\documentclass[11pt]{article}
\pdfoutput=1

\usepackage{amssymb}
\usepackage{amsmath}
\usepackage{bbold}
\usepackage{color}
\usepackage{colordvi}
\usepackage{dsfont}
\usepackage{fancybox}
\usepackage[footnotesize]{caption2}
\usepackage{graphicx}
\usepackage[center,footnotesize,hang]{subfigure}
\usepackage{bbm}
\usepackage{url}
\usepackage{multirow}
\usepackage{array}
\usepackage{arydshln}
\usepackage{pifont}
\usepackage{mathrsfs}
\usepackage{fancybox}
\usepackage{cite}
\usepackage[colorlinks]{hyperref}
\usepackage{enumitem}
\usepackage{longtable}
\newcommand{\ignore}[1]{}
\usepackage{enumitem}

\newcommand{\PreserveBackslash}[1]{\let\temp=\\#1\let\\=\temp}
\newcolumntype{C}[1]{>{\PreserveBackslash\centering}p{#1}}
\newcolumntype{R}[1]{>{\PreserveBackslash\raggedleft}p{#1}}
\newcolumntype{L}[1]{>{\PreserveBackslash\raggedright}p{#1}}
\allowdisplaybreaks
\newcommand{\bq}{\begin{eqnarray}}
\newcommand{\nq}{\end{eqnarray}}

\newcommand{\f}{ }

\def\bvec#1{\raise1.5ex\hbox{$\rightarrow$}\mkern-16.5mu #1}

\begin{document}
\title{
\begin{flushright}
\hfill\mbox{{\small\tt USTC-ICTS/PCFT-21-06 }} \\[5mm]
\begin{minipage}{0.2\linewidth}
\normalsize
\end{minipage}
\end{flushright}
{\Large \bf
$SU(5)$ GUTs with $A_4$ modular symmetry
\\[2mm]}}
\date{}

\author{
Peng Chen$^{1}$\footnote{E-mail: {\tt
pche@mail.ustc.edu.cn}},  \
Gui-Jun~Ding$^{2,3}$\footnote{E-mail: {\tt
dinggj@ustc.edu.cn}},  \
Stephen~F.~King$^{4}$\footnote{E-mail: {\tt king@soton.ac.uk}}, \
\\*[20pt]
\centerline{
\begin{minipage}{\linewidth}
\begin{center}
$^1${\it \small College of Information Science and Engineering,Ocean University of China, Qingdao 266100, China} \\[2mm]
$^2${\it \small Peng Huanwu Center for Fundamental Theory, Hefei, Anhui 230026, China} \\[2mm]
$^3${\it \small
Interdisciplinary Center for Theoretical Study and  Department of Modern Physics,\\
University of Science and Technology of China, Hefei, Anhui 230026, China}\\[2mm]
$^4${\it \small
Physics and Astronomy, University of Southampton, Southampton, SO17 1BJ, U.K.}
\end{center}
\end{minipage}}
\\[10mm]}
\maketitle
\thispagestyle{empty}

\begin{abstract}

We combine $SU(5)$ Grand Unified Theories (GUTs) with $A_4$ modular symmetry and present a comprehensive analysis of the
resulting quark and lepton mass matrices for all the simplest cases. Classifying the models according to the representation assignments of the matter fields under $A_4$, we find that there are seven types of $SU(5)$ models with $A_4$ modular symmetry. We present 53 benchmark models with the fewest free parameters. The parameter space of each model is scanned to optimize the agreement between predictions and experimental data, and predictions for the masses and mixing parameters of quarks and leptons are given at the best fitting points. The best fit predictions for the leptonic CP violating Dirac phase, the lightest neutrino mass and the neutrinoless double beta decay parameter when displayed graphically are observed to cover a wide range of possible values, but are clustered around particular regions, allowing future neutrino experiments to discriminate between the different types of models.

\end{abstract}
\newpage

\section{Introduction}

The standard model (SM) of electroweak interactions based on $SU(2)_L\times U(1)_Y$ is sixty years old this year~\cite{Glashow:1961tr}, although the inclusion of the Higgs mechanism and QCD required a further decade of work as has been well documented elsewhere.
It is by now well established and has successfully passed many precision tests at various energy levels. After the discovery of the Higgs boson at the LHC, all particles predicted by the three family SM have been observed. Although there is no concrete experimental evidence for new physics beyond the SM, apart from the neutrino masses and mixing, most physicists believe that the SM should be a low energy effective theory rather than a fundamental theory. One essential question of the SM is how to understand the hierarchical mass spectrum and flavor mixing patterns of quarks and leptons, including neutrinos whose masses lie beyond the SM. The masses of quarks and leptons span about 13 orders of magnitude. The upper limit of the lightest neutrino mass is smaller than $0.04$ eV while the top quark mass is around $173$ GeV. The three quark mixing angles are hierarchical and the largest one is the Cabibbo angle $\theta^{q}_{12}\simeq13.09^{\circ}$, and the CP phase of the quark sector is $\delta_{CP}^q\simeq68.53^\circ$~\cite{Zyla:2020zbs}.
In contrast to the quark sector, the solar mixing angle $\theta_{12}^l\simeq33.44^\circ$ and the atmospheric mixing angle $\theta_{23}^l\simeq49.2^\circ$ are large, and the reactor mixing angle is small with $\theta_{13}^l\simeq8.57^\circ$, while the CP violation of the lepton sector has not been confirmed yet~\cite{Esteban:2020cvm}. The quark and charged lepton masses are described by different interaction strengths with the Higgs doublets within the SM, but their values cannot be predicted.
Understanding the flavor structure of the SM from the first principles is one of the greatest challenges in particle physics. Much effort has been devoted to
addressing the flavor puzzle, and symmetry has been an important guiding principle. It is found that the non-Abelian discrete flavor symmetry is particularly suitable to explain the large lepton mixing angles $\theta^{l}_{12}$ and $\theta^{l}_{23}$~\cite{King:2013eh}.
In conventional flavor symmetry models,
a complicated vacuum alignment is frequently required, since flavons are generally necessary and their vacuum expectation values (VEVs) should be oriented along certain directions in flavor space.

Recently modular invariance has been suggested to play the role of flavor symmetry, especially in the neutrino sector~\cite{Feruglio:2017spp}, an approach which is inspired by superstring theory with compactified extra dimensions. The finite discrete flavor symmetry groups arise as the quotient group of modular group $SL(2,Z)$ over the principal congruence subgroups. The quark and lepton fields transform nontrivially under the finite modular groups and are assigned to various modular weights, thus modular invariance requires the Yukawa couplings are modular forms which are holomorphic functions of the complex modulus $\tau$. The flavon fields other than the modulus might not be needed and the flavor symmetry can be entirely broken by the vacuum expectation value (VEV) of the modulus $\tau$. Moreover, all higher dimensional operators in the superpotential are completely fixed by the modular invariance.
The modular form of level $N$ and integral weight $k$ can be arranged into some modular multiplets of the homogeneous finite modular group $\Gamma'_N\equiv\Gamma/\Gamma(N)$~\cite{Liu:2019khw}, and they can be organized into modular multiplets of the inhomogeneous finite modular group $\Gamma_N\equiv\overline{\Gamma}/\overline{\Gamma}(N)$ if $k$ is an even number~\cite{Feruglio:2017spp}. The inhomogeneous finite modular group $\Gamma_N$ of lower levels
$N=2$~\cite{Kobayashi:2018vbk,Kobayashi:2018wkl,Kobayashi:2019rzp,Okada:2019xqk}, $N=3$~\cite{Feruglio:2017spp,Criado:2018thu,Kobayashi:2018vbk,Kobayashi:2018scp,deAnda:2018ecu,Okada:2018yrn,Kobayashi:2018wkl,Novichkov:2018yse,Nomura:2019jxj,Okada:2019uoy,Nomura:2019yft,Ding:2019zxk,Okada:2019mjf,Nomura:2019lnr,Kobayashi:2019xvz,Asaka:2019vev,Gui-JunDing:2019wap,Zhang:2019ngf,Nomura:2019xsb,Wang:2019xbo,Kobayashi:2019gtp,King:2020qaj,Ding:2020yen,Okada:2020rjb,Nomura:2020opk,Okada:2020brs,Yao:2020qyy,Feruglio:2021dte}, $N=4$~\cite{Penedo:2018nmg,Novichkov:2018ovf,deMedeirosVarzielas:2019cyj,Kobayashi:2019mna,King:2019vhv,Criado:2019tzk,Wang:2019ovr,Gui-JunDing:2019wap,Wang:2020dbp}, $N=5$~\cite{Novichkov:2018nkm,Ding:2019xna,Criado:2019tzk} and $N=7$~\cite{Ding:2020msi} have been considered and a large number of models have been constructed. All the modular forms of integral weights can be generated from the tensor products of weight one modular forms and
the odd weight modular forms are in the representations with $\rho_{\mathbf{r}}(S^2)=-1$. The homogeneous finite modular groups $\Gamma'_N$ provide richer structure of modular forms for flavor model building,
and the small groups $\Gamma'_3\cong T'$~\cite{Liu:2019khw,Lu:2019vgm}, $\Gamma'_4\cong S'_4$~\cite{Liu:2020akv,Novichkov:2020eep} and $\Gamma'_5\cong A'_5$~\cite{Wang:2020lxk} have been studied.
If the modular weight $k$ of the operator is not an integer, $(c\tau+d)^k$ is not the automorphy factor anymore and some multiplier is needed, consequently the modular group should be extended to its metaplectic covering~\cite{Liu:2020msy}. The framework of modular invariance have been extended to include the modular forms of rational weights $k/2$~\cite{Liu:2020msy} and $k/5$~\cite{Yao:2020zml}.
The generalized CP symmetry can be consistently imposed in the context of modular symmetry, the modulus is determined to transform as $\tau\rightarrow-\tau^{*}$ under the action of CP~\cite{Novichkov:2019sqv,Baur:2019kwi,Baur:2019iai}. The CP transformation matrix is completely fixed by the consistency condition up to an overall phase, and it is exactly the canonical CP transformation in the symmetric basis~\cite{Novichkov:2019sqv}. The more fundamental theory such as string theory sometimes requires several compact space with more than one modulus parametrizing its shape. The modular invariance approach has been extended to incorporate several factorizable~\cite{deMedeirosVarzielas:2019cyj} and non-factorizable moduli~\cite{Ding:2020zxw}.

Grand unified theories (GUTs) are amongst the most well motivated theories beyond the SM, realising the elegant aspiration to unify the three gauge interactions of the SM into a simple gauge group~\cite{Georgi:1974sy}. The fermion representations of the SM are unified into a smaller number of multiplets under the GUT gauge group, with quarks and leptons being assigned to the same GUT multiplet, thereby providing an explanation for charge quantisation.
Imposing a family symmetry together with GUTs can help to address the problem of quark and lepton mass and mixing hierarchies~\cite{King:2017guk}.
Among many different GUT scenarios, $SU(5)$~\cite{Georgi:1974sy} is the minimal choice, being
the smallest simple group which can accommodate the gauge symmetry of the SM.
Although neutrino masses are not predicted by $SU(5)$, they can easily be accommodated as singlet representations of the GUT group.
Among the many choices of family symmetry, $A_4$ is the minimal choice which admits triplet representations~\cite{Ma:2001dn}.
Combining $A_4$ family symmetry with $SU(5)$ GUTs~\cite{Bjorkeroth:2015ora} also requires vacuum alignment of the flavons in order to break the $A_4$, and so there is a strong motivation for introducing modular symmetry in such frameworks.
Indeed, modular symmetry in the context of $SU(5)$ GUTs was first studied in an $(\Gamma_3\simeq A_4)\times SU(5)$ model
in~\cite{deAnda:2018ecu}. Other modular GUT models were subsequently constructed based on
$(\Gamma_2\simeq S_3)\times SU(5)$~\cite{Kobayashi:2019rzp,Du:2020ylx}, and $(\Gamma_4\simeq S_4)\times SU(5)$~\cite{Zhao:2021jxg}.

In this paper we shall perform a comprehensive study of the $\Gamma_3\simeq A_4$ modular symmetry in the framework of supersymmetric (SUSY) $SU(5)$ GUTs. It is known that the fifteen matter fields in each generation are embedded into two chiral supermultiplets $\mathbf{\overline{5}}$ and $\mathbf{10}$. The $SU(5)$ gauge symmetry is spontaneously broken down to the SM gauge group $SU(3)_c\times SU(2)\times U(1)_Y$ by the VEV of the Higgs $H_{\mathbf{24}}$ in the adjoint representation of $SU(5)$. In the minimal $SU(5)$ model, two Higgs multiplets $H_{\mathbf{5}}$ and $H_{\overline{\mathbf{5}}}$ in the fundamental representation $\mathbf{5}$ and antifundamental representation $\overline{\mathbf{5}}$ of $SU(5)$ further break the SM gauge symmetry into $SU(3)_c\times U(1)_{\text{EM}}$.
The charged leptons and down-type quarks are in the same GUT multiplets, the down type quark mass matrix is the transpose of the charged lepton mass matrix in the minimal $SU(5)$ such that the masses of the charged leptons and down quarks would be identical.
In order to account the different masses of down quarks and charged leptons, additional Higgs multiplet $H_{\mathbf{\overline{45}}}$ is introduced in our models~\cite{Georgi:1979df}.
The neutrinos are massless in the minimal $SU(5)$ GUT because of the absence of right-handed neutrinos. In the present work, we extend the matter contents by including right-handed neutrinos which are $SU(5)$ singlets, and the neutrino masses are generated by the type-I seesaw mechanism~\cite{Minkowski:1977sc,Yanagida:1979as,GellMann:1980vs,Mohapatra:1979ia,Schechter:1980gr}. The most minimal version of the seesaw mechanism involves two additional right-handed neutrinos~\cite{King:1999mb,Frampton:2002qc}, consequently both versions of seesaw models with two and three right-handed neutrinos are considered. The $A_4$ group has three singlet representations $\mathbf{1}$, $\mathbf{1}'$, $\mathbf{1}''$ and a triplet representation $\mathbf{3}$. We assume the three Higgs fields are modular invariants and their modular weights are vanishing, and the three generations of matter fields are assigned to transform as singlets or triplet of $A_4$. The purpose of this work is to find phenomenologically viable $SU(5)$ GUT models based on $A_4$ modular symmetry with less parameters. We classify all possible $A_4$ modular $SU(5)$ GUT models into seven different types according to transformation properties of the matter superfields under $A_4$. Intensive numerical analysis is performed for each model and we have optimized the free parameters of the models in order to match the experimental data.

This paper is organized as follows. In section~\ref{modular-review}, we briefly review modular symmetry and modular forms, and modular forms of level 3 are listed. In section~\ref{sec:lassification}, we present the most general form of the $A_4$ modular quark and lepton mass matrices in $SU(5)$, and all possible $A_4$ representation assignments of matter fields are considered.
In section~\ref{sec:benchmark-models}, according to the representation assignments of matter fields, we find that there are seven different types of $SU(5)$ GUT models with $A_4$ modular symmetry, and some benchmark models are presented. We don't consider the cases in which no matter fields are assigned to triplet of $A_4$ since generally more free parameters would be involved.
In section~\ref{sec:numerical-results}, we show the results of a scan of the parameter space for each benchmark model, we search for the minimum of the $\chi^2$ function to optimize the agreement between predictions and experimental data. Finally we draw the conclusions in section~\ref{sec:conclusion}. Appendix~\ref{sec:appendix-best-fit} gives the best fit values of the input parameters and the predictions for the masses, mixing angles and CP violating phases of quarks and leptons.

\section{\label{modular-review}Modular symmetry and modular forms of level $N=3$}

The modular group $SL(2, Z)$ consists of matrices with integer entries and determinant 1, it acts on the upper half plane as fractional linear(Mobius) transformations:
\begin{equation}
\tau\mapsto\gamma\tau=\frac{a\tau+b}{c\tau+d},~~~\begin{pmatrix}
a  ~&  b  \\
c  ~& d
\end{pmatrix}\in SL(2, Z),~~~\texttt{Im}(\tau)>0\,.
\end{equation}
It is easily verified that the kernel of this action is the center $C=\{I, -I\}$ where $I$ denotes the two dimensional unit matrix. Therefore the group of fractional linear transformations is isomorphic to the projective special linear group $PSL(2, Z)\cong SL(2, Z)/C=\overline{\Gamma}$ which is the quotient of $SL(2, Z)$ by its center $C$. Although the group $PSL(2, Z)$ is of infinite order, it can be generated by two transformations $S$ and $T$,
\begin{eqnarray}
\nonumber&& S=\begin{pmatrix}
0  ~&~  1  \\
-1  ~&~  0
\end{pmatrix}: \tau\rightarrow -\frac{1}{\tau}\,, \\
&& T=\begin{pmatrix}
1  ~&~  1  \\
0  ~&~  1
\end{pmatrix}~~: \tau\rightarrow \tau+1\,,
\end{eqnarray}
Taking the quotient of $PSL(2, Z)$ over the principal congruence subgroup of level $N$, we can obtain the (inhomogeneous) finite modular group $\Gamma_N\cong \overline{\Gamma}/\overline{\Gamma}(N)$ with
\begin{equation}
\Gamma(N)=\left\{\begin{pmatrix}
a  ~&  b \\
c  ~& d
\end{pmatrix}\in SL(2, Z),~~\begin{pmatrix}
a  ~&  b \\
c  ~& d
\end{pmatrix} (\texttt{mod}~N)=\begin{pmatrix}
1  ~& 0 \\
0  ~& 1
\end{pmatrix}
\right\}\,,
\end{equation}
and $\overline{\Gamma}(N)=\Gamma(N)$ for $N>2$ and $\overline{\Gamma}(N)=\Gamma(N)/\{I, -I\}$ for $N=1, 2$. Obviously $T^N\in \Gamma(N)$, consequently the finite modular group $\Gamma_N$ can be generated by the generators $S$ and $T$ satisfying
\begin{equation}
S^2=(ST)^3=T^{N}=1\,.
\end{equation}
Additional relations are needed to render the group finite make for $N>5$~\cite{Ding:2020msi}. In the present we are interested in the minimal finite modular group $\Gamma_3=\overline{\Gamma}/\overline{\Gamma}(3)\cong A_4$ which can be seen as the symmetry group of the tetrahedron.
There are 4 inequivalent irreducible representations of $A_4$: three singlets $\mathbf{1}$, $\mathbf{1}'$, $\mathbf{1}''$ and a triplet $\mathbf{3}$.
In the three singlet representations, the generators $S$ and $T$ are
\begin{eqnarray}
\nonumber  \mathbf{1}:&&~~ S=1, ~~~ T=1 \,,  \\
\nonumber \mathbf{1}':&&~~ S=1,~~~ T=\omega \,,  \\
\mathbf{1}'':&&~~S=1, ~~~ T=\omega^{2} \,,
\end{eqnarray}
where $\omega=e^{2\pi i/3}$ refers to a cubic root of unity. We shall denote ${\mathbf1}\equiv{\mathbf1}^{0}$,
${\mathbf1}'\equiv{\mathbf1}^{1}$  and ${\mathbf1}''\equiv{\mathbf1}^{2}$
when constructing models in the following.
The triplet representation $\mathbf{3}$ in the bassi where $T$ is diagonal is given by
\begin{equation}
\label{eq:rep-gen}S=\frac{1}{3}\begin{pmatrix}
-1 ~& 2  ~& 2  \\
2  ~& -1 ~& 2 \\
2  ~& 2  ~& -1
\end{pmatrix}, ~\quad~
T=\begin{pmatrix}
1 ~&~ 0 ~&~ 0 \\
0 ~&~ \omega ~&~ 0 \\
0 ~&~ 0 ~&~ \omega^{2}
\end{pmatrix} \,,
\end{equation}
The decompositions of the direct product of $A_4$ representations are
\begin{eqnarray}
\nonumber&& \mathbf{1}'\otimes\mathbf{1}'=\mathbf{1}'',~~~ \mathbf{1}'\otimes\mathbf{1}''=\mathbf{1},~~~ \mathbf{1}''\otimes\mathbf{1}''=\mathbf{1}'\,,\\
&&\mathbf{3}\otimes \mathbf{3}= \mathbf{1}\oplus \mathbf{1'}\oplus \mathbf{1''}\oplus \mathbf{3}\oplus \mathbf{3}\,.
\end{eqnarray}
Given two triplets $\alpha=(\alpha_1,\alpha_2,\alpha_3)$ and  $\beta=(\beta_1,\beta_2,\beta_3)$, the irreducible representations obtained from their product are:
\begin{eqnarray}
\nonumber &&\mathbf{1}=\alpha_1\beta_1+\alpha_2\beta_3+\alpha_3\beta_2\,, \\
\nonumber &&\mathbf{1}'=\alpha_3\beta_3+\alpha_1\beta_2+\alpha_2\beta_1\,, \\
\nonumber &&\mathbf{1}''=\alpha_2\beta_2+\alpha_1\beta_3+\alpha_3\beta_1\,, \\
\nonumber &&\mathbf{3}_S=(
2\alpha_1\beta_1-\alpha_2\beta_3-\alpha_3\beta_2,
2\alpha_3\beta_3-\alpha_1\beta_2-\alpha_2\beta_1,
2\alpha_2\beta_2-\alpha_1\beta_3-\alpha_3\beta_1)\,, \\
\label{eq:decomp-rules} &&\mathbf{3}_A=(
\alpha_2\beta_3-\alpha_3\beta_2,
\alpha_1\beta_2-\alpha_2\beta_1,
\alpha_3\beta_1-\alpha_1\beta_3)\,.
\end{eqnarray}
where $\mathbf{3}_{S(A)}$ denotes the symmetric (antisymmetric) combination.
The modular forms of level $N$ and even weight $k$ span a linear space of finite dimension, and they can be organized into some modular multiplets $Y^{(k)}_{\mathbf{r}}(\tau)$ transforming in the irreducible representation $\mathbf{r}$ of $\Gamma_N$ up to the automorphy factor,
\begin{equation}
Y^{(k)}_{\mathbf{r}}(\gamma\tau)=(c\tau+d)^{k}\rho_{\mathbf{r}}(\gamma)Y^{(k)}_{\mathbf{r}}(\tau),~~~\forall\gamma=\begin{pmatrix}
a  ~&  b  \\
c  ~&  d
\end{pmatrix}\in\overline{\Gamma}\,.
\end{equation}
The modular forms of level 3 has been constructed ~\cite{Feruglio:2017spp}, and it turns out that there are only three linearly independent modular forms $Y_{1,2,3}(\tau)$ of weight 2 and level 3,
\begin{eqnarray}
Y_1(\tau) &=& \frac{i}{2\pi}\left[ \frac{\eta'(\tau/3)}{\eta(\tau/3)}  +\frac{\eta'((\tau +1)/3)}{\eta((\tau+1)/3)}
+\frac{\eta'((\tau +2)/3)}{\eta((\tau+2)/3)} - \frac{27\eta'(3\tau)}{\eta(3\tau)}  \right], \nonumber \\
Y_2(\tau) &=& \frac{-i}{\pi}\left[ \frac{\eta'(\tau/3)}{\eta(\tau/3)}  +\omega^2\frac{\eta'((\tau +1)/3)}{\eta((\tau+1)/3)}
+\omega \frac{\eta'((\tau +2)/3)}{\eta((\tau+2)/3)}  \right] ,\nonumber \\
Y_3(\tau) &=& \frac{-i}{\pi}\left[ \frac{\eta'(\tau/3)}{\eta(\tau/3)}  +\omega\frac{\eta'((\tau +1)/3)}{\eta((\tau+1)/3)}
+\omega^2 \frac{\eta'((\tau +2)/3)}{\eta((\tau+2)/3)} \right]\,,
\end{eqnarray}
where $\eta(\tau)$ is the famous Dedekind eta-function,
\begin{equation}
\eta(\tau)=q^{1/24} \prod_{n =1}^\infty (1-q^n), ~~~  q=e^{2\pi i\tau}\,.
\end{equation}
Thus the $q-$expansion of $Y_{1,2,3}(\tau)$ reads
\begin{eqnarray}
\nonumber&&Y_1(\tau)=1 + 12q + 36q^2 + 12q^3 + 84q^4 + 72q^5 +\dots\,,\\
\nonumber&&Y_2(\tau)=-6q^{1/3}(1 + 7q + 8q^2 + 18q^3 + 14q^4 +\dots)\,,\\
&&Y_2(\tau)=-18q^{2/3}(1 + 2q + 5q^2 + 4q^3 + 8q^4 +\dots)\,.
\end{eqnarray}
The above three modular forms can be arranged into a triplet $Y^{(2)}_{\mathbf{3}}=\left(Y_1, Y_2, Y_3\right)^{T}$ which transforms as a three-dimensional irreducible representation of $A_4$ up to the automorphy factor:
\begin{equation}
Y^{(2)}_{\mathbf{3}}(\tau)\xrightarrow{S}Y^{(2)}_{\mathbf{3}}(-1/\tau)=\tau^2 \rho_{\mathbf{3}}(S)Y^{(2)}_{\mathbf{3}}(\tau),~~~~
Y^{(2)}_{\mathbf{3}}(\tau)\xrightarrow{T}Y^{(2)}_{\mathbf{3}}(\tau+1)=\rho_{\mathbf{3}}(T)Y^{(2)}_{\mathbf{3}}(\tau)\,,
\end{equation}
where $\rho_{\mathbf{3}}(S)$ and $\rho_{\mathbf{3}}(T)$ are the representation matrices of $S$ and $T$ respectively given in Eq.~\eqref{eq:rep-gen}. The weight $2k$ modular forms of level 3 are homogeneous polynomials $Y_{i_1}Y_{i_2}\ldots Y_{i_k}$ of degree $k$ in $Y_i(\tau)$. There are five linearly independent weight four modular forms and they decompose as $\mathbf{3}\oplus\mathbf{1}\oplus\mathbf{1}'$ under $A_4$,
\begin{eqnarray}
\nonumber Y^{(4)}_{\mathbf{3}}&=&\frac{1}{2}(Y^{(2)}_{\mathbf{3}}Y^{(2)}_{\mathbf{3}})_{\mathbf{3}}=
\begin{pmatrix}
Y_1^2-Y_2 Y_3\\
Y_3^2-Y_1 Y_2\\
Y_2^2-Y_1 Y_3
\end{pmatrix}\,, \\
\nonumber Y^{(4)}_{\mathbf{1}}&=&(Y^{(2)}_{\mathbf{3}}Y^{(2)}_{\mathbf{3}})_{\mathbf{1}}=Y_1^2+2 Y_2 Y_3\,, \\
Y^{(4)}_{\mathbf{1}'}&=&(Y^{(2)}_{\mathbf{3}}Y^{(2)}_{\mathbf{3}})_{\mathbf{1}'}=Y_3^2+2 Y_1 Y_2\,.
\end{eqnarray}
The tensor products of weight 2 and weight 4 modular forms give rise to weight 6 modular forms,
\begin{eqnarray}
Y^{(6)}_{\mathbf{1}}&=&(Y^{(2)}_{\mathbf{3}}Y^{(4)}_{\mathbf{3}})_{\mathbf{1}}=Y_1^3+Y_2^3+Y_3^3-3 Y_1 Y_2 Y_3\,,\nonumber\\
Y^{(6)}_{\mathbf{3}I}&=&Y^{(2)}_{\mathbf{3}}Y^{(4)}_{\mathbf{1}}=(Y_1^2+2Y_2Y_3)\begin{pmatrix}
Y_1\\
Y_2\\
Y_3
\end{pmatrix}\,,\nonumber\\
Y^{(6)}_{\mathbf{3}II}&=&Y^{(2)}_{\mathbf{3}}Y^{(4)}_{\mathbf{1}'}=
(Y_3^2+2 Y_1Y_2)\begin{pmatrix}
Y_3\\
Y_1\\
Y_2
\end{pmatrix}\,.
\end{eqnarray}
The weight 8 modular forms which can be arranged into two triplets and three singlets of $A_4$
\begin{eqnarray}
\nonumber Y^{(8)}_{\mathbf{1}}&=&(Y^{(2)}_{\mathbf{3}}Y^{(6)}_{\mathbf{3}I})_{\mathbf{1}}=(Y_1^2+2 Y_2 Y_3)^2\,,\\
\nonumber Y^{(8)}_{\mathbf{1'}}&=&(Y^{(2)}_{\mathbf{3}}Y^{(6)}_{\mathbf{3}I})_{\mathbf{1'}}=(Y_1^2+2 Y_2 Y_3)(Y_3^2+2 Y_1 Y_2)\,, \\
\nonumber Y^{(8)}_{\mathbf{1''}}&=&(Y^{(2)}_{\mathbf{3}}Y^{(6)}_{\mathbf{3}II})_{\mathbf{1''}}=(Y_3^2+2 Y_1 Y_2)^2\,,\\
\nonumber Y^{(8)}_{\mathbf{3}I}&=&Y^{(2)}_{\mathbf{3}}Y^{(6)}_{\mathbf{1}}=(Y_1^3+Y_2^3+Y_3^3-3 Y_1 Y_2 Y_3)\begin{pmatrix}
Y_1 \\
Y_2\\
Y_3
\end{pmatrix}\,,\\
Y^{(8)}_{\mathbf{3}II}&=&(Y^{(2)}_{\mathbf{3}}Y^{(6)}_{\mathbf{3}II})_{\mathbf{3}_A}=(Y_3^2+2 Y_1Y_2)\begin{pmatrix}
Y^2_2-Y_1Y_3\\
Y^2_1-Y_2Y_3\\
Y^2_3-Y_1Y_2
\end{pmatrix}\,.
\end{eqnarray}
Notice that modular forms in all irreducible representations of $A_4$ appears at weight 8, both $Y^{(6)}_{\mathbf{3}I}(\tau)$ and $Y^{(8)}_{\mathbf{3}I}(\tau)$ are proportional to $Y^{(2)}_{\mathbf{3}}(\tau)$. We summarize the even weight modular forms of level 3 and their transformation under $A_4$ in table~\ref{Tab:Level3_MF}.

\begin{table}[t!]
\centering
\begin{tabular}{|c|c|}
\hline  \hline

weight $k$ & representation $\mathbf{r}$ \\ \hline

$2$ & $\mathbf{3}$\\ \hline

$4$ & $\mathbf{1}, \mathbf{1}', \mathbf{3}$\\ \hline

$6$ & $\mathbf{1}, \mathbf{3}, \mathbf{3}$\\ \hline

$8$ & $\mathbf{1}, \mathbf{1}', \mathbf{1}'', \mathbf{3}, \mathbf{3}$\\ \hline \hline
\end{tabular}
\caption{\label{Tab:Level3_MF}
The transformation properties of the weight $k$ and the level 3 modular forms $Y^{(k)}_{\mathbf{r}}$ under $A_4$.
}
\end{table}

\section{\label{sec:lassification}General form of fermion mass matrices }

We formulate our models in the framework of supersymmetric $SU(5)$ GUT. The neutrino masses are generated by the type-I seesaw mechanism~\cite{Minkowski:1977sc,Yanagida:1979as,GellMann:1980vs,Mohapatra:1979ia,Schechter:1980gr}, consequently three (or two) right-handed neutrinos $N$ are introduced and they are $SU(5)$ singlets. Thus all left-handed quark and lepton superfields within each family are embedded into
three $SU(5)$ multiplets $\mathbf{1}$, $\overline{\mathbf{5}}$ and $\mathbf{10}$ with
\begin{eqnarray}
N=\nu^c\,,\qquad
\overline{F}=
\left(\begin{array}{c}
d^c_r\\
d^c_g\\
d^c_b\\
e\\
-\nu
\end{array}\right)\,,\qquad
T=
\left(\begin{array}{ccccc}
0~&u^c_b~&-u^c_g~&-u_r~&-d_r\\
-u^c_b~&0~&u^c_r~&-u_g~&-d_g\\
u^c_g~&-u^c_r~&0~&-u_b~&-d_b\\
u_r~&u_g~&u_b~&0~&e^c\\
d_r~&d_g~&d_b~&-e^c~&0
\end{array}\right)\,,
\label{eq:su5multiplets}
\end{eqnarray}
where the superscript $c$ denotes charge conjugation of the right-handed superfields. The weight and transformation of the above matter superfields under $A_4$ are denoted as $k_{\psi}$ and $\rho_{\psi}$ respectively with $\psi=N, \overline{F}, T$. Moreover, the $SU(5)$ Higgs fields $H_{5}$ and $H_{\overline{5}}$, $H_{\overline{45}}$ is considered to account for the mass differences of down quarks and charged leptons. All the three Higgs fields are assumed to be invariant under $A_4$ and their modular weights are zero. Each of these GUT Higgs representations contains an $SU(2)_L$ Higgs doublet, the low energy doublet $H_u$ in the minimal supersymmetric standard model arises from $H_5$, and $H_d$ originates from a linear combination of $H_{\overline{5}}$ and $H_{\overline{45}}$.
The three generations of $F$, $T$ and $N$ can transform as either a triplet $\mathbf{3}$ or three singlets under $A_4$ modular symmetry. The most general superpotential for quark and lepton masses is of the following form,
\begin{equation}
\label{eq:mass-SU5}{\cal W} =NN f_M(Y) + N\overline{F}H_{5} f_N(Y)
 + \overline{F}TH_{\overline{5}}f_{D}(Y) + \overline{F}T H_{\overline{45}}f'_{D}(Y)+TTH_{5}f_{U}(Y) \,,
\end{equation}
where $f_M(Y)$, $f_N(Y)$, $f_{D}(Y)$, $f'_{D}(Y)$ and $f_{U}(Y)$ are functions of modular forms, and they can be determined by the requirement that each term of $\mathcal{W}$ is modular invariant with total vanishing modular weight.
Because both $H_{\overline{5}}$ and $H_{\overline{45}}$ are assumed to be $A_4$ invariant singlet and their modular weights are vanishing, the modular functions $f_{D}(Y)$ and $f'_{D}(Y)$ are of the same form. In the following, we shall present the most general form of the mass matrices of quarks and leptons in $SU(5)$ models with $A_4$ modular symmetry. We shall consider modular forms up to weight 8 in the following, and extension to higher weight modular forms is straightforward.

\subsection{Majorana mass matrix of the right-handed neutrinos}

The first term $NN f_M(Y)$ in Eq.~\eqref{eq:mass-SU5} gives rise to Majorana mass matrix of the right-handed neutrinos, and the modular function $f_M(Y)$ is completely fixed by the weight and representation of $N$.

\begin{itemize}
\item{$\rho_N=\mathbf{3}$}

Using the contraction rules in Eq.~\eqref{eq:decomp-rules}, we can read off the mass matrix of the right-handed neutrinos for different values of $N$'s weight $k_N$,
\begin{eqnarray}
\begin{aligned}
k_N=0&~~:~~ M_N=\Lambda\,S^{(0)}_{\mathbf1}\,,\\
k_N=1&~~:~~ M_N=\Lambda\,S^{(2)}_{\mathbf3}\,,\\
k_N=2&~~:~~ M_N=\Lambda_1 S^{(4)}_{\mathbf3}+\Lambda_2\,S^{(4)}_{\mathbf1}+\Lambda_3\,S^{(4)}_{\mathbf1'}\,,\\
k_N=3&~~:~~ M_N=\Lambda_1 S^{(6)}_{\mathbf{3}I}+\Lambda_2 S^{(6)}_{\mathbf{3}II}+\Lambda_3 S^{(6)}_{\mathbf1}\,,\\
k_N=4&~~:~~ M_N=\Lambda_1 S^{(8)}_{\mathbf{3}I}+\Lambda_2 S^{(8)}_{\mathbf{3}II}+\Lambda_3 S^{(8)}_{\mathbf1}+\Lambda_4 S^{(8)}_{\mathbf1'}+\Lambda_5 S^{(8)}_{\mathbf1''}\,,
\end{aligned}
\end{eqnarray}
where $\Lambda_i$ are the characteristic scale of
flavour dynamics.
For simplicity we have defined
\begin{eqnarray}
S^{(k)}_{\mathbf1} =
Y_{\mathbf1}^{(k)}(\tau)\left(
\begin{matrix}
 1 ~& 0 ~& 0 \\
 0 ~& 0 ~& 1 \\
 0 ~& 1 ~& 0 \\
\end{matrix}
\right)\,,\quad
S^{(k)}_{\mathbf1'} =
Y_{\mathbf1'}^{(k)}(\tau)\left(
\begin{matrix}
 0 ~& 0 ~& 1 \\
 0 ~& 1 ~& 0 \\
 1 ~& 0 ~& 0 \\
\end{matrix}
\right)\,,\quad
S^{(k)}_{\mathbf1''} =
Y_{\mathbf1''}^{(k)}(\tau)\left(
\begin{matrix}
 0 ~& 1 ~& 0 \\
 1 ~& 0 ~& 0 \\
 0 ~& 0 ~& 1 \\
\end{matrix}
\right)\,,
\end{eqnarray}
and
\begin{eqnarray}
S^{(k)}_{{\mathbf3}}(\tau) =
\left(\begin{matrix}
 2Y^{(k)}_{{{\mathbf3}},1}(\tau) ~& -Y^{(k)}_{{{\mathbf3}},3}(\tau)  ~& -Y^{(k)}_{{{\mathbf3}},2}(\tau) \\
 -Y^{(k)}_{{{\mathbf3}},3}(\tau) ~& 2Y^{(k)}_{{{\mathbf3}},2}(\tau)  ~& -Y^{(k)}_{{{\mathbf3}},1}(\tau) \\
 -Y^{(k)}_{{{\mathbf3}},2}(\tau) ~& -Y^{(k)}_{{{\mathbf3}},1}(\tau)  ~& 2Y^{(k)}_{{{\mathbf3}},3}(\tau)
\end{matrix}
\right)\,,
\end{eqnarray}
where we denote $Y_{\mathbf1}^{(0)}(\tau)=1$ and ${{\mathbf3}}$ stands for
${\mathbf3}$, ${\mathbf3}I$ or ${\mathbf3}II$.

\item{$\rho_N=\mathbf{1}, \mathbf{1}', \mathbf{1}''$ }

In this case, $f_M(Y)$ has to be modular form transforming as singlet under $A_4$. It is known that the mass matrix $M_N$ of right-handed neutrinos is a symmetric matrix. For the modular weight $k_{N_i}+k_{N_j}\le8$, the possible nonzero elements depend on their weights and $A_4$ kronecker products
\begin{eqnarray}
\nonumber&&k_{N_i}+k_{N_j}=0~~:~~(M_N)_{ij} = \alpha^{M}_{ij} \Lambda ~~\text{for}~~\rho_{N_i}\otimes\rho_{N_j}={\mathbf1}\,,\\
\nonumber&&k_{N_i}+k_{N_j}=4~~:~~(M_N)_{ij} = \left\{\begin{array}{cc}\alpha^{M}_{ij} \Lambda Y^{(4)}_{\mathbf1}(\tau) &~~\text{for}~~\rho_{N_i}\otimes\rho_{N_j}={\mathbf1}\,,\\[0.1in]
 \alpha^{M}_{ij} \Lambda Y^{(4)}_{\mathbf1'}(\tau)&~~\text{for}~~\rho_{N_i}\otimes\rho_{N_j}={\mathbf1''}\,,
\end{array}\right.\\
\nonumber&&k_{N_i}+k_{N_j}=6~~:~~(M_N)_{ij} =\alpha^{M}_{ij} \Lambda Y^{(6)}_{\mathbf1}(\tau)~~\text{for}~~\rho_{N_i}\otimes\rho_{N_j}={\mathbf1}\,,\\
&&k_{N_i}+k_{N_j}=8~~:~~(M_N)_{ij} = \left\{\begin{array}{cc} \alpha^{M}_{ij} \Lambda Y^{(8)}_{\mathbf1}(\tau) &~~\text{for}~~\rho_{N_i}\otimes\rho_{N_j}={\mathbf1}\,,\\[0.1in]
\alpha^{M}_{ij} \Lambda Y^{(8)}_{\mathbf1''}(\tau) &~~\text{for}~~\rho_{N_i}\otimes\rho_{N_j}={\mathbf1'}\,,\\[0.1in]
\alpha^{M}_{ij}\Lambda Y^{(8)}_{\mathbf1'}(\tau)&~~\text{for}~~\rho_{N_i}\otimes\rho_{N_j}={\mathbf1''}\,,
\end{array}\right.
\end{eqnarray}
where $\alpha^{M}_{ij}$ are coupling constants.

\end{itemize}

\subsection{Dirac neutrino Yukawa coupling}

The Dirac neutrino Yukawa coupling arises from the term $N\overline{F}H_{5} f_N(Y)$ in Eq.~\eqref{eq:mass-SU5}. In the following, we report the Dirac neutrino Yukawa coupling for different weight and representation assignments of $N$ and $\overline{F}$.

\begin{itemize}

\item{$\rho_N=\rho_{\overline{F}}=\mathbf{3}$}

For different values of $k_N$ and $k_{\overline{F}}$, we can read out the neutrino Yukawa couplings ${\cal Y}^\nu_5$ as follows.
\begin{eqnarray}
\nonumber k_N+k_{\overline{F}}=0&&~:~ {\cal Y}^\nu_5= \,S^{(0)}_{\mathbf1}\,,\\
\nonumber k_N+k_{\overline{F}}=2&&~:~ {\cal Y}^\nu_5= \,S^{(2)}_{\mathbf3} + A^{(2)}_{\mathbf3}\,,\\
\nonumber k_N+k_{\overline{F}}=4&&~:~ {\cal Y}^\nu_5= \,S^{(4)}_{\mathbf3} + A^{(4)}_{\mathbf3} + S^{(4)}_{\mathbf1} + S^{(4)}_{\mathbf1'}\,,\\
\nonumber k_N+k_{\overline{F}}=6&&~:~ {\cal Y}^\nu_5= \,S^{(6)}_{\mathbf{3}I} + A^{(6)}_{\mathbf{3}I} + S^{(6)}_{\mathbf{3}II} + A^{(6)}_{\mathbf{3}II} + S^{(6)}_{\mathbf1}\,,\\
k_N+k_{\overline{F}}=8&&~:~ {\cal Y}^\nu_5= \,S^{(8)}_{\mathbf{3}I} + A^{(8)}_{\mathbf{3}I} + S^{(8)}_{\mathbf{3}II} + A^{(8)}_{\mathbf{3}II} + S^{(8)}_{\mathbf1} + S^{(8)}_{\mathbf1'} + S^{(8)}_{\mathbf1''}\,,
\end{eqnarray}
where we have omitted the coupling coefficients of each independent term and
\begin{eqnarray}
A^{(k)}_{{\mathbf3}}(\tau) =
\left(\begin{matrix}
 0 ~& Y^{(k)}_{{{\mathbf3}},3}(\tau)  ~&~ -Y^{(k)}_{{{\mathbf3}},2}(\tau) \\
 -Y^{(k)}_{{{\mathbf3}},3}(\tau) ~& 0  ~& Y^{(k)}_{{{\mathbf3}},1}(\tau) \\
 Y^{(k)}_{{{\mathbf3}},2}(\tau) ~& -Y^{(k)}_{{{\mathbf3}},1}(\tau)  ~& 0
\end{matrix}\right)\,.
\end{eqnarray}

\item{$\rho_N=\mathbf{3}$, $\rho_{\overline{F}}=\mathbf{1}, \mathbf{1}', \mathbf{1}''$}

In this case, the modular function $f^N(Y)$ should be a modular form in the representation $\mathbf{3}$ of $A_4$. For notational simplicity, we denote ${\mathbf1}^{0,1,2}={\mathbf1},{\mathbf1}',{\mathbf1}''$ and  $(\rho_{\overline{F}_1}\,,\rho_{\overline{F}_2}\,,\rho_{\overline{F}_3})=({\mathbf1}^{a_{\overline{F}_1}}\,,{\mathbf1}^{a_{\overline{F}_2}}\,,{\mathbf1}^{a_{\overline{F}_3}})$ with $a_{\overline{F}_j}=0$, $1$, $2$. Then the $(ij)$ element of the Dirac neutrino Yukawa coupling matrix is given by,
\begin{eqnarray}
\label{eq:contraction-triMF}\begin{aligned}
k_{N}+k_{\overline{F}_j}=2,4~&:~~~({\cal Y}^\nu_5)_{ij} =\alpha_{\nu_j} Y_{{\mathbf3},\langle i+a_{\overline{F}_j}\rangle}^{(k_{N}+k_{\overline{F}_j})}\,,\\
k_{N}+k_{\overline{F}_j}=6,8~&:~~~({\cal Y}^\nu_5)_{ij} =\alpha_{\nu_j} Y_{\mathbf{3}I,\langle i+a_{\overline{F}_j}\rangle}^{(k_{N}+k_{\overline{F}_j})} + \alpha_{\nu_j}^\prime Y_{\mathbf{3}II,\langle i+a_{\overline{F}_j}\rangle}^{(k_{N}+k_{\overline{F}_j})} \,,
\end{aligned}
\end{eqnarray}
with $\langle 1\rangle=\langle4\rangle=1$,
$\langle 2\rangle=\langle5\rangle=3$ and
$\langle 0\rangle=\langle3\rangle=2$. Here $i$ and $j$ are the generation indices of $N$ and $\overline{F}$ respectively. The neutrino Yukawa couplings can be compactly written as
\begin{eqnarray}
{\cal Y}^\nu_5 &=&  \left(C_{a_{\overline{F}_1}}^{k_{N}+k_{\overline{F}_1}}\,,C_{a_{\overline{F}_2}}^{k_{N}+k_{\overline{F}_2}}\,,C_{a_{\overline{F}_3}}^{k_{N}+k_{\overline{F}_3}}\right)\,,
\end{eqnarray}
with
\begin{equation}
C_{a_{\overline{F}_j}}^{k_{N}+k_{\overline{F}_j}}=\begin{pmatrix}
Y_{{\mathbf3},\langle 1+a_{\overline{F}_j}\rangle}^{(k_{N}+k_{\overline{F}_j})}(\tau)  \\
Y_{{\mathbf3},\langle 2+a_{\overline{F}_j}\rangle}^{(k_{N}+k_{\overline{F}_j})}(\tau) \\
Y_{{\mathbf3},\langle 3+a_{\overline{F}_j}\rangle}^{(k_{N}+k_{\overline{F}_j})}(\tau)
\end{pmatrix}\,.
\end{equation}
To be more specific, we have

\begin{eqnarray}
\nonumber&& k_{N}+k_{\overline{F}_j}=2,4~:~~C_{a_{\overline{F}_j}}^{k_{N}+k_{\overline{F}_j}}=\left\{\begin{array}{cc}
\Big(Y_{{\mathbf3},1}^{(k_{N}+k_{\overline{F}_j})}\,,Y_{{\mathbf3},3}^{(k_{N}+k_{\overline{F}_j})}\,,Y_{{\mathbf3},2}^{(k_{N}+k_{\overline{F}_j})}\Big)^T &~~\text{for}~~a_{\overline{F}_j}=0\,,\\[0.1in]
\Big(Y_{{\mathbf3},3}^{(k_{N}+k_{\overline{F}_j})}\,,Y_{{\mathbf3},2}^{(k_{N}+k_{\overline{F}_j})}\,,Y_{{\mathbf3},1}^{(k_{N}+k_{\overline{F}_j})}\Big)^T
&~~\text{for}~~a_{\overline{F}_j}=1\,, \\[0.1in]
\Big(Y_{{\mathbf3},2}^{(k_{N}+k_{\overline{F}_j})}\,,Y_{{\mathbf3},1}^{(k_{N}+k_{\overline{F}_j})}\,,Y_{{\mathbf3},3}^{(k_{N}+k_{\overline{F}_j})}\Big)^T &~~\text{for}~~a_{\overline{F}_j}=2\,,
\end{array}
\right. \\
\nonumber&& k_{N}+k_{\overline{F}_j}=6,8~:~~C_{a_{\overline{F}_j}}^{k_{N}+k_{\overline{F}_j}}=\left\{\begin{array}{cc}
\Big(Y_{{\mathbf{3}I},1}^{(k_{N}+k_{\overline{F}_j})}\,,Y_{{\mathbf{3}I},3}^{(k_{N}+k_{\overline{F}_j})}\,,Y_{{\mathbf{3}I},2}^{(k_{N}+k_{\overline{F}_j})}\Big)^T & \\
+\Big(Y_{{\mathbf{3}II},1}^{(k_{N}+k_{\overline{F}_j})}\,,Y_{{\mathbf{3}II},3}^{(k_{N}+k_{\overline{F}_j})}\,,Y_{{\mathbf{3}II},2}^{(k_{N}+k_{\overline{F}_j})}\Big)^T &~~\text{for}~~a_{\overline{F}_j}=0\,,  \\[0.15in]
\Big(Y_{{\mathbf{3}I},3}^{(k_{N}+k_{\overline{F}_j})}\,,Y_{{\mathbf{3}I},2}^{(k_{N}+k_{\overline{F}_j})}\,,Y_{{\mathbf{3}I},1}^{(k_{N}+k_{\overline{F}_j})}\Big)^T &  \\
+\Big(Y_{{\mathbf{3}II},3}^{(k_{N}+k_{\overline{F}_j})}\,,Y_{{\mathbf{3}II},2}^{(k_{N}+k_{\overline{F}_j})}\,,Y_{{\mathbf{3}II},1}^{(k_{N}+k_{\overline{F}_j})}\Big)^T &~~\text{for}~~a_{\overline{F}_j}=1\,,  \\[0.15in]
\Big(Y_{{\mathbf{3}I},2}^{(k_{N}+k_{\overline{F}_j})}\,,Y_{{\mathbf{3}I},1}^{(k_{N}+k_{\overline{F}_j})}\,,Y_{{\mathbf{3}I},3}^{(k_{N}+k_{\overline{F}_j})}\Big)^T & \\
+\Big(Y_{{\mathbf{3}II},2}^{(k_{N}+k_{\overline{F}_j})}\,,Y_{{\mathbf{3}II},1}^{(k_{N}+k_{\overline{F}_j})}\,,Y_{{\mathbf{3}II},3}^{(k_{N}+k_{\overline{F}_j})}\Big)^T
&~~\text{for}~~a_{\overline{F}_j}=2\,,
\end{array}
\right.
\end{eqnarray}
where we have omitted the overall coefficient associated with each independent contraction for every column.

\item{$\rho_N=\mathbf{1}, \mathbf{1}', \mathbf{1}''$, $\rho_{\overline{F}}=\mathbf{3}$ }

Similar to previous case,  we denote $(\rho_{N_1}\,,\rho_{N_2}\,,\rho_{N_3})=({\mathbf1}^{a_{N_1}}\,,{\mathbf1}^{a_{N_2}}\,,{\mathbf1}^{a_{N_3}})$ with $a_{N_i}=0$, $1$, $2$ and the neutrino Yukawa couplings read as
\begin{eqnarray}
\nonumber k_{N_i}+k_{\overline{F}}=2,4~&&:~~~({\cal Y}^\nu_5)_{ij} =  \alpha_{\nu_i} Y_{{\mathbf3},\langle j+a_{N_i}\rangle}^{(k_{N_i}+k_{\overline{F}})}\,,\\
k_{N_i}+k_{\overline{F}}=6,8~&&:~~~({\cal Y}^\nu_5)_{ij} =  \alpha_{\nu_i} Y_{\mathbf{3}I,\langle j+a_{N_i}\rangle}^{(k_{N_i}+k_{\overline{F}})} + \alpha_{\nu_i}^\prime Y_{\mathbf{3}II,\langle j+a_{N_i}\rangle}^{(k_{N_i}+k_{\overline{F}})} \,,
\end{eqnarray}
where $i$ and $j$ are the flavor indices of $N$ and $\overline{F}$ respectively, $\alpha_{\nu_i}$ and $\alpha_{\nu_i}^\prime$ are free coupling constants.
We can also write
\begin{eqnarray}
{\cal Y}^\nu_5 =  \left(C_{a_{N_1}}^{k_{N_1}+k_{\overline{F}}}\,,C_{a_{N_2}}^{k_{N_2}+k_{\overline{F}}}\,,C_{a_{N_3}}^{k_{N_3}+k_{\overline{F}}}\right)^T\,.
\end{eqnarray}

\end{itemize}

\subsection{The charged lepton and down quark sectors }

The masses of the charged lepton and down type quarks are described by the terms $\overline{F}TH_{\overline{5}}f_{d}(Y)$ and $\overline{F}T H_{\overline{45}}f'_{d}(Y)$ in Eq.~\eqref{eq:mass-SU5}. Modular invariance strongly constrains the form of the modular functions $f_{D}(Y)$ and $f'_{D}(Y)$.

\begin{itemize}
\item{$\rho_{\overline{F}}=\rho_{T}=\mathbf{3}$ }

The Yukawa coupling with the Higgs $H_{\overline{5}}$ is determined by the modular weight of matter fields.
\begin{eqnarray}
\nonumber k_{\overline{F}}+k_T=0~~&&:~~
{\cal Y}^d_{\overline{5}} = S^{(0)}_{\mathbf1}\,, \\
\nonumber k_{\overline{F}}+k_T=2~~&&:~~
{\cal Y}^d_{\overline{5}}  = S^{(2)}_{\mathbf3} + A^{(2)}_{\mathbf3}\,,\\
\nonumber k_{\overline{F}}+k_T=4~~&&:~~
{\cal Y}^d_{\overline{5}}  = S^{(4)}_{\mathbf3} + A^{(4)}_{\mathbf3} + S^{(4)}_{\mathbf1} + S^{(4)}_{\mathbf1'}\,,\\
\nonumber k_{\overline{F}}+k_T=6~~&&:~~
{\cal Y}^d_{\overline{5}}  = S^{(6)}_{\mathbf{3}I} + A^{(6)}_{\mathbf{3}I} + S^{(6)}_{\mathbf{3}II} + A^{(6)}_{\mathbf{3}II} + S^{(6)}_{\mathbf1}\,,\\
 k_{\overline{F}}+k_T=8~~&&:~~
{\cal Y}^d_{\overline{5}}  =
S^{(8)}_{\mathbf{3}I}
+A^{(8)}_{\mathbf{3}I}
+S^{(8)}_{\mathbf{3}II}
+A^{(8)}_{\mathbf{3}II}
+S^{(8)}_{\mathbf1}
+S^{(8)}_{\mathbf1'}
+S^{(8)}_{\mathbf1''}\,,
\end{eqnarray}
where the dimensionless coefficient in front of each term has been omitted. Analogously the Yukawa coupling ${\cal Y}^d_{\overline{45}}$ with $H_{\overline{45}}$ can be obtained, and it is found to be of the same form as ${\cal Y}^d_{\overline{5}}$ because both $H_{\overline{5}}$ and $H_{\overline{45}}$ are assumed to be $A_4$ singlets with vanishing modular weight.

\item{$\rho_{\overline{F}}=\mathbf{3}$, $\rho_T=\mathbf{1}, \mathbf{1}', \mathbf{1}''$ }

The charged lepton and down quark Yukawa couplings depend on the $A_4$
representation of the ten-dimensional matter fields $T$ and their weights. We can read out
\begin{eqnarray}
\nonumber k_{\overline F}+k_{T_j}=2,4~&&:~~
({\cal Y}^d_{\overline{5}} )_{ij} = \alpha_{d_j} Y_{{\mathbf3},\langle i+a_{T_j}\rangle}^{(k_{\overline F}+k_{T_j})}\,,\\
k_{\overline F}+k_{T_j}=6,8~&&:~~
({\cal Y}^d_{\overline{5}} )_{ij} = \alpha_{d_j} Y_{\mathbf{3}I,\langle i+a_{T_j}\rangle}^{(k_{\overline F}+k_{T_j})}
+ \alpha_{d_j}^\prime Y_{\mathbf{3}II,\langle i+a_{T_j}\rangle}^{(k_{\overline F}+k_{T_j})}\,,
\end{eqnarray}
where $\alpha_{d_j}$ and $\alpha_{d_j}^\prime$ are coupling constants, and $i$ and $j$ stand for the flavor indices of $\overline{F}$ and $T$ respectively. Note ${\cal Y}^d_{\overline{45}}$ is of the same form as ${\cal Y}^d_{\overline{5}}$.

\item{$\rho_{\overline{F}}=\mathbf{1}, \mathbf{1}', \mathbf{1}''$, $\rho_T=\mathbf{3}$ }

For this assignment, the $(ij)$ element of the Yukawa coupling matrix is found to be
\begin{eqnarray}
\nonumber k_{\overline{F}_i} + k_T=2,4~&&:~~
({\cal Y}^d_{\overline{5}} )_{ij} = \alpha_{d_i} Y_{{\mathbf3},\langle j+a_{\overline{F}_i}\rangle}^{(k_{\overline{F}_i} + k_T)}\,,\\
k_{\overline{F}_i} + k_T=6,8~&&:~~
({\cal Y}^d_{\overline{5}} )_{ij} = \alpha_{d_i} Y_{\mathbf{3}I,\langle j+a_{\overline{F}_i}\rangle}^{(k_{\overline{F}_i} + k_T)}
 + \alpha_{d_i}^\prime Y_{\mathbf{3}II,\langle j+a_{\overline{F}_i}\rangle}^{(k_{\overline{F}_i} + k_T)}\,,
\end{eqnarray}
The electroweak Higgs field $H_d$ is a linear combination of the doublet components of the Higgs fields $H_{\overline{5}}$ and
$H_{\overline{45}}$, and the mixing angle can be absorbed into the coupling constants. Hence the charged lepton and down quark mass matrices are of the following form:
\begin{equation}
\label{eq:Me-Md}M_e=(\mathcal Y^d_{\overline{5}}-3\mathcal Y^d_{\overline{45}})^T v_d,~~~~M_d=(\mathcal Y^d_{\overline{5}}+\mathcal Y^d_{\overline{45}}) v_d\,.
\end{equation}
The Georgi-Jarlskog factor ``$-3$'' can account for the mass differences of charged leptons and down quarks~\cite{Georgi:1979df}.

\end{itemize}

\subsection{The up quark sector }

The up quark mass matrix is symmetric in the $SU(5)$ GUT theory and it is predicted to be of the same form as the right-handed neutrino mass matrix $M_N$ in the current context of $A_4$ modular symmetry.

\begin{itemize}

\item{$\rho_{T}=\mathbf{3}$ }

We can read out the up quark Yukawa coupling matrix for different values of modular weight $k_{T}$ as follows,
\begin{eqnarray}
\nonumber k_T=0~&&:~~ {\cal Y}^u_5=S^{(0)}_{\mathbf1}\,,\\
\nonumber k_T=1~&&:~~  {\cal Y}^u_5=S^{(2)}_{\mathbf3}\,,\\
\nonumber k_T=2~&&:~~  {\cal Y}^u_5= S^{(4)}_{\mathbf3}+S^{(4)}_{\mathbf1}+S^{(4)}_{\mathbf1'}\,,\\
\nonumber k_T=3~&&:~~  {\cal Y}^u_5= S^{(6)}_{\mathbf{3}I}+S^{(6)}_{\mathbf{3}II}+S^{(6)}_{\mathbf1} \,,\\
 k_T=4~&&:~~ {\cal Y}^u_5= S^{(8)}_{\mathbf{3}I}+S^{(8)}_{\mathbf{3}II}+S^{(8)}_{\mathbf1}+S^{(8)}_{\mathbf1'}+S^{(8)}_{\mathbf1''}\,.
\end{eqnarray}

\item{$\rho_T=\mathbf{1}, \mathbf{1}', \mathbf{1}''$ }

For the modular weight $k_{T_i}+k_{T_j}\le8$, the possible non-vanishing elements of ${\cal Y}^u_5$ are
\begin{eqnarray}
\nonumber&&k_{T_i}+k_{T_j}=0~~:~~({\cal Y}^u_5)_{ij} = \alpha^u_{ij}  ~~\text{for}~~\rho_{T_i}\otimes\rho_{T_j}={\mathbf1}\,, \\
\nonumber&&k_{T_i}+k_{T_j}=4~~:~~({\cal Y}^u_5)_{ij} = \left\{\begin{array}{cc}\alpha^{u}_{ij}  Y^{(4)}_{\mathbf1}(\tau) &~~\text{for}~~\rho_{T_i}\otimes\rho_{T_j}={\mathbf1}\,,\\[0.1in]
 \alpha^{u}_{ij} Y^{(4)}_{\mathbf1'}(\tau)&~~\text{for}~~\rho_{T_i}\otimes\rho_{T_j}={\mathbf1''}\,,
\end{array}\right.\\
\nonumber&&k_{T_i}+k_{T_j}=6~~:~~({\cal Y}^u_5)_{ij} =\alpha^{u}_{ij} Y^{(6)}_{\mathbf1}(\tau)~~\text{for}~~\rho_{T_i}\otimes\rho_{T_j}={\mathbf1}\,,\\
&&k_{T_i}+k_{T_j}=8~~:~~({\cal Y}^u_5)_{ij} = \left\{\begin{array}{cc} \alpha^{u}_{ij} Y^{(8)}_{\mathbf1}(\tau) &~~\text{for}~~\rho_{T_i}\otimes\rho_{T_j}={\mathbf1}\,,\\[0.1in]
\alpha^{u}_{ij} Y^{(8)}_{\mathbf1''}(\tau) &~~\text{for}~~\rho_{T_i}\otimes\rho_{T_j}={\mathbf1'}\,,\\[0.1in]
\alpha^{u}_{ij} Y^{(8)}_{\mathbf1'}(\tau)&~~\text{for}~~\rho_{T_i}\otimes\rho_{T_j}={\mathbf1''}\,.
\end{array}\right.
\end{eqnarray}

\end{itemize}

\section{\label{sec:benchmark-models}Benchmark models}

We can classify all the possible $A_4$ modular $SU(5)$ models according to the transformation properties of the matter fields $N$, $\overline{F}$ and $T$ under $A_4$. We don't consider the cases in which all matter fields are assigned to $A_4$ singlets, since the Yukawa superpotentials would be less constrained by modular symmetry and generally more free parameters would be involved. If three right-handed neutrinos are introduced in the seesaw mechanism, we have the following five types of models:
\begin{eqnarray}
\label{eq:type1}\text{Type-I}~\f&:&\f\quad~N\sim\mathbf{3}\,,\qquad\qquad\qquad~~
\overline{F}\sim\mathbf{3}\,,\qquad\qquad\qquad~~
T_{1,2,3}\sim\mathbf{1}(\mathbf{1}',\mathbf{1}'')\,,\\
\label{eq:type2}\text{Type-II}~\f&:&\f\quad~N\sim\mathbf{3}\,,\qquad\qquad\qquad~~ \overline{F}_{1,2,3}\sim\mathbf{1}(\mathbf{1}',\mathbf{1}'')\,,\qquad T\sim\mathbf{3}\,,\\
\label{eq:type3}\text{Type-III}~\f&:&\f\quad~N_{1,2,3}\sim\mathbf{1}(\mathbf{1}',\mathbf{1}'')\,,\qquad \overline{F}\sim\mathbf{3}\,,\qquad\qquad\qquad~~ T\sim\mathbf{3}\,,\\
\label{eq:type4}\text{Type-IV}~\f&:&\f\quad~
N\sim\mathbf{3}\,,\qquad\qquad\qquad~~
\overline{F}\sim\mathbf{3}\,,\qquad\qquad\qquad~~
T\sim\mathbf{3}\,,\\
\label{eq:type5}\text{Type-V}~\f&:&\f\quad~N_{1,2,3}\sim\mathbf{1}(\mathbf{1}',\mathbf{1}'')\,,\qquad
\overline{F}\sim\mathbf{3}\,,\qquad\qquad\qquad~~
T_{1,2,3}\sim\mathbf{1}(\mathbf{1}',\mathbf{1}'')\,.
\end{eqnarray}
As shown in Ref.~\cite{Ding:2019zxk},
there are ten possibilities for the singlet assignment of the three generations of matter fields,
\begin{eqnarray}
\begin{aligned}
&
C_{1}~:~ \{\mathbf1, \mathbf1, \mathbf1\}\,,\quad~~~
C_{2} ~:~ \{\mathbf1^{\prime}, \mathbf1^{\prime}, \mathbf1^{\prime}\}\,,\quad~
C_{3} ~:~ \{\mathbf1^{\prime\prime}, \mathbf1^{\prime\prime}, \mathbf1^{\prime\prime}\}\,,\\
&
C_{4} ~:~ \{\mathbf1, \mathbf1, \mathbf1^{\prime}\}\,,\quad~~
C_{5} ~:~ \{\mathbf1, \mathbf1, \mathbf1^{\prime\prime}\}\,,\quad~~
C_{6} ~:~ \{\mathbf1^{\prime}, \mathbf1^{\prime}, \mathbf1\}\,,\\
&
C_{7} ~:~ \{\mathbf1^{\prime}, \mathbf1^{\prime}, \mathbf1^{\prime\prime}\}\,,\quad
C_{8}~:~ \{\mathbf1^{\prime\prime}, \mathbf1^{\prime\prime}, \mathbf1\}\,, \quad\,
C_{9} ~:~ \{\mathbf1^{\prime\prime}, \mathbf1^{\prime\prime}, \mathbf1^{\prime}\}\,,\quad
C_{10} ~:~\{\mathbf1, \mathbf1^{\prime}, \mathbf1^{\prime\prime}\}\,.
\end{aligned}
\label{eq:assignments}
\end{eqnarray}
The three generations of matter fields are distinguished by their modular weights and transformation rules under $A_4$. If any two generations of fermion fields are assigned to the same singlet representation of $A_4$, their modular weight should be different. Once can also permute the above singlet assignments together with modular weights, for instance, the three generations of matter fields can be assigned to transform as $\mathbf{1}$, $\mathbf{1^{\prime}}$ and $\mathbf1^{\prime\prime}$ or $\mathbf{1^{\prime\prime}}$, $\mathbf{1}$, $\mathbf{1^{\prime}}$, this amounts to a redefinition of the fields. As a consequence, the predictions for the masses and mixing matrices of the quarks and leptons would keep invariant.

Moreover, if neutrino masses are generated by the seesaw mechanism with two right-handed neutrinos which transform as singlets under $A_4$, then we have another two kinds of models:
\begin{eqnarray}
\label{eq:type6}\text{Type-VI}~\f&:&\f\quad~
N_{1,2}\sim\mathbf{1}(\mathbf{1}',\mathbf{1}'')\,,\qquad\quad
\overline{F}\sim\mathbf{3}\,,\qquad\quad
T\sim\mathbf{3}\,,\\
\label{eq:type7}\text{Type-VII}~\f&:&\f\quad~
N_{1,2}\sim\mathbf{1}(\mathbf{1}',\mathbf{1}'')\,,\qquad\quad
\overline{F}\sim\mathbf{3}\,,\qquad\quad
T_{1,2,3}\sim\mathbf{1}(\mathbf{1}',\mathbf{1}'')\,.
\end{eqnarray}
As is generally true in two right-handed neutrino models, the lightest neutrino would be massless. The right-handed neutrinos $N_1$ and $N_2$ can transform in either the same or different manner under $A_4$,
\begin{eqnarray}
\nonumber&&D_{1}~:~ \{\mathbf1, \mathbf1\}\,,\quad~~~D_{2} ~:~ \{\mathbf1^{\prime}, \mathbf1^{\prime}\}\,,\quad~~~D_{3} ~:~ \{\mathbf1^{\prime\prime}, \mathbf1^{\prime\prime}\}\,,\\
&&D_{4}~:~ \{\mathbf1, \mathbf1^{\prime}\}\,,\quad~~~D_{5} ~:~ \{\mathbf1, \mathbf1^{\prime\prime}\}\,,\quad~~~D_{6} ~:~ \{\mathbf1^{\prime}, \mathbf1^{\prime\prime}\}\,.
\end{eqnarray}
Note that $N_1$ and $N_2$ are assigned to the same $A_4$ singlet for the cases of $D_{1, 2, 3}$ and they should carry different modular weights to be distinguished. For the Type-VII models, the three generations of the ten-plets $T_1$, $T_2$ and $T_3$ have ten possible representation assignments as shown in Eq.~\eqref{eq:assignments}.
For each type of models, we comprehensively scan over the weights and  representations of $N$, $\overline{F}$ and $T$. The requirement that each term of the superpotential $\mathcal{W}$ has vanishing total weight entails the modular weights of the matter fields $N$, $\overline{F}$ and $T$ are integers. We neglect these cases that one column or one row of the fermion mass matrix is vanishing, since at least one fermion would be massless. A numerical analysis is performed for each model, and the strategy of numerical analysis is discussed in section~\ref{sec:numerical-results}. In the following, we present some benchmark models which can give acceptable masses and mixing parameters of both quark and leptons for certain values of the input parameters. The assignments of the matter fields for these viable models are summarized in table~\ref{tab:models-summary}.

\tabcolsep=0.1cm
\renewcommand\arraystretch{0.65}
\begin{table}[hptb]
\centering
  \begin{tabular}{|c|c|c|c|c|c|c|c|c|}
\hline\hline
Type-I& \#P& $(\rho_{N},\rho_{F},\rho_{T_{1}},\rho_{T_{2}},\rho_{T_{3}})$& $k_{N}$& $k_F$& $(k_{T_{1}},k_{T_{2}},k_{T_{3}})$ \\\hline\hline

 $\mathcal{I}_{1}$  & 24 & $(\mathbf{3} , \mathbf{3}, \mathbf{1^{\prime\prime}},\mathbf{1},\mathbf{1})$ & 1 &1& (1,1,3) \\ \hline
 $\mathcal{I}_{2}$  & 24 & $(\mathbf{3} , \mathbf{3}, \mathbf{1^{\prime\prime}},\mathbf{1^{\prime}},\mathbf{1^{\prime}})$ & 0 &2& (2,0,4) \\ \hline
 $\mathcal{I}_{3}$  & 24 & $(\mathbf{3} , \mathbf{3}, \mathbf{1^{\prime\prime}},\mathbf{1^{\prime\prime}},\mathbf{1^{\prime}})$ & 0 &2& (0,2,4) \\ \hline
 $\mathcal{I}_{4}$  & 24 & $(\mathbf{3} , \mathbf{3}, \mathbf{1^{\prime\prime}},\mathbf{1},\mathbf{1})$ & 0 &2& (0,0,4) \\ \hline
 $\mathcal{I}_{5}$  & 24 & $(\mathbf{3} , \mathbf{3}, \mathbf{1^{\prime\prime}},\mathbf{1},\mathbf{1})$ & 0 &2& (0,2,4) \\ \hline
 $\mathcal{I}_{6}$  & 24 & $(\mathbf{3} , \mathbf{3}, \mathbf{1^{\prime\prime}},\mathbf{1},\mathbf{1})$ & 0 &2& (2,2,4) \\ \hline
 $\mathcal{I}_{7}$  & 24 & $(\mathbf{3} , \mathbf{3}, \mathbf{1^{\prime}},\mathbf{1},\mathbf{1^{\prime\prime}})$ & 0 &2& (0,0,4) \\ \hline
 $\mathcal{I}_{8}$  & 24 & $(\mathbf{3} , \mathbf{3}, \mathbf{1^{\prime\prime}},\mathbf{1},\mathbf{1^{\prime}})$ & 2 &0& (2,2,2) \\ \hline
 $\mathcal{I}_{9}$  & 24 & $(\mathbf{3} , \mathbf{3}, \mathbf{1^{\prime\prime}},\mathbf{1},\mathbf{1^{\prime}})$ & 0 &2& (2,2,4) \\ \hline
 $\mathcal{I}_{10}$  & 24 & $(\mathbf{3} , \mathbf{3}, \mathbf{1^{\prime}},\mathbf{1^{\prime}},\mathbf{1^{\prime\prime}})$ & 0 &2& (0,2,4) \\ \hline
 $\mathcal{I}_{11}$  & 24 & $(\mathbf{3} , \mathbf{3}, \mathbf{1^{\prime\prime}},\mathbf{1^{\prime}},\mathbf{1^{\prime}})$ & 0 &2& (2,2,4) \\ \hline
 $\mathcal{I}_{12}$  & 24 & $(\mathbf{3} , \mathbf{3}, \mathbf{1^{\prime\prime}},\mathbf{1^{\prime\prime}},\mathbf{1^{\prime}})$ & 0 &2& (2,4,2) \\ \hline
 $\mathcal{I}_{13}$  & 24 & $(\mathbf{3} , \mathbf{3}, \mathbf{1^{\prime\prime}},\mathbf{1},\mathbf{1^{\prime}})$ & 2 &0& (2,2,4) \\ \hline
 $\mathcal{I}_{14}$  & 24 & $(\mathbf{3} , \mathbf{3}, \mathbf{1^{\prime\prime}},\mathbf{1},\mathbf{1^{\prime}})$ & 2 &0& (2,4,4) \\ \hline
 $\mathcal{I}_{15}$  & 24 & $(\mathbf{3} , \mathbf{3}, \mathbf{1^{\prime\prime}},\mathbf{1^{\prime}},\mathbf{1^{\prime}})$ & 2 &0& (2,2,4) \\ \hline
 $\mathcal{I}_{16}$  & 24 & $(\mathbf{3} , \mathbf{3}, \mathbf{1^{\prime\prime}},\mathbf{1^{\prime\prime}},\mathbf{1^{\prime}})$ & 2 &0& (2,4,2) \\ \hline
 $\mathcal{I}_{17}$  & 24 & $(\mathbf{3} , \mathbf{3}, \mathbf{1^{\prime\prime}},\mathbf{1^{\prime\prime}},\mathbf{1^{\prime}})$ & 2 &0& (2,4,4) \\ \hline
\hline
Type-II&\#P& $(\rho_{N},\rho_{T},\rho_{F_1},\rho_{F_2},\rho_{F_3})$& $k_{N}$& $(k_{F_1},k_{F_2},k_{F_3})$& $k_{T}$ \\\hline\hline

 $\mathcal{II}_{1}$  & 24 & $(\mathbf{3} , \mathbf{3} , \mathbf{1},\mathbf{1^{\prime\prime}},\mathbf{1^{\prime}})$ & 1& (1,3,1) & 3 \\ \hline
 $\mathcal{II}_{2}$  & 24 & $(\mathbf{3} , \mathbf{3} , \mathbf{1^{\prime\prime}},\mathbf{1^{\prime\prime}},\mathbf{1})$ & 1& (1,3,1) & 3 \\ \hline
\hline
Type-V&\#P& $(\rho_{F},\rho_{N_1},\rho_{N_2},\rho_{N_3},\rho_{T_{1}},\rho_{T_{2}},\rho_{T_{3}})$& $(k_{N_1},k_{N_2},k_{N_3})$& $k_F$& $(k_{T_{1}},k_{T_{2}},k_{T_{3}})$ \\\hline\hline

 $\mathcal{V}_{1}$  & 24 & $(\mathbf{3}, \mathbf{1},\mathbf{1},\mathbf{1^{\prime\prime}},\mathbf{1^{\prime\prime}},\mathbf{1},\mathbf{1})$ & (1,3,3) &1& (1,1,3) \\ \hline
 $\mathcal{V}_{2}$  & 24 & $(\mathbf{3}, \mathbf{1},\mathbf{1^{\prime\prime}},\mathbf{1^{\prime}},\mathbf{1^{\prime\prime}},\mathbf{1},\mathbf{1})$ & (3,1,3) &1& (1,1,3) \\ \hline
 $\mathcal{V}_{3}$  & 24 & $(\mathbf{3}, \mathbf{1},\mathbf{1^{\prime\prime}},\mathbf{1^{\prime}},\mathbf{1^{\prime\prime}},\mathbf{1},\mathbf{1^{\prime}})$ & (0,2,2) &2& (2,2,2) \\ \hline
 $\mathcal{V}_{4}$  & 24 & $(\mathbf{3}, \mathbf{1},\mathbf{1^{\prime\prime}},\mathbf{1^{\prime}},\mathbf{1^{\prime\prime}},\mathbf{1},\mathbf{1^{\prime}})$ & (4,2,2) &0& (2,2,2) \\ \hline
 $\mathcal{V}_{5}$  & 24 & $(\mathbf{3}, \mathbf{1},\mathbf{1^{\prime\prime}},\mathbf{1^{\prime}},\mathbf{1^{\prime\prime}},\mathbf{1^{\prime}},\mathbf{1^{\prime}})$ & (4,2,2) &0& (2,2,4) \\ \hline
 $\mathcal{V}_{6}$  & 24 & $(\mathbf{3}, \mathbf{1},\mathbf{1^{\prime\prime}},\mathbf{1^{\prime}},\mathbf{1^{\prime\prime}},\mathbf{1^{\prime\prime}},\mathbf{1^{\prime}})$ & (4,2,2) &0& (2,4,2) \\ \hline
 $\mathcal{V}_{7}$  & 24 & $(\mathbf{3}, \mathbf{1},\mathbf{1^{\prime\prime}},\mathbf{1^{\prime}},\mathbf{1^{\prime\prime}},\mathbf{1^{\prime\prime}},\mathbf{1^{\prime}})$ & (4,2,2) &0& (2,4,4) \\ \hline
 $\mathcal{V}_{8}$  & 24 & $(\mathbf{3}, \mathbf{1^{\prime}},\mathbf{1^{\prime}},\mathbf{1},\mathbf{1^{\prime\prime}},\mathbf{1},\mathbf{1^{\prime}})$ & (2,4,2) &0& (2,2,2) \\ \hline
 $\mathcal{V}_{9}$  & 24 & $(\mathbf{3}, \mathbf{1^{\prime}},\mathbf{1^{\prime}},\mathbf{1},\mathbf{1^{\prime\prime}},\mathbf{1^{\prime}},\mathbf{1^{\prime}})$ & (2,4,2) &0& (2,2,4) \\ \hline
 $\mathcal{V}_{10}$  & 24 & $(\mathbf{3}, \mathbf{1^{\prime}},\mathbf{1^{\prime}},\mathbf{1},\mathbf{1^{\prime\prime}},\mathbf{1^{\prime\prime}},\mathbf{1^{\prime}})$ & (2,4,2) &0& (2,4,2) \\ \hline
 $\mathcal{V}_{11}$  & 24 & $(\mathbf{3}, \mathbf{1^{\prime}},\mathbf{1^{\prime}},\mathbf{1},\mathbf{1^{\prime\prime}},\mathbf{1^{\prime\prime}},\mathbf{1^{\prime}})$ & (2,4,2) &0& (2,4,4) \\ \hline
 $\mathcal{V}_{12}$  & 24 & $(\mathbf{3}, \mathbf{1^{\prime\prime}},\mathbf{1^{\prime\prime}},\mathbf{1},\mathbf{1^{\prime\prime}},\mathbf{1},\mathbf{1^{\prime}})$ & (2,4,2) &0& (2,2,2) \\ \hline
 $\mathcal{V}_{13}$  & 24 & $(\mathbf{3}, \mathbf{1^{\prime\prime}},\mathbf{1^{\prime\prime}},\mathbf{1},\mathbf{1^{\prime\prime}},\mathbf{1},\mathbf{1^{\prime}})$ & (2,4,2) &0& (2,4,4) \\ \hline
 $\mathcal{V}_{14}$  & 24 & $(\mathbf{3}, \mathbf{1^{\prime\prime}},\mathbf{1^{\prime\prime}},\mathbf{1},\mathbf{1^{\prime\prime}},\mathbf{1},\mathbf{1})$ & (2,4,2) &0& (2,2,4) \\ \hline
 $\mathcal{V}_{15}$  & 24 & $(\mathbf{3}, \mathbf{1},\mathbf{1^{\prime\prime}},\mathbf{1^{\prime}},\mathbf{1^{\prime}},\mathbf{1^{\prime\prime}},\mathbf{1^{\prime\prime}})$ & (0,0,0) &2& (0,0,4) \\ \hline
 $\mathcal{V}_{16}$  & 24 & $(\mathbf{3}, \mathbf{1},\mathbf{1^{\prime\prime}},\mathbf{1^{\prime}},\mathbf{1^{\prime}},\mathbf{1^{\prime\prime}},\mathbf{1})$ & (0,0,0) &2& (0,0,4) \\ \hline
 $\mathcal{V}_{17}$  & 24 & $(\mathbf{3}, \mathbf{1},\mathbf{1^{\prime\prime}},\mathbf{1^{\prime}},\mathbf{1^{\prime}},\mathbf{1^{\prime\prime}},\mathbf{1})$ & (0,0,0) &4& (0,0,4) \\ \hline
 $\mathcal{V}_{18}$  & 24 & $(\mathbf{3}, \mathbf{1},\mathbf{1^{\prime\prime}},\mathbf{1^{\prime}},\mathbf{1^{\prime}},\mathbf{1^{\prime\prime}},\mathbf{1})$ & (2,0,0) &2& (0,0,4) \\ \hline
 $\mathcal{V}_{19}$  & 24 & $(\mathbf{3}, \mathbf{1},\mathbf{1^{\prime\prime}},\mathbf{1^{\prime}},\mathbf{1^{\prime}},\mathbf{1^{\prime\prime}},\mathbf{1^{\prime\prime}})$ & (0,0,0) &4& (0,0,4) \\ \hline
 $\mathcal{V}_{20}$  & 24 & $(\mathbf{3}, \mathbf{1},\mathbf{1^{\prime\prime}},\mathbf{1^{\prime}},\mathbf{1^{\prime}},\mathbf{1^{\prime\prime}},\mathbf{1^{\prime\prime}})$ & (2,0,0) &2& (0,0,4) \\ \hline
 $\mathcal{V}_{21}$  & 24 & $(\mathbf{3}, \mathbf{1},\mathbf{1^{\prime\prime}},\mathbf{1^{\prime}},\mathbf{1^{\prime\prime}},\mathbf{1^{\prime}},\mathbf{1^{\prime}})$ & (0,0,0) &2& (2,0,4) \\ \hline
 $\mathcal{V}_{22}$  & 24 & $(\mathbf{3}, \mathbf{1},\mathbf{1^{\prime\prime}},\mathbf{1^{\prime}},\mathbf{1^{\prime\prime}},\mathbf{1^{\prime}},\mathbf{1^{\prime}})$ & (2,0,0) &2& (2,0,4) \\ \hline
 $\mathcal{V}_{23}$  & 24 & $(\mathbf{3}, \mathbf{1},\mathbf{1^{\prime\prime}},\mathbf{1^{\prime}},\mathbf{1^{\prime\prime}},\mathbf{1^{\prime\prime}},\mathbf{1^{\prime}})$ & (0,0,0) &2& (0,2,4) \\ \hline
 $\mathcal{V}_{24}$  & 24 & $(\mathbf{3}, \mathbf{1},\mathbf{1^{\prime\prime}},\mathbf{1^{\prime}},\mathbf{1^{\prime\prime}},\mathbf{1^{\prime\prime}},\mathbf{1^{\prime}})$ & (2,0,0) &2& (0,2,4) \\ \hline
 $\mathcal{V}_{25}$  & 24 & $(\mathbf{3}, \mathbf{1^{\prime}},\mathbf{1^{\prime}},\mathbf{1^{\prime\prime}},\mathbf{1^{\prime}},\mathbf{1^{\prime\prime}},\mathbf{1})$ & (0,2,0) &2& (0,0,4) \\ \hline
 $\mathcal{V}_{26}$  & 24 & $(\mathbf{3}, \mathbf{1^{\prime}},\mathbf{1^{\prime}},\mathbf{1^{\prime\prime}},\mathbf{1^{\prime}},\mathbf{1^{\prime\prime}},\mathbf{1^{\prime\prime}})$ & (0,2,0) &2& (0,0,4) \\ \hline
 $\mathcal{V}_{27}$  & 24 & $(\mathbf{3}, \mathbf{1^{\prime}},\mathbf{1^{\prime}},\mathbf{1^{\prime\prime}},\mathbf{1^{\prime\prime}},\mathbf{1^{\prime}},\mathbf{1^{\prime}})$ & (0,2,0) &2& (2,0,4) \\ \hline
 $\mathcal{V}_{28}$  & 24 & $(\mathbf{3}, \mathbf{1^{\prime}},\mathbf{1^{\prime}},\mathbf{1^{\prime\prime}},\mathbf{1^{\prime\prime}},\mathbf{1^{\prime\prime}},\mathbf{1^{\prime}})$ & (0,2,0) &2& (0,2,4) \\ \hline
\hline
Type-VII&\#P& $(\rho_{F},\rho_{N_1},\rho_{N_2},\rho_{T_{1}},\rho_{T_{2}},\rho_{T_{3}})$& $(k_{N_1},k_{N_2})$& $k_F$& $(k_{T_{1}},k_{T_{2}},k_{T_{3}})$ \\\hline\hline

 $\mathcal{VII}_{1}$  & 23 & $(\mathbf{3} , \mathbf{1},\mathbf{1^{\prime\prime}},\mathbf{1^{\prime\prime}},\mathbf{1},\mathbf{1})$ & (1,3) &3& (1,1,3) \\ \hline
 $\mathcal{VII}_{2}$  & 23 & $(\mathbf{3} , \mathbf{1^{\prime}},\mathbf{1^{\prime}},\mathbf{1^{\prime\prime}},\mathbf{1},\mathbf{1})$ & (1,3) &3& (1,1,3) \\ \hline
 $\mathcal{VII}_{3}$  & 23 & $(\mathbf{3} , \mathbf{1^{\prime}},\mathbf{1^{\prime\prime}},\mathbf{1^{\prime\prime}},\mathbf{1},\mathbf{1})$ & (1,3) &3& (1,1,3) \\ \hline
 $\mathcal{VII}_{4}$  & 23 & $(\mathbf{3} , \mathbf{1},\mathbf{1},\mathbf{1^{\prime\prime}},\mathbf{1^{\prime\prime}},\mathbf{1^{\prime}})$ & (0,2) &2& (0,2,4) \\ \hline
 $\mathcal{VII}_{5}$  & 23 & $(\mathbf{3} , \mathbf{1},\mathbf{1^{\prime}},\mathbf{1^{\prime\prime}},\mathbf{1^{\prime\prime}},\mathbf{1^{\prime}})$ & (0,2) &2& (0,2,4) \\ \hline
 $\mathcal{VII}_{6}$  & 23 & $(\mathbf{3} , \mathbf{1},\mathbf{1^{\prime}},\mathbf{1^{\prime\prime}},\mathbf{1^{\prime\prime}},\mathbf{1^{\prime}})$ & (2,2) &2& (0,2,4) \\ \hline
\hline
\end{tabular}
\caption{\label{tab:models-summary} Summary of phenomenologically viable $SU(5)$ GUT models based on $A_4$ modular symmetry with less free parameters. Here the real and imaginary part of $\tau$ are taken as free parameters. }
\end{table}

\subsection{Type-I}
\begin{itemize}[leftmargin=1.0em]
\item{~Model ${\cal I}_{1}$~:~$(\rho_N,\rho_{F},\rho_{T_1},\rho_{T_2},\rho_{T_3})=(\mathbf{3},\mathbf{3},\mathbf{1^{\prime\prime}},\mathbf{1},\mathbf{1})\,,~ (k_{N},k_{F},k_{T_1},k_{T_2},k_{T_3})=(1,1,1,1,3)$}

The modular invariant superpotentials for quark and lepton masses are of the following form,
\begin{eqnarray}
\nonumber {\cal W}_\nu &=&
   \alpha_{N_1} \Lambda (NN)_{\mathbf{3}_S}Y_{\mathbf{3}}^{(2)}
 + \alpha_{\nu_1} (NF)_{\mathbf{3}_S}Y_{\mathbf{3}}^{(2)} H_{5}
 + \alpha_{\nu_2} (NF)_{\mathbf{3}_A}Y_{\mathbf{3}}^{(2)} H_{5}
\,,\\
\nonumber {\cal W}_d &=&
   \alpha_{d_1} (FT_1)_{\mathbf{3}}Y_{\mathbf{3}}^{(2)} H_{\overline{5}}
 + \alpha_{d_2} (FT_2)_{\mathbf{3}}Y_{\mathbf{3}}^{(2)} H_{\overline{5}}
 + \alpha_{d_3} (FT_3)_{\mathbf{3}}Y_{\mathbf{3}}^{(4)} H_{\overline{5}}
\\\nonumber &&
 + \alpha^\prime_{d_1} (FT_1)_{\mathbf{3}}Y_{\mathbf{3}}^{(2)} H_{\overline{45}}
 + \alpha^\prime_{d_2} (FT_2)_{\mathbf{3}}Y_{\mathbf{3}}^{(2)} H_{\overline{45}}
 + \alpha^\prime_{d_3} (FT_3)_{\mathbf{3}}Y_{\mathbf{3}}^{(4)} H_{\overline{45}}
\,,\\
{\cal W}_u &=&
   \alpha_{u_1}  (T_1T_3)_{\mathbf{1^{\prime\prime}}}Y_{\mathbf{1^\prime}}^{(4)} H_{5}
 + \alpha_{u_2}  (T_2T_3)_{\mathbf{1}}Y_{\mathbf{1}}^{(4)} H_{5}
 + \alpha_{u_3}  (T_3T_3)_{\mathbf{1}}Y_{\mathbf{1}}^{(6)} H_{5}
\,.
\end{eqnarray}
The right-handed neutrino mass matrix and the Yukawa matrix are given by
\begin{eqnarray}
m_M^{\nu} &=& \Lambda  \left(
\begin{array}{ccc}
 2 \alpha_{N_1}Y_{\mathbf{3},1}^{(2)} ~&~ -\alpha_{N_1}Y_{\mathbf{3},3}^{(2)} ~&~ -\alpha_{N_1}Y_{\mathbf{3},2}^{(2)} \\
 -\alpha_{N_1}Y_{\mathbf{3},3}^{(2)} ~&~ 2 \alpha_{N_1}Y_{\mathbf{3},2}^{(2)} ~&~ -\alpha_{N_1}Y_{\mathbf{3},1}^{(2)} \\
 -\alpha_{N_1}Y_{\mathbf{3},2}^{(2)} ~&~ -\alpha_{N_1}Y_{\mathbf{3},1}^{(2)} ~&~ 2 \alpha_{N_1}Y_{\mathbf{3},3}^{(2)} \\
\end{array}
\right)\,,\nonumber\\
{\cal Y}_5^{\nu} &=& \left(
\begin{array}{ccc}
 2 \alpha_{\nu_1}Y_{\mathbf{3},1}^{(2)} ~&~ -\left(\alpha_{\nu_1}-\alpha_{\nu_2}\right)Y_{\mathbf{3},3}^{(2)} ~&~ -\left(\alpha_{\nu_1}+\alpha_{\nu_2}\right)Y_{\mathbf{3},2}^{(2)} \\
 -\left(\alpha_{\nu_1}+\alpha_{\nu_2}\right)Y_{\mathbf{3},3}^{(2)} ~&~ 2 \alpha_{\nu_1}Y_{\mathbf{3},2}^{(2)} ~&~ -\left(\alpha_{\nu_1}-\alpha_{\nu_2}\right)Y_{\mathbf{3},1}^{(2)} \\
 -\left(\alpha_{\nu_1}-\alpha_{\nu_2}\right)Y_{\mathbf{3},2}^{(2)} ~&~ -\left(\alpha_{\nu_1}+\alpha_{\nu_2}\right)Y_{\mathbf{3},1}^{(2)} ~&~ 2 \alpha_{\nu_1}Y_{\mathbf{3},3}^{(2)} \\
\end{array}
\right)\,,\nonumber\\
{\cal Y}_{\overline{5}}^d &=& \left(
\begin{array}{ccc}
 \alpha_{d_1}Y_{\mathbf{3},2}^{(2)} ~&~ \alpha_{d_2}Y_{\mathbf{3},1}^{(2)} ~&~ \alpha_{d_3}Y_{\mathbf{3},1}^{(4)} \\
 \alpha_{d_1}Y_{\mathbf{3},1}^{(2)}  ~&~ \alpha_{d_2}Y_{\mathbf{3},3}^{(2)} ~&~ \alpha_{d_3}Y_{\mathbf{3},3}^{(4)} \\
 \alpha_{d_1}Y_{\mathbf{3},3}^{(2)} ~&~ \alpha_{d_2}Y_{\mathbf{3},2}^{(2)} ~&~ \alpha_{d_3}Y_{\mathbf{3},2}^{(4)} \\
\end{array}
\right)\,,\nonumber\\
{\cal Y}_5^u &=& \left(
\begin{array}{ccc}
 0 ~&~ 0 ~&~ \alpha_{u_1}Y_{\mathbf{1}^\prime}^{(4)} \\
 0 ~&~ 0 ~&~ \alpha_{u_2}Y_\mathbf{1}^{(4)} \\
 \alpha_{u_1}Y_{\mathbf{1}^\prime}^{(4)} ~&~ \alpha_{u_2}Y_\mathbf{1}^{(4)} ~&~ \alpha_{u_3}Y_\mathbf{1}^{(6)} \\
\end{array}
\right)\,.
\end{eqnarray}
Since both $H_{\overline{5}}$ and $H_{\overline{45}}$ are assumed to be invariant under $A_4$ and their modular weights are vanishing, ${\cal Y}_{\overline{5}}^d$ and ${\cal Y}_{\overline{45}}^d$ are of the same form except that the couplings $\alpha_{d_i}$ are replaced by $\alpha'_{d_i}$. Hence we shall not give the explicit form of the Yukawa coupling ${\cal Y}_{\overline{45}}^d$ here and below.
The parameters $\alpha_{N_1}$, $\alpha_{\nu_1}$, $\alpha_{d_1}$, $\alpha_{d_2}$, $\alpha_{d_3}$, $\alpha_{d_1}^\prime$ and $\alpha_{u_3}$ can be taken real by field redefinition, and $\alpha_{\nu_2}$, $\alpha_{d_2}^\prime$, $\alpha_{d_3}^\prime$, $\alpha_{u_1}$ and $\alpha_{u_2}$ are generically complex.
\item{~Model ${\cal I}_{2}$~:~$(\rho_N,\rho_{F},\rho_{T_1},\rho_{T_2},\rho_{T_3})=(\mathbf{3},\mathbf{3},\mathbf{1^{\prime\prime}},\mathbf{1^{\prime}},\mathbf{1^{\prime}})\,,~ (k_{N},k_{F},k_{T_1},k_{T_2},k_{T_3})=(0,2,2,0,4)$}

We can read out the superpotential for quark and leptons as follows,
\begin{eqnarray}
\nonumber {\cal W}_\nu &=&
   \alpha_{N_1} \Lambda (NN)_{\mathbf{1}}
 + \alpha_{\nu_1} (NF)_{\mathbf{3}_S}Y_{\mathbf{3}}^{(2)} H_{5}
 + \alpha_{\nu_2} (NF)_{\mathbf{3}_A}Y_{\mathbf{3}}^{(2)} H_{5}
\,,\\
\nonumber {\cal W}_d &=&
   \alpha_{d_1} (FT_1)_{\mathbf{3}}Y_{\mathbf{3}}^{(4)} H_{\overline{5}}
 + \alpha_{d_2} (FT_2)_{\mathbf{3}}Y_{\mathbf{3}}^{(2)} H_{\overline{5}}
 + \alpha_{d_3} (FT_3)_{\mathbf{3}}Y_{\mathbf{3},I}^{(6)} H_{\overline{5}}
 + \alpha_{d_4} (FT_3)_{\mathbf{3}}Y_{\mathbf{3},II}^{(6)} H_{\overline{5}}
\\\nonumber &&
 + \alpha^\prime_{d_1} (FT_1)_{\mathbf{3}}Y_{\mathbf{3}}^{(4)} H_{\overline{45}}
 + \alpha^\prime_{d_2} (FT_2)_{\mathbf{3}}Y_{\mathbf{3}}^{(2)} H_{\overline{45}}
 + \alpha^\prime_{d_3} (FT_3)_{\mathbf{3}}Y_{\mathbf{3},I}^{(6)} H_{\overline{45}}
 + \alpha^\prime_{d_4} (FT_3)_{\mathbf{3}}Y_{\mathbf{3},II}^{(6)} H_{\overline{45}}
\,,\\
{\cal W}_u &=&
   \alpha_{u_1}  (T_1T_3)_{\mathbf{1}}Y_{\mathbf{1}}^{(6)} H_{5}
 + \alpha_{u_2}  (T_2T_3)_{\mathbf{1^{\prime\prime}}}Y_{\mathbf{1^\prime}}^{(4)} H_{5}
 + \alpha_{u_3}  (T_3T_3)_{\mathbf{1^{\prime\prime}}}Y_{\mathbf{1^\prime}}^{(8)} H_{5}
\,.
\end{eqnarray}
The right-handed neutrino mass matrix and the Yukawa matrix are given by
\begin{eqnarray}
m_M^{\nu} &=& \Lambda  \left(
\begin{array}{ccc}
 \alpha_{N_1} ~&~ 0 ~&~ 0 \\
 0 ~&~ 0 ~&~ \alpha_{N_1} \\
 0 ~&~ \alpha_{N_1} ~&~ 0 \\
\end{array}
\right)\,,\nonumber\\
{\cal Y}_{\overline{5}}^d &=& \left(
\begin{array}{ccc}
 \alpha_{d_1}Y_{\mathbf{3},2}^{(4)} ~&~ \alpha_{d_2}Y_{\mathbf{3},3}^{(2)} ~&~ \alpha_{d_3}Y_{\mathbf{3}I,3}^{(6)}+\alpha_{d_4}Y_{\mathbf{3}II,3}^{(6)} \\
 \alpha_{d_1}Y_{\mathbf{3},1}^{(4)} ~&~ \alpha_{d_2}Y_{\mathbf{3},2}^{(2)} ~&~ \alpha_{d_3}Y_{\mathbf{3}I,2}^{(6)}+\alpha_{d_4}Y_{\mathbf{3}II,2}^{(6)} \\
 \alpha_{d_1}Y_{\mathbf{3},3}^{(4)} ~&~ \alpha_{d_2}Y_{\mathbf{3},1}^{(2)} ~&~ \alpha_{d_3}Y_{\mathbf{3}I,1}^{(6)}+\alpha_{d_4}Y_{\mathbf{3}II,1}^{(6)} \\
\end{array}
\right)\,,\nonumber\\
{\cal Y}_5^u &=& \left(
\begin{array}{ccc}
 0 ~&~ 0 ~&~ \alpha_{u_1}Y_\mathbf{1}^{(6)} \\
 0 ~&~ 0 ~&~ \alpha_{u_2}Y_{\mathbf{1}^\prime}^{(4)} \\
 \alpha_{u_1}Y_\mathbf{1}^{(6)} ~&~ \alpha_{u_2}Y_{\mathbf{1}^\prime}^{(4)} ~&~ \alpha_{u_3}Y_{\mathbf{1}^\prime}^{(8)} \\
\end{array}
\right)\,.
\end{eqnarray}
The ${\cal Y}_5^{\nu}$ matrix is in common with that of the model ${\cal I}_{1}$.
The parameters $\alpha_{N_1}$, $\alpha_{\nu_1}$, $\alpha_{d_1}$, $\alpha_{d_2}$, $\alpha_{d_3}$, $\alpha_{d_1}^\prime$ and $\alpha_{u_3}$ can be taken real by field redefinition, and $\alpha_{\nu_2}$, $\alpha_{d_4}$, $\alpha_{d_2}^\prime$, $\alpha_{d_3}^\prime$, $\alpha_{d_4}^\prime$, $\alpha_{u_1}$ and $\alpha_{u_2}$ are generically complex.
\item{~Model ${\cal I}_{3}$~:~$(\rho_N,\rho_{F},\rho_{T_1},\rho_{T_2},\rho_{T_3})=(\mathbf{3},\mathbf{3},\mathbf{1^{\prime\prime}},\mathbf{1^{\prime\prime}},\mathbf{1^{\prime}})\,,~ (k_{N},k_{F},k_{T_1},k_{T_2},k_{T_3})=(0,2,0,2,4)$}

The modular invariant superpotential is of the following form:
\begin{eqnarray}
\nonumber {\cal W}_\nu &=&
   \alpha_{N_1} \Lambda (NN)_{\mathbf{1}}
 + \alpha_{\nu_1} (NF)_{\mathbf{3}_S}Y_{\mathbf{3}}^{(2)} H_{5}
 + \alpha_{\nu_2} (NF)_{\mathbf{3}_A}Y_{\mathbf{3}}^{(2)} H_{5}
\,,\\
\nonumber {\cal W}_d &=&
   \alpha_{d_1} (FT_1)_{\mathbf{3}}Y_{\mathbf{3}}^{(2)} H_{\overline{5}}
 + \alpha_{d_2} (FT_2)_{\mathbf{3}}Y_{\mathbf{3}}^{(4)} H_{\overline{5}}
 + \alpha_{d_3} (FT_3)_{\mathbf{3}}Y_{\mathbf{3},I}^{(6)} H_{\overline{5}}
 + \alpha_{d_4} (FT_3)_{\mathbf{3}}Y_{\mathbf{3},II}^{(6)} H_{\overline{5}}
\\\nonumber &&
 + \alpha^\prime_{d_1} (FT_1)_{\mathbf{3}}Y_{\mathbf{3}}^{(2)} H_{\overline{45}}
 + \alpha^\prime_{d_2} (FT_2)_{\mathbf{3}}Y_{\mathbf{3}}^{(4)} H_{\overline{45}}
 + \alpha^\prime_{d_3} (FT_3)_{\mathbf{3}}Y_{\mathbf{3},I}^{(6)} H_{\overline{45}}
 + \alpha^\prime_{d_4} (FT_3)_{\mathbf{3}}Y_{\mathbf{3},II}^{(6)} H_{\overline{45}}
\,,\\
{\cal W}_u &=&
   \alpha_{u_1}  (T_1T_3)_{\mathbf{1}}Y_{\mathbf{1}}^{(4)} H_{5}
 + \alpha_{u_2}  (T_2T_3)_{\mathbf{1}}Y_{\mathbf{1}}^{(6)} H_{5}
 + \alpha_{u_3}  (T_3T_3)_{\mathbf{1^{\prime\prime}}}Y_{\mathbf{1^\prime}}^{(8)} H_{5}
\,,
\end{eqnarray}
which leads to
\begin{eqnarray}
{\cal Y}_{\overline{5}}^d &=& \left(
\begin{array}{ccc}
 \alpha_{d_1}Y_{\mathbf{3},2}^{(2)} ~&~ \alpha_{d_2}Y_{\mathbf{3},2}^{(4)} ~&~ \alpha_{d_3}Y_{\mathbf{3}I,3}^{(6)}+\alpha_{d_4}Y_{\mathbf{3}II,3}^{(6)} \\
 \alpha_{d_1}Y_{\mathbf{3},1}^{(2)}  ~&~ \alpha_{d_2}Y_{\mathbf{3},1}^{(4)} ~&~ \alpha_{d_3}Y_{\mathbf{3}I,2}^{(6)}+\alpha_{d_4}Y_{\mathbf{3}II,2}^{(6)} \\
 \alpha_{d_1}Y_{\mathbf{3},3}^{(2)} ~&~ \alpha_{d_2}Y_{\mathbf{3},3}^{(4)} ~&~ \alpha_{d_3}Y_{\mathbf{3}I,1}^{(6)}+\alpha_{d_4}Y_{\mathbf{3}II,1}^{(6)} \\
\end{array}
\right)\,,\nonumber\\
{\cal Y}_5^u &=& \left(
\begin{array}{ccc}
 0 ~&~ 0 ~&~ \alpha_{u_1}Y_\mathbf{1}^{(4)} \\
 0 ~&~ 0 ~&~ \alpha_{u_2}Y_\mathbf{1}^{(6)} \\
 \alpha_{u_1}Y_\mathbf{1}^{(4)} ~&~ \alpha_{u_2}Y_\mathbf{1}^{(6)} ~&~ \alpha_{u_3}Y_{\mathbf{1}^\prime}^{(8)} \\
\end{array}
\right)\,.
\end{eqnarray}
The right-handed neutrino mass matrix $M_N$ matrix is the same as that of the model $\mathcal{I}_{2}$, and ${\cal Y}_5^{\nu}$ is the same as that of the model $\mathcal{I}_{1}$.
The phases of the parameters $\alpha_{N_1}$, $\alpha_{\nu_1}$, $\alpha_{d_1}$, $\alpha_{d_2}$, $\alpha_{d_3}$, $\alpha_{d_1}^\prime$ and $\alpha_{u_3}$ can be removed by field redefinition, and the remaining parameters $\alpha_{\nu_2}$, $\alpha_{d_4}$, $\alpha_{d_2}^\prime$, $\alpha_{d_3}^\prime$, $\alpha_{d_4}^\prime$, $\alpha_{u_1}$ and $\alpha_{u_2}$ are complex.
\item{~Model ${\cal I}_{4}$~:~$(\rho_N,\rho_{F},\rho_{T_1},\rho_{T_2},\rho_{T_3})=(\mathbf{3},\mathbf{3},\mathbf{1^{\prime\prime}},\mathbf{1},\mathbf{1})\,,~ (k_{N},k_{F},k_{T_1},k_{T_2},k_{T_3})=(0,2,0,0,4)$}

The quark and lepton masses are described by the following superpotential:
\begin{eqnarray}
\nonumber {\cal W}_\nu &=&
   \alpha_{N_1} \Lambda (NN)_{\mathbf{1}}
 + \alpha_{\nu_1} (NF)_{\mathbf{3}_S}Y_{\mathbf{3}}^{(2)} H_{5}
 + \alpha_{\nu_2} (NF)_{\mathbf{3}_A}Y_{\mathbf{3}}^{(2)} H_{5}
\,,\\
\nonumber {\cal W}_d &=&
   \alpha_{d_1} (FT_1)_{\mathbf{3}}Y_{\mathbf{3}}^{(2)} H_{\overline{5}}
 + \alpha_{d_2} (FT_2)_{\mathbf{3}}Y_{\mathbf{3}}^{(2)} H_{\overline{5}}
 + \alpha_{d_3} (FT_3)_{\mathbf{3}}Y_{\mathbf{3},I}^{(6)} H_{\overline{5}}
 + \alpha_{d_4} (FT_3)_{\mathbf{3}}Y_{\mathbf{3},II}^{(6)} H_{\overline{5}}
\\\nonumber &&
 + \alpha^\prime_{d_1} (FT_1)_{\mathbf{3}}Y_{\mathbf{3}}^{(2)} H_{\overline{45}}
 + \alpha^\prime_{d_2} (FT_2)_{\mathbf{3}}Y_{\mathbf{3}}^{(2)} H_{\overline{45}}
 + \alpha^\prime_{d_3} (FT_3)_{\mathbf{3}}Y_{\mathbf{3},I}^{(6)} H_{\overline{45}}
 + \alpha^\prime_{d_4} (FT_3)_{\mathbf{3}}Y_{\mathbf{3},II}^{(6)} H_{\overline{45}}
\,,\\
{\cal W}_u &=&
   \alpha_{u_1}  (T_1T_3)_{\mathbf{1^{\prime\prime}}}Y_{\mathbf{1^\prime}}^{(4)} H_{5}
 + \alpha_{u_2}  (T_2T_2)_{\mathbf{1}}   H_{5}
 + \alpha_{u_3}  (T_2T_3)_{\mathbf{1}}Y_{\mathbf{1}}^{(4)} H_{5}
 + \alpha_{u_4}  (T_3T_3)_{\mathbf{1}}Y_{\mathbf{1}}^{(8)} H_{5}
\,.
\end{eqnarray}
The right-handed neutrino mass matrix and the Yukawa matrix are given by
\begin{eqnarray}
{\cal Y}_{\overline{5}}^d &=& \left(
\begin{array}{ccc}
 \alpha_{d_1}Y_{\mathbf{3},2}^{(2)} ~&~ \alpha_{d_2}Y_{\mathbf{3},1}^{(2)} ~&~ \alpha_{d_3}Y_{\mathbf{3}I,1}^{(6)}+\alpha_{d_4}Y_{\mathbf{3}II,1}^{(6)} \\
 \alpha_{d_1}Y_{\mathbf{3},1}^{(2)}  ~&~ \alpha_{d_2}Y_{\mathbf{3},3}^{(2)} ~&~ \alpha_{d_3}Y_{\mathbf{3}I,3}^{(6)}+\alpha_{d_4}Y_{\mathbf{3}II,3}^{(6)} \\
 \alpha_{d_1}Y_{\mathbf{3},3}^{(2)} ~&~ \alpha_{d_2}Y_{\mathbf{3},2}^{(2)} ~&~ \alpha_{d_3}Y_{\mathbf{3}I,2}^{(6)}+\alpha_{d_4}Y_{\mathbf{3}II,2}^{(6)} \\
\end{array}
\right)\,,\nonumber\\
{\cal Y}_5^u &=& \left(
\begin{array}{ccc}
 0 ~&~ 0 ~&~ \alpha_{u_1}Y_{\mathbf{1}^\prime}^{(4)} \\
 0 ~&~ \alpha_{u_2} ~&~ \alpha_{u_3}Y_\mathbf{1}^{(4)} \\
 \alpha_{u_1}Y_{\mathbf{1}^\prime}^{(4)} ~&~ \alpha_{u_3}Y_\mathbf{1}^{(4)} ~&~ \alpha_{u_4}Y_\mathbf{1}^{(8)} \\
\end{array}
\right)\,.
\end{eqnarray}
The matrices $M_N$ and ${\cal Y}_5^{\nu}$ coincide with those of the model ${\cal I}_{3}$ as well.
The phases of the parameters $\alpha_{N_1}$, $\alpha_{\nu_1}$, $\alpha_{d_1}$, $\alpha_{d_2}$, $\alpha_{d_3}$, $\alpha_{d_1}^\prime$ and $\alpha_{u_3}$ can be removed by field redefinition, and the other parameters are complex.
\item{~Model ${\cal I}_{5}$~:~$(\rho_N,\rho_{F},\rho_{T_1},\rho_{T_2},\rho_{T_3})=(\mathbf{3},\mathbf{3},\mathbf{1^{\prime\prime}},\mathbf{1},\mathbf{1})\,,~ (k_{N},k_{F},k_{T_1},k_{T_2},k_{T_3})=(0,2,0,2,4)$}

The modular invariant superpotentials for quark and lepton masses are of the following form,
\begin{eqnarray}
\nonumber {\cal W}_\nu &=&
   \alpha_{N_1} \Lambda (NN)_{\mathbf{1}}
 + \alpha_{\nu_1} (NF)_{\mathbf{3}_S}Y_{\mathbf{3}}^{(2)} H_{5}
 + \alpha_{\nu_2} (NF)_{\mathbf{3}_A}Y_{\mathbf{3}}^{(2)} H_{5}
\,,\\
\nonumber {\cal W}_d &=&
   \alpha_{d_1} (FT_1)_{\mathbf{3}}Y_{\mathbf{3}}^{(2)} H_{\overline{5}}
 + \alpha_{d_2} (FT_2)_{\mathbf{3}}Y_{\mathbf{3}}^{(4)} H_{\overline{5}}
 + \alpha_{d_3} (FT_3)_{\mathbf{3}}Y_{\mathbf{3},I}^{(6)} H_{\overline{5}}
 + \alpha_{d_4} (FT_3)_{\mathbf{3}}Y_{\mathbf{3},II}^{(6)} H_{\overline{5}}
\\\nonumber &&
 + \alpha^\prime_{d_1} (FT_1)_{\mathbf{3}}Y_{\mathbf{3}}^{(2)} H_{\overline{45}}
 + \alpha^\prime_{d_2} (FT_2)_{\mathbf{3}}Y_{\mathbf{3}}^{(4)} H_{\overline{45}}
 + \alpha^\prime_{d_3} (FT_3)_{\mathbf{3}}Y_{\mathbf{3},I}^{(6)} H_{\overline{45}}
 + \alpha^\prime_{d_4} (FT_3)_{\mathbf{3}}Y_{\mathbf{3},II}^{(6)} H_{\overline{45}}
\,,\\
{\cal W}_u &=&
   \alpha_{u_1}  (T_1T_3)_{\mathbf{1^{\prime\prime}}}Y_{\mathbf{1^\prime}}^{(4)} H_{5}
 + \alpha_{u_2}  (T_2T_2)_{\mathbf{1}}Y_{\mathbf{1}}^{(4)} H_{5}
 + \alpha_{u_3}  (T_2T_3)_{\mathbf{1}}Y_{\mathbf{1}}^{(6)} H_{5}
 + \alpha_{u_4}  (T_3T_3)_{\mathbf{1}}Y_{\mathbf{1}}^{(8)} H_{5}
\,.
\end{eqnarray}
The right-handed neutrino mass matrix and the Yukawa coupling matrices are predicted to be
\begin{eqnarray}
{\cal Y}_{\overline{5}}^d &=& \left(
\begin{array}{ccc}
 \alpha_{d_1}Y_{\mathbf{3},2}^{(2)} ~&~ \alpha_{d_2}Y_{\mathbf{3},1}^{(4)} ~&~ \alpha_{d_3}Y_{\mathbf{3}I,1}^{(6)}+\alpha_{d_4}Y_{\mathbf{3}II,1}^{(6)} \\
 \alpha_{d_1}Y_{\mathbf{3},1}^{(2)}  ~&~ \alpha_{d_2}Y_{\mathbf{3},3}^{(4)} ~&~ \alpha_{d_3}Y_{\mathbf{3}I,3}^{(6)}+\alpha_{d_4}Y_{\mathbf{3}II,3}^{(6)} \\
 \alpha_{d_1}Y_{\mathbf{3},3}^{(2)} ~&~ \alpha_{d_2}Y_{\mathbf{3},2}^{(4)} ~&~ \alpha_{d_3}Y_{\mathbf{3}I,2}^{(6)}+\alpha_{d_4}Y_{\mathbf{3}II,2}^{(6)} \\
\end{array}
\right)\,,\nonumber\\
{\cal Y}_5^u &=& \left(
\begin{array}{ccc}
 0 ~&~ 0 ~&~ \alpha_{u_1}Y_{\mathbf{1}^\prime}^{(4)} \\
 0 ~&~ \alpha_{u_2}Y_\mathbf{1}^{(4)} ~&~ \alpha_{u_3}Y_\mathbf{1}^{(6)} \\
 \alpha_{u_1}Y_{\mathbf{1}^\prime}^{(4)} ~&~ \alpha_{u_3}Y_\mathbf{1}^{(6)} ~&~ \alpha_{u_4}Y_\mathbf{1}^{(8)} \\
\end{array}
\right)\,.
\end{eqnarray}
The matrices $M_N$ and ${\cal Y}_5^{\nu}$ are the same as the corresponding ones of the model ${\cal I}_{3}$ as well.
The parameters $\alpha_{N_1}$, $\alpha_{\nu_1}$, $\alpha_{d_1}$, $\alpha_{d_2}$, $\alpha_{d_3}$, $\alpha_{d_1}^\prime$ and $\alpha_{u_3}$ can be taken real by field redefinition, and the other parameters are complex.
\item{~Model ${\cal I}_{6}$~:~$(\rho_N,\rho_{F},\rho_{T_1},\rho_{T_2},\rho_{T_3})=(\mathbf{3},\mathbf{3},\mathbf{1^{\prime\prime}},\mathbf{1},\mathbf{1})\,,~ (k_{N},k_{F},k_{T_1},k_{T_2},k_{T_3})=(0,2,2,2,4)$}

The quark and lepton masses are described by the following superpotential:
\begin{eqnarray}
\nonumber {\cal W}_\nu &=&
   \alpha_{N_1} \Lambda (NN)_{\mathbf{1}}
 + \alpha_{\nu_1} (NF)_{\mathbf{3}_S}Y_{\mathbf{3}}^{(2)} H_{5}
 + \alpha_{\nu_2} (NF)_{\mathbf{3}_A}Y_{\mathbf{3}}^{(2)} H_{5}
\,,\\
\nonumber {\cal W}_d &=&
   \alpha_{d_1} (FT_1)_{\mathbf{3}}Y_{\mathbf{3}}^{(4)} H_{\overline{5}}
 + \alpha_{d_2} (FT_2)_{\mathbf{3}}Y_{\mathbf{3}}^{(4)} H_{\overline{5}}
 + \alpha_{d_3} (FT_3)_{\mathbf{3}}Y_{\mathbf{3},I}^{(6)} H_{\overline{5}}
 + \alpha_{d_4} (FT_3)_{\mathbf{3}}Y_{\mathbf{3},II}^{(6)} H_{\overline{5}}
\\\nonumber &&
 + \alpha^\prime_{d_1} (FT_1)_{\mathbf{3}}Y_{\mathbf{3}}^{(4)} H_{\overline{45}}
 + \alpha^\prime_{d_2} (FT_2)_{\mathbf{3}}Y_{\mathbf{3}}^{(4)} H_{\overline{45}}
 + \alpha^\prime_{d_3} (FT_3)_{\mathbf{3}}Y_{\mathbf{3},I}^{(6)} H_{\overline{45}}
 + \alpha^\prime_{d_4} (FT_3)_{\mathbf{3}}Y_{\mathbf{3},II}^{(6)} H_{\overline{45}}
\,,\\
{\cal W}_u &=&
   \alpha_{u_1}  (T_1T_2)_{\mathbf{1^{\prime\prime}}}Y_{\mathbf{1^\prime}}^{(4)} H_{5}
 + \alpha_{u_2}  (T_2T_2)_{\mathbf{1}}Y_{\mathbf{1}}^{(4)} H_{5}
 + \alpha_{u_3}  (T_2T_3)_{\mathbf{1}}Y_{\mathbf{1}}^{(6)} H_{5}
 + \alpha_{u_4}  (T_3T_3)_{\mathbf{1}}Y_{\mathbf{1}}^{(8)} H_{5}
\,,
\end{eqnarray}
which gives rise to
\begin{eqnarray}
{\cal Y}_{\overline{5}}^d &=& \left(
\begin{array}{ccc}
 \alpha_{d_1}Y_{\mathbf{3},2}^{(4)} ~&~ \alpha_{d_2}Y_{\mathbf{3},1}^{(4)} ~&~ \alpha_{d_3}Y_{\mathbf{3}I,1}^{(6)}+\alpha_{d_4}Y_{\mathbf{3}II,1}^{(6)} \\
 \alpha_{d_1}Y_{\mathbf{3},1}^{(4)} ~&~ \alpha_{d_2}Y_{\mathbf{3},3}^{(4)} ~&~ \alpha_{d_3}Y_{\mathbf{3}I,3}^{(6)}+\alpha_{d_4}Y_{\mathbf{3}II,3}^{(6)} \\
 \alpha_{d_1}Y_{\mathbf{3},3}^{(4)} ~&~ \alpha_{d_2}Y_{\mathbf{3},2}^{(4)} ~&~ \alpha_{d_3}Y_{\mathbf{3}I,2}^{(6)}+\alpha_{d_4}Y_{\mathbf{3}II,2}^{(6)} \\
\end{array}
\right)\,,\nonumber\\
{\cal Y}_5^u &=& \left(
\begin{array}{ccc}
 0 ~&~ \alpha_{u_1}Y_{\mathbf{1}^\prime}^{(4)} ~&~ 0 \\
 \alpha_{u_1}Y_{\mathbf{1}^\prime}^{(4)} ~&~ \alpha_{u_2}Y_\mathbf{1}^{(4)} ~&~ \alpha_{u_3}Y_\mathbf{1}^{(6)} \\
 0 ~&~ \alpha_{u_3}Y_\mathbf{1}^{(6)} ~&~ \alpha_{u_4}Y_\mathbf{1}^{(8)} \\
\end{array}
\right)\,.
\end{eqnarray}
The matrices $M_N$ and ${\cal Y}_5^{\nu}$ are in common with those of the model ${\cal I}_{3}$ as well.
The parameters $\alpha_{N_1}$, $\alpha_{\nu_1}$, $\alpha_{d_1}$, $\alpha_{d_2}$, $\alpha_{d_3}$, $\alpha_{d_1}^\prime$ and $\alpha_{u_3}$ can be taken real by field redefinition, and $\alpha_{\nu_2}$, $\alpha_{d_4}$, $\alpha_{d_2}^\prime$, $\alpha_{d_3}^\prime$, $\alpha_{d_4}^\prime$, $\alpha_{u_1}$, $\alpha_{u_2}$ and $\alpha_{u_4}$ are complex.
\item{~Model ${\cal I}_{7}$~:~$(\rho_N,\rho_{F},\rho_{T_1},\rho_{T_2},\rho_{T_3})=(\mathbf{3},\mathbf{3},\mathbf{1^{\prime}},\mathbf{1},\mathbf{1^{\prime\prime}})\,,~ (k_{N},k_{F},k_{T_1},k_{T_2},k_{T_3})=(0,2,0,0,4)$}

The modular invariant superpotential is of the following form:
\begin{eqnarray}
\nonumber {\cal W}_\nu &=&
   \alpha_{N_1} \Lambda (NN)_{\mathbf{1}}
 + \alpha_{\nu_1} (NF)_{\mathbf{3}_S}Y_{\mathbf{3}}^{(2)} H_{5}
 + \alpha_{\nu_2} (NF)_{\mathbf{3}_A}Y_{\mathbf{3}}^{(2)} H_{5}
\,,\\
\nonumber {\cal W}_d &=&
   \alpha_{d_1} (FT_1)_{\mathbf{3}}Y_{\mathbf{3}}^{(2)} H_{\overline{5}}
 + \alpha_{d_2} (FT_2)_{\mathbf{3}}Y_{\mathbf{3}}^{(2)} H_{\overline{5}}
 + \alpha_{d_3} (FT_3)_{\mathbf{3}}Y_{\mathbf{3},I}^{(6)} H_{\overline{5}}
 + \alpha_{d_4} (FT_3)_{\mathbf{3}}Y_{\mathbf{3},II}^{(6)} H_{\overline{5}}
\\\nonumber &&
 + \alpha^\prime_{d_1} (FT_1)_{\mathbf{3}}Y_{\mathbf{3}}^{(2)} H_{\overline{45}}
 + \alpha^\prime_{d_2} (FT_2)_{\mathbf{3}}Y_{\mathbf{3}}^{(2)} H_{\overline{45}}
 + \alpha^\prime_{d_3} (FT_3)_{\mathbf{3}}Y_{\mathbf{3},I}^{(6)} H_{\overline{45}}
 + \alpha^\prime_{d_4} (FT_3)_{\mathbf{3}}Y_{\mathbf{3},II}^{(6)} H_{\overline{45}}
\,,\\
{\cal W}_u &=&
   \alpha_{u_1}  (T_1T_3)_{\mathbf{1}}Y_{\mathbf{1}}^{(4)} H_{5}
 + \alpha_{u_2}  (T_2T_2)_{\mathbf{1}}   H_{5}
 + \alpha_{u_3}  (T_2T_3)_{\mathbf{1^{\prime\prime}}}Y_{\mathbf{1^\prime}}^{(4)} H_{5}
 + \alpha_{u_4}  (T_3T_3)_{\mathbf{1^{\prime}}}Y_{\mathbf{1^{\prime\prime}}}^{(8)} H_{5}
\,.
\end{eqnarray}
The right-handed neutrino mass matrix and the Yukawa matrix are given by
\begin{eqnarray}
{\cal Y}_{\overline{5}}^d &=& \left(
\begin{array}{ccc}
 \alpha_{d_1}Y_{\mathbf{3},3}^{(2)} ~&~ \alpha_{d_2}Y_{\mathbf{3},1}^{(2)} ~&~ \alpha_{d_3}Y_{\mathbf{3}I,2}^{(6)}+\alpha_{d_4}Y_{\mathbf{3}II,2}^{(6)} \\
 \alpha_{d_1}Y_{\mathbf{3},2}^{(2)} ~&~ \alpha_{d_2}Y_{\mathbf{3},3}^{(2)} ~&~ \alpha_{d_3}Y_{\mathbf{3}I,1}^{(6)}+\alpha_{d_4}Y_{\mathbf{3}II,1}^{(6)} \\
 \alpha_{d_1}Y_{\mathbf{3},1}^{(2)}  ~&~ \alpha_{d_2}Y_{\mathbf{3},2}^{(2)} ~&~ \alpha_{d_3}Y_{\mathbf{3}I,3}^{(6)}+\alpha_{d_4}Y_{\mathbf{3}II,3}^{(6)} \\
\end{array}
\right)\,,\nonumber\\
{\cal Y}_5^u &=& \left(
\begin{array}{ccc}
 0 ~&~ 0 ~&~ \alpha_{u_1}Y_\mathbf{1}^{(4)} \\
 0 ~&~ \alpha_{u_2} ~&~ \alpha_{u_3}Y_{\mathbf{1}^\prime}^{(4)} \\
 \alpha_{u_1}Y_\mathbf{1}^{(4)} ~&~ \alpha_{u_3}Y_{\mathbf{1}^\prime}^{(4)} ~&~ \alpha_{u_4}Y_{\mathbf{1}^{\prime\prime}}^{(8)} \\
\end{array}
\right)\,.
\end{eqnarray}
The matrices $M_N$ and ${\cal Y}_5^{\nu}$ coincide with those of the model ${\cal I}_{3}$ as well.
The parameters $\alpha_{N_1}$, $\alpha_{\nu_1}$, $\alpha_{d_1}$, $\alpha_{d_2}$, $\alpha_{d_3}$, $\alpha_{d_1}^\prime$ and $\alpha_{u_3}$ can be taken real by field redefinition, and the other parameters are complex.
\item{~Model ${\cal I}_{8}$~:~$(\rho_N,\rho_{F},\rho_{T_1},\rho_{T_2},\rho_{T_3})=(\mathbf{3},\mathbf{3},\mathbf{1^{\prime\prime}},\mathbf{1},\mathbf{1^{\prime}})\,,~ (k_{N},k_{F},k_{T_1},k_{T_2},k_{T_3})=(2,0,2,2,2)$}

The modular invariant superpotential is of the following form:
\begin{eqnarray}
\nonumber {\cal W}_\nu &=&
   \alpha_{N_1} \Lambda (NN)_{\mathbf{1}}Y_{\mathbf{1}}^{(4)}
 + \alpha_{N_2} \Lambda (NN)_{\mathbf{1^{\prime\prime}}}Y_{\mathbf{1^\prime}}^{(4)}
 + \alpha_{N_3} \Lambda (NN)_{\mathbf{3}_S}Y_{\mathbf{3}}^{(4)}
 + \alpha_{\nu_1} (NF)_{\mathbf{3}_S}Y_{\mathbf{3}}^{(2)} H_{5}
\\
\nonumber&&
 + \alpha_{\nu_2} (NF)_{\mathbf{3}_A}Y_{\mathbf{3}}^{(2)} H_{5}
\,,\\
\nonumber {\cal W}_d &=&
   \alpha_{d_1} (FT_1)_{\mathbf{3}}Y_{\mathbf{3}}^{(2)} H_{\overline{5}}
 + \alpha_{d_2} (FT_2)_{\mathbf{3}}Y_{\mathbf{3}}^{(2)} H_{\overline{5}}
 + \alpha_{d_3} (FT_3)_{\mathbf{3}}Y_{\mathbf{3}}^{(2)} H_{\overline{5}}
\\\nonumber &&
 + \alpha^\prime_{d_1} (FT_1)_{\mathbf{3}}Y_{\mathbf{3}}^{(2)} H_{\overline{45}}
 + \alpha^\prime_{d_2} (FT_2)_{\mathbf{3}}Y_{\mathbf{3}}^{(2)} H_{\overline{45}}
 + \alpha^\prime_{d_3} (FT_3)_{\mathbf{3}}Y_{\mathbf{3}}^{(2)} H_{\overline{45}}
\,,\\
{\cal W}_u &=&
   \alpha_{u_1}  (T_1T_2)_{\mathbf{1^{\prime\prime}}}Y_{\mathbf{1^\prime}}^{(4)} H_{5}
 + \alpha_{u_2}  (T_1T_3)_{\mathbf{1}}Y_{\mathbf{1}}^{(4)} H_{5}
 + \alpha_{u_3}  (T_2T_2)_{\mathbf{1}}Y_{\mathbf{1}}^{(4)} H_{5}
 + \alpha_{u_4}  (T_3T_3)_{\mathbf{1^{\prime\prime}}}Y_{\mathbf{1^\prime}}^{(4)} H_{5}
\,.
\end{eqnarray}
The right-handed neutrino mass matrix and the Yukawa matrix are given by
\begin{eqnarray}
m_M^{\nu} &=& \Lambda  \left(
\begin{array}{ccc}
 \alpha_{N_1}Y_\mathbf{1}^{(4)}+2 \alpha_{N_3}Y_{\mathbf{3},1}^{(4)} ~&~ -\alpha_{N_3}Y_{\mathbf{3},3}^{(4)} ~&~ \alpha_{N_2}Y_{\mathbf{1}^\prime}^{(4)}-\alpha_{N_3}Y_{\mathbf{3},2}^{(4)} \\
 -\alpha_{N_3}Y_{\mathbf{3},3}^{(4)} ~&~ \alpha_{N_2}Y_{\mathbf{1}^\prime}^{(4)}+2 \alpha_{N_3}Y_{\mathbf{3},2}^{(4)} ~&~ \alpha_{N_1}Y_\mathbf{1}^{(4)}-\alpha_{N_3}Y_{\mathbf{3},1}^{(4)} \\
 \alpha_{N_2}Y_{\mathbf{1}^\prime}^{(4)}-\alpha_{N_3}Y_{\mathbf{3},2}^{(4)} ~&~ \alpha_{N_1}Y_\mathbf{1}^{(4)}-\alpha_{N_3}Y_{\mathbf{3},1}^{(4)} ~&~ 2 \alpha_{N_3}Y_{\mathbf{3},3}^{(4)} \\
\end{array}
\right)\,,\nonumber\\
{\cal Y}_{\overline{5}}^d &=& \left(
\begin{array}{ccc}
 \alpha_{d_1}Y_{\mathbf{3},2}^{(2)} ~&~ \alpha_{d_2}Y_{\mathbf{3},1}^{(2)} ~&~ \alpha_{d_3}Y_{\mathbf{3},3}^{(2)} \\
 \alpha_{d_1}Y_{\mathbf{3},1}^{(2)}  ~&~ \alpha_{d_2}Y_{\mathbf{3},3}^{(2)} ~&~ \alpha_{d_3}Y_{\mathbf{3},2}^{(2)} \\
 \alpha_{d_1}Y_{\mathbf{3},3}^{(2)} ~&~ \alpha_{d_2}Y_{\mathbf{3},2}^{(2)} ~&~ \alpha_{d_3}Y_{\mathbf{3},1}^{(2)}  \\
\end{array}
\right)\,,\nonumber\\
{\cal Y}_5^u &=& \left(
\begin{array}{ccc}
 0 ~&~ \alpha_{u_1}Y_{\mathbf{1}^\prime}^{(4)} ~&~ \alpha_{u_2}Y_\mathbf{1}^{(4)} \\
 \alpha_{u_1}Y_{\mathbf{1}^\prime}^{(4)} ~&~ \alpha_{u_3}Y_\mathbf{1}^{(4)} ~&~ 0 \\
 \alpha_{u_2}Y_\mathbf{1}^{(4)} ~&~ 0 ~&~ \alpha_{u_4}Y_{\mathbf{1}^\prime}^{(4)} \\
\end{array}
\right)\,.
\end{eqnarray}
The ${\cal Y}_5^{\nu}$ matrix coincide with those of the model ${\cal I}_{2}$ as well.
The phases of the parameters $\alpha_{N_1}$, $\alpha_{\nu_1}$, $\alpha_{d_1}$, $\alpha_{d_2}$, $\alpha_{d_3}$, $\alpha_{d_1}^\prime$ and $\alpha_{u_3}$ are unphysical, and the remaining parameters are complex.
\item{~Model ${\cal I}_{9}$~:~$(\rho_N,\rho_{F},\rho_{T_1},\rho_{T_2},\rho_{T_3})=(\mathbf{3},\mathbf{3},\mathbf{1^{\prime\prime}},\mathbf{1},\mathbf{1^{\prime}})\,,~ (k_{N},k_{F},k_{T_1},k_{T_2},k_{T_3})=(0,2,2,2,4)$}

We can read out the superpotentials relevant to quark and lepton masses
\begin{eqnarray}
\nonumber {\cal W}_\nu &=&
   \alpha_{N_1} \Lambda (NN)_{\mathbf{1}}
 + \alpha_{\nu_1} (NF)_{\mathbf{3}_S}Y_{\mathbf{3}}^{(2)} H_{5}
 + \alpha_{\nu_2} (NF)_{\mathbf{3}_A}Y_{\mathbf{3}}^{(2)} H_{5}
\,,\\
\nonumber {\cal W}_d &=&
   \alpha_{d_1} (FT_1)_{\mathbf{3}}Y_{\mathbf{3}}^{(4)} H_{\overline{5}}
 + \alpha_{d_2} (FT_2)_{\mathbf{3}}Y_{\mathbf{3}}^{(4)} H_{\overline{5}}
 + \alpha_{d_3} (FT_3)_{\mathbf{3}}Y_{\mathbf{3},I}^{(6)} H_{\overline{5}}
 + \alpha_{d_4} (FT_3)_{\mathbf{3}}Y_{\mathbf{3},II}^{(6)} H_{\overline{5}}
\\\nonumber &&
 + \alpha^\prime_{d_1} (FT_1)_{\mathbf{3}}Y_{\mathbf{3}}^{(4)} H_{\overline{45}}
 + \alpha^\prime_{d_2} (FT_2)_{\mathbf{3}}Y_{\mathbf{3}}^{(4)} H_{\overline{45}}
 + \alpha^\prime_{d_3} (FT_3)_{\mathbf{3}}Y_{\mathbf{3},I}^{(6)} H_{\overline{45}}
 + \alpha^\prime_{d_4} (FT_3)_{\mathbf{3}}Y_{\mathbf{3},II}^{(6)} H_{\overline{45}}
\,,\\
{\cal W}_u &=&
   \alpha_{u_1}  (T_1T_2)_{\mathbf{1^{\prime\prime}}}Y_{\mathbf{1^\prime}}^{(4)} H_{5}
 + \alpha_{u_2}  (T_1T_3)_{\mathbf{1}}Y_{\mathbf{1}}^{(6)} H_{5}
 + \alpha_{u_3}  (T_2T_2)_{\mathbf{1}}Y_{\mathbf{1}}^{(4)} H_{5}
 + \alpha_{u_4}  (T_3T_3)_{\mathbf{1^{\prime\prime}}}Y_{\mathbf{1^\prime}}^{(8)} H_{5}
\,,
\end{eqnarray}
which lead to
\begin{eqnarray}
{\cal Y}_{\overline{5}}^d &=& \left(
\begin{array}{ccc}
 \alpha_{d_1}Y_{\mathbf{3},2}^{(4)} ~&~ \alpha_{d_2}Y_{\mathbf{3},1}^{(4)} ~&~ \alpha_{d_3}Y_{\mathbf{3}I,3}^{(6)}+\alpha_{d_4}Y_{\mathbf{3}II,3}^{(6)} \\
 \alpha_{d_1}Y_{\mathbf{3},1}^{(4)} ~&~ \alpha_{d_2}Y_{\mathbf{3},3}^{(4)} ~&~ \alpha_{d_3}Y_{\mathbf{3}I,2}^{(6)}+\alpha_{d_4}Y_{\mathbf{3}II,2}^{(6)} \\
 \alpha_{d_1}Y_{\mathbf{3},3}^{(4)} ~&~ \alpha_{d_2}Y_{\mathbf{3},2}^{(4)} ~&~ \alpha_{d_3}Y_{\mathbf{3}I,1}^{(6)}+\alpha_{d_4}Y_{\mathbf{3}II,1}^{(6)} \\
\end{array}
\right)\,,\nonumber\\
{\cal Y}_5^u &=& \left(
\begin{array}{ccc}
 0 ~&~ \alpha_{u_1}Y_{\mathbf{1}^\prime}^{(4)} ~&~ \alpha_{u_2}Y_\mathbf{1}^{(6)} \\
 \alpha_{u_1}Y_{\mathbf{1}^\prime}^{(4)} ~&~ \alpha_{u_3}Y_\mathbf{1}^{(4)} ~&~ 0 \\
 \alpha_{u_2}Y_\mathbf{1}^{(6)} ~&~ 0 ~&~ \alpha_{u_4}Y_{\mathbf{1}^\prime}^{(8)} \\
\end{array}
\right)\,.
\end{eqnarray}
The matrices $M_N$ and ${\cal Y}_5^{\nu}$ are identical with those of the model ${\cal I}_{3}$ as well.
The phases of the parameters $\alpha_{N_1}$, $\alpha_{\nu_1}$, $\alpha_{d_1}$, $\alpha_{d_2}$, $\alpha_{d_3}$, $\alpha_{d_1}^\prime$ and $\alpha_{u_3}$ are unphysical, and the other parameters are complex.
\item{~Model ${\cal I}_{10}$~:~$(\rho_N,\rho_{F},\rho_{T_1},\rho_{T_2},\rho_{T_3})=(\mathbf{3},\mathbf{3},\mathbf{1^{\prime}},\mathbf{1^{\prime}},\mathbf{1^{\prime\prime}})\,,~ (k_{N},k_{F},k_{T_1},k_{T_2},k_{T_3})=(0,2,0,2,4)$}

We can read out the superpotential for quark and leptons as follows,
\begin{eqnarray}
\nonumber {\cal W}_\nu &=&
   \alpha_{N_1} \Lambda (NN)_{\mathbf{1}}
 + \alpha_{\nu_1} (NF)_{\mathbf{3}_S}Y_{\mathbf{3}}^{(2)} H_{5}
 + \alpha_{\nu_2} (NF)_{\mathbf{3}_A}Y_{\mathbf{3}}^{(2)} H_{5}
\,,\\
\nonumber {\cal W}_d &=&
   \alpha_{d_1} (FT_1)_{\mathbf{3}}Y_{\mathbf{3}}^{(2)} H_{\overline{5}}
 + \alpha_{d_2} (FT_2)_{\mathbf{3}}Y_{\mathbf{3}}^{(4)} H_{\overline{5}}
 + \alpha_{d_3} (FT_3)_{\mathbf{3}}Y_{\mathbf{3},I}^{(6)} H_{\overline{5}}
 + \alpha_{d_4} (FT_3)_{\mathbf{3}}Y_{\mathbf{3},II}^{(6)} H_{\overline{5}}
\\\nonumber &&
 + \alpha^\prime_{d_1} (FT_1)_{\mathbf{3}}Y_{\mathbf{3}}^{(2)} H_{\overline{45}}
 + \alpha^\prime_{d_2} (FT_2)_{\mathbf{3}}Y_{\mathbf{3}}^{(4)} H_{\overline{45}}
 + \alpha^\prime_{d_3} (FT_3)_{\mathbf{3}}Y_{\mathbf{3},I}^{(6)} H_{\overline{45}}
 + \alpha^\prime_{d_4} (FT_3)_{\mathbf{3}}Y_{\mathbf{3},II}^{(6)} H_{\overline{45}}
\,,\\
{\cal W}_u &=&
   \alpha_{u_1}  (T_1T_3)_{\mathbf{1}}Y_{\mathbf{1}}^{(4)} H_{5}
 + \alpha_{u_2}  (T_2T_2)_{\mathbf{1^{\prime\prime}}}Y_{\mathbf{1^\prime}}^{(4)} H_{5}
 + \alpha_{u_3}  (T_2T_3)_{\mathbf{1}}Y_{\mathbf{1}}^{(6)} H_{5}
 + \alpha_{u_4}  (T_3T_3)_{\mathbf{1^{\prime}}}Y_{\mathbf{1^{\prime\prime}}}^{(8)} H_{5}
\,.
\end{eqnarray}
The right-handed neutrino mass matrix and the Yukawa coupling matrices are predicted to be
\begin{eqnarray}
{\cal Y}_{\overline{5}}^d &=& \left(
\begin{array}{ccc}
 \alpha_{d_1}Y_{\mathbf{3},3}^{(2)} ~&~ \alpha_{d_2}Y_{\mathbf{3},3}^{(4)} ~&~ \alpha_{d_3}Y_{\mathbf{3}I,2}^{(6)}+\alpha_{d_4}Y_{\mathbf{3}II,2}^{(6)} \\
 \alpha_{d_1}Y_{\mathbf{3},2}^{(2)} ~&~ \alpha_{d_2}Y_{\mathbf{3},2}^{(4)} ~&~ \alpha_{d_3}Y_{\mathbf{3}I,1}^{(6)}+\alpha_{d_4}Y_{\mathbf{3}II,1}^{(6)} \\
 \alpha_{d_1}Y_{\mathbf{3},1}^{(2)}  ~&~ \alpha_{d_2}Y_{\mathbf{3},1}^{(4)} ~&~ \alpha_{d_3}Y_{\mathbf{3}I,3}^{(6)}+\alpha_{d_4}Y_{\mathbf{3}II,3}^{(6)} \\
\end{array}
\right)\,,\nonumber\\
{\cal Y}_5^u &=& \left(
\begin{array}{ccc}
 0 ~&~ 0 ~&~ \alpha_{u_1}Y_\mathbf{1}^{(4)} \\
 0 ~&~ \alpha_{u_2}Y_{\mathbf{1}^\prime}^{(4)} ~&~ \alpha_{u_3}Y_\mathbf{1}^{(6)} \\
 \alpha_{u_1}Y_\mathbf{1}^{(4)} ~&~ \alpha_{u_3}Y_\mathbf{1}^{(6)} ~&~ \alpha_{u_4}Y_{\mathbf{1}^{\prime\prime}}^{(8)} \\
\end{array}
\right)\,.
\end{eqnarray}
The matrices $M_N$ and ${\cal Y}_5^{\nu}$ are the same as the corresponding ones of the model ${\cal I}_{3}$ as well.
The parameters $\alpha_{N_1}$, $\alpha_{\nu_1}$, $\alpha_{d_1}$, $\alpha_{d_2}$, $\alpha_{d_3}$, $\alpha_{d_1}^\prime$ and $\alpha_{u_3}$ are real, and the remaining parameters are complex.
\item{~Model ${\cal I}_{11}$~:~$(\rho_N,\rho_{F},\rho_{T_1},\rho_{T_2},\rho_{T_3})=(\mathbf{3},\mathbf{3},\mathbf{1^{\prime\prime}},\mathbf{1^{\prime}},\mathbf{1^{\prime}})\,,~ (k_{N},k_{F},k_{T_1},k_{T_2},k_{T_3})=(0,2,2,2,4)$}

The modular invariant superpotential for quark and lepton mass is given by
\begin{eqnarray}
\nonumber {\cal W}_\nu &=&
   \alpha_{N_1} \Lambda (NN)_{\mathbf{1}}
 + \alpha_{\nu_1} (NF)_{\mathbf{3}_S}Y_{\mathbf{3}}^{(2)} H_{5}
 + \alpha_{\nu_2} (NF)_{\mathbf{3}_A}Y_{\mathbf{3}}^{(2)} H_{5}
\,,\\
\nonumber {\cal W}_d &=&
   \alpha_{d_1} (FT_1)_{\mathbf{3}}Y_{\mathbf{3}}^{(4)} H_{\overline{5}}
 + \alpha_{d_2} (FT_2)_{\mathbf{3}}Y_{\mathbf{3}}^{(4)} H_{\overline{5}}
 + \alpha_{d_3} (FT_3)_{\mathbf{3}}Y_{\mathbf{3},I}^{(6)} H_{\overline{5}}
 + \alpha_{d_4} (FT_3)_{\mathbf{3}}Y_{\mathbf{3},II}^{(6)} H_{\overline{5}}
\\\nonumber &&
 + \alpha^\prime_{d_1} (FT_1)_{\mathbf{3}}Y_{\mathbf{3}}^{(4)} H_{\overline{45}}
 + \alpha^\prime_{d_2} (FT_2)_{\mathbf{3}}Y_{\mathbf{3}}^{(4)} H_{\overline{45}}
 + \alpha^\prime_{d_3} (FT_3)_{\mathbf{3}}Y_{\mathbf{3},I}^{(6)} H_{\overline{45}}
 + \alpha^\prime_{d_4} (FT_3)_{\mathbf{3}}Y_{\mathbf{3},II}^{(6)} H_{\overline{45}}
\,,\\
{\cal W}_u &=&
   \alpha_{u_1}  (T_1T_2)_{\mathbf{1}}Y_{\mathbf{1}}^{(4)} H_{5}
 + \alpha_{u_2}  (T_1T_3)_{\mathbf{1}}Y_{\mathbf{1}}^{(6)} H_{5}
 + \alpha_{u_3}  (T_2T_2)_{\mathbf{1^{\prime\prime}}}Y_{\mathbf{1^\prime}}^{(4)} H_{5}
 + \alpha_{u_4}  (T_3T_3)_{\mathbf{1^{\prime\prime}}}Y_{\mathbf{1^\prime}}^{(8)} H_{5}
\,.
\end{eqnarray}
The right-handed neutrino mass matrix and the Yukawa matrix are given by
\begin{eqnarray}
{\cal Y}_{\overline{5}}^d &=& \left(
\begin{array}{ccc}
 \alpha_{d_1}Y_{\mathbf{3},2}^{(4)} ~&~ \alpha_{d_2}Y_{\mathbf{3},3}^{(4)} ~&~ \alpha_{d_3}Y_{\mathbf{3}I,3}^{(6)}+\alpha_{d_4}Y_{\mathbf{3}II,3}^{(6)} \\
 \alpha_{d_1}Y_{\mathbf{3},1}^{(4)} ~&~ \alpha_{d_2}Y_{\mathbf{3},2}^{(4)} ~&~ \alpha_{d_3}Y_{\mathbf{3}I,2}^{(6)}+\alpha_{d_4}Y_{\mathbf{3}II,2}^{(6)} \\
 \alpha_{d_1}Y_{\mathbf{3},3}^{(4)} ~&~ \alpha_{d_2}Y_{\mathbf{3},1}^{(4)} ~&~ \alpha_{d_3}Y_{\mathbf{3}I,1}^{(6)}+\alpha_{d_4}Y_{\mathbf{3}II,1}^{(6)} \\
\end{array}
\right)\,,\nonumber\\
{\cal Y}_5^u &=& \left(
\begin{array}{ccc}
 0 ~&~ \alpha_{u_1}Y_\mathbf{1}^{(4)} ~&~ \alpha_{u_2}Y_\mathbf{1}^{(6)} \\
 \alpha_{u_1}Y_\mathbf{1}^{(4)} ~&~ \alpha_{u_3}Y_{\mathbf{1}^\prime}^{(4)} ~&~ 0 \\
 \alpha_{u_2}Y_\mathbf{1}^{(6)} ~&~ 0 ~&~ \alpha_{u_4}Y_{\mathbf{1}^\prime}^{(8)} \\
\end{array}
\right)\,.
\end{eqnarray}
The matrices $M_N$ and ${\cal Y}_5^{\nu}$ are identical with those of the model ${\cal I}_{3}$ as well.
The parameters $\alpha_{N_1}$, $\alpha_{\nu_1}$, $\alpha_{d_1}$, $\alpha_{d_2}$, $\alpha_{d_3}$, $\alpha_{d_1}^\prime$ and $\alpha_{u_3}$ can be taken real by field redefinition, and the remaining parameters are complex.
\item{~Model ${\cal I}_{12}$~:~$(\rho_N,\rho_{F},\rho_{T_1},\rho_{T_2},\rho_{T_3})=(\mathbf{3},\mathbf{3},\mathbf{1^{\prime\prime}},\mathbf{1^{\prime\prime}},\mathbf{1^{\prime}})\,,~ (k_{N},k_{F},k_{T_1},k_{T_2},k_{T_3})=(0,2,2,4,2)$}

The quark and lepton masses are described by the following superpotential:
\begin{eqnarray}
\nonumber {\cal W}_\nu &=&
   \alpha_{N_1} \Lambda (NN)_{\mathbf{1}}
 + \alpha_{\nu_1} (NF)_{\mathbf{3}_S}Y_{\mathbf{3}}^{(2)} H_{5}
 + \alpha_{\nu_2} (NF)_{\mathbf{3}_A}Y_{\mathbf{3}}^{(2)} H_{5}
\,,\\
\nonumber {\cal W}_d &=&
   \alpha_{d_1} (FT_1)_{\mathbf{3}}Y_{\mathbf{3}}^{(4)} H_{\overline{5}}
 + \alpha_{d_2} (FT_2)_{\mathbf{3}}Y_{\mathbf{3},I}^{(6)} H_{\overline{5}}
 + \alpha_{d_4} (FT_2)_{\mathbf{3}}Y_{\mathbf{3},II}^{(6)} H_{\overline{5}}
 + \alpha_{d_3} (FT_3)_{\mathbf{3}}Y_{\mathbf{3}}^{(4)} H_{\overline{5}}
\\\nonumber &&
 + \alpha^\prime_{d_1} (FT_1)_{\mathbf{3}}Y_{\mathbf{3}}^{(4)} H_{\overline{45}}
 + \alpha^\prime_{d_2} (FT_2)_{\mathbf{3}}Y_{\mathbf{3},I}^{(6)} H_{\overline{45}}
 + \alpha^\prime_{d_4} (FT_2)_{\mathbf{3}}Y_{\mathbf{3},II}^{(6)} H_{\overline{45}}
 + \alpha^\prime_{d_3} (FT_3)_{\mathbf{3}}Y_{\mathbf{3}}^{(4)} H_{\overline{45}}
\,,\\
{\cal W}_u &=&
   \alpha_{u_1}  (T_1T_3)_{\mathbf{1}}Y_{\mathbf{1}}^{(4)} H_{5}
 + \alpha_{u_2}  (T_2T_2)_{\mathbf{1^{\prime}}}Y_{\mathbf{1^{\prime\prime}}}^{(8)} H_{5}
 + \alpha_{u_3}  (T_2T_3)_{\mathbf{1}}Y_{\mathbf{1}}^{(6)} H_{5}
 + \alpha_{u_4}  (T_3T_3)_{\mathbf{1^{\prime\prime}}}Y_{\mathbf{1^\prime}}^{(4)} H_{5}
\,,
\end{eqnarray}
which gives rise to
\begin{eqnarray}
{\cal Y}_{\overline{5}}^d &=& \left(
\begin{array}{ccc}
 \alpha_{d_1}Y_{\mathbf{3},2}^{(4)} ~&~ \alpha_{d_2}Y_{\mathbf{3}I,2}^{(6)}+\alpha_{d_4}Y_{\mathbf{3}II,2}^{(6)} ~&~ \alpha_{d_3}Y_{\mathbf{3},3}^{(4)} \\
 \alpha_{d_1}Y_{\mathbf{3},1}^{(4)} ~&~ \alpha_{d_2}Y_{\mathbf{3}I,1}^{(6)}+\alpha_{d_4}Y_{\mathbf{3}II,1}^{(6)} ~&~ \alpha_{d_3}Y_{\mathbf{3},2}^{(4)} \\
 \alpha_{d_1}Y_{\mathbf{3},3}^{(4)} ~&~ \alpha_{d_2}Y_{\mathbf{3}I,3}^{(6)}+\alpha_{d_4}Y_{\mathbf{3}II,3}^{(6)} ~&~ \alpha_{d_3}Y_{\mathbf{3},1}^{(4)} \\
\end{array}
\right)\,,\nonumber\\
{\cal Y}_5^u &=& \left(
\begin{array}{ccc}
 0 ~&~ 0 ~&~ \alpha_{u_1}Y_\mathbf{1}^{(4)} \\
 0 ~&~ \alpha_{u_2}Y_{\mathbf{1}^{\prime\prime}}^{(8)} ~&~ \alpha_{u_3}Y_\mathbf{1}^{(6)} \\
 \alpha_{u_1}Y_\mathbf{1}^{(4)} ~&~ \alpha_{u_3}Y_\mathbf{1}^{(6)} ~&~ \alpha_{u_4}Y_{\mathbf{1}^\prime}^{(4)} \\
\end{array}
\right)\,.
\end{eqnarray}
The matrices $M_N$ and ${\cal Y}_5^{\nu}$ are identical with the corresponding ones of the model ${\cal I}_{3}$ as well.
The phases of the parameters $\alpha_{N_1}$, $\alpha_{\nu_1}$, $\alpha_{d_1}$, $\alpha_{d_2}$, $\alpha_{d_3}$, $\alpha_{d_1}^\prime$ and $\alpha_{u_3}$ can be removed by field redefinition, and the remaining parameters are complex.
\item{~Model ${\cal I}_{13}$~:~$(\rho_N,\rho_{F},\rho_{T_1},\rho_{T_2},\rho_{T_3})=(\mathbf{3},\mathbf{3},\mathbf{1^{\prime\prime}},\mathbf{1},\mathbf{1^{\prime}})\,,~ (k_{N},k_{F},k_{T_1},k_{T_2},k_{T_3})=(2,0,2,2,4)$}

The modular invariant superpotential is of the following form:
\begin{eqnarray}
\nonumber {\cal W}_\nu &=&
   \alpha_{N_1} \Lambda (NN)_{\mathbf{1}}Y_{\mathbf{1}}^{(4)}
 + \alpha_{N_2} \Lambda (NN)_{\mathbf{1^{\prime\prime}}}Y_{\mathbf{1^\prime}}^{(4)}
 + \alpha_{N_3} \Lambda (NN)_{\mathbf{3}_S}Y_{\mathbf{3}}^{(4)}
 + \alpha_{\nu_1} (NF)_{\mathbf{3}_S}Y_{\mathbf{3}}^{(2)} H_{5}
\\
\nonumber&&
 + \alpha_{\nu_2} (NF)_{\mathbf{3}_A}Y_{\mathbf{3}}^{(2)} H_{5}
\,,\\
\nonumber {\cal W}_d &=&
   \alpha_{d_1} (FT_1)_{\mathbf{3}}Y_{\mathbf{3}}^{(2)} H_{\overline{5}}
 + \alpha_{d_2} (FT_2)_{\mathbf{3}}Y_{\mathbf{3}}^{(2)} H_{\overline{5}}
 + \alpha_{d_3} (FT_3)_{\mathbf{3}}Y_{\mathbf{3}}^{(4)} H_{\overline{5}}
\\\nonumber &&
 + \alpha^\prime_{d_1} (FT_1)_{\mathbf{3}}Y_{\mathbf{3}}^{(2)} H_{\overline{45}}
 + \alpha^\prime_{d_2} (FT_2)_{\mathbf{3}}Y_{\mathbf{3}}^{(2)} H_{\overline{45}}
 + \alpha^\prime_{d_3} (FT_3)_{\mathbf{3}}Y_{\mathbf{3}}^{(4)} H_{\overline{45}}
\,,\\
{\cal W}_u &=&
   \alpha_{u_1}  (T_1T_2)_{\mathbf{1^{\prime\prime}}}Y_{\mathbf{1^\prime}}^{(4)} H_{5}
 + \alpha_{u_2}  (T_1T_3)_{\mathbf{1}}Y_{\mathbf{1}}^{(6)} H_{5}
 + \alpha_{u_3}  (T_2T_2)_{\mathbf{1}}Y_{\mathbf{1}}^{(4)} H_{5}
 + \alpha_{u_4}  (T_3T_3)_{\mathbf{1^{\prime\prime}}}Y_{\mathbf{1^\prime}}^{(8)} H_{5}
\,.
\end{eqnarray}
which gives rise to
\begin{eqnarray}
{\cal Y}_{\overline{5}}^d &=& \left(
\begin{array}{ccc}
 \alpha_{d_1}Y_{\mathbf{3},2}^{(2)} ~&~ \alpha_{d_2}Y_{\mathbf{3},1}^{(2)} ~&~ \alpha_{d_3}Y_{\mathbf{3},3}^{(4)} \\
 \alpha_{d_1}Y_{\mathbf{3},1}^{(2)}  ~&~ \alpha_{d_2}Y_{\mathbf{3},3}^{(2)} ~&~ \alpha_{d_3}Y_{\mathbf{3},2}^{(4)} \\
 \alpha_{d_1}Y_{\mathbf{3},3}^{(2)} ~&~ \alpha_{d_2}Y_{\mathbf{3},2}^{(2)} ~&~ \alpha_{d_3}Y_{\mathbf{3},1}^{(4)} \\
\end{array}
\right)\,.
\end{eqnarray}
The $M_N$ matrix coincides with that of the model ${\cal I}_{8}$.
The ${\cal Y}_5^{\nu}$ matrix coincides with that of the model ${\cal I}_{1}$.
The ${\cal Y}_5^u$ matrix is identical with the corresponding one of the model ${\cal I}_{9}$.
The parameters $\alpha_{N_3}$, $\alpha_{\nu_1}$, $\alpha_{d_1}$, $\alpha_{d_2}$, $\alpha_{d_3}$, $\alpha_{d_1}^\prime$ and $\alpha_{u_3}$ are real, and the remaining parameters $\alpha_{N_1}$, $\alpha_{N_2}$, $\alpha_{\nu_2}$, $\alpha_{d_2}^\prime$, $\alpha_{d_3}^\prime$, $\alpha_{u_1}$, $\alpha_{u_2}$ and $\alpha_{u_4}$ are complex.
\item{~Model ${\cal I}_{14}$~:~$(\rho_N,\rho_{F},\rho_{T_1},\rho_{T_2},\rho_{T_3})=(\mathbf{3},\mathbf{3},\mathbf{1^{\prime\prime}},\mathbf{1},\mathbf{1^{\prime}})\,,~ (k_{N},k_{F},k_{T_1},k_{T_2},k_{T_3})=(2,0,2,4,4)$}

We can read out the the superpotential for the fermion masses as,
\begin{eqnarray}
\nonumber {\cal W}_\nu &=&
   \alpha_{N_1} \Lambda (NN)_{\mathbf{1}}Y_{\mathbf{1}}^{(4)}
 + \alpha_{N_2} \Lambda (NN)_{\mathbf{1^{\prime\prime}}}Y_{\mathbf{1^\prime}}^{(4)}
 + \alpha_{N_3} \Lambda (NN)_{\mathbf{3}_S}Y_{\mathbf{3}}^{(4)}
 + \alpha_{\nu_1} (NF)_{\mathbf{3}_S}Y_{\mathbf{3}}^{(2)} H_{5}
\\
\nonumber&&
 + \alpha_{\nu_2} (NF)_{\mathbf{3}_A}Y_{\mathbf{3}}^{(2)} H_{5}
\,,\\
\nonumber {\cal W}_d &=&
   \alpha_{d_1} (FT_1)_{\mathbf{3}}Y_{\mathbf{3}}^{(2)} H_{\overline{5}}
 + \alpha_{d_2} (FT_2)_{\mathbf{3}}Y_{\mathbf{3}}^{(4)} H_{\overline{5}}
 + \alpha_{d_3} (FT_3)_{\mathbf{3}}Y_{\mathbf{3}}^{(4)} H_{\overline{5}}
\\\nonumber &&
 + \alpha^\prime_{d_1} (FT_1)_{\mathbf{3}}Y_{\mathbf{3}}^{(2)} H_{\overline{45}}
 + \alpha^\prime_{d_2} (FT_2)_{\mathbf{3}}Y_{\mathbf{3}}^{(4)} H_{\overline{45}}
 + \alpha^\prime_{d_3} (FT_3)_{\mathbf{3}}Y_{\mathbf{3}}^{(4)} H_{\overline{45}}
\,,\\
{\cal W}_u &=&
   \alpha_{u_1}  (T_1T_3)_{\mathbf{1}}Y_{\mathbf{1}}^{(6)} H_{5}
 + \alpha_{u_2}  (T_2T_2)_{\mathbf{1}}Y_{\mathbf{1}}^{(8)} H_{5}
 + \alpha_{u_3}  (T_2T_3)_{\mathbf{1^{\prime}}}Y_{\mathbf{1^{\prime\prime}}}^{(8)} H_{5}
 + \alpha_{u_4}  (T_3T_3)_{\mathbf{1^{\prime\prime}}}Y_{\mathbf{1^\prime}}^{(8)} H_{5}
\,.
\end{eqnarray}
which gives rise to
\begin{eqnarray}
{\cal Y}_{\overline{5}}^d &=& \left(
\begin{array}{ccc}
 \alpha_{d_1}Y_{\mathbf{3},2}^{(2)} ~&~ \alpha_{d_2}Y_{\mathbf{3},1}^{(4)} ~&~ \alpha_{d_3}Y_{\mathbf{3},3}^{(4)} \\
 \alpha_{d_1}Y_{\mathbf{3},1}^{(2)}  ~&~ \alpha_{d_2}Y_{\mathbf{3},3}^{(4)} ~&~ \alpha_{d_3}Y_{\mathbf{3},2}^{(4)} \\
 \alpha_{d_1}Y_{\mathbf{3},3}^{(2)} ~&~ \alpha_{d_2}Y_{\mathbf{3},2}^{(4)} ~&~ \alpha_{d_3}Y_{\mathbf{3},1}^{(4)} \\
\end{array}
\right)\,,\nonumber\\
{\cal Y}_5^u &=& \left(
\begin{array}{ccc}
 0 ~&~ 0 ~&~ \alpha_{u_1}Y_\mathbf{1}^{(6)} \\
 0 ~&~ \alpha_{u_2}Y_\mathbf{1}^{(8)} ~&~ \alpha_{u_3}Y_{\mathbf{1}^{\prime\prime}}^{(8)} \\
 \alpha_{u_1}Y_\mathbf{1}^{(6)} ~&~ \alpha_{u_3}Y_{\mathbf{1}^{\prime\prime}}^{(8)} ~&~ \alpha_{u_4}Y_{\mathbf{1}^\prime}^{(8)} \\
\end{array}
\right)\,.
\end{eqnarray}
The matrices $M_N$ and ${\cal Y}_5^{\nu}$ are identical with those of the models ${\cal I}_{8}$ and ${\cal I}_{1}$ respectively.
The parameters $\alpha_{N_3}$, $\alpha_{\nu_1}$, $\alpha_{d_1}$, $\alpha_{d_2}$, $\alpha_{d_3}$, $\alpha_{d_1}^\prime$ and $\alpha_{u_3}$ can taken to be real without loss of generality, while $\alpha_{N_1}$, $\alpha_{N_2}$, $\alpha_{\nu_2}$, $\alpha_{d_2}^\prime$, $\alpha_{d_3}^\prime$, $\alpha_{u_1}$, $\alpha_{u_2}$ and $\alpha_{u_4}$ are generically complex.
\item{~Model ${\cal I}_{15}$~:~$(\rho_N,\rho_{F},\rho_{T_1},\rho_{T_2},\rho_{T_3})=(\mathbf{3},\mathbf{3},\mathbf{1^{\prime\prime}},\mathbf{1^{\prime}},\mathbf{1^{\prime}})\,,~ (k_{N},k_{F},k_{T_1},k_{T_2},k_{T_3})=(2,0,2,2,4)$}

The modular invariant superpotential for quark and lepton mass is given by
\begin{eqnarray}
\nonumber {\cal W}_\nu &=&
   \alpha_{N_1} \Lambda (NN)_{\mathbf{1}}Y_{\mathbf{1}}^{(4)}
 + \alpha_{N_2} \Lambda (NN)_{\mathbf{1^{\prime\prime}}}Y_{\mathbf{1^\prime}}^{(4)}
 + \alpha_{N_3} \Lambda (NN)_{\mathbf{3}_S}Y_{\mathbf{3}}^{(4)}
 + \alpha_{\nu_1} (NF)_{\mathbf{3}_S}Y_{\mathbf{3}}^{(2)} H_{5}
\\
\nonumber&&
 + \alpha_{\nu_2} (NF)_{\mathbf{3}_A}Y_{\mathbf{3}}^{(2)} H_{5}
\,,\\
\nonumber {\cal W}_d &=&
   \alpha_{d_1} (FT_1)_{\mathbf{3}}Y_{\mathbf{3}}^{(2)} H_{\overline{5}}
 + \alpha_{d_2} (FT_2)_{\mathbf{3}}Y_{\mathbf{3}}^{(2)} H_{\overline{5}}
 + \alpha_{d_3} (FT_3)_{\mathbf{3}}Y_{\mathbf{3}}^{(4)} H_{\overline{5}}
\\\nonumber &&
 + \alpha^\prime_{d_1} (FT_1)_{\mathbf{3}}Y_{\mathbf{3}}^{(2)} H_{\overline{45}}
 + \alpha^\prime_{d_2} (FT_2)_{\mathbf{3}}Y_{\mathbf{3}}^{(2)} H_{\overline{45}}
 + \alpha^\prime_{d_3} (FT_3)_{\mathbf{3}}Y_{\mathbf{3}}^{(4)} H_{\overline{45}}
\,,\\
{\cal W}_u &=&
   \alpha_{u_1}  (T_1T_2)_{\mathbf{1}}Y_{\mathbf{1}}^{(4)} H_{5}
 + \alpha_{u_2}  (T_1T_3)_{\mathbf{1}}Y_{\mathbf{1}}^{(6)} H_{5}
 + \alpha_{u_3}  (T_2T_2)_{\mathbf{1^{\prime\prime}}}Y_{\mathbf{1^\prime}}^{(4)} H_{5}
 + \alpha_{u_4}  (T_3T_3)_{\mathbf{1^{\prime\prime}}}Y_{\mathbf{1^\prime}}^{(8)} H_{5}
\,,
\end{eqnarray}
which leads to
\begin{eqnarray}
{\cal Y}_{\overline{5}}^d &=& \left(
\begin{array}{ccc}
 \alpha_{d_1}Y_{\mathbf{3},2}^{(2)} ~&~ \alpha_{d_2}Y_{\mathbf{3},3}^{(2)} ~&~ \alpha_{d_3}Y_{\mathbf{3},3}^{(4)} \\
 \alpha_{d_1}Y_{\mathbf{3},1}^{(2)}  ~&~ \alpha_{d_2}Y_{\mathbf{3},2}^{(2)} ~&~ \alpha_{d_3}Y_{\mathbf{3},2}^{(4)} \\
 \alpha_{d_1}Y_{\mathbf{3},3}^{(2)} ~&~ \alpha_{d_2}Y_{\mathbf{3},1}^{(2)} ~&~ \alpha_{d_3}Y_{\mathbf{3},1}^{(4)} \\
\end{array}
\right)\,.
\end{eqnarray}
The $M_N$ matrix coincides with the corresponding one of the model ${\cal I}_{8}$.
The ${\cal Y}_5^{\nu}$ matrix coincides with that of the model ${\cal I}_{1}$.
The ${\cal Y}_5^u$ matrix is identical with that of the model ${\cal I}_{11}$.
The parameters $\alpha_{N_3}$, $\alpha_{\nu_1}$, $\alpha_{d_1}$, $\alpha_{d_2}$, $\alpha_{d_3}$, $\alpha_{d_1}^\prime$ and $\alpha_{u_3}$ can be taken real by field redefinition, and the remaining parameters are complex.
\item{~Model ${\cal I}_{16}$~:~$(\rho_N,\rho_{F},\rho_{T_1},\rho_{T_2},\rho_{T_3})=(\mathbf{3},\mathbf{3},\mathbf{1^{\prime\prime}},\mathbf{1^{\prime\prime}},\mathbf{1^{\prime}})\,,~ (k_{N},k_{F},k_{T_1},k_{T_2},k_{T_3})=(2,0,2,4,2)$}

We can read out the superpotentials relevant to quark and lepton masses
\begin{eqnarray}
\nonumber {\cal W}_\nu &=&
   \alpha_{N_1} \Lambda (NN)_{\mathbf{1}}Y_{\mathbf{1}}^{(4)}
 + \alpha_{N_2} \Lambda (NN)_{\mathbf{1^{\prime\prime}}}Y_{\mathbf{1^\prime}}^{(4)}
 + \alpha_{N_3} \Lambda (NN)_{\mathbf{3}_S}Y_{\mathbf{3}}^{(4)}
 + \alpha_{\nu_1} (NF)_{\mathbf{3}_S}Y_{\mathbf{3}}^{(2)} H_{5}
\\
\nonumber&&
 + \alpha_{\nu_2} (NF)_{\mathbf{3}_A}Y_{\mathbf{3}}^{(2)} H_{5}
\,,\\
\nonumber {\cal W}_d &=&
   \alpha_{d_1} (FT_1)_{\mathbf{3}}Y_{\mathbf{3}}^{(2)} H_{\overline{5}}
 + \alpha_{d_2} (FT_2)_{\mathbf{3}}Y_{\mathbf{3}}^{(4)} H_{\overline{5}}
 + \alpha_{d_3} (FT_3)_{\mathbf{3}}Y_{\mathbf{3}}^{(2)} H_{\overline{5}}
\\\nonumber &&
 + \alpha^\prime_{d_1} (FT_1)_{\mathbf{3}}Y_{\mathbf{3}}^{(2)} H_{\overline{45}}
 + \alpha^\prime_{d_2} (FT_2)_{\mathbf{3}}Y_{\mathbf{3}}^{(4)} H_{\overline{45}}
 + \alpha^\prime_{d_3} (FT_3)_{\mathbf{3}}Y_{\mathbf{3}}^{(2)} H_{\overline{45}}
\,,\\
{\cal W}_u &=&
   \alpha_{u_1}  (T_1T_3)_{\mathbf{1}}Y_{\mathbf{1}}^{(4)} H_{5}
 + \alpha_{u_2}  (T_2T_2)_{\mathbf{1^{\prime}}}Y_{\mathbf{1^{\prime\prime}}}^{(8)} H_{5}
 + \alpha_{u_3}  (T_2T_3)_{\mathbf{1}}Y_{\mathbf{1}}^{(6)} H_{5}
 + \alpha_{u_4}  (T_3T_3)_{\mathbf{1^{\prime\prime}}}Y_{\mathbf{1^\prime}}^{(4)} H_{5}
\,.
\end{eqnarray}
The right-handed neutrino mass matrix and the Yukawa matrix are given by
\begin{eqnarray}
{\cal Y}_{\overline{5}}^d &=& \left(
\begin{array}{ccc}
 \alpha_{d_1}Y_{\mathbf{3},2}^{(2)} ~&~ \alpha_{d_2}Y_{\mathbf{3},2}^{(4)} ~&~ \alpha_{d_3}Y_{\mathbf{3},3}^{(2)} \\
 \alpha_{d_1}Y_{\mathbf{3},1}^{(2)}  ~&~ \alpha_{d_2}Y_{\mathbf{3},1}^{(4)} ~&~ \alpha_{d_3}Y_{\mathbf{3},2}^{(2)} \\
 \alpha_{d_1}Y_{\mathbf{3},3}^{(2)} ~&~ \alpha_{d_2}Y_{\mathbf{3},3}^{(4)} ~&~ \alpha_{d_3}Y_{\mathbf{3},1}^{(2)}  \\
\end{array}
\right)\,.
\end{eqnarray}
The $M_N$ matrix is the same as the corresponding one of the model ${\cal I}_{8}$.
The ${\cal Y}_5^{\nu}$ matrix is in common with the corresponding one of the model ${\cal I}_{1}$.
The ${\cal Y}_5^u$ matrix coincides with the corresponding one of the model ${\cal I}_{12}$.
The parameters $\alpha_{N_3}$, $\alpha_{\nu_1}$, $\alpha_{d_1}$, $\alpha_{d_2}$, $\alpha_{d_3}$, $\alpha_{d_1}^\prime$ and $\alpha_{u_3}$ can taken to be real without loss of generality, while $\alpha_{N_1}$, $\alpha_{N_2}$, $\alpha_{\nu_2}$, $\alpha_{d_2}^\prime$, $\alpha_{d_3}^\prime$, $\alpha_{u_1}$, $\alpha_{u_2}$ and $\alpha_{u_4}$ are generically complex.
\item{~Model ${\cal I}_{17}$~:~$(\rho_N,\rho_{F},\rho_{T_1},\rho_{T_2},\rho_{T_3})=(\mathbf{3},\mathbf{3},\mathbf{1^{\prime\prime}},\mathbf{1^{\prime\prime}},\mathbf{1^{\prime}})\,,~ (k_{N},k_{F},k_{T_1},k_{T_2},k_{T_3})=(2,0,2,4,4)$}

We can read out the the superpotential for the fermion masses as,
\begin{eqnarray}
\nonumber {\cal W}_\nu &=&
   \alpha_{N_1} \Lambda (NN)_{\mathbf{1}}Y_{\mathbf{1}}^{(4)}
 + \alpha_{N_2} \Lambda (NN)_{\mathbf{1^{\prime\prime}}}Y_{\mathbf{1^\prime}}^{(4)}
 + \alpha_{N_3} \Lambda (NN)_{\mathbf{3}_S}Y_{\mathbf{3}}^{(4)}
 + \alpha_{\nu_1} (NF)_{\mathbf{3}_S}Y_{\mathbf{3}}^{(2)} H_{5}
\\
\nonumber&&
 + \alpha_{\nu_2} (NF)_{\mathbf{3}_A}Y_{\mathbf{3}}^{(2)} H_{5}
\,,\\
\nonumber {\cal W}_d &=&
   \alpha_{d_1} (FT_1)_{\mathbf{3}}Y_{\mathbf{3}}^{(2)} H_{\overline{5}}
 + \alpha_{d_2} (FT_2)_{\mathbf{3}}Y_{\mathbf{3}}^{(4)} H_{\overline{5}}
 + \alpha_{d_3} (FT_3)_{\mathbf{3}}Y_{\mathbf{3}}^{(4)} H_{\overline{5}}
\\\nonumber &&
 + \alpha^\prime_{d_1} (FT_1)_{\mathbf{3}}Y_{\mathbf{3}}^{(2)} H_{\overline{45}}
 + \alpha^\prime_{d_2} (FT_2)_{\mathbf{3}}Y_{\mathbf{3}}^{(4)} H_{\overline{45}}
 + \alpha^\prime_{d_3} (FT_3)_{\mathbf{3}}Y_{\mathbf{3}}^{(4)} H_{\overline{45}}
\,,\\
{\cal W}_u &=&
   \alpha_{u_1}  (T_1T_3)_{\mathbf{1}}Y_{\mathbf{1}}^{(6)} H_{5}
 + \alpha_{u_2}  (T_2T_2)_{\mathbf{1^{\prime}}}Y_{\mathbf{1^{\prime\prime}}}^{(8)} H_{5}
 + \alpha_{u_3}  (T_2T_3)_{\mathbf{1}}Y_{\mathbf{1}}^{(8)} H_{5}
 + \alpha_{u_4}  (T_3T_3)_{\mathbf{1^{\prime\prime}}}Y_{\mathbf{1^\prime}}^{(8)} H_{5}
\,.
\end{eqnarray}
The right-handed neutrino mass matrix and the Yukawa matrix are given by
\begin{eqnarray}
{\cal Y}_{\overline{5}}^d &=& \left(
\begin{array}{ccc}
 \alpha_{d_1}Y_{\mathbf{3},2}^{(2)} ~&~ \alpha_{d_2}Y_{\mathbf{3},2}^{(4)} ~&~ \alpha_{d_3}Y_{\mathbf{3},3}^{(4)} \\
 \alpha_{d_1}Y_{\mathbf{3},1}^{(2)}  ~&~ \alpha_{d_2}Y_{\mathbf{3},1}^{(4)} ~&~ \alpha_{d_3}Y_{\mathbf{3},2}^{(4)} \\
 \alpha_{d_1}Y_{\mathbf{3},3}^{(2)} ~&~ \alpha_{d_2}Y_{\mathbf{3},3}^{(4)} ~&~ \alpha_{d_3}Y_{\mathbf{3},1}^{(4)} \\
\end{array}
\right)\,,\nonumber\\
{\cal Y}_5^u &=& \left(
\begin{array}{ccc}
 0 ~&~ 0 ~&~ \alpha_{u_1}Y_\mathbf{1}^{(6)} \\
 0 ~&~ \alpha_{u_2}Y_{\mathbf{1}^{\prime\prime}}^{(8)} ~&~ \alpha_{u_3}Y_\mathbf{1}^{(8)} \\
 \alpha_{u_1}Y_\mathbf{1}^{(6)} ~&~ \alpha_{u_3}Y_\mathbf{1}^{(8)} ~&~ \alpha_{u_4}Y_{\mathbf{1}^\prime}^{(8)} \\
\end{array}
\right)\,.
\end{eqnarray}
The matrices $M_N$ and ${\cal Y}_5^{\nu}$ coincide with the corresponding ones of the model ${\cal I}_{14}$ as well.
The parameters $\alpha_{N_3}$, $\alpha_{\nu_1}$, $\alpha_{d_1}$, $\alpha_{d_2}$, $\alpha_{d_3}$, $\alpha_{d_1}^\prime$ and $\alpha_{u_3}$ can be taken real by field redefinition, and the other parameters are complex.
\end{itemize}
\subsection{Type-II}
\begin{itemize}[leftmargin=1.0em]
\item{~Model ${\cal II}_{1}$~:~$(\rho_N,\rho_{F_1},\rho_{F_2},\rho_{F_3},\rho_{T})=(\mathbf{3},\mathbf{1},\mathbf{1^{\prime\prime}},\mathbf{1^{\prime}},\mathbf{3})\,,~ (k_{N},k_{F_1},k_{F_2},k_{F_3},k_{T})=(1,1,3,1,3)$}

The quark and lepton masses are described by the following superpotential:
\begin{eqnarray}
\nonumber {\cal W}_\nu &=&
   \alpha_{N_1} \Lambda (NN)_{\mathbf{3}_S}Y_{\mathbf{3}}^{(2)}
 + \alpha_{\nu_1} (NF_1)_{\mathbf{3}}Y_{\mathbf{3}}^{(2)} H_{5}
 + \alpha_{\nu_2} (NF_2)_{\mathbf{3}}Y_{\mathbf{3}}^{(4)} H_{5}
 + \alpha_{\nu_3} (NF_3)_{\mathbf{3}}Y_{\mathbf{3}}^{(2)} H_{5}
\,,\\
\nonumber {\cal W}_d &=&
   \alpha_{d_1} (F_1T)_{\mathbf{3}}Y_{\mathbf{3}}^{(4)} H_{\overline{5}}
 + \alpha_{d_2} (F_2T)_{\mathbf{3}}Y_{\mathbf{3},I}^{(6)} H_{\overline{5}}
 + \alpha_{d_4} (F_2T)_{\mathbf{3}}Y_{\mathbf{3},II}^{(6)} H_{\overline{5}}
 + \alpha_{d_3} (F_3T)_{\mathbf{3}}Y_{\mathbf{3}}^{(4)} H_{\overline{5}}
\\\nonumber &&
 + \alpha^\prime_{d_1} (F_1T)_{\mathbf{3}}Y_{\mathbf{3}}^{(4)} H_{\overline{45}}
 + \alpha^\prime_{d_2} (F_2T)_{\mathbf{3}}Y_{\mathbf{3},I}^{(6)} H_{\overline{45}}
 + \alpha^\prime_{d_4} (F_2T)_{\mathbf{3}}Y_{\mathbf{3},II}^{(6)} H_{\overline{45}}
 + \alpha^\prime_{d_3} (F_3T)_{\mathbf{3}}Y_{\mathbf{3}}^{(4)} H_{\overline{45}}
\,,\\
{\cal W}_u &=&
   \alpha_{u_1}  (TT)_{\mathbf{1}}Y_{\mathbf{1}}^{(6)} H_{5}
 + \alpha_{u_2}  (TT)_{\mathbf{3}_S}Y_{\mathbf{3},I}^{(6)} H_{5}
 + \alpha_{u_3}  (TT)_{\mathbf{3}_S}Y_{\mathbf{3},II}^{(6)} H_{5}
\,.
\end{eqnarray}
The right-handed neutrino mass matrix and the Yukawa matrix are given by
\begin{eqnarray}
{\cal Y}_5^{\nu} &=& \left(
\begin{array}{ccc}
 \alpha_{\nu_1}Y_{\mathbf{3},1}^{(2)} ~&~ \alpha_{\nu_2}Y_{\mathbf{3},2}^{(4)} ~&~ \alpha_{\nu_3}Y_{\mathbf{3},3}^{(2)} \\
 \alpha_{\nu_1}Y_{\mathbf{3},3}^{(2)} ~&~ \alpha_{\nu_2}Y_{\mathbf{3},1}^{(4)} ~&~ \alpha_{\nu_3}Y_{\mathbf{3},2}^{(2)} \\
 \alpha_{\nu_1}Y_{\mathbf{3},2}^{(2)} ~&~ \alpha_{\nu_2}Y_{\mathbf{3},3}^{(4)}  ~&~ \alpha_{\nu_3}Y_{\mathbf{3},1}^{(2)} \\
\end{array}
\right)\,,\nonumber\\
{\cal Y}_{\overline{5}}^d &=& \left(
\begin{array}{ccc}
 \alpha_{d_1}Y_{\mathbf{3},1}^{(4)} ~&~ \alpha_{d_1}Y_{\mathbf{3},3}^{(4)} ~&~ \alpha_{d_1}Y_{\mathbf{3},2}^{(4)} \\
 \alpha_{d_2}Y_{\mathbf{3}I,2}^{(6)}+\alpha_{d_4}Y_{\mathbf{3}II,2}^{(6)} ~&~ \alpha_{d_2}Y_{\mathbf{3}I,1}^{(6)}+\alpha_{d_4}Y_{\mathbf{3}II,1}^{(6)} ~&~ \alpha_{d_2}Y_{\mathbf{3}I,3}^{(6)}+\alpha_{d_4}Y_{\mathbf{3}II,3}^{(6)} \\
 \alpha_{d_3}Y_{\mathbf{3},3}^{(4)} ~&~ \alpha_{d_3}Y_{\mathbf{3},2}^{(4)} ~&~ \alpha_{d_3}Y_{\mathbf{3},1}^{(4)} \\
\end{array}
\right)\,,\nonumber\\
{\cal Y}_5^u &=& \left(
\begin{array}{ccc}
 \alpha_{u_1}Y_\mathbf{1}^{(6)}+2 \alpha_{u_2}Y_{\mathbf{3}I,1}^{(6)}+2 \alpha_{u_3}Y_{\mathbf{3}II,1}^{(6)} ~&~ -\alpha_{u_2}Y_{\mathbf{3}I,3}^{(6)}-\alpha_{u_3}Y_{\mathbf{3}II,3}^{(6)} ~&~ -\alpha_{u_2}Y_{\mathbf{3}I,2}^{(6)}-\alpha_{u_3}Y_{\mathbf{3}II,2}^{(6)} \\
 -\alpha_{u_2}Y_{\mathbf{3}I,3}^{(6)}-\alpha_{u_3}Y_{\mathbf{3}II,3}^{(6)} ~&~ 2 \left(\alpha_{u_2}Y_{\mathbf{3}I,2}^{(6)}+\alpha_{u_3}Y_{\mathbf{3}II,2}^{(6)}\right) ~&~ \alpha_{u_1}Y_\mathbf{1}^{(6)}-\alpha_{u_2}Y_{\mathbf{3}I,1}^{(6)}-\alpha_{u_3}Y_{\mathbf{3}II,1}^{(6)}  \\
 -\alpha_{u_2}Y_{\mathbf{3}I,2}^{(6)}-\alpha_{u_3}Y_{\mathbf{3}II,2}^{(6)} ~&~ \alpha_{u_1}Y_\mathbf{1}^{(6)}-\alpha_{u_2}Y_{\mathbf{3}I,1}^{(6)}-\alpha_{u_3}Y_{\mathbf{3}II,1}^{(6)}  ~&~ 2 \left(\alpha_{u_2}Y_{\mathbf{3}I,3}^{(6)}+\alpha_{u_3}Y_{\mathbf{3}II,3}^{(6)}\right) \\
\end{array}
\right)\,.
\end{eqnarray}
The $M_N$ matrix coincides with the corresponding one of the model ${\cal I}_{1}$.
The parameters $\alpha_{N_1}$, $\alpha_{\nu_1}$, $\alpha_{\nu_2}$, $\alpha_{\nu_3}$, $\alpha_{d_1}$, $\alpha_{d_2}$, $\alpha_{d_3}$, $\alpha_{d_1}^\prime$ and $\alpha_{u_3}$ can be taken real by field redefinition, and the remaining parameters are complex.
\item{~Model ${\cal II}_{2}$~:~$(\rho_N,\rho_{F_1},\rho_{F_2},\rho_{F_3},\rho_{T})=(\mathbf{3},\mathbf{1^{\prime\prime}},\mathbf{1^{\prime\prime}},\mathbf{1},\mathbf{3})\,,~ (k_{N},k_{F_1},k_{F_2},k_{F_3},k_{T})=(1,1,3,1,3)$}

The modular invariant superpotential for quark and lepton mass is given by
\begin{eqnarray}
\nonumber {\cal W}_\nu &=&
   \alpha_{N_1} \Lambda (NN)_{\mathbf{3}_S}Y_{\mathbf{3}}^{(2)}
 + \alpha_{\nu_1} (NF_1)_{\mathbf{3}}Y_{\mathbf{3}}^{(2)} H_{5}
 + \alpha_{\nu_2} (NF_2)_{\mathbf{3}}Y_{\mathbf{3}}^{(4)} H_{5}
 + \alpha_{\nu_3} (NF_3)_{\mathbf{3}}Y_{\mathbf{3}}^{(2)} H_{5}
\,,\\
\nonumber {\cal W}_d &=&
   \alpha_{d_1} (F_1T)_{\mathbf{3}}Y_{\mathbf{3}}^{(4)} H_{\overline{5}}
 + \alpha_{d_2} (F_2T)_{\mathbf{3}}Y_{\mathbf{3},I}^{(6)} H_{\overline{5}}
 + \alpha_{d_4} (F_2T)_{\mathbf{3}}Y_{\mathbf{3},II}^{(6)} H_{\overline{5}}
 + \alpha_{d_3} (F_3T)_{\mathbf{3}}Y_{\mathbf{3}}^{(4)} H_{\overline{5}}
\\\nonumber &&
 + \alpha^\prime_{d_1} (F_1T)_{\mathbf{3}}Y_{\mathbf{3}}^{(4)} H_{\overline{45}}
 + \alpha^\prime_{d_2} (F_2T)_{\mathbf{3}}Y_{\mathbf{3},I}^{(6)} H_{\overline{45}}
 + \alpha^\prime_{d_4} (F_2T)_{\mathbf{3}}Y_{\mathbf{3},II}^{(6)} H_{\overline{45}}
 + \alpha^\prime_{d_3} (F_3T)_{\mathbf{3}}Y_{\mathbf{3}}^{(4)} H_{\overline{45}}
\,,\\
{\cal W}_u &=&
   \alpha_{u_1}  (TT)_{\mathbf{1}}Y_{\mathbf{1}}^{(6)} H_{5}
 + \alpha_{u_2}  (TT)_{\mathbf{3}_S}Y_{\mathbf{3},I}^{(6)} H_{5}
 + \alpha_{u_3}  (TT)_{\mathbf{3}_S}Y_{\mathbf{3},II}^{(6)} H_{5}
\,.
\end{eqnarray}
The right-handed neutrino mass matrix and the Yukawa matrix are given by
\begin{eqnarray}
{\cal Y}_5^{\nu} &=& \left(
\begin{array}{ccc}
 \alpha_{\nu_1}Y_{\mathbf{3},2}^{(2)} ~&~ \alpha_{\nu_2}Y_{\mathbf{3},2}^{(4)} ~&~ \alpha_{\nu_3}Y_{\mathbf{3},1}^{(2)} \\
 \alpha_{\nu_1}Y_{\mathbf{3},1}^{(2)} ~&~ \alpha_{\nu_2}Y_{\mathbf{3},1}^{(4)} ~&~ \alpha_{\nu_3}Y_{\mathbf{3},3}^{(2)} \\
 \alpha_{\nu_1}Y_{\mathbf{3},3}^{(2)} ~&~ \alpha_{\nu_2}Y_{\mathbf{3},3}^{(4)}  ~&~ \alpha_{\nu_3}Y_{\mathbf{3},2}^{(2)} \\
\end{array}
\right)\,,\nonumber\\
{\cal Y}_{\overline{5}}^d &=& \left(
\begin{array}{ccc}
 \alpha_{d_1}Y_{\mathbf{3},2}^{(4)} ~&~ \alpha_{d_1}Y_{\mathbf{3},1}^{(4)} ~&~ \alpha_{d_1}Y_{\mathbf{3},3}^{(4)} \\
 \alpha_{d_2}Y_{\mathbf{3}I,2}^{(6)}+\alpha_{d_4}Y_{\mathbf{3}II,2}^{(6)} ~&~ \alpha_{d_2}Y_{\mathbf{3}I,1}^{(6)}+\alpha_{d_4}Y_{\mathbf{3}II,1}^{(6)} ~&~ \alpha_{d_2}Y_{\mathbf{3}I,3}^{(6)}+\alpha_{d_4}Y_{\mathbf{3}II,3}^{(6)} \\
 \alpha_{d_3}Y_{\mathbf{3},1}^{(4)} ~&~ \alpha_{d_3}Y_{\mathbf{3},3}^{(4)} ~&~ \alpha_{d_3}Y_{\mathbf{3},2}^{(4)} \\
\end{array}
\right)\,.
\end{eqnarray}
The matrices $M_N$ and ${\cal Y}_5^u$ are identical with those of the models ${\cal I}_{1}$ and ${\cal II}_{1}$ respectively.
The phases of the parameters $\alpha_{N_1}$, $\alpha_{\nu_1}$, $\alpha_{\nu_2}$, $\alpha_{\nu_3}$, $\alpha_{d_1}$, $\alpha_{d_2}$, $\alpha_{d_3}$, $\alpha_{d_1}^\prime$ and $\alpha_{u_3}$ are unphysical, and $\alpha_{d_4}$, $\alpha_{d_2}^\prime$, $\alpha_{d_4}^\prime$, $\alpha_{d_3}^\prime$, $\alpha_{u_1}$ and $\alpha_{u_2}$ are generically complex.
\end{itemize}
\subsection{Type-V}
\begin{itemize}[leftmargin=1.0em]
\item{~Model ${\cal V}_{1}$~:~$\begin{aligned}
&(\rho_{N_1},\rho_{N_2},\rho_{N_3},\rho_F,\rho_{T_1},\rho_{T_2},\rho_{T_3})=(\mathbf{1},\mathbf{1},\mathbf{1^{\prime\prime}},\mathbf{3},\mathbf{1^{\prime\prime}},\mathbf{1},\mathbf{1})\,,\\
& (k_{N_1},k_{N_2},k_{N_3},k_{F},k_{T_1},k_{T_2},k_{T_3})=(1,3,3,1,1,1,3)
\end{aligned}$}

We can read out the superpotential for quark and leptons as follows,
\begin{eqnarray}
\nonumber {\cal W}_\nu &=&
   \alpha_{N_1} \Lambda (N_1N_2)_{\mathbf{1}}Y_{\mathbf{1}}^{(4)}
 + \alpha_{N_2} \Lambda (N_1N_3)_{\mathbf{1^{\prime\prime}}}Y_{\mathbf{1^\prime}}^{(4)}
 + \alpha_{N_3} \Lambda (N_2N_2)_{\mathbf{1}}Y_{\mathbf{1}}^{(6)}
 + \alpha_{\nu_1} (N_1F)_{\mathbf{3}}Y_{\mathbf{3}}^{(2)} H_{5}
\\
\nonumber&&
 + \alpha_{\nu_2} (N_2F)_{\mathbf{3}}Y_{\mathbf{3}}^{(4)} H_{5}
 + \alpha_{\nu_3} (N_3F)_{\mathbf{3}}Y_{\mathbf{3}}^{(4)} H_{5}
\,,\\
\nonumber {\cal W}_d &=&
   \alpha_{d_1} (FT_1)_{\mathbf{3}}Y_{\mathbf{3}}^{(2)} H_{\overline{5}}
 + \alpha_{d_2} (FT_2)_{\mathbf{3}}Y_{\mathbf{3}}^{(2)} H_{\overline{5}}
 + \alpha_{d_3} (FT_3)_{\mathbf{3}}Y_{\mathbf{3}}^{(4)} H_{\overline{5}}
\\\nonumber &&
 + \alpha^\prime_{d_1} (FT_1)_{\mathbf{3}}Y_{\mathbf{3}}^{(2)} H_{\overline{45}}
 + \alpha^\prime_{d_2} (FT_2)_{\mathbf{3}}Y_{\mathbf{3}}^{(2)} H_{\overline{45}}
 + \alpha^\prime_{d_3} (FT_3)_{\mathbf{3}}Y_{\mathbf{3}}^{(4)} H_{\overline{45}}
\,,\\
{\cal W}_u &=&
   \alpha_{u_1}  (T_1T_3)_{\mathbf{1^{\prime\prime}}}Y_{\mathbf{1^\prime}}^{(4)} H_{5}
 + \alpha_{u_2}  (T_2T_3)_{\mathbf{1}}Y_{\mathbf{1}}^{(4)} H_{5}
 + \alpha_{u_3}  (T_3T_3)_{\mathbf{1}}Y_{\mathbf{1}}^{(6)} H_{5}
\,.
\end{eqnarray}
We can straightforwardly read out the right-handed neutrino mass matrix and the Yukawa matrices as follows,
\begin{eqnarray}
m_M^{\nu} &=& \Lambda  \left(
\begin{array}{ccc}
 0 ~&~ \alpha_{N_1}Y_\mathbf{1}^{(4)} ~&~ \alpha_{N_2}Y_{\mathbf{1}^\prime}^{(4)} \\
 \alpha_{N_1}Y_\mathbf{1}^{(4)} ~&~ Y_\mathbf{1}^{(6)} \alpha_{N_3} ~&~ 0 \\
 \alpha_{N_2}Y_{\mathbf{1}^\prime}^{(4)} ~&~ 0 ~&~ 0 \\
\end{array}
\right)\,,\nonumber\\
{\cal Y}_5^{\nu} &=& \left(
\begin{array}{ccc}
 \alpha_{\nu_1}Y_{\mathbf{3},1}^{(2)} ~&~ \alpha_{\nu_1}Y_{\mathbf{3},3}^{(2)} ~&~ \alpha_{\nu_1}Y_{\mathbf{3},2}^{(2)} \\
 \alpha_{\nu_2}Y_{\mathbf{3},1}^{(4)} ~&~ \alpha_{\nu_2}Y_{\mathbf{3},3}^{(4)}  ~&~ \alpha_{\nu_2}Y_{\mathbf{3},2}^{(4)} \\
 \alpha_{\nu_3}Y_{\mathbf{3},2}^{(4)} ~&~ \alpha_{\nu_3}Y_{\mathbf{3},1}^{(4)} ~&~ \alpha_{\nu_3}Y_{\mathbf{3},3}^{(4)}  \\
\end{array}
\right)\,.
\end{eqnarray}
The matrices ${\cal Y}_{\overline{5}}^d$ and ${\cal Y}_5^u$ are the same as those of the model ${\cal I}_{1}$ as well.
The parameters $\alpha_{N_1}$, $\alpha_{\nu_1}$, $\alpha_{\nu_2}$, $\alpha_{\nu_3}$, $\alpha_{d_1}$, $\alpha_{d_2}$, $\alpha_{d_3}$, $\alpha_{d_1}^\prime$ and $\alpha_{u_3}$ can be taken real by field redefinition, and $\alpha_{N_2}$, $\alpha_{N_3}$, $\alpha_{d_2}^\prime$, $\alpha_{d_3}^\prime$, $\alpha_{u_1}$ and $\alpha_{u_2}$ are complex.
\item{~Model ${\cal V}_{2}$~:~$\begin{aligned}
&(\rho_{N_1},\rho_{N_2},\rho_{N_3},\rho_F,\rho_{T_1},\rho_{T_2},\rho_{T_3})=(\mathbf{1},\mathbf{1^{\prime\prime}},\mathbf{1^{\prime}},\mathbf{3},\mathbf{1^{\prime\prime}},\mathbf{1},\mathbf{1})\,,\\
& (k_{N_1},k_{N_2},k_{N_3},k_{F},k_{T_1},k_{T_2},k_{T_3})=(3,1,3,1,1,1,3)
\end{aligned}$}

The modular invariant superpotentials for quark and lepton masses are of the following form,
\begin{eqnarray}
\nonumber {\cal W}_\nu &=&
   \alpha_{N_1} \Lambda (N_1N_1)_{\mathbf{1}}Y_{\mathbf{1}}^{(6)}
 + \alpha_{N_2} \Lambda (N_1N_2)_{\mathbf{1^{\prime\prime}}}Y_{\mathbf{1^\prime}}^{(4)}
 + \alpha_{N_3} \Lambda (N_2N_3)_{\mathbf{1}}Y_{\mathbf{1}}^{(4)}
 + \alpha_{\nu_1} (N_1F)_{\mathbf{3}}Y_{\mathbf{3}}^{(4)} H_{5}
\\
\nonumber&&
 + \alpha_{\nu_2} (N_2F)_{\mathbf{3}}Y_{\mathbf{3}}^{(2)} H_{5}
 + \alpha_{\nu_3} (N_3F)_{\mathbf{3}}Y_{\mathbf{3}}^{(4)} H_{5}
\,,\\
\nonumber {\cal W}_d &=&
   \alpha_{d_1} (FT_1)_{\mathbf{3}}Y_{\mathbf{3}}^{(2)} H_{\overline{5}}
 + \alpha_{d_2} (FT_2)_{\mathbf{3}}Y_{\mathbf{3}}^{(2)} H_{\overline{5}}
 + \alpha_{d_3} (FT_3)_{\mathbf{3}}Y_{\mathbf{3}}^{(4)} H_{\overline{5}}
\\\nonumber &&
 + \alpha^\prime_{d_1} (FT_1)_{\mathbf{3}}Y_{\mathbf{3}}^{(2)} H_{\overline{45}}
 + \alpha^\prime_{d_2} (FT_2)_{\mathbf{3}}Y_{\mathbf{3}}^{(2)} H_{\overline{45}}
 + \alpha^\prime_{d_3} (FT_3)_{\mathbf{3}}Y_{\mathbf{3}}^{(4)} H_{\overline{45}}
\,,\\
{\cal W}_u &=&
   \alpha_{u_1}  (T_1T_3)_{\mathbf{1^{\prime\prime}}}Y_{\mathbf{1^\prime}}^{(4)} H_{5}
 + \alpha_{u_2}  (T_2T_3)_{\mathbf{1}}Y_{\mathbf{1}}^{(4)} H_{5}
 + \alpha_{u_3}  (T_3T_3)_{\mathbf{1}}Y_{\mathbf{1}}^{(6)} H_{5}
\,.
\end{eqnarray}
We can straightforwardly read out the right-handed neutrino mass matrix and the Yukawa matrices as follows,
\begin{eqnarray}
m_M^{\nu} &=& \Lambda  \left(
\begin{array}{ccc}
 Y_\mathbf{1}^{(6)} \alpha_{N_1} ~&~ \alpha_{N_2}Y_{\mathbf{1}^\prime}^{(4)} ~&~ 0 \\
 \alpha_{N_2}Y_{\mathbf{1}^\prime}^{(4)} ~&~ 0 ~&~ \alpha_{N_3}Y_\mathbf{1}^{(4)} \\
 0 ~&~ \alpha_{N_3}Y_\mathbf{1}^{(4)} ~&~ 0 \\
\end{array}
\right)\,,\nonumber\\
{\cal Y}_5^{\nu} &=& \left(
\begin{array}{ccc}
 \alpha_{\nu_1}Y_{\mathbf{3},1}^{(4)} ~&~ \alpha_{\nu_1}Y_{\mathbf{3},3}^{(4)} ~&~ \alpha_{\nu_1}Y_{\mathbf{3},2}^{(4)} \\
 \alpha_{\nu_2}Y_{\mathbf{3},2}^{(2)} ~&~ \alpha_{\nu_2}Y_{\mathbf{3},1}^{(2)} ~&~ \alpha_{\nu_2}Y_{\mathbf{3},3}^{(2)} \\
 \alpha_{\nu_3}Y_{\mathbf{3},3}^{(4)}  ~&~ \alpha_{\nu_3}Y_{\mathbf{3},2}^{(4)} ~&~ \alpha_{\nu_3}Y_{\mathbf{3},1}^{(4)} \\
\end{array}
\right)\,.
\end{eqnarray}
The matrices ${\cal Y}_{\overline{5}}^d$ and ${\cal Y}_5^u$ coincide with the corresponding ones of the model ${\cal V}_{1}$ as well.
The parameters $\alpha_{N_1}$, $\alpha_{\nu_1}$, $\alpha_{\nu_2}$, $\alpha_{\nu_3}$, $\alpha_{d_1}$, $\alpha_{d_2}$, $\alpha_{d_3}$, $\alpha_{d_1}^\prime$ and $\alpha_{u_3}$ can be taken real by field redefinition, and $\alpha_{N_2}$, $\alpha_{N_3}$, $\alpha_{d_2}^\prime$, $\alpha_{d_3}^\prime$, $\alpha_{u_1}$ and $\alpha_{u_2}$ are generically complex.
\item{~Model ${\cal V}_{3}$~:~$\begin{aligned}
&(\rho_{N_1},\rho_{N_2},\rho_{N_3},\rho_F,\rho_{T_1},\rho_{T_2},\rho_{T_3})=(\mathbf{1},\mathbf{1^{\prime\prime}},\mathbf{1^{\prime}},\mathbf{3},\mathbf{1^{\prime\prime}},\mathbf{1},\mathbf{1^{\prime}})\,,\\
& (k_{N_1},k_{N_2},k_{N_3},k_{F},k_{T_1},k_{T_2},k_{T_3})=(0,2,2,2,2,2,2)
\end{aligned}$}

The modular invariant superpotential for quark and lepton mass is given by
\begin{eqnarray}
\nonumber {\cal W}_\nu &=&
   \alpha_{N_1} \Lambda (N_1N_1)_{\mathbf{1}}
 + \alpha_{N_2} \Lambda (N_2N_3)_{\mathbf{1}}Y_{\mathbf{1}}^{(4)}
 + \alpha_{N_3} \Lambda (N_3N_3)_{\mathbf{1^{\prime\prime}}}Y_{\mathbf{1^\prime}}^{(4)}
 + \alpha_{\nu_1} (N_1F)_{\mathbf{3}}Y_{\mathbf{3}}^{(2)} H_{5}
\\
\nonumber&&
 + \alpha_{\nu_2} (N_2F)_{\mathbf{3}}Y_{\mathbf{3}}^{(4)} H_{5}
 + \alpha_{\nu_3} (N_3F)_{\mathbf{3}}Y_{\mathbf{3}}^{(4)} H_{5}
\,,\\
\nonumber {\cal W}_d &=&
   \alpha_{d_1} (FT_1)_{\mathbf{3}}Y_{\mathbf{3}}^{(4)} H_{\overline{5}}
 + \alpha_{d_2} (FT_2)_{\mathbf{3}}Y_{\mathbf{3}}^{(4)} H_{\overline{5}}
 + \alpha_{d_3} (FT_3)_{\mathbf{3}}Y_{\mathbf{3}}^{(4)} H_{\overline{5}}
\\\nonumber &&
 + \alpha^\prime_{d_1} (FT_1)_{\mathbf{3}}Y_{\mathbf{3}}^{(4)} H_{\overline{45}}
 + \alpha^\prime_{d_2} (FT_2)_{\mathbf{3}}Y_{\mathbf{3}}^{(4)} H_{\overline{45}}
 + \alpha^\prime_{d_3} (FT_3)_{\mathbf{3}}Y_{\mathbf{3}}^{(4)} H_{\overline{45}}
\,,\\
{\cal W}_u &=&
   \alpha_{u_1}  (T_1T_2)_{\mathbf{1^{\prime\prime}}}Y_{\mathbf{1^\prime}}^{(4)} H_{5}
 + \alpha_{u_2}  (T_1T_3)_{\mathbf{1}}Y_{\mathbf{1}}^{(4)} H_{5}
 + \alpha_{u_3}  (T_2T_2)_{\mathbf{1}}Y_{\mathbf{1}}^{(4)} H_{5}
 + \alpha_{u_4}  (T_3T_3)_{\mathbf{1^{\prime\prime}}}Y_{\mathbf{1^\prime}}^{(4)} H_{5}
\,,
\end{eqnarray}
which lead to
\begin{eqnarray}
m_M^{\nu} &=& \Lambda  \left(
\begin{array}{ccc}
 \alpha_{N_1} ~&~ 0 ~&~ 0 \\
 0 ~&~ 0 ~&~ \alpha_{N_2}Y_\mathbf{1}^{(4)} \\
 0 ~&~ \alpha_{N_2}Y_\mathbf{1}^{(4)} ~&~ \alpha_{N_3}Y_{\mathbf{1}^\prime}^{(4)} \\
\end{array}
\right)\,,\nonumber\\
{\cal Y}_5^{\nu} &=& \left(
\begin{array}{ccc}
 \alpha_{\nu_1}Y_{\mathbf{3},1}^{(2)} ~&~ \alpha_{\nu_1}Y_{\mathbf{3},3}^{(2)} ~&~ \alpha_{\nu_1}Y_{\mathbf{3},2}^{(2)} \\
 \alpha_{\nu_2}Y_{\mathbf{3},2}^{(4)} ~&~ \alpha_{\nu_2}Y_{\mathbf{3},1}^{(4)} ~&~ \alpha_{\nu_2}Y_{\mathbf{3},3}^{(4)}  \\
 \alpha_{\nu_3}Y_{\mathbf{3},3}^{(4)}  ~&~ \alpha_{\nu_3}Y_{\mathbf{3},2}^{(4)} ~&~ \alpha_{\nu_3}Y_{\mathbf{3},1}^{(4)} \\
\end{array}
\right)\,,\nonumber\\
{\cal Y}_{\overline{5}}^d &=& \left(
\begin{array}{ccc}
 \alpha_{d_1}Y_{\mathbf{3},2}^{(4)} ~&~ \alpha_{d_2}Y_{\mathbf{3},1}^{(4)} ~&~ \alpha_{d_3}Y_{\mathbf{3},3}^{(4)} \\
 \alpha_{d_1}Y_{\mathbf{3},1}^{(4)} ~&~ \alpha_{d_2}Y_{\mathbf{3},3}^{(4)} ~&~ \alpha_{d_3}Y_{\mathbf{3},2}^{(4)} \\
 \alpha_{d_1}Y_{\mathbf{3},3}^{(4)} ~&~ \alpha_{d_2}Y_{\mathbf{3},2}^{(4)} ~&~ \alpha_{d_3}Y_{\mathbf{3},1}^{(4)} \\
\end{array}
\right)\,.
\end{eqnarray}
The ${\cal Y}_5^u$ matrix coincides with the corresponding one of the model ${\cal I}_{8}$.
The parameters $\alpha_{N_1}$, $\alpha_{\nu_1}$, $\alpha_{\nu_2}$, $\alpha_{\nu_3}$, $\alpha_{d_1}$, $\alpha_{d_2}$, $\alpha_{d_3}$, $\alpha_{d_1}^\prime$ and $\alpha_{u_3}$ can be taken real by field redefinition, and $\alpha_{N_2}$, $\alpha_{N_3}$, $\alpha_{d_2}^\prime$, $\alpha_{d_3}^\prime$, $\alpha_{u_1}$, $\alpha_{u_2}$ and $\alpha_{u_4}$ are complex.
\item{~Model ${\cal V}_{4}$~:~$\begin{aligned}
&(\rho_{N_1},\rho_{N_2},\rho_{N_3},\rho_F,\rho_{T_1},\rho_{T_2},\rho_{T_3})=(\mathbf{1},\mathbf{1^{\prime\prime}},\mathbf{1^{\prime}},\mathbf{3},\mathbf{1^{\prime\prime}},\mathbf{1},\mathbf{1^{\prime}})\,,\\
& (k_{N_1},k_{N_2},k_{N_3},k_{F},k_{T_1},k_{T_2},k_{T_3})=(4,2,2,0,2,2,2)
\end{aligned}$}

We can read out the superpotential for quark and leptons as follows,
\begin{eqnarray}
\nonumber {\cal W}_\nu &=&
   \alpha_{N_1} \Lambda (N_1N_1)_{\mathbf{1}}Y_{\mathbf{1}}^{(8)}
 + \alpha_{N_2} \Lambda (N_2N_3)_{\mathbf{1}}Y_{\mathbf{1}}^{(4)}
 + \alpha_{N_3} \Lambda (N_3N_3)_{\mathbf{1^{\prime\prime}}}Y_{\mathbf{1^\prime}}^{(4)}
 + \alpha_{\nu_1} (N_1F)_{\mathbf{3}}Y_{\mathbf{3}}^{(4)} H_{5}
\\
\nonumber&&
 + \alpha_{\nu_2} (N_2F)_{\mathbf{3}}Y_{\mathbf{3}}^{(2)} H_{5}
 + \alpha_{\nu_3} (N_3F)_{\mathbf{3}}Y_{\mathbf{3}}^{(2)} H_{5}
\,,\\
\nonumber {\cal W}_d &=&
   \alpha_{d_1} (FT_1)_{\mathbf{3}}Y_{\mathbf{3}}^{(2)} H_{\overline{5}}
 + \alpha_{d_2} (FT_2)_{\mathbf{3}}Y_{\mathbf{3}}^{(2)} H_{\overline{5}}
 + \alpha_{d_3} (FT_3)_{\mathbf{3}}Y_{\mathbf{3}}^{(2)} H_{\overline{5}}
\\\nonumber &&
 + \alpha^\prime_{d_1} (FT_1)_{\mathbf{3}}Y_{\mathbf{3}}^{(2)} H_{\overline{45}}
 + \alpha^\prime_{d_2} (FT_2)_{\mathbf{3}}Y_{\mathbf{3}}^{(2)} H_{\overline{45}}
 + \alpha^\prime_{d_3} (FT_3)_{\mathbf{3}}Y_{\mathbf{3}}^{(2)} H_{\overline{45}}
\,,\\
{\cal W}_u &=&
   \alpha_{u_1}  (T_1T_2)_{\mathbf{1^{\prime\prime}}}Y_{\mathbf{1^\prime}}^{(4)} H_{5}
 + \alpha_{u_2}  (T_1T_3)_{\mathbf{1}}Y_{\mathbf{1}}^{(4)} H_{5}
 + \alpha_{u_3}  (T_2T_2)_{\mathbf{1}}Y_{\mathbf{1}}^{(4)} H_{5}
 + \alpha_{u_4}  (T_3T_3)_{\mathbf{1^{\prime\prime}}}Y_{\mathbf{1^\prime}}^{(4)} H_{5}
\,.
\end{eqnarray}
which gives rise to
\begin{eqnarray}
m_M^{\nu} &=& \Lambda  \left(
\begin{array}{ccc}
 \alpha_{N_1}Y_\mathbf{1}^{(8)} ~&~ 0 ~&~ 0 \\
 0 ~&~ 0 ~&~ \alpha_{N_2}Y_\mathbf{1}^{(4)} \\
 0 ~&~ \alpha_{N_2}Y_\mathbf{1}^{(4)} ~&~ \alpha_{N_3}Y_{\mathbf{1}^\prime}^{(4)} \\
\end{array}
\right)\,,\nonumber\\
{\cal Y}_5^{\nu} &=& \left(
\begin{array}{ccc}
 \alpha_{\nu_1}Y_{\mathbf{3},1}^{(4)} ~&~ \alpha_{\nu_1}Y_{\mathbf{3},3}^{(4)} ~&~ \alpha_{\nu_1}Y_{\mathbf{3},2}^{(4)} \\
 \alpha_{\nu_2}Y_{\mathbf{3},2}^{(2)} ~&~ \alpha_{\nu_2}Y_{\mathbf{3},1}^{(2)} ~&~ \alpha_{\nu_2}Y_{\mathbf{3},3}^{(2)} \\
 \alpha_{\nu_3}Y_{\mathbf{3},3}^{(2)} ~&~ \alpha_{\nu_3}Y_{\mathbf{3},2}^{(2)} ~&~ \alpha_{\nu_3}Y_{\mathbf{3},1}^{(2)} \\
\end{array}
\right)\,.
\end{eqnarray}
The matrices ${\cal Y}_{\overline{5}}^d$ and ${\cal Y}_5^u$ are the same as those of the model ${\cal I}_{8}$ as well.
The parameters $\alpha_{N_1}$, $\alpha_{\nu_1}$, $\alpha_{\nu_2}$, $\alpha_{\nu_3}$, $\alpha_{d_1}$, $\alpha_{d_2}$, $\alpha_{d_3}$, $\alpha_{d_1}^\prime$ and $\alpha_{u_3}$ can taken to be real without loss of generality, while $\alpha_{N_2}$, $\alpha_{N_3}$, $\alpha_{d_2}^\prime$, $\alpha_{d_3}^\prime$, $\alpha_{u_1}$, $\alpha_{u_2}$ and $\alpha_{u_4}$ are generically complex.
\item{~Model ${\cal V}_{5}$~:~$\begin{aligned}
&(\rho_{N_1},\rho_{N_2},\rho_{N_3},\rho_F,\rho_{T_1},\rho_{T_2},\rho_{T_3})=(\mathbf{1},\mathbf{1^{\prime\prime}},\mathbf{1^{\prime}},\mathbf{3},\mathbf{1^{\prime\prime}},\mathbf{1^{\prime}},\mathbf{1^{\prime}})\,,\\
& (k_{N_1},k_{N_2},k_{N_3},k_{F},k_{T_1},k_{T_2},k_{T_3})=(4,2,2,0,2,2,4)
\end{aligned}$}

The quark and lepton masses are described by the following superpotential:
\begin{eqnarray}
\nonumber {\cal W}_\nu &=&
   \alpha_{N_1} \Lambda (N_1N_1)_{\mathbf{1}}Y_{\mathbf{1}}^{(8)}
 + \alpha_{N_2} \Lambda (N_2N_3)_{\mathbf{1}}Y_{\mathbf{1}}^{(4)}
 + \alpha_{N_3} \Lambda (N_3N_3)_{\mathbf{1^{\prime\prime}}}Y_{\mathbf{1^\prime}}^{(4)}
 + \alpha_{\nu_1} (N_1F)_{\mathbf{3}}Y_{\mathbf{3}}^{(4)} H_{5}
\\
\nonumber&&
 + \alpha_{\nu_2} (N_2F)_{\mathbf{3}}Y_{\mathbf{3}}^{(2)} H_{5}
 + \alpha_{\nu_3} (N_3F)_{\mathbf{3}}Y_{\mathbf{3}}^{(2)} H_{5}
\,,\\
\nonumber {\cal W}_d &=&
   \alpha_{d_1} (FT_1)_{\mathbf{3}}Y_{\mathbf{3}}^{(2)} H_{\overline{5}}
 + \alpha_{d_2} (FT_2)_{\mathbf{3}}Y_{\mathbf{3}}^{(2)} H_{\overline{5}}
 + \alpha_{d_3} (FT_3)_{\mathbf{3}}Y_{\mathbf{3}}^{(4)} H_{\overline{5}}
\\\nonumber &&
 + \alpha^\prime_{d_1} (FT_1)_{\mathbf{3}}Y_{\mathbf{3}}^{(2)} H_{\overline{45}}
 + \alpha^\prime_{d_2} (FT_2)_{\mathbf{3}}Y_{\mathbf{3}}^{(2)} H_{\overline{45}}
 + \alpha^\prime_{d_3} (FT_3)_{\mathbf{3}}Y_{\mathbf{3}}^{(4)} H_{\overline{45}}
\,,\\
{\cal W}_u &=&
   \alpha_{u_1}  (T_1T_2)_{\mathbf{1}}Y_{\mathbf{1}}^{(4)} H_{5}
 + \alpha_{u_2}  (T_1T_3)_{\mathbf{1}}Y_{\mathbf{1}}^{(6)} H_{5}
 + \alpha_{u_3}  (T_2T_2)_{\mathbf{1^{\prime\prime}}}Y_{\mathbf{1^\prime}}^{(4)} H_{5}
 + \alpha_{u_4}  (T_3T_3)_{\mathbf{1^{\prime\prime}}}Y_{\mathbf{1^\prime}}^{(8)} H_{5}
\,.
\end{eqnarray}
The matrices $M_N$ and ${\cal Y}_5^{\nu}$ are the same as those of the model ${\cal V}_{4}$ as well.
The ${\cal Y}_{\overline{5}}^d$ matrix is identical with the corresponding one of the model ${\cal I}_{15}$.
The ${\cal Y}_5^u$ matrix is in common with that of the model ${\cal I}_{11}$.
The parameters $\alpha_{N_1}$, $\alpha_{\nu_1}$, $\alpha_{\nu_2}$, $\alpha_{\nu_3}$, $\alpha_{d_1}$, $\alpha_{d_2}$, $\alpha_{d_3}$, $\alpha_{d_1}^\prime$ and $\alpha_{u_3}$ can be taken real by field redefinition, and the other parameters are complex.
\item{~Model ${\cal V}_{6}$~:~$\begin{aligned}
&(\rho_{N_1},\rho_{N_2},\rho_{N_3},\rho_F,\rho_{T_1},\rho_{T_2},\rho_{T_3})=(\mathbf{1},\mathbf{1^{\prime\prime}},\mathbf{1^{\prime}},\mathbf{3},\mathbf{1^{\prime\prime}},\mathbf{1^{\prime\prime}},\mathbf{1^{\prime}})\,,\\
& (k_{N_1},k_{N_2},k_{N_3},k_{F},k_{T_1},k_{T_2},k_{T_3})=(4,2,2,0,2,4,2)
\end{aligned}$}

The modular invariant superpotential for quark and lepton mass is given by
\begin{eqnarray}
\nonumber {\cal W}_\nu &=&
   \alpha_{N_1} \Lambda (N_1N_1)_{\mathbf{1}}Y_{\mathbf{1}}^{(8)}
 + \alpha_{N_2} \Lambda (N_2N_3)_{\mathbf{1}}Y_{\mathbf{1}}^{(4)}
 + \alpha_{N_3} \Lambda (N_3N_3)_{\mathbf{1^{\prime\prime}}}Y_{\mathbf{1^\prime}}^{(4)}
 + \alpha_{\nu_1} (N_1F)_{\mathbf{3}}Y_{\mathbf{3}}^{(4)} H_{5}
\\
\nonumber&&
 + \alpha_{\nu_2} (N_2F)_{\mathbf{3}}Y_{\mathbf{3}}^{(2)} H_{5}
 + \alpha_{\nu_3} (N_3F)_{\mathbf{3}}Y_{\mathbf{3}}^{(2)} H_{5}
\,,\\
\nonumber {\cal W}_d &=&
   \alpha_{d_1} (FT_1)_{\mathbf{3}}Y_{\mathbf{3}}^{(2)} H_{\overline{5}}
 + \alpha_{d_2} (FT_2)_{\mathbf{3}}Y_{\mathbf{3}}^{(4)} H_{\overline{5}}
 + \alpha_{d_3} (FT_3)_{\mathbf{3}}Y_{\mathbf{3}}^{(2)} H_{\overline{5}}
\\\nonumber &&
 + \alpha^\prime_{d_1} (FT_1)_{\mathbf{3}}Y_{\mathbf{3}}^{(2)} H_{\overline{45}}
 + \alpha^\prime_{d_2} (FT_2)_{\mathbf{3}}Y_{\mathbf{3}}^{(4)} H_{\overline{45}}
 + \alpha^\prime_{d_3} (FT_3)_{\mathbf{3}}Y_{\mathbf{3}}^{(2)} H_{\overline{45}}
\,,\\
{\cal W}_u &=&
   \alpha_{u_1}  (T_1T_3)_{\mathbf{1}}Y_{\mathbf{1}}^{(4)} H_{5}
 + \alpha_{u_2}  (T_2T_2)_{\mathbf{1^{\prime}}}Y_{\mathbf{1^{\prime\prime}}}^{(8)} H_{5}
 + \alpha_{u_3}  (T_2T_3)_{\mathbf{1}}Y_{\mathbf{1}}^{(6)} H_{5}
 + \alpha_{u_4}  (T_3T_3)_{\mathbf{1^{\prime\prime}}}Y_{\mathbf{1^\prime}}^{(4)} H_{5}
\,.
\end{eqnarray}
The matrices $M_N$ and ${\cal Y}_5^{\nu}$ are identical with the corresponding ones of the model ${\cal V}_{4}$ as well.
The ${\cal Y}_{\overline{5}}^d$ matrix coincides with that of the model ${\cal I}_{16}$.
The ${\cal Y}_5^u$ matrix is identical with that of the model ${\cal I}_{12}$.
The phases of the parameters $\alpha_{N_1}$, $\alpha_{\nu_1}$, $\alpha_{\nu_2}$, $\alpha_{\nu_3}$, $\alpha_{d_1}$, $\alpha_{d_2}$, $\alpha_{d_3}$, $\alpha_{d_1}^\prime$ and $\alpha_{u_3}$ can be removed by field redefinition, and the remaining parameters are complex.
\item{~Model ${\cal V}_{7}$~:~$\begin{aligned}
&(\rho_{N_1},\rho_{N_2},\rho_{N_3},\rho_F,\rho_{T_1},\rho_{T_2},\rho_{T_3})=(\mathbf{1},\mathbf{1^{\prime\prime}},\mathbf{1^{\prime}},\mathbf{3},\mathbf{1^{\prime\prime}},\mathbf{1^{\prime\prime}},\mathbf{1^{\prime}})\,,\\
& (k_{N_1},k_{N_2},k_{N_3},k_{F},k_{T_1},k_{T_2},k_{T_3})=(4,2,2,0,2,4,4)
\end{aligned}$}

The modular invariant superpotentials for quark and lepton masses are of the following form,
\begin{eqnarray}
\nonumber {\cal W}_\nu &=&
   \alpha_{N_1} \Lambda (N_1N_1)_{\mathbf{1}}Y_{\mathbf{1}}^{(8)}
 + \alpha_{N_2} \Lambda (N_2N_3)_{\mathbf{1}}Y_{\mathbf{1}}^{(4)}
 + \alpha_{N_3} \Lambda (N_3N_3)_{\mathbf{1^{\prime\prime}}}Y_{\mathbf{1^\prime}}^{(4)}
 + \alpha_{\nu_1} (N_1F)_{\mathbf{3}}Y_{\mathbf{3}}^{(4)} H_{5}
\\
\nonumber&&
 + \alpha_{\nu_2} (N_2F)_{\mathbf{3}}Y_{\mathbf{3}}^{(2)} H_{5}
 + \alpha_{\nu_3} (N_3F)_{\mathbf{3}}Y_{\mathbf{3}}^{(2)} H_{5}
\,,\\
\nonumber {\cal W}_d &=&
   \alpha_{d_1} (FT_1)_{\mathbf{3}}Y_{\mathbf{3}}^{(2)} H_{\overline{5}}
 + \alpha_{d_2} (FT_2)_{\mathbf{3}}Y_{\mathbf{3}}^{(4)} H_{\overline{5}}
 + \alpha_{d_3} (FT_3)_{\mathbf{3}}Y_{\mathbf{3}}^{(4)} H_{\overline{5}}
\\\nonumber &&
 + \alpha^\prime_{d_1} (FT_1)_{\mathbf{3}}Y_{\mathbf{3}}^{(2)} H_{\overline{45}}
 + \alpha^\prime_{d_2} (FT_2)_{\mathbf{3}}Y_{\mathbf{3}}^{(4)} H_{\overline{45}}
 + \alpha^\prime_{d_3} (FT_3)_{\mathbf{3}}Y_{\mathbf{3}}^{(4)} H_{\overline{45}}
\,,\\
{\cal W}_u &=&
   \alpha_{u_1}  (T_1T_3)_{\mathbf{1}}Y_{\mathbf{1}}^{(6)} H_{5}
 + \alpha_{u_2}  (T_2T_2)_{\mathbf{1^{\prime}}}Y_{\mathbf{1^{\prime\prime}}}^{(8)} H_{5}
 + \alpha_{u_3}  (T_2T_3)_{\mathbf{1}}Y_{\mathbf{1}}^{(8)} H_{5}
 + \alpha_{u_4}  (T_3T_3)_{\mathbf{1^{\prime\prime}}}Y_{\mathbf{1^\prime}}^{(8)} H_{5}
\,.
\end{eqnarray}
The matrices $M_N$ and ${\cal Y}_5^{\nu}$ are identical with those of the model ${\cal V}_{4}$ as well.
The matrices ${\cal Y}_{\overline{5}}^d$ and ${\cal Y}_5^u$ are the same as those of the model ${\cal I}_{17}$ as well.
The parameters $\alpha_{N_1}$, $\alpha_{\nu_1}$, $\alpha_{\nu_2}$, $\alpha_{\nu_3}$, $\alpha_{d_1}$, $\alpha_{d_2}$, $\alpha_{d_3}$, $\alpha_{d_1}^\prime$ and $\alpha_{u_3}$ can be taken real by field redefinition, and $\alpha_{N_2}$, $\alpha_{N_3}$, $\alpha_{d_2}^\prime$, $\alpha_{d_3}^\prime$, $\alpha_{u_1}$, $\alpha_{u_2}$ and $\alpha_{u_4}$ are generically complex.
\item{~Model ${\cal V}_{8}$~:~$\begin{aligned}
&(\rho_{N_1},\rho_{N_2},\rho_{N_3},\rho_F,\rho_{T_1},\rho_{T_2},\rho_{T_3})=(\mathbf{1^{\prime}},\mathbf{1^{\prime}},\mathbf{1},\mathbf{3},\mathbf{1^{\prime\prime}},\mathbf{1},\mathbf{1^{\prime}})\,,\\
& (k_{N_1},k_{N_2},k_{N_3},k_{F},k_{T_1},k_{T_2},k_{T_3})=(2,4,2,0,2,2,2)
\end{aligned}$}

We can read out the the superpotential for the fermion masses as,
\begin{eqnarray}
\nonumber {\cal W}_\nu &=&
   \alpha_{N_1} \Lambda (N_1N_1)_{\mathbf{1^{\prime\prime}}}Y_{\mathbf{1^\prime}}^{(4)}
 + \alpha_{N_2} \Lambda (N_2N_2)_{\mathbf{1^{\prime\prime}}}Y_{\mathbf{1^\prime}}^{(8)}
 + \alpha_{N_3} \Lambda (N_3N_3)_{\mathbf{1}}Y_{\mathbf{1}}^{(4)}
 + \alpha_{\nu_1} (N_1F)_{\mathbf{3}}Y_{\mathbf{3}}^{(2)} H_{5}
\\
\nonumber&&
 + \alpha_{\nu_2} (N_2F)_{\mathbf{3}}Y_{\mathbf{3}}^{(4)} H_{5}
 + \alpha_{\nu_3} (N_3F)_{\mathbf{3}}Y_{\mathbf{3}}^{(2)} H_{5}
\,,\\
\nonumber {\cal W}_d &=&
   \alpha_{d_1} (FT_1)_{\mathbf{3}}Y_{\mathbf{3}}^{(2)} H_{\overline{5}}
 + \alpha_{d_2} (FT_2)_{\mathbf{3}}Y_{\mathbf{3}}^{(2)} H_{\overline{5}}
 + \alpha_{d_3} (FT_3)_{\mathbf{3}}Y_{\mathbf{3}}^{(2)} H_{\overline{5}}
\\\nonumber &&
 + \alpha^\prime_{d_1} (FT_1)_{\mathbf{3}}Y_{\mathbf{3}}^{(2)} H_{\overline{45}}
 + \alpha^\prime_{d_2} (FT_2)_{\mathbf{3}}Y_{\mathbf{3}}^{(2)} H_{\overline{45}}
 + \alpha^\prime_{d_3} (FT_3)_{\mathbf{3}}Y_{\mathbf{3}}^{(2)} H_{\overline{45}}
\,,\\
{\cal W}_u &=&
   \alpha_{u_1}  (T_1T_2)_{\mathbf{1^{\prime\prime}}}Y_{\mathbf{1^\prime}}^{(4)} H_{5}
 + \alpha_{u_2}  (T_1T_3)_{\mathbf{1}}Y_{\mathbf{1}}^{(4)} H_{5}
 + \alpha_{u_3}  (T_2T_2)_{\mathbf{1}}Y_{\mathbf{1}}^{(4)} H_{5}
 + \alpha_{u_4}  (T_3T_3)_{\mathbf{1^{\prime\prime}}}Y_{\mathbf{1^\prime}}^{(4)} H_{5}
\,.
\end{eqnarray}
We can straightforwardly read out the right-handed neutrino mass matrix and the Yukawa matrices as follows,
\begin{eqnarray}
m_M^{\nu} &=& \Lambda  \left(
\begin{array}{ccc}
 \alpha_{N_1}Y_{\mathbf{1}^\prime}^{(4)} ~&~ 0 ~&~ 0 \\
 0 ~&~ \alpha_{N_2}Y_{\mathbf{1}^\prime}^{(8)} ~&~ 0 \\
 0 ~&~ 0 ~&~ \alpha_{N_3}Y_\mathbf{1}^{(4)} \\
\end{array}
\right)\,,\nonumber\\
{\cal Y}_5^{\nu} &=& \left(
\begin{array}{ccc}
 \alpha_{\nu_1}Y_{\mathbf{3},3}^{(2)} ~&~ \alpha_{\nu_1}Y_{\mathbf{3},2}^{(2)} ~&~ \alpha_{\nu_1}Y_{\mathbf{3},1}^{(2)} \\
 \alpha_{\nu_2}Y_{\mathbf{3},3}^{(4)}  ~&~ \alpha_{\nu_2}Y_{\mathbf{3},2}^{(4)} ~&~ \alpha_{\nu_2}Y_{\mathbf{3},1}^{(4)} \\
 \alpha_{\nu_3}Y_{\mathbf{3},1}^{(2)} ~&~ \alpha_{\nu_3}Y_{\mathbf{3},3}^{(2)} ~&~ \alpha_{\nu_3}Y_{\mathbf{3},2}^{(2)} \\
\end{array}
\right)\,.
\end{eqnarray}
The matrices ${\cal Y}_{\overline{5}}^d$ and ${\cal Y}_5^u$ are identical with the corresponding ones of the model ${\cal V}_{4}$ as well.
The parameters $\alpha_{N_1}$, $\alpha_{\nu_1}$, $\alpha_{\nu_2}$, $\alpha_{\nu_3}$, $\alpha_{d_1}$, $\alpha_{d_2}$, $\alpha_{d_3}$, $\alpha_{d_1}^\prime$ and $\alpha_{u_3}$ can be taken real by field redefinition, and $\alpha_{N_2}$, $\alpha_{N_3}$, $\alpha_{d_2}^\prime$, $\alpha_{d_3}^\prime$, $\alpha_{u_1}$, $\alpha_{u_2}$ and $\alpha_{u_4}$ are generically complex.
\item{~Model ${\cal V}_{9}$~:~$\begin{aligned}
&(\rho_{N_1},\rho_{N_2},\rho_{N_3},\rho_F,\rho_{T_1},\rho_{T_2},\rho_{T_3})=(\mathbf{1^{\prime}},\mathbf{1^{\prime}},\mathbf{1},\mathbf{3},\mathbf{1^{\prime\prime}},\mathbf{1^{\prime}},\mathbf{1^{\prime}})\,,\\
& (k_{N_1},k_{N_2},k_{N_3},k_{F},k_{T_1},k_{T_2},k_{T_3})=(2,4,2,0,2,2,4)
\end{aligned}$}

The superpotentials relevant to quark and lepton masses are of the form,
\begin{eqnarray}
\nonumber {\cal W}_\nu &=&
   \alpha_{N_1} \Lambda (N_1N_1)_{\mathbf{1^{\prime\prime}}}Y_{\mathbf{1^\prime}}^{(4)}
 + \alpha_{N_2} \Lambda (N_2N_2)_{\mathbf{1^{\prime\prime}}}Y_{\mathbf{1^\prime}}^{(8)}
 + \alpha_{N_3} \Lambda (N_3N_3)_{\mathbf{1}}Y_{\mathbf{1}}^{(4)}
 + \alpha_{\nu_1} (N_1F)_{\mathbf{3}}Y_{\mathbf{3}}^{(2)} H_{5}
\\
\nonumber&&
 + \alpha_{\nu_2} (N_2F)_{\mathbf{3}}Y_{\mathbf{3}}^{(4)} H_{5}
 + \alpha_{\nu_3} (N_3F)_{\mathbf{3}}Y_{\mathbf{3}}^{(2)} H_{5}
\,,\\
\nonumber {\cal W}_d &=&
   \alpha_{d_1} (FT_1)_{\mathbf{3}}Y_{\mathbf{3}}^{(2)} H_{\overline{5}}
 + \alpha_{d_2} (FT_2)_{\mathbf{3}}Y_{\mathbf{3}}^{(2)} H_{\overline{5}}
 + \alpha_{d_3} (FT_3)_{\mathbf{3}}Y_{\mathbf{3}}^{(4)} H_{\overline{5}}
\\\nonumber &&
 + \alpha^\prime_{d_1} (FT_1)_{\mathbf{3}}Y_{\mathbf{3}}^{(2)} H_{\overline{45}}
 + \alpha^\prime_{d_2} (FT_2)_{\mathbf{3}}Y_{\mathbf{3}}^{(2)} H_{\overline{45}}
 + \alpha^\prime_{d_3} (FT_3)_{\mathbf{3}}Y_{\mathbf{3}}^{(4)} H_{\overline{45}}
\,,\\
{\cal W}_u &=&
   \alpha_{u_1}  (T_1T_2)_{\mathbf{1}}Y_{\mathbf{1}}^{(4)} H_{5}
 + \alpha_{u_2}  (T_1T_3)_{\mathbf{1}}Y_{\mathbf{1}}^{(6)} H_{5}
 + \alpha_{u_3}  (T_2T_2)_{\mathbf{1^{\prime\prime}}}Y_{\mathbf{1^\prime}}^{(4)} H_{5}
 + \alpha_{u_4}  (T_3T_3)_{\mathbf{1^{\prime\prime}}}Y_{\mathbf{1^\prime}}^{(8)} H_{5}
\,.
\end{eqnarray}
The matrices $M_N$ and ${\cal Y}_5^{\nu}$ are identical with the corresponding ones of the model ${\cal V}_{8}$ as well.
The ${\cal Y}_{\overline{5}}^d$ matrix is the same as that of the model ${\cal I}_{15}$.
The ${\cal Y}_5^u$ matrix is identical with the corresponding one of the model ${\cal I}_{11}$.
The parameters $\alpha_{N_1}$, $\alpha_{\nu_1}$, $\alpha_{\nu_2}$, $\alpha_{\nu_3}$, $\alpha_{d_1}$, $\alpha_{d_2}$, $\alpha_{d_3}$, $\alpha_{d_1}^\prime$ and $\alpha_{u_3}$ can be taken real by field redefinition, and $\alpha_{N_2}$, $\alpha_{N_3}$, $\alpha_{d_2}^\prime$, $\alpha_{d_3}^\prime$, $\alpha_{u_1}$, $\alpha_{u_2}$ and $\alpha_{u_4}$ are complex.
\item{~Model ${\cal V}_{10}$~:~$\begin{aligned}
&(\rho_{N_1},\rho_{N_2},\rho_{N_3},\rho_F,\rho_{T_1},\rho_{T_2},\rho_{T_3})=(\mathbf{1^{\prime}},\mathbf{1^{\prime}},\mathbf{1},\mathbf{3},\mathbf{1^{\prime\prime}},\mathbf{1^{\prime\prime}},\mathbf{1^{\prime}})\,,\\
& (k_{N_1},k_{N_2},k_{N_3},k_{F},k_{T_1},k_{T_2},k_{T_3})=(2,4,2,0,2,4,2)
\end{aligned}$}

We can read out the superpotential for quark and leptons as follows,
\begin{eqnarray}
\nonumber {\cal W}_\nu &=&
   \alpha_{N_1} \Lambda (N_1N_1)_{\mathbf{1^{\prime\prime}}}Y_{\mathbf{1^\prime}}^{(4)}
 + \alpha_{N_2} \Lambda (N_2N_2)_{\mathbf{1^{\prime\prime}}}Y_{\mathbf{1^\prime}}^{(8)}
 + \alpha_{N_3} \Lambda (N_3N_3)_{\mathbf{1}}Y_{\mathbf{1}}^{(4)}
 + \alpha_{\nu_1} (N_1F)_{\mathbf{3}}Y_{\mathbf{3}}^{(2)} H_{5}
\\
\nonumber&&
 + \alpha_{\nu_2} (N_2F)_{\mathbf{3}}Y_{\mathbf{3}}^{(4)} H_{5}
 + \alpha_{\nu_3} (N_3F)_{\mathbf{3}}Y_{\mathbf{3}}^{(2)} H_{5}
\,,\\
\nonumber {\cal W}_d &=&
   \alpha_{d_1} (FT_1)_{\mathbf{3}}Y_{\mathbf{3}}^{(2)} H_{\overline{5}}
 + \alpha_{d_2} (FT_2)_{\mathbf{3}}Y_{\mathbf{3}}^{(4)} H_{\overline{5}}
 + \alpha_{d_3} (FT_3)_{\mathbf{3}}Y_{\mathbf{3}}^{(2)} H_{\overline{5}}
\\\nonumber &&
 + \alpha^\prime_{d_1} (FT_1)_{\mathbf{3}}Y_{\mathbf{3}}^{(2)} H_{\overline{45}}
 + \alpha^\prime_{d_2} (FT_2)_{\mathbf{3}}Y_{\mathbf{3}}^{(4)} H_{\overline{45}}
 + \alpha^\prime_{d_3} (FT_3)_{\mathbf{3}}Y_{\mathbf{3}}^{(2)} H_{\overline{45}}
\,,\\
{\cal W}_u &=&
   \alpha_{u_1}  (T_1T_3)_{\mathbf{1}}Y_{\mathbf{1}}^{(4)} H_{5}
 + \alpha_{u_2}  (T_2T_2)_{\mathbf{1^{\prime}}}Y_{\mathbf{1^{\prime\prime}}}^{(8)} H_{5}
 + \alpha_{u_3}  (T_2T_3)_{\mathbf{1}}Y_{\mathbf{1}}^{(6)} H_{5}
 + \alpha_{u_4}  (T_3T_3)_{\mathbf{1^{\prime\prime}}}Y_{\mathbf{1^\prime}}^{(4)} H_{5}
\,.
\end{eqnarray}
The matrices $M_N$ and ${\cal Y}_5^{\nu}$ are the same as those of the model ${\cal V}_{8}$ as well.
The ${\cal Y}_{\overline{5}}^d$ matrix coincides with the corresponding one of the model ${\cal I}_{16}$.
The ${\cal Y}_5^u$ matrix is in common with that of the model ${\cal I}_{12}$.
The parameters $\alpha_{N_1}$, $\alpha_{\nu_1}$, $\alpha_{\nu_2}$, $\alpha_{\nu_3}$, $\alpha_{d_1}$, $\alpha_{d_2}$, $\alpha_{d_3}$, $\alpha_{d_1}^\prime$ and $\alpha_{u_3}$ can taken to be real without loss of generality, while the other parameters are complex.
\item{~Model ${\cal V}_{11}$~:~$\begin{aligned}
&(\rho_{N_1},\rho_{N_2},\rho_{N_3},\rho_F,\rho_{T_1},\rho_{T_2},\rho_{T_3})=(\mathbf{1^{\prime}},\mathbf{1^{\prime}},\mathbf{1},\mathbf{3},\mathbf{1^{\prime\prime}},\mathbf{1^{\prime\prime}},\mathbf{1^{\prime}})\,,\\
& (k_{N_1},k_{N_2},k_{N_3},k_{F},k_{T_1},k_{T_2},k_{T_3})=(2,4,2,0,2,4,4)
\end{aligned}$}

We can read out the superpotential for quark and leptons as follows,
\begin{eqnarray}
\nonumber {\cal W}_\nu &=&
   \alpha_{N_1} \Lambda (N_1N_1)_{\mathbf{1^{\prime\prime}}}Y_{\mathbf{1^\prime}}^{(4)}
 + \alpha_{N_2} \Lambda (N_2N_2)_{\mathbf{1^{\prime\prime}}}Y_{\mathbf{1^\prime}}^{(8)}
 + \alpha_{N_3} \Lambda (N_3N_3)_{\mathbf{1}}Y_{\mathbf{1}}^{(4)}
 + \alpha_{\nu_1} (N_1F)_{\mathbf{3}}Y_{\mathbf{3}}^{(2)} H_{5}
\\
\nonumber&&
 + \alpha_{\nu_2} (N_2F)_{\mathbf{3}}Y_{\mathbf{3}}^{(4)} H_{5}
 + \alpha_{\nu_3} (N_3F)_{\mathbf{3}}Y_{\mathbf{3}}^{(2)} H_{5}
\,,\\
\nonumber {\cal W}_d &=&
   \alpha_{d_1} (FT_1)_{\mathbf{3}}Y_{\mathbf{3}}^{(2)} H_{\overline{5}}
 + \alpha_{d_2} (FT_2)_{\mathbf{3}}Y_{\mathbf{3}}^{(4)} H_{\overline{5}}
 + \alpha_{d_3} (FT_3)_{\mathbf{3}}Y_{\mathbf{3}}^{(4)} H_{\overline{5}}
\\\nonumber &&
 + \alpha^\prime_{d_1} (FT_1)_{\mathbf{3}}Y_{\mathbf{3}}^{(2)} H_{\overline{45}}
 + \alpha^\prime_{d_2} (FT_2)_{\mathbf{3}}Y_{\mathbf{3}}^{(4)} H_{\overline{45}}
 + \alpha^\prime_{d_3} (FT_3)_{\mathbf{3}}Y_{\mathbf{3}}^{(4)} H_{\overline{45}}
\,,\\
{\cal W}_u &=&
   \alpha_{u_1}  (T_1T_3)_{\mathbf{1}}Y_{\mathbf{1}}^{(6)} H_{5}
 + \alpha_{u_2}  (T_2T_2)_{\mathbf{1^{\prime}}}Y_{\mathbf{1^{\prime\prime}}}^{(8)} H_{5}
 + \alpha_{u_3}  (T_2T_3)_{\mathbf{1}}Y_{\mathbf{1}}^{(8)} H_{5}
 + \alpha_{u_4}  (T_3T_3)_{\mathbf{1^{\prime\prime}}}Y_{\mathbf{1^\prime}}^{(8)} H_{5}
\,.
\end{eqnarray}
The matrices $M_N$ and ${\cal Y}_5^{\nu}$ are identical with the corresponding ones of the model ${\cal V}_{8}$ as well.
The matrices ${\cal Y}_{\overline{5}}^d$ and ${\cal Y}_5^u$ are identical with the corresponding ones of the model ${\cal I}_{17}$ as well.
The parameters $\alpha_{N_1}$, $\alpha_{\nu_1}$, $\alpha_{\nu_2}$, $\alpha_{\nu_3}$, $\alpha_{d_1}$, $\alpha_{d_2}$, $\alpha_{d_3}$, $\alpha_{d_1}^\prime$ and $\alpha_{u_3}$ can be taken real by field redefinition, and the remaining parameters are complex.
\item{~Model ${\cal V}_{12}$~:~$\begin{aligned}
&(\rho_{N_1},\rho_{N_2},\rho_{N_3},\rho_F,\rho_{T_1},\rho_{T_2},\rho_{T_3})=(\mathbf{1^{\prime\prime}},\mathbf{1^{\prime\prime}},\mathbf{1},\mathbf{3},\mathbf{1^{\prime\prime}},\mathbf{1},\mathbf{1^{\prime}})\,,\\
& (k_{N_1},k_{N_2},k_{N_3},k_{F},k_{T_1},k_{T_2},k_{T_3})=(2,4,2,0,2,2,2)
\end{aligned}$}

We can read out the the superpotential for the fermion masses as,
\begin{eqnarray}
\nonumber {\cal W}_\nu &=&
   \alpha_{N_1} \Lambda (N_1N_3)_{\mathbf{1^{\prime\prime}}}Y_{\mathbf{1^\prime}}^{(4)}
 + \alpha_{N_2} \Lambda (N_2N_2)_{\mathbf{1^{\prime}}}Y_{\mathbf{1^{\prime\prime}}}^{(8)}
 + \alpha_{N_3} \Lambda (N_3N_3)_{\mathbf{1}}Y_{\mathbf{1}}^{(4)}
 + \alpha_{\nu_1} (N_1F)_{\mathbf{3}}Y_{\mathbf{3}}^{(2)} H_{5}
\\
\nonumber&&
 + \alpha_{\nu_2} (N_2F)_{\mathbf{3}}Y_{\mathbf{3}}^{(4)} H_{5}
 + \alpha_{\nu_3} (N_3F)_{\mathbf{3}}Y_{\mathbf{3}}^{(2)} H_{5}
\,,\\
\nonumber {\cal W}_d &=&
   \alpha_{d_1} (FT_1)_{\mathbf{3}}Y_{\mathbf{3}}^{(2)} H_{\overline{5}}
 + \alpha_{d_2} (FT_2)_{\mathbf{3}}Y_{\mathbf{3}}^{(2)} H_{\overline{5}}
 + \alpha_{d_3} (FT_3)_{\mathbf{3}}Y_{\mathbf{3}}^{(2)} H_{\overline{5}}
\\\nonumber &&
 + \alpha^\prime_{d_1} (FT_1)_{\mathbf{3}}Y_{\mathbf{3}}^{(2)} H_{\overline{45}}
 + \alpha^\prime_{d_2} (FT_2)_{\mathbf{3}}Y_{\mathbf{3}}^{(2)} H_{\overline{45}}
 + \alpha^\prime_{d_3} (FT_3)_{\mathbf{3}}Y_{\mathbf{3}}^{(2)} H_{\overline{45}}
\,,\\
{\cal W}_u &=&
   \alpha_{u_1}  (T_1T_2)_{\mathbf{1^{\prime\prime}}}Y_{\mathbf{1^\prime}}^{(4)} H_{5}
 + \alpha_{u_2}  (T_1T_3)_{\mathbf{1}}Y_{\mathbf{1}}^{(4)} H_{5}
 + \alpha_{u_3}  (T_2T_2)_{\mathbf{1}}Y_{\mathbf{1}}^{(4)} H_{5}
 + \alpha_{u_4}  (T_3T_3)_{\mathbf{1^{\prime\prime}}}Y_{\mathbf{1^\prime}}^{(4)} H_{5}
\,,
\end{eqnarray}
which lead to
\begin{eqnarray}
m_M^{\nu} &=& \Lambda  \left(
\begin{array}{ccc}
 0 ~&~ 0 ~&~ \alpha_{N_1}Y_{\mathbf{1}^\prime}^{(4)} \\
 0 ~&~ \alpha_{N_2}Y_{\mathbf{1}^{\prime\prime}}^{(8)} ~&~ 0 \\
 \alpha_{N_1}Y_{\mathbf{1}^\prime}^{(4)} ~&~ 0 ~&~ \alpha_{N_3}Y_\mathbf{1}^{(4)} \\
\end{array}
\right)\,,\nonumber\\
{\cal Y}_5^{\nu} &=& \left(
\begin{array}{ccc}
 \alpha_{\nu_1}Y_{\mathbf{3},2}^{(2)} ~&~ \alpha_{\nu_1}Y_{\mathbf{3},1}^{(2)} ~&~ \alpha_{\nu_1}Y_{\mathbf{3},3}^{(2)} \\
 \alpha_{\nu_2}Y_{\mathbf{3},2}^{(4)} ~&~ \alpha_{\nu_2}Y_{\mathbf{3},1}^{(4)} ~&~ \alpha_{\nu_2}Y_{\mathbf{3},3}^{(4)}  \\
 \alpha_{\nu_3}Y_{\mathbf{3},1}^{(2)} ~&~ \alpha_{\nu_3}Y_{\mathbf{3},3}^{(2)} ~&~ \alpha_{\nu_3}Y_{\mathbf{3},2}^{(2)} \\
\end{array}
\right)\,.
\end{eqnarray}
The matrices ${\cal Y}_{\overline{5}}^d$ and ${\cal Y}_5^u$ are in common with the corresponding ones of the model ${\cal V}_{4}$ as well.
The parameters $\alpha_{N_1}$, $\alpha_{\nu_1}$, $\alpha_{\nu_2}$, $\alpha_{\nu_3}$, $\alpha_{d_1}$, $\alpha_{d_2}$, $\alpha_{d_3}$, $\alpha_{d_1}^\prime$ and $\alpha_{u_3}$ can taken to be real without loss of generality, while the remaining parameters are complex.
\item{~Model ${\cal V}_{13}$~:~$\begin{aligned}
&(\rho_{N_1},\rho_{N_2},\rho_{N_3},\rho_F,\rho_{T_1},\rho_{T_2},\rho_{T_3})=(\mathbf{1^{\prime\prime}},\mathbf{1^{\prime\prime}},\mathbf{1},\mathbf{3},\mathbf{1^{\prime\prime}},\mathbf{1},\mathbf{1^{\prime}})\,,\\
& (k_{N_1},k_{N_2},k_{N_3},k_{F},k_{T_1},k_{T_2},k_{T_3})=(2,4,2,0,2,4,4)
\end{aligned}$}

We can read out the the superpotential for the fermion masses as,
\begin{eqnarray}
\nonumber {\cal W}_\nu &=&
   \alpha_{N_1} \Lambda (N_1N_3)_{\mathbf{1^{\prime\prime}}}Y_{\mathbf{1^\prime}}^{(4)}
 + \alpha_{N_2} \Lambda (N_2N_2)_{\mathbf{1^{\prime}}}Y_{\mathbf{1^{\prime\prime}}}^{(8)}
 + \alpha_{N_3} \Lambda (N_3N_3)_{\mathbf{1}}Y_{\mathbf{1}}^{(4)}
 + \alpha_{\nu_1} (N_1F)_{\mathbf{3}}Y_{\mathbf{3}}^{(2)} H_{5}
\\
\nonumber&&
 + \alpha_{\nu_2} (N_2F)_{\mathbf{3}}Y_{\mathbf{3}}^{(4)} H_{5}
 + \alpha_{\nu_3} (N_3F)_{\mathbf{3}}Y_{\mathbf{3}}^{(2)} H_{5}
\,,\\
\nonumber {\cal W}_d &=&
   \alpha_{d_1} (FT_1)_{\mathbf{3}}Y_{\mathbf{3}}^{(2)} H_{\overline{5}}
 + \alpha_{d_2} (FT_2)_{\mathbf{3}}Y_{\mathbf{3}}^{(4)} H_{\overline{5}}
 + \alpha_{d_3} (FT_3)_{\mathbf{3}}Y_{\mathbf{3}}^{(4)} H_{\overline{5}}
\\\nonumber &&
 + \alpha^\prime_{d_1} (FT_1)_{\mathbf{3}}Y_{\mathbf{3}}^{(2)} H_{\overline{45}}
 + \alpha^\prime_{d_2} (FT_2)_{\mathbf{3}}Y_{\mathbf{3}}^{(4)} H_{\overline{45}}
 + \alpha^\prime_{d_3} (FT_3)_{\mathbf{3}}Y_{\mathbf{3}}^{(4)} H_{\overline{45}}
\,,\\
{\cal W}_u &=&
   \alpha_{u_1}  (T_1T_3)_{\mathbf{1}}Y_{\mathbf{1}}^{(6)} H_{5}
 + \alpha_{u_2}  (T_2T_2)_{\mathbf{1}}Y_{\mathbf{1}}^{(8)} H_{5}
 + \alpha_{u_3}  (T_2T_3)_{\mathbf{1^{\prime}}}Y_{\mathbf{1^{\prime\prime}}}^{(8)} H_{5}
 + \alpha_{u_4}  (T_3T_3)_{\mathbf{1^{\prime\prime}}}Y_{\mathbf{1^\prime}}^{(8)} H_{5}
\,.
\end{eqnarray}
The matrices $M_N$ and ${\cal Y}_5^{\nu}$ are the same as those of the model ${\cal V}_{12}$ as well.
The matrices ${\cal Y}_{\overline{5}}^d$ and ${\cal Y}_5^u$ are identical with the corresponding ones of the model ${\cal I}_{14}$ as well.
The phases of the parameters $\alpha_{N_1}$, $\alpha_{\nu_1}$, $\alpha_{\nu_2}$, $\alpha_{\nu_3}$, $\alpha_{d_1}$, $\alpha_{d_2}$, $\alpha_{d_3}$, $\alpha_{d_1}^\prime$ and $\alpha_{u_3}$ can be removed by field redefinition, and the remaining parameters are complex.
\item{~Model ${\cal V}_{14}$~:~$\begin{aligned}
&(\rho_{N_1},\rho_{N_2},\rho_{N_3},\rho_F,\rho_{T_1},\rho_{T_2},\rho_{T_3})=(\mathbf{1^{\prime\prime}},\mathbf{1^{\prime\prime}},\mathbf{1},\mathbf{3},\mathbf{1^{\prime\prime}},\mathbf{1},\mathbf{1})\,,\\
& (k_{N_1},k_{N_2},k_{N_3},k_{F},k_{T_1},k_{T_2},k_{T_3})=(2,4,2,0,2,2,4)
\end{aligned}$}

The modular invariant superpotential for quark and lepton mass is given by
\begin{eqnarray}
\nonumber {\cal W}_\nu &=&
   \alpha_{N_1} \Lambda (N_1N_3)_{\mathbf{1^{\prime\prime}}}Y_{\mathbf{1^\prime}}^{(4)}
 + \alpha_{N_2} \Lambda (N_2N_2)_{\mathbf{1^{\prime}}}Y_{\mathbf{1^{\prime\prime}}}^{(8)}
 + \alpha_{N_3} \Lambda (N_3N_3)_{\mathbf{1}}Y_{\mathbf{1}}^{(4)}
 + \alpha_{\nu_1} (N_1F)_{\mathbf{3}}Y_{\mathbf{3}}^{(2)} H_{5}
\\
\nonumber&&
 + \alpha_{\nu_2} (N_2F)_{\mathbf{3}}Y_{\mathbf{3}}^{(4)} H_{5}
 + \alpha_{\nu_3} (N_3F)_{\mathbf{3}}Y_{\mathbf{3}}^{(2)} H_{5}
\,,\\
\nonumber {\cal W}_d &=&
   \alpha_{d_1} (FT_1)_{\mathbf{3}}Y_{\mathbf{3}}^{(2)} H_{\overline{5}}
 + \alpha_{d_2} (FT_2)_{\mathbf{3}}Y_{\mathbf{3}}^{(2)} H_{\overline{5}}
 + \alpha_{d_3} (FT_3)_{\mathbf{3}}Y_{\mathbf{3}}^{(4)} H_{\overline{5}}
\\\nonumber &&
 + \alpha^\prime_{d_1} (FT_1)_{\mathbf{3}}Y_{\mathbf{3}}^{(2)} H_{\overline{45}}
 + \alpha^\prime_{d_2} (FT_2)_{\mathbf{3}}Y_{\mathbf{3}}^{(2)} H_{\overline{45}}
 + \alpha^\prime_{d_3} (FT_3)_{\mathbf{3}}Y_{\mathbf{3}}^{(4)} H_{\overline{45}}
\,,\\
{\cal W}_u &=&
   \alpha_{u_1}  (T_1T_2)_{\mathbf{1^{\prime\prime}}}Y_{\mathbf{1^\prime}}^{(4)} H_{5}
 + \alpha_{u_2}  (T_2T_2)_{\mathbf{1}}Y_{\mathbf{1}}^{(4)} H_{5}
 + \alpha_{u_3}  (T_2T_3)_{\mathbf{1}}Y_{\mathbf{1}}^{(6)} H_{5}
 + \alpha_{u_4}  (T_3T_3)_{\mathbf{1}}Y_{\mathbf{1}}^{(8)} H_{5}
\,.
\end{eqnarray}
The matrices $M_N$ and ${\cal Y}_5^{\nu}$ are the same as those of the model ${\cal V}_{12}$ as well.
The ${\cal Y}_{\overline{5}}^d$ matrix is the same as the corresponding one of the model ${\cal I}_{1}$.
The ${\cal Y}_5^u$ matrix is identical with that of the model ${\cal I}_{6}$.
The phases of the parameters $\alpha_{N_1}$, $\alpha_{\nu_1}$, $\alpha_{\nu_2}$, $\alpha_{\nu_3}$, $\alpha_{d_1}$, $\alpha_{d_2}$, $\alpha_{d_3}$, $\alpha_{d_1}^\prime$ and $\alpha_{u_3}$ are unphysical, and the other parameters are complex.
\item{~Model ${\cal V}_{15}$~:~$\begin{aligned}
&(\rho_{N_1},\rho_{N_2},\rho_{N_3},\rho_F,\rho_{T_1},\rho_{T_2},\rho_{T_3})=(\mathbf{1},\mathbf{1^{\prime\prime}},\mathbf{1^{\prime}},\mathbf{3},\mathbf{1^{\prime}},\mathbf{1^{\prime\prime}},\mathbf{1^{\prime\prime}})\,,\\
& (k_{N_1},k_{N_2},k_{N_3},k_{F},k_{T_1},k_{T_2},k_{T_3})=(0,0,0,2,0,0,4)
\end{aligned}$}

The quark and lepton masses are described by the following superpotential:
\begin{eqnarray}
\nonumber {\cal W}_\nu &=&
   \alpha_{N_1} \Lambda (N_1N_1)_{\mathbf{1}}
 + \alpha_{N_2} \Lambda (N_2N_3)_{\mathbf{1}}
 + \alpha_{\nu_1} (N_1F)_{\mathbf{3}}Y_{\mathbf{3}}^{(2)} H_{5}
 + \alpha_{\nu_2} (N_2F)_{\mathbf{3}}Y_{\mathbf{3}}^{(2)} H_{5}
\\
\nonumber&&
 + \alpha_{\nu_3} (N_3F)_{\mathbf{3}}Y_{\mathbf{3}}^{(2)} H_{5}
\,,\\
\nonumber {\cal W}_d &=&
   \alpha_{d_1} (FT_1)_{\mathbf{3}}Y_{\mathbf{3}}^{(2)} H_{\overline{5}}
 + \alpha_{d_2} (FT_2)_{\mathbf{3}}Y_{\mathbf{3}}^{(2)} H_{\overline{5}}
 + \alpha_{d_3} (FT_3)_{\mathbf{3}}Y_{\mathbf{3},I}^{(6)} H_{\overline{5}}
 + \alpha_{d_4} (FT_3)_{\mathbf{3}}Y_{\mathbf{3},II}^{(6)} H_{\overline{5}}
\\\nonumber &&
 + \alpha^\prime_{d_1} (FT_1)_{\mathbf{3}}Y_{\mathbf{3}}^{(2)} H_{\overline{45}}
 + \alpha^\prime_{d_2} (FT_2)_{\mathbf{3}}Y_{\mathbf{3}}^{(2)} H_{\overline{45}}
 + \alpha^\prime_{d_3} (FT_3)_{\mathbf{3}}Y_{\mathbf{3},I}^{(6)} H_{\overline{45}}
 + \alpha^\prime_{d_4} (FT_3)_{\mathbf{3}}Y_{\mathbf{3},II}^{(6)} H_{\overline{45}}
\,,\\
{\cal W}_u &=&
   \alpha_{u_1}  (T_1T_2)_{\mathbf{1}}   H_{5}
 + \alpha_{u_2}  (T_1T_3)_{\mathbf{1}}Y_{\mathbf{1}}^{(4)} H_{5}
 + \alpha_{u_3}  (T_3T_3)_{\mathbf{1^{\prime}}}Y_{\mathbf{1^{\prime\prime}}}^{(8)} H_{5}
\,,
\end{eqnarray}
which lead to
\begin{eqnarray}
m_M^{\nu} &=& \Lambda  \left(
\begin{array}{ccc}
 \alpha_{N_1} ~&~ 0 ~&~ 0 \\
 0 ~&~ 0 ~&~ \alpha_{N_2} \\
 0 ~&~ \alpha_{N_2} ~&~ 0 \\
\end{array}
\right)\,,\nonumber\\
{\cal Y}_5^{\nu} &=& \left(
\begin{array}{ccc}
 \alpha_{\nu_1}Y_{\mathbf{3},1}^{(2)} ~&~ \alpha_{\nu_1}Y_{\mathbf{3},3}^{(2)} ~&~ \alpha_{\nu_1}Y_{\mathbf{3},2}^{(2)} \\
 \alpha_{\nu_2}Y_{\mathbf{3},2}^{(2)} ~&~ \alpha_{\nu_2}Y_{\mathbf{3},1}^{(2)} ~&~ \alpha_{\nu_2}Y_{\mathbf{3},3}^{(2)} \\
 \alpha_{\nu_3}Y_{\mathbf{3},3}^{(2)} ~&~ \alpha_{\nu_3}Y_{\mathbf{3},2}^{(2)} ~&~ \alpha_{\nu_3}Y_{\mathbf{3},1}^{(2)} \\
\end{array}
\right)\,,\nonumber\\
{\cal Y}_{\overline{5}}^d &=& \left(
\begin{array}{ccc}
 \alpha_{d_1}Y_{\mathbf{3},3}^{(2)} ~&~ \alpha_{d_2}Y_{\mathbf{3},2}^{(2)} ~&~ \alpha_{d_3}Y_{\mathbf{3}I,2}^{(6)}+\alpha_{d_4}Y_{\mathbf{3}II,2}^{(6)} \\
 \alpha_{d_1}Y_{\mathbf{3},2}^{(2)} ~&~ \alpha_{d_2}Y_{\mathbf{3},1}^{(2)} ~&~ \alpha_{d_3}Y_{\mathbf{3}I,1}^{(6)}+\alpha_{d_4}Y_{\mathbf{3}II,1}^{(6)} \\
 \alpha_{d_1}Y_{\mathbf{3},1}^{(2)}  ~&~ \alpha_{d_2}Y_{\mathbf{3},3}^{(2)} ~&~ \alpha_{d_3}Y_{\mathbf{3}I,3}^{(6)}+\alpha_{d_4}Y_{\mathbf{3}II,3}^{(6)} \\
\end{array}
\right)\,,\nonumber\\
{\cal Y}_5^u &=& \left(
\begin{array}{ccc}
 0 ~&~ \alpha_{u_1} ~&~ \alpha_{u_2}Y_\mathbf{1}^{(4)} \\
 \alpha_{u_1} ~&~ 0 ~&~ 0 \\
 \alpha_{u_2}Y_\mathbf{1}^{(4)} ~&~ 0 ~&~ \alpha_{u_3}Y_{\mathbf{1}^{\prime\prime}}^{(8)} \\
\end{array}
\right)\,.
\end{eqnarray}
The phases of the parameters $\alpha_{N_1}$, $\alpha_{\nu_1}$, $\alpha_{\nu_2}$, $\alpha_{\nu_3}$, $\alpha_{d_1}$, $\alpha_{d_2}$, $\alpha_{d_3}$, $\alpha_{d_1}^\prime$ and $\alpha_{u_3}$ are unphysical, and the remaining parameters are complex.
\item{~Model ${\cal V}_{16}$~:~$\begin{aligned}
&(\rho_{N_1},\rho_{N_2},\rho_{N_3},\rho_F,\rho_{T_1},\rho_{T_2},\rho_{T_3})=(\mathbf{1},\mathbf{1^{\prime\prime}},\mathbf{1^{\prime}},\mathbf{3},\mathbf{1^{\prime}},\mathbf{1^{\prime\prime}},\mathbf{1})\,,\\
& (k_{N_1},k_{N_2},k_{N_3},k_{F},k_{T_1},k_{T_2},k_{T_3})=(0,0,0,2,0,0,4)
\end{aligned}$}

The quark and lepton masses are described by the following superpotential:
\begin{eqnarray}
\nonumber {\cal W}_\nu &=&
   \alpha_{N_1} \Lambda (N_1N_1)_{\mathbf{1}}
 + \alpha_{N_2} \Lambda (N_2N_3)_{\mathbf{1}}
 + \alpha_{\nu_1} (N_1F)_{\mathbf{3}}Y_{\mathbf{3}}^{(2)} H_{5}
 + \alpha_{\nu_2} (N_2F)_{\mathbf{3}}Y_{\mathbf{3}}^{(2)} H_{5}
\\
\nonumber&&
 + \alpha_{\nu_3} (N_3F)_{\mathbf{3}}Y_{\mathbf{3}}^{(2)} H_{5}
\,,\\
\nonumber {\cal W}_d &=&
   \alpha_{d_1} (FT_1)_{\mathbf{3}}Y_{\mathbf{3}}^{(2)} H_{\overline{5}}
 + \alpha_{d_2} (FT_2)_{\mathbf{3}}Y_{\mathbf{3}}^{(2)} H_{\overline{5}}
 + \alpha_{d_3} (FT_3)_{\mathbf{3}}Y_{\mathbf{3},I}^{(6)} H_{\overline{5}}
 + \alpha_{d_4} (FT_3)_{\mathbf{3}}Y_{\mathbf{3},II}^{(6)} H_{\overline{5}}
\\\nonumber &&
 + \alpha^\prime_{d_1} (FT_1)_{\mathbf{3}}Y_{\mathbf{3}}^{(2)} H_{\overline{45}}
 + \alpha^\prime_{d_2} (FT_2)_{\mathbf{3}}Y_{\mathbf{3}}^{(2)} H_{\overline{45}}
 + \alpha^\prime_{d_3} (FT_3)_{\mathbf{3}}Y_{\mathbf{3},I}^{(6)} H_{\overline{45}}
 + \alpha^\prime_{d_4} (FT_3)_{\mathbf{3}}Y_{\mathbf{3},II}^{(6)} H_{\overline{45}}
\,,\\
{\cal W}_u &=&
   \alpha_{u_1}  (T_1T_2)_{\mathbf{1}}   H_{5}
 + \alpha_{u_2}  (T_2T_3)_{\mathbf{1^{\prime\prime}}}Y_{\mathbf{1^\prime}}^{(4)} H_{5}
 + \alpha_{u_3}  (T_3T_3)_{\mathbf{1}}Y_{\mathbf{1}}^{(8)} H_{5}
\,.
\end{eqnarray}
The right-handed neutrino mass matrix and the Yukawa matrix are given by
\begin{eqnarray}
{\cal Y}_{\overline{5}}^d &=& \left(
\begin{array}{ccc}
 \alpha_{d_1}Y_{\mathbf{3},3}^{(2)} ~&~ \alpha_{d_2}Y_{\mathbf{3},2}^{(2)} ~&~ \alpha_{d_3}Y_{\mathbf{3}I,1}^{(6)}+\alpha_{d_4}Y_{\mathbf{3}II,1}^{(6)} \\
 \alpha_{d_1}Y_{\mathbf{3},2}^{(2)} ~&~ \alpha_{d_2}Y_{\mathbf{3},1}^{(2)} ~&~ \alpha_{d_3}Y_{\mathbf{3}I,3}^{(6)}+\alpha_{d_4}Y_{\mathbf{3}II,3}^{(6)} \\
 \alpha_{d_1}Y_{\mathbf{3},1}^{(2)}  ~&~ \alpha_{d_2}Y_{\mathbf{3},3}^{(2)} ~&~ \alpha_{d_3}Y_{\mathbf{3}I,2}^{(6)}+\alpha_{d_4}Y_{\mathbf{3}II,2}^{(6)} \\
\end{array}
\right)\,,\nonumber\\
{\cal Y}_5^u &=& \left(
\begin{array}{ccc}
 0 ~&~ \alpha_{u_1} ~&~ 0 \\
 \alpha_{u_1} ~&~ 0 ~&~ \alpha_{u_2}Y_{\mathbf{1}^\prime}^{(4)} \\
 0 ~&~ \alpha_{u_2}Y_{\mathbf{1}^\prime}^{(4)} ~&~ \alpha_{u_3}Y_\mathbf{1}^{(8)} \\
\end{array}
\right)\,.
\end{eqnarray}
The matrices $M_N$ and ${\cal Y}_5^{\nu}$ are identical with the corresponding ones of the model ${\cal V}_{15}$ as well.
The parameters $\alpha_{N_1}$, $\alpha_{\nu_1}$, $\alpha_{\nu_2}$, $\alpha_{\nu_3}$, $\alpha_{d_1}$, $\alpha_{d_2}$, $\alpha_{d_3}$, $\alpha_{d_1}^\prime$ and $\alpha_{u_3}$ can be taken real by field redefinition, and the other parameters are complex.
\item{~Model ${\cal V}_{17}$~:~$\begin{aligned}
&(\rho_{N_1},\rho_{N_2},\rho_{N_3},\rho_F,\rho_{T_1},\rho_{T_2},\rho_{T_3})=(\mathbf{1},\mathbf{1^{\prime\prime}},\mathbf{1^{\prime}},\mathbf{3},\mathbf{1^{\prime}},\mathbf{1^{\prime\prime}},\mathbf{1})\,,\\
& (k_{N_1},k_{N_2},k_{N_3},k_{F},k_{T_1},k_{T_2},k_{T_3})=(0,0,0,4,0,0,4)
\end{aligned}$}

We can read out the superpotentials relevant to quark and lepton masses
\begin{eqnarray}
\nonumber {\cal W}_\nu &=&
   \alpha_{N_1} \Lambda (N_1N_1)_{\mathbf{1}}
 + \alpha_{N_2} \Lambda (N_2N_3)_{\mathbf{1}}
 + \alpha_{\nu_1} (N_1F)_{\mathbf{3}}Y_{\mathbf{3}}^{(4)} H_{5}
 + \alpha_{\nu_2} (N_2F)_{\mathbf{3}}Y_{\mathbf{3}}^{(4)} H_{5}
\\
\nonumber&&
 + \alpha_{\nu_3} (N_3F)_{\mathbf{3}}Y_{\mathbf{3}}^{(4)} H_{5}
\,,\\
\nonumber {\cal W}_d &=&
   \alpha_{d_1} (FT_1)_{\mathbf{3}}Y_{\mathbf{3}}^{(4)} H_{\overline{5}}
 + \alpha_{d_2} (FT_2)_{\mathbf{3}}Y_{\mathbf{3}}^{(4)} H_{\overline{5}}
 + \alpha_{d_3} (FT_3)_{\mathbf{3}}Y_{\mathbf{3},I}^{(8)} H_{\overline{5}}
 + \alpha_{d_4} (FT_3)_{\mathbf{3}}Y_{\mathbf{3},II}^{(8)} H_{\overline{5}}
\\\nonumber &&
 + \alpha^\prime_{d_1} (FT_1)_{\mathbf{3}}Y_{\mathbf{3}}^{(4)} H_{\overline{45}}
 + \alpha^\prime_{d_2} (FT_2)_{\mathbf{3}}Y_{\mathbf{3}}^{(4)} H_{\overline{45}}
 + \alpha^\prime_{d_3} (FT_3)_{\mathbf{3}}Y_{\mathbf{3},I}^{(8)} H_{\overline{45}}
 + \alpha^\prime_{d_4} (FT_3)_{\mathbf{3}}Y_{\mathbf{3},II}^{(8)} H_{\overline{45}}
\,,\\
{\cal W}_u &=&
   \alpha_{u_1}  (T_1T_2)_{\mathbf{1}}   H_{5}
 + \alpha_{u_2}  (T_2T_3)_{\mathbf{1^{\prime\prime}}}Y_{\mathbf{1^\prime}}^{(4)} H_{5}
 + \alpha_{u_3}  (T_3T_3)_{\mathbf{1}}Y_{\mathbf{1}}^{(8)} H_{5}
\,.
\end{eqnarray}
The right-handed neutrino mass matrix and the Yukawa matrix are given by
\begin{eqnarray}
{\cal Y}_5^{\nu} &=& \left(
\begin{array}{ccc}
 \alpha_{\nu_1}Y_{\mathbf{3},1}^{(4)} ~&~ \alpha_{\nu_1}Y_{\mathbf{3},3}^{(4)} ~&~ \alpha_{\nu_1}Y_{\mathbf{3},2}^{(4)} \\
 \alpha_{\nu_2}Y_{\mathbf{3},2}^{(4)} ~&~ \alpha_{\nu_2}Y_{\mathbf{3},1}^{(4)} ~&~ \alpha_{\nu_2}Y_{\mathbf{3},3}^{(4)}  \\
 \alpha_{\nu_3}Y_{\mathbf{3},3}^{(4)}  ~&~ \alpha_{\nu_3}Y_{\mathbf{3},2}^{(4)} ~&~ \alpha_{\nu_3}Y_{\mathbf{3},1}^{(4)} \\
\end{array}
\right)\,,\nonumber\\
{\cal Y}_{\overline{5}}^d &=& \left(
\begin{array}{ccc}
 \alpha_{d_1}Y_{\mathbf{3},3}^{(4)} ~&~ \alpha_{d_2}Y_{\mathbf{3},2}^{(4)} ~&~ \alpha_{d_3}Y_{\mathbf{3}I,1}^{(8)}+\alpha_{d_4}Y_{\mathbf{3}II,1}^{(8)} \\
 \alpha_{d_1}Y_{\mathbf{3},2}^{(4)} ~&~ \alpha_{d_2}Y_{\mathbf{3},1}^{(4)} ~&~ \alpha_{d_3}Y_{\mathbf{3}I,3}^{(8)}+\alpha_{d_4}Y_{\mathbf{3}II,3}^{(8)} \\
 \alpha_{d_1}Y_{\mathbf{3},1}^{(4)} ~&~ \alpha_{d_2}Y_{\mathbf{3},3}^{(4)} ~&~ \alpha_{d_3}Y_{\mathbf{3}I,2}^{(8)}+\alpha_{d_4}Y_{\mathbf{3}II,2}^{(8)} \\
\end{array}
\right)\,.
\end{eqnarray}
The matrices $M_N$ and ${\cal Y}_5^u$ are identical with those of the models ${\cal V}_{15}$ and ${\cal V}_{16}$ respectively.
The parameters $\alpha_{N_1}$, $\alpha_{\nu_1}$, $\alpha_{\nu_2}$, $\alpha_{\nu_3}$, $\alpha_{d_1}$, $\alpha_{d_2}$, $\alpha_{d_3}$, $\alpha_{d_1}^\prime$ and $\alpha_{u_3}$ can taken to be real without loss of generality, while the remaining parameters $\alpha_{N_2}$, $\alpha_{d_4}$, $\alpha_{d_2}^\prime$, $\alpha_{d_3}^\prime$, $\alpha_{d_4}^\prime$, $\alpha_{u_1}$ and $\alpha_{u_2}$ are complex.
\item{~Model ${\cal V}_{18}$~:~$\begin{aligned}
&(\rho_{N_1},\rho_{N_2},\rho_{N_3},\rho_F,\rho_{T_1},\rho_{T_2},\rho_{T_3})=(\mathbf{1},\mathbf{1^{\prime\prime}},\mathbf{1^{\prime}},\mathbf{3},\mathbf{1^{\prime}},\mathbf{1^{\prime\prime}},\mathbf{1})\,,\\
& (k_{N_1},k_{N_2},k_{N_3},k_{F},k_{T_1},k_{T_2},k_{T_3})=(2,0,0,2,0,0,4)
\end{aligned}$}

The quark and lepton masses are described by the following superpotential:
\begin{eqnarray}
\nonumber {\cal W}_\nu &=&
   \alpha_{N_1} \Lambda (N_1N_1)_{\mathbf{1}}Y_{\mathbf{1}}^{(4)}
 + \alpha_{N_2} \Lambda (N_2N_3)_{\mathbf{1}}
 + \alpha_{\nu_1} (N_1F)_{\mathbf{3}}Y_{\mathbf{3}}^{(4)} H_{5}
 + \alpha_{\nu_2} (N_2F)_{\mathbf{3}}Y_{\mathbf{3}}^{(2)} H_{5}
\\
\nonumber&&
 + \alpha_{\nu_3} (N_3F)_{\mathbf{3}}Y_{\mathbf{3}}^{(2)} H_{5}
\,,\\
\nonumber {\cal W}_d &=&
   \alpha_{d_1} (FT_1)_{\mathbf{3}}Y_{\mathbf{3}}^{(2)} H_{\overline{5}}
 + \alpha_{d_2} (FT_2)_{\mathbf{3}}Y_{\mathbf{3}}^{(2)} H_{\overline{5}}
 + \alpha_{d_3} (FT_3)_{\mathbf{3}}Y_{\mathbf{3},I}^{(6)} H_{\overline{5}}
 + \alpha_{d_4} (FT_3)_{\mathbf{3}}Y_{\mathbf{3},II}^{(6)} H_{\overline{5}}
\\\nonumber &&
 + \alpha^\prime_{d_1} (FT_1)_{\mathbf{3}}Y_{\mathbf{3}}^{(2)} H_{\overline{45}}
 + \alpha^\prime_{d_2} (FT_2)_{\mathbf{3}}Y_{\mathbf{3}}^{(2)} H_{\overline{45}}
 + \alpha^\prime_{d_3} (FT_3)_{\mathbf{3}}Y_{\mathbf{3},I}^{(6)} H_{\overline{45}}
 + \alpha^\prime_{d_4} (FT_3)_{\mathbf{3}}Y_{\mathbf{3},II}^{(6)} H_{\overline{45}}
\,,\\
{\cal W}_u &=&
   \alpha_{u_1}  (T_1T_2)_{\mathbf{1}}   H_{5}
 + \alpha_{u_2}  (T_2T_3)_{\mathbf{1^{\prime\prime}}}Y_{\mathbf{1^\prime}}^{(4)} H_{5}
 + \alpha_{u_3}  (T_3T_3)_{\mathbf{1}}Y_{\mathbf{1}}^{(8)} H_{5}
\,,
\end{eqnarray}
which gives rise to
\begin{eqnarray}
m_M^{\nu} &=& \Lambda  \left(
\begin{array}{ccc}
 \alpha_{N_1}Y_\mathbf{1}^{(4)} ~&~ 0 ~&~ 0 \\
 0 ~&~ 0 ~&~ \alpha_{N_2} \\
 0 ~&~ \alpha_{N_2} ~&~ 0 \\
\end{array}
\right)\,.
\end{eqnarray}
The ${\cal Y}_5^{\nu}$ matrix is identical with the corresponding one of the model ${\cal V}_{4}$.
The matrices ${\cal Y}_{\overline{5}}^d$ and ${\cal Y}_5^u$ are identical with those of the model ${\cal V}_{16}$ as well.
The parameters $\alpha_{N_1}$, $\alpha_{\nu_1}$, $\alpha_{\nu_2}$, $\alpha_{\nu_3}$, $\alpha_{d_1}$, $\alpha_{d_2}$, $\alpha_{d_3}$, $\alpha_{d_1}^\prime$ and $\alpha_{u_3}$ can be taken real by field redefinition, and the other parameters are complex.
\item{~Model ${\cal V}_{19}$~:~$\begin{aligned}
&(\rho_{N_1},\rho_{N_2},\rho_{N_3},\rho_F,\rho_{T_1},\rho_{T_2},\rho_{T_3})=(\mathbf{1},\mathbf{1^{\prime\prime}},\mathbf{1^{\prime}},\mathbf{3},\mathbf{1^{\prime}},\mathbf{1^{\prime\prime}},\mathbf{1^{\prime\prime}})\,,\\
& (k_{N_1},k_{N_2},k_{N_3},k_{F},k_{T_1},k_{T_2},k_{T_3})=(0,0,0,4,0,0,4)
\end{aligned}$}

The modular invariant superpotential for quark and lepton mass is given by
\begin{eqnarray}
\nonumber {\cal W}_\nu &=&
   \alpha_{N_1} \Lambda (N_1N_1)_{\mathbf{1}}
 + \alpha_{N_2} \Lambda (N_2N_3)_{\mathbf{1}}
 + \alpha_{\nu_1} (N_1F)_{\mathbf{3}}Y_{\mathbf{3}}^{(4)} H_{5}
 + \alpha_{\nu_2} (N_2F)_{\mathbf{3}}Y_{\mathbf{3}}^{(4)} H_{5}
\\
\nonumber&&
 + \alpha_{\nu_3} (N_3F)_{\mathbf{3}}Y_{\mathbf{3}}^{(4)} H_{5}
\,,\\
\nonumber {\cal W}_d &=&
   \alpha_{d_1} (FT_1)_{\mathbf{3}}Y_{\mathbf{3}}^{(4)} H_{\overline{5}}
 + \alpha_{d_2} (FT_2)_{\mathbf{3}}Y_{\mathbf{3}}^{(4)} H_{\overline{5}}
 + \alpha_{d_3} (FT_3)_{\mathbf{3}}Y_{\mathbf{3},I}^{(8)} H_{\overline{5}}
 + \alpha_{d_4} (FT_3)_{\mathbf{3}}Y_{\mathbf{3},II}^{(8)} H_{\overline{5}}
\\\nonumber &&
 + \alpha^\prime_{d_1} (FT_1)_{\mathbf{3}}Y_{\mathbf{3}}^{(4)} H_{\overline{45}}
 + \alpha^\prime_{d_2} (FT_2)_{\mathbf{3}}Y_{\mathbf{3}}^{(4)} H_{\overline{45}}
 + \alpha^\prime_{d_3} (FT_3)_{\mathbf{3}}Y_{\mathbf{3},I}^{(8)} H_{\overline{45}}
 + \alpha^\prime_{d_4} (FT_3)_{\mathbf{3}}Y_{\mathbf{3},II}^{(8)} H_{\overline{45}}
\,,\\
{\cal W}_u &=&
   \alpha_{u_1}  (T_1T_2)_{\mathbf{1}}   H_{5}
 + \alpha_{u_2}  (T_1T_3)_{\mathbf{1}}Y_{\mathbf{1}}^{(4)} H_{5}
 + \alpha_{u_3}  (T_3T_3)_{\mathbf{1^{\prime}}}Y_{\mathbf{1^{\prime\prime}}}^{(8)} H_{5}
\,.
\end{eqnarray}
The right-handed neutrino mass matrix and the Yukawa matrix are given by
\begin{eqnarray}
{\cal Y}_{\overline{5}}^d &=& \left(
\begin{array}{ccc}
 \alpha_{d_1}Y_{\mathbf{3},3}^{(4)} ~&~ \alpha_{d_2}Y_{\mathbf{3},2}^{(4)} ~&~ \alpha_{d_3}Y_{\mathbf{3}I,2}^{(8)}+\alpha_{d_4}Y_{\mathbf{3}II,2}^{(8)} \\
 \alpha_{d_1}Y_{\mathbf{3},2}^{(4)} ~&~ \alpha_{d_2}Y_{\mathbf{3},1}^{(4)} ~&~ \alpha_{d_3}Y_{\mathbf{3}I,1}^{(8)}+\alpha_{d_4}Y_{\mathbf{3}II,1}^{(8)} \\
 \alpha_{d_1}Y_{\mathbf{3},1}^{(4)} ~&~ \alpha_{d_2}Y_{\mathbf{3},3}^{(4)} ~&~ \alpha_{d_3}Y_{\mathbf{3}I,3}^{(8)}+\alpha_{d_4}Y_{\mathbf{3}II,3}^{(8)} \\
\end{array}
\right)\,.
\end{eqnarray}
The matrices $M_N$ and ${\cal Y}_5^u$ are identical with the corresponding ones of the model ${\cal V}_{15}$ as well.
The ${\cal Y}_5^{\nu}$ matrix is identical with that of the model ${\cal V}_{17}$.
The parameters $\alpha_{N_1}$, $\alpha_{\nu_1}$, $\alpha_{\nu_2}$, $\alpha_{\nu_3}$, $\alpha_{d_1}$, $\alpha_{d_2}$, $\alpha_{d_3}$, $\alpha_{d_1}^\prime$ and $\alpha_{u_3}$ can taken to be real without loss of generality, while the other parameters are complex.
\item{~Model ${\cal V}_{20}$~:~$\begin{aligned}
&(\rho_{N_1},\rho_{N_2},\rho_{N_3},\rho_F,\rho_{T_1},\rho_{T_2},\rho_{T_3})=(\mathbf{1},\mathbf{1^{\prime\prime}},\mathbf{1^{\prime}},\mathbf{3},\mathbf{1^{\prime}},\mathbf{1^{\prime\prime}},\mathbf{1^{\prime\prime}})\,,\\
& (k_{N_1},k_{N_2},k_{N_3},k_{F},k_{T_1},k_{T_2},k_{T_3})=(2,0,0,2,0,0,4)
\end{aligned}$}

The modular invariant superpotential is of the following form:
\begin{eqnarray}
\nonumber {\cal W}_\nu &=&
   \alpha_{N_1} \Lambda (N_1N_1)_{\mathbf{1}}Y_{\mathbf{1}}^{(4)}
 + \alpha_{N_2} \Lambda (N_2N_3)_{\mathbf{1}}
 + \alpha_{\nu_1} (N_1F)_{\mathbf{3}}Y_{\mathbf{3}}^{(4)} H_{5}
 + \alpha_{\nu_2} (N_2F)_{\mathbf{3}}Y_{\mathbf{3}}^{(2)} H_{5}
\\
\nonumber&&
 + \alpha_{\nu_3} (N_3F)_{\mathbf{3}}Y_{\mathbf{3}}^{(2)} H_{5}
\,,\\
\nonumber {\cal W}_d &=&
   \alpha_{d_1} (FT_1)_{\mathbf{3}}Y_{\mathbf{3}}^{(2)} H_{\overline{5}}
 + \alpha_{d_2} (FT_2)_{\mathbf{3}}Y_{\mathbf{3}}^{(2)} H_{\overline{5}}
 + \alpha_{d_3} (FT_3)_{\mathbf{3}}Y_{\mathbf{3},I}^{(6)} H_{\overline{5}}
 + \alpha_{d_4} (FT_3)_{\mathbf{3}}Y_{\mathbf{3},II}^{(6)} H_{\overline{5}}
\\\nonumber &&
 + \alpha^\prime_{d_1} (FT_1)_{\mathbf{3}}Y_{\mathbf{3}}^{(2)} H_{\overline{45}}
 + \alpha^\prime_{d_2} (FT_2)_{\mathbf{3}}Y_{\mathbf{3}}^{(2)} H_{\overline{45}}
 + \alpha^\prime_{d_3} (FT_3)_{\mathbf{3}}Y_{\mathbf{3},I}^{(6)} H_{\overline{45}}
 + \alpha^\prime_{d_4} (FT_3)_{\mathbf{3}}Y_{\mathbf{3},II}^{(6)} H_{\overline{45}}
\,,\\
{\cal W}_u &=&
   \alpha_{u_1}  (T_1T_2)_{\mathbf{1}}   H_{5}
 + \alpha_{u_2}  (T_1T_3)_{\mathbf{1}}Y_{\mathbf{1}}^{(4)} H_{5}
 + \alpha_{u_3}  (T_3T_3)_{\mathbf{1^{\prime}}}Y_{\mathbf{1^{\prime\prime}}}^{(8)} H_{5}
\,.
\end{eqnarray}
The $M_N$ matrix coincides with that of the model ${\cal V}_{18}$.
The ${\cal Y}_5^{\nu}$ matrix is the same as that of the model ${\cal V}_{4}$.
The matrices ${\cal Y}_{\overline{5}}^d$ and ${\cal Y}_5^u$ are identical with those of the model ${\cal V}_{15}$ as well.
The parameters $\alpha_{N_1}$, $\alpha_{\nu_1}$, $\alpha_{\nu_2}$, $\alpha_{\nu_3}$, $\alpha_{d_1}$, $\alpha_{d_2}$, $\alpha_{d_3}$, $\alpha_{d_1}^\prime$ and $\alpha_{u_3}$ can be taken real by field redefinition, and $\alpha_{N_2}$, $\alpha_{d_4}$, $\alpha_{d_2}^\prime$, $\alpha_{d_3}^\prime$, $\alpha_{d_4}^\prime$, $\alpha_{u_1}$ and $\alpha_{u_2}$ are complex.
\item{~Model ${\cal V}_{21}$~:~$\begin{aligned}
&(\rho_{N_1},\rho_{N_2},\rho_{N_3},\rho_F,\rho_{T_1},\rho_{T_2},\rho_{T_3})=(\mathbf{1},\mathbf{1^{\prime\prime}},\mathbf{1^{\prime}},\mathbf{3},\mathbf{1^{\prime\prime}},\mathbf{1^{\prime}},\mathbf{1^{\prime}})\,,\\
& (k_{N_1},k_{N_2},k_{N_3},k_{F},k_{T_1},k_{T_2},k_{T_3})=(0,0,0,2,2,0,4)
\end{aligned}$}

The modular invariant superpotential is of the following form:
\begin{eqnarray}
\nonumber {\cal W}_\nu &=&
   \alpha_{N_1} \Lambda (N_1N_1)_{\mathbf{1}}
 + \alpha_{N_2} \Lambda (N_2N_3)_{\mathbf{1}}
 + \alpha_{\nu_1} (N_1F)_{\mathbf{3}}Y_{\mathbf{3}}^{(2)} H_{5}
 + \alpha_{\nu_2} (N_2F)_{\mathbf{3}}Y_{\mathbf{3}}^{(2)} H_{5}
\\
\nonumber&&
 + \alpha_{\nu_3} (N_3F)_{\mathbf{3}}Y_{\mathbf{3}}^{(2)} H_{5}
\,,\\
\nonumber {\cal W}_d &=&
   \alpha_{d_1} (FT_1)_{\mathbf{3}}Y_{\mathbf{3}}^{(4)} H_{\overline{5}}
 + \alpha_{d_2} (FT_2)_{\mathbf{3}}Y_{\mathbf{3}}^{(2)} H_{\overline{5}}
 + \alpha_{d_3} (FT_3)_{\mathbf{3}}Y_{\mathbf{3},I}^{(6)} H_{\overline{5}}
 + \alpha_{d_4} (FT_3)_{\mathbf{3}}Y_{\mathbf{3},II}^{(6)} H_{\overline{5}}
\\\nonumber &&
 + \alpha^\prime_{d_1} (FT_1)_{\mathbf{3}}Y_{\mathbf{3}}^{(4)} H_{\overline{45}}
 + \alpha^\prime_{d_2} (FT_2)_{\mathbf{3}}Y_{\mathbf{3}}^{(2)} H_{\overline{45}}
 + \alpha^\prime_{d_3} (FT_3)_{\mathbf{3}}Y_{\mathbf{3},I}^{(6)} H_{\overline{45}}
 + \alpha^\prime_{d_4} (FT_3)_{\mathbf{3}}Y_{\mathbf{3},II}^{(6)} H_{\overline{45}}
\,,\\
{\cal W}_u &=&
   \alpha_{u_1}  (T_1T_3)_{\mathbf{1}}Y_{\mathbf{1}}^{(6)} H_{5}
 + \alpha_{u_2}  (T_2T_3)_{\mathbf{1^{\prime\prime}}}Y_{\mathbf{1^\prime}}^{(4)} H_{5}
 + \alpha_{u_3}  (T_3T_3)_{\mathbf{1^{\prime\prime}}}Y_{\mathbf{1^\prime}}^{(8)} H_{5}
\,.
\end{eqnarray}
The matrices $M_N$ and ${\cal Y}_5^{\nu}$ are identical with the corresponding ones of the model ${\cal V}_{15}$ as well.
The matrices ${\cal Y}_{\overline{5}}^d$ and ${\cal Y}_5^u$ are the same as those of the model ${\cal I}_{2}$ as well.
The phases of the parameters $\alpha_{N_1}$, $\alpha_{\nu_1}$, $\alpha_{\nu_2}$, $\alpha_{\nu_3}$, $\alpha_{d_1}$, $\alpha_{d_2}$, $\alpha_{d_3}$, $\alpha_{d_1}^\prime$ and $\alpha_{u_3}$ are unphysical, and the remaining parameters $\alpha_{N_2}$, $\alpha_{d_4}$, $\alpha_{d_2}^\prime$, $\alpha_{d_3}^\prime$, $\alpha_{d_4}^\prime$, $\alpha_{u_1}$ and $\alpha_{u_2}$ are complex.
\item{~Model ${\cal V}_{22}$~:~$\begin{aligned}
&(\rho_{N_1},\rho_{N_2},\rho_{N_3},\rho_F,\rho_{T_1},\rho_{T_2},\rho_{T_3})=(\mathbf{1},\mathbf{1^{\prime\prime}},\mathbf{1^{\prime}},\mathbf{3},\mathbf{1^{\prime\prime}},\mathbf{1^{\prime}},\mathbf{1^{\prime}})\,,\\
& (k_{N_1},k_{N_2},k_{N_3},k_{F},k_{T_1},k_{T_2},k_{T_3})=(2,0,0,2,2,0,4)
\end{aligned}$}

The modular invariant superpotential for quark and lepton mass is given by
\begin{eqnarray}
\nonumber {\cal W}_\nu &=&
   \alpha_{N_1} \Lambda (N_1N_1)_{\mathbf{1}}Y_{\mathbf{1}}^{(4)}
 + \alpha_{N_2} \Lambda (N_2N_3)_{\mathbf{1}}
 + \alpha_{\nu_1} (N_1F)_{\mathbf{3}}Y_{\mathbf{3}}^{(4)} H_{5}
 + \alpha_{\nu_2} (N_2F)_{\mathbf{3}}Y_{\mathbf{3}}^{(2)} H_{5}
\\
\nonumber&&
 + \alpha_{\nu_3} (N_3F)_{\mathbf{3}}Y_{\mathbf{3}}^{(2)} H_{5}
\,,\\
\nonumber {\cal W}_d &=&
   \alpha_{d_1} (FT_1)_{\mathbf{3}}Y_{\mathbf{3}}^{(4)} H_{\overline{5}}
 + \alpha_{d_2} (FT_2)_{\mathbf{3}}Y_{\mathbf{3}}^{(2)} H_{\overline{5}}
 + \alpha_{d_3} (FT_3)_{\mathbf{3}}Y_{\mathbf{3},I}^{(6)} H_{\overline{5}}
 + \alpha_{d_4} (FT_3)_{\mathbf{3}}Y_{\mathbf{3},II}^{(6)} H_{\overline{5}}
\\\nonumber &&
 + \alpha^\prime_{d_1} (FT_1)_{\mathbf{3}}Y_{\mathbf{3}}^{(4)} H_{\overline{45}}
 + \alpha^\prime_{d_2} (FT_2)_{\mathbf{3}}Y_{\mathbf{3}}^{(2)} H_{\overline{45}}
 + \alpha^\prime_{d_3} (FT_3)_{\mathbf{3}}Y_{\mathbf{3},I}^{(6)} H_{\overline{45}}
 + \alpha^\prime_{d_4} (FT_3)_{\mathbf{3}}Y_{\mathbf{3},II}^{(6)} H_{\overline{45}}
\,,\\
{\cal W}_u &=&
   \alpha_{u_1}  (T_1T_3)_{\mathbf{1}}Y_{\mathbf{1}}^{(6)} H_{5}
 + \alpha_{u_2}  (T_2T_3)_{\mathbf{1^{\prime\prime}}}Y_{\mathbf{1^\prime}}^{(4)} H_{5}
 + \alpha_{u_3}  (T_3T_3)_{\mathbf{1^{\prime\prime}}}Y_{\mathbf{1^\prime}}^{(8)} H_{5}
\,.
\end{eqnarray}
The $M_N$ matrix is identical with the corresponding one of the model ${\cal V}_{18}$.
The ${\cal Y}_5^{\nu}$ matrix is the same as that of the model ${\cal V}_{4}$.
The matrices ${\cal Y}_{\overline{5}}^d$ and ${\cal Y}_5^u$ are identical with those of the model ${\cal I}_{2}$ as well.
The parameters $\alpha_{N_1}$, $\alpha_{\nu_1}$, $\alpha_{\nu_2}$, $\alpha_{\nu_3}$, $\alpha_{d_1}$, $\alpha_{d_2}$, $\alpha_{d_3}$, $\alpha_{d_1}^\prime$ and $\alpha_{u_3}$ can be taken real by field redefinition, and the remaining parameters $\alpha_{N_2}$, $\alpha_{d_4}$, $\alpha_{d_2}^\prime$, $\alpha_{d_3}^\prime$, $\alpha_{d_4}^\prime$, $\alpha_{u_1}$ and $\alpha_{u_2}$ are complex.
\item{~Model ${\cal V}_{23}$~:~$\begin{aligned}
&(\rho_{N_1},\rho_{N_2},\rho_{N_3},\rho_F,\rho_{T_1},\rho_{T_2},\rho_{T_3})=(\mathbf{1},\mathbf{1^{\prime\prime}},\mathbf{1^{\prime}},\mathbf{3},\mathbf{1^{\prime\prime}},\mathbf{1^{\prime\prime}},\mathbf{1^{\prime}})\,,\\
& (k_{N_1},k_{N_2},k_{N_3},k_{F},k_{T_1},k_{T_2},k_{T_3})=(0,0,0,2,0,2,4)
\end{aligned}$}

The modular invariant superpotentials for quark and lepton masses are of the following form,
\begin{eqnarray}
\nonumber {\cal W}_\nu &=&
   \alpha_{N_1} \Lambda (N_1N_1)_{\mathbf{1}}
 + \alpha_{N_2} \Lambda (N_2N_3)_{\mathbf{1}}
 + \alpha_{\nu_1} (N_1F)_{\mathbf{3}}Y_{\mathbf{3}}^{(2)} H_{5}
 + \alpha_{\nu_2} (N_2F)_{\mathbf{3}}Y_{\mathbf{3}}^{(2)} H_{5}
\\
\nonumber&&
 + \alpha_{\nu_3} (N_3F)_{\mathbf{3}}Y_{\mathbf{3}}^{(2)} H_{5}
\,,\\
\nonumber {\cal W}_d &=&
   \alpha_{d_1} (FT_1)_{\mathbf{3}}Y_{\mathbf{3}}^{(2)} H_{\overline{5}}
 + \alpha_{d_2} (FT_2)_{\mathbf{3}}Y_{\mathbf{3}}^{(4)} H_{\overline{5}}
 + \alpha_{d_3} (FT_3)_{\mathbf{3}}Y_{\mathbf{3},I}^{(6)} H_{\overline{5}}
 + \alpha_{d_4} (FT_3)_{\mathbf{3}}Y_{\mathbf{3},II}^{(6)} H_{\overline{5}}
\\\nonumber &&
 + \alpha^\prime_{d_1} (FT_1)_{\mathbf{3}}Y_{\mathbf{3}}^{(2)} H_{\overline{45}}
 + \alpha^\prime_{d_2} (FT_2)_{\mathbf{3}}Y_{\mathbf{3}}^{(4)} H_{\overline{45}}
 + \alpha^\prime_{d_3} (FT_3)_{\mathbf{3}}Y_{\mathbf{3},I}^{(6)} H_{\overline{45}}
 + \alpha^\prime_{d_4} (FT_3)_{\mathbf{3}}Y_{\mathbf{3},II}^{(6)} H_{\overline{45}}
\,,\\
{\cal W}_u &=&
   \alpha_{u_1}  (T_1T_3)_{\mathbf{1}}Y_{\mathbf{1}}^{(4)} H_{5}
 + \alpha_{u_2}  (T_2T_3)_{\mathbf{1}}Y_{\mathbf{1}}^{(6)} H_{5}
 + \alpha_{u_3}  (T_3T_3)_{\mathbf{1^{\prime\prime}}}Y_{\mathbf{1^\prime}}^{(8)} H_{5}
\,.
\end{eqnarray}
The matrices $M_N$ and ${\cal Y}_5^{\nu}$ are identical with those of the model ${\cal V}_{15}$ as well.
The matrices ${\cal Y}_{\overline{5}}^d$ and ${\cal Y}_5^u$ are identical with those of the model ${\cal I}_{3}$ as well.
The phases of the parameters $\alpha_{N_1}$, $\alpha_{\nu_1}$, $\alpha_{\nu_2}$, $\alpha_{\nu_3}$, $\alpha_{d_1}$, $\alpha_{d_2}$, $\alpha_{d_3}$, $\alpha_{d_1}^\prime$ and $\alpha_{u_3}$ can be removed by field redefinition, and the remaining parameters are complex.
\item{~Model ${\cal V}_{24}$~:~$\begin{aligned}
&(\rho_{N_1},\rho_{N_2},\rho_{N_3},\rho_F,\rho_{T_1},\rho_{T_2},\rho_{T_3})=(\mathbf{1},\mathbf{1^{\prime\prime}},\mathbf{1^{\prime}},\mathbf{3},\mathbf{1^{\prime\prime}},\mathbf{1^{\prime\prime}},\mathbf{1^{\prime}})\,,\\
& (k_{N_1},k_{N_2},k_{N_3},k_{F},k_{T_1},k_{T_2},k_{T_3})=(2,0,0,2,0,2,4)
\end{aligned}$}

The superpotentials relevant to quark and lepton masses are of the form,
\begin{eqnarray}
\nonumber {\cal W}_\nu &=&
   \alpha_{N_1} \Lambda (N_1N_1)_{\mathbf{1}}Y_{\mathbf{1}}^{(4)}
 + \alpha_{N_2} \Lambda (N_2N_3)_{\mathbf{1}}
 + \alpha_{\nu_1} (N_1F)_{\mathbf{3}}Y_{\mathbf{3}}^{(4)} H_{5}
 + \alpha_{\nu_2} (N_2F)_{\mathbf{3}}Y_{\mathbf{3}}^{(2)} H_{5}
\\
\nonumber&&
 + \alpha_{\nu_3} (N_3F)_{\mathbf{3}}Y_{\mathbf{3}}^{(2)} H_{5}
\,,\\
\nonumber {\cal W}_d &=&
   \alpha_{d_1} (FT_1)_{\mathbf{3}}Y_{\mathbf{3}}^{(2)} H_{\overline{5}}
 + \alpha_{d_2} (FT_2)_{\mathbf{3}}Y_{\mathbf{3}}^{(4)} H_{\overline{5}}
 + \alpha_{d_3} (FT_3)_{\mathbf{3}}Y_{\mathbf{3},I}^{(6)} H_{\overline{5}}
 + \alpha_{d_4} (FT_3)_{\mathbf{3}}Y_{\mathbf{3},II}^{(6)} H_{\overline{5}}
\\\nonumber &&
 + \alpha^\prime_{d_1} (FT_1)_{\mathbf{3}}Y_{\mathbf{3}}^{(2)} H_{\overline{45}}
 + \alpha^\prime_{d_2} (FT_2)_{\mathbf{3}}Y_{\mathbf{3}}^{(4)} H_{\overline{45}}
 + \alpha^\prime_{d_3} (FT_3)_{\mathbf{3}}Y_{\mathbf{3},I}^{(6)} H_{\overline{45}}
 + \alpha^\prime_{d_4} (FT_3)_{\mathbf{3}}Y_{\mathbf{3},II}^{(6)} H_{\overline{45}}
\,,\\
{\cal W}_u &=&
   \alpha_{u_1}  (T_1T_3)_{\mathbf{1}}Y_{\mathbf{1}}^{(4)} H_{5}
 + \alpha_{u_2}  (T_2T_3)_{\mathbf{1}}Y_{\mathbf{1}}^{(6)} H_{5}
 + \alpha_{u_3}  (T_3T_3)_{\mathbf{1^{\prime\prime}}}Y_{\mathbf{1^\prime}}^{(8)} H_{5}
\,.
\end{eqnarray}
The $M_N$ matrix is identical with that of the model ${\cal V}_{18}$.
The ${\cal Y}_5^{\nu}$ matrix coincides with that of the model ${\cal V}_{4}$.
The matrices ${\cal Y}_{\overline{5}}^d$ and ${\cal Y}_5^u$ are identical with those of the model ${\cal I}_{3}$ as well.
The phases of the parameters $\alpha_{N_1}$, $\alpha_{\nu_1}$, $\alpha_{\nu_2}$, $\alpha_{\nu_3}$, $\alpha_{d_1}$, $\alpha_{d_2}$, $\alpha_{d_3}$, $\alpha_{d_1}^\prime$ and $\alpha_{u_3}$ can be removed by field redefinition, and the other parameters are complex.
\item{~Model ${\cal V}_{25}$~:~$\begin{aligned}
&(\rho_{N_1},\rho_{N_2},\rho_{N_3},\rho_F,\rho_{T_1},\rho_{T_2},\rho_{T_3})=(\mathbf{1^{\prime}},\mathbf{1^{\prime}},\mathbf{1^{\prime\prime}},\mathbf{3},\mathbf{1^{\prime}},\mathbf{1^{\prime\prime}},\mathbf{1})\,,\\
& (k_{N_1},k_{N_2},k_{N_3},k_{F},k_{T_1},k_{T_2},k_{T_3})=(0,2,0,2,0,0,4)
\end{aligned}$}

The modular invariant superpotentials for quark and lepton masses are of the following form,
\begin{eqnarray}
\nonumber {\cal W}_\nu &=&
   \alpha_{N_1} \Lambda (N_1N_3)_{\mathbf{1}}
 + \alpha_{N_2} \Lambda (N_2N_2)_{\mathbf{1^{\prime\prime}}}Y_{\mathbf{1^\prime}}^{(4)}
 + \alpha_{\nu_1} (N_1F)_{\mathbf{3}}Y_{\mathbf{3}}^{(2)} H_{5}
 + \alpha_{\nu_2} (N_2F)_{\mathbf{3}}Y_{\mathbf{3}}^{(4)} H_{5}
\\
\nonumber&&
 + \alpha_{\nu_3} (N_3F)_{\mathbf{3}}Y_{\mathbf{3}}^{(2)} H_{5}
\,,\\
\nonumber {\cal W}_d &=&
   \alpha_{d_1} (FT_1)_{\mathbf{3}}Y_{\mathbf{3}}^{(2)} H_{\overline{5}}
 + \alpha_{d_2} (FT_2)_{\mathbf{3}}Y_{\mathbf{3}}^{(2)} H_{\overline{5}}
 + \alpha_{d_3} (FT_3)_{\mathbf{3}}Y_{\mathbf{3},I}^{(6)} H_{\overline{5}}
 + \alpha_{d_4} (FT_3)_{\mathbf{3}}Y_{\mathbf{3},II}^{(6)} H_{\overline{5}}
\\\nonumber &&
 + \alpha^\prime_{d_1} (FT_1)_{\mathbf{3}}Y_{\mathbf{3}}^{(2)} H_{\overline{45}}
 + \alpha^\prime_{d_2} (FT_2)_{\mathbf{3}}Y_{\mathbf{3}}^{(2)} H_{\overline{45}}
 + \alpha^\prime_{d_3} (FT_3)_{\mathbf{3}}Y_{\mathbf{3},I}^{(6)} H_{\overline{45}}
 + \alpha^\prime_{d_4} (FT_3)_{\mathbf{3}}Y_{\mathbf{3},II}^{(6)} H_{\overline{45}}
\,,\\
{\cal W}_u &=&
   \alpha_{u_1}  (T_1T_2)_{\mathbf{1}}   H_{5}
 + \alpha_{u_2}  (T_2T_3)_{\mathbf{1^{\prime\prime}}}Y_{\mathbf{1^\prime}}^{(4)} H_{5}
 + \alpha_{u_3}  (T_3T_3)_{\mathbf{1}}Y_{\mathbf{1}}^{(8)} H_{5}
\,.
\end{eqnarray}
The right-handed neutrino mass matrix and the Yukawa matrix are given by
\begin{eqnarray}
m_M^{\nu} &=& \Lambda  \left(
\begin{array}{ccc}
 0 ~&~ 0 ~&~ \alpha_{N_1} \\
 0 ~&~ \alpha_{N_2}Y_{\mathbf{1}^\prime}^{(4)} ~&~ 0 \\
 \alpha_{N_1} ~&~ 0 ~&~ 0 \\
\end{array}
\right)\,,\nonumber\\
{\cal Y}_5^{\nu} &=& \left(
\begin{array}{ccc}
 \alpha_{\nu_1}Y_{\mathbf{3},3}^{(2)} ~&~ \alpha_{\nu_1}Y_{\mathbf{3},2}^{(2)} ~&~ \alpha_{\nu_1}Y_{\mathbf{3},1}^{(2)} \\
 \alpha_{\nu_2}Y_{\mathbf{3},3}^{(4)}  ~&~ \alpha_{\nu_2}Y_{\mathbf{3},2}^{(4)} ~&~ \alpha_{\nu_2}Y_{\mathbf{3},1}^{(4)} \\
 \alpha_{\nu_3}Y_{\mathbf{3},2}^{(2)} ~&~ \alpha_{\nu_3}Y_{\mathbf{3},1}^{(2)} ~&~ \alpha_{\nu_3}Y_{\mathbf{3},3}^{(2)} \\
\end{array}
\right)\,.
\end{eqnarray}
The matrices ${\cal Y}_{\overline{5}}^d$ and ${\cal Y}_5^u$ are identical with those of the model ${\cal V}_{16}$ as well.
The parameters $\alpha_{N_1}$, $\alpha_{\nu_1}$, $\alpha_{\nu_2}$, $\alpha_{\nu_3}$, $\alpha_{d_1}$, $\alpha_{d_2}$, $\alpha_{d_3}$, $\alpha_{d_1}^\prime$ and $\alpha_{u_3}$ can be taken real by field redefinition, and $\alpha_{N_2}$, $\alpha_{d_4}$, $\alpha_{d_2}^\prime$, $\alpha_{d_3}^\prime$, $\alpha_{d_4}^\prime$, $\alpha_{u_1}$ and $\alpha_{u_2}$ are generically complex.
\item{~Model ${\cal V}_{26}$~:~$\begin{aligned}
&(\rho_{N_1},\rho_{N_2},\rho_{N_3},\rho_F,\rho_{T_1},\rho_{T_2},\rho_{T_3})=(\mathbf{1^{\prime}},\mathbf{1^{\prime}},\mathbf{1^{\prime\prime}},\mathbf{3},\mathbf{1^{\prime}},\mathbf{1^{\prime\prime}},\mathbf{1^{\prime\prime}})\,,\\
& (k_{N_1},k_{N_2},k_{N_3},k_{F},k_{T_1},k_{T_2},k_{T_3})=(0,2,0,2,0,0,4)
\end{aligned}$}

The modular invariant superpotentials for quark and lepton masses are of the following form,
\begin{eqnarray}
\nonumber {\cal W}_\nu &=&
   \alpha_{N_1} \Lambda (N_1N_3)_{\mathbf{1}}
 + \alpha_{N_2} \Lambda (N_2N_2)_{\mathbf{1^{\prime\prime}}}Y_{\mathbf{1^\prime}}^{(4)}
 + \alpha_{\nu_1} (N_1F)_{\mathbf{3}}Y_{\mathbf{3}}^{(2)} H_{5}
 + \alpha_{\nu_2} (N_2F)_{\mathbf{3}}Y_{\mathbf{3}}^{(4)} H_{5}
\\
\nonumber&&
 + \alpha_{\nu_3} (N_3F)_{\mathbf{3}}Y_{\mathbf{3}}^{(2)} H_{5}
\,,\\
\nonumber {\cal W}_d &=&
   \alpha_{d_1} (FT_1)_{\mathbf{3}}Y_{\mathbf{3}}^{(2)} H_{\overline{5}}
 + \alpha_{d_2} (FT_2)_{\mathbf{3}}Y_{\mathbf{3}}^{(2)} H_{\overline{5}}
 + \alpha_{d_3} (FT_3)_{\mathbf{3}}Y_{\mathbf{3},I}^{(6)} H_{\overline{5}}
 + \alpha_{d_4} (FT_3)_{\mathbf{3}}Y_{\mathbf{3},II}^{(6)} H_{\overline{5}}
\\\nonumber &&
 + \alpha^\prime_{d_1} (FT_1)_{\mathbf{3}}Y_{\mathbf{3}}^{(2)} H_{\overline{45}}
 + \alpha^\prime_{d_2} (FT_2)_{\mathbf{3}}Y_{\mathbf{3}}^{(2)} H_{\overline{45}}
 + \alpha^\prime_{d_3} (FT_3)_{\mathbf{3}}Y_{\mathbf{3},I}^{(6)} H_{\overline{45}}
 + \alpha^\prime_{d_4} (FT_3)_{\mathbf{3}}Y_{\mathbf{3},II}^{(6)} H_{\overline{45}}
\,,\\
{\cal W}_u &=&
   \alpha_{u_1}  (T_1T_2)_{\mathbf{1}}   H_{5}
 + \alpha_{u_2}  (T_1T_3)_{\mathbf{1}}Y_{\mathbf{1}}^{(4)} H_{5}
 + \alpha_{u_3}  (T_3T_3)_{\mathbf{1^{\prime}}}Y_{\mathbf{1^{\prime\prime}}}^{(8)} H_{5}
\,.
\end{eqnarray}
The matrices $M_N$ and ${\cal Y}_5^{\nu}$ are identical with those of the model ${\cal V}_{25}$ as well.
The matrices ${\cal Y}_{\overline{5}}^d$ and ${\cal Y}_5^u$ are identical with those of the model ${\cal V}_{15}$ as well.
The parameters $\alpha_{N_1}$, $\alpha_{\nu_1}$, $\alpha_{\nu_2}$, $\alpha_{\nu_3}$, $\alpha_{d_1}$, $\alpha_{d_2}$, $\alpha_{d_3}$, $\alpha_{d_1}^\prime$ and $\alpha_{u_3}$ can be taken real by field redefinition, and $\alpha_{N_2}$, $\alpha_{d_4}$, $\alpha_{d_2}^\prime$, $\alpha_{d_3}^\prime$, $\alpha_{d_4}^\prime$, $\alpha_{u_1}$ and $\alpha_{u_2}$ are generically complex.
\item{~Model ${\cal V}_{27}$~:~$\begin{aligned}
&(\rho_{N_1},\rho_{N_2},\rho_{N_3},\rho_F,\rho_{T_1},\rho_{T_2},\rho_{T_3})=(\mathbf{1^{\prime}},\mathbf{1^{\prime}},\mathbf{1^{\prime\prime}},\mathbf{3},\mathbf{1^{\prime\prime}},\mathbf{1^{\prime}},\mathbf{1^{\prime}})\,,\\
& (k_{N_1},k_{N_2},k_{N_3},k_{F},k_{T_1},k_{T_2},k_{T_3})=(0,2,0,2,2,0,4)
\end{aligned}$}

We can read out the the superpotential for the fermion masses as,
\begin{eqnarray}
\nonumber {\cal W}_\nu &=&
   \alpha_{N_1} \Lambda (N_1N_3)_{\mathbf{1}}
 + \alpha_{N_2} \Lambda (N_2N_2)_{\mathbf{1^{\prime\prime}}}Y_{\mathbf{1^\prime}}^{(4)}
 + \alpha_{\nu_1} (N_1F)_{\mathbf{3}}Y_{\mathbf{3}}^{(2)} H_{5}
 + \alpha_{\nu_2} (N_2F)_{\mathbf{3}}Y_{\mathbf{3}}^{(4)} H_{5}
\\
\nonumber&&
 + \alpha_{\nu_3} (N_3F)_{\mathbf{3}}Y_{\mathbf{3}}^{(2)} H_{5}
\,,\\
\nonumber {\cal W}_d &=&
   \alpha_{d_1} (FT_1)_{\mathbf{3}}Y_{\mathbf{3}}^{(4)} H_{\overline{5}}
 + \alpha_{d_2} (FT_2)_{\mathbf{3}}Y_{\mathbf{3}}^{(2)} H_{\overline{5}}
 + \alpha_{d_3} (FT_3)_{\mathbf{3}}Y_{\mathbf{3},I}^{(6)} H_{\overline{5}}
 + \alpha_{d_4} (FT_3)_{\mathbf{3}}Y_{\mathbf{3},II}^{(6)} H_{\overline{5}}
\\\nonumber &&
 + \alpha^\prime_{d_1} (FT_1)_{\mathbf{3}}Y_{\mathbf{3}}^{(4)} H_{\overline{45}}
 + \alpha^\prime_{d_2} (FT_2)_{\mathbf{3}}Y_{\mathbf{3}}^{(2)} H_{\overline{45}}
 + \alpha^\prime_{d_3} (FT_3)_{\mathbf{3}}Y_{\mathbf{3},I}^{(6)} H_{\overline{45}}
 + \alpha^\prime_{d_4} (FT_3)_{\mathbf{3}}Y_{\mathbf{3},II}^{(6)} H_{\overline{45}}
\,,\\
{\cal W}_u &=&
   \alpha_{u_1}  (T_1T_3)_{\mathbf{1}}Y_{\mathbf{1}}^{(6)} H_{5}
 + \alpha_{u_2}  (T_2T_3)_{\mathbf{1^{\prime\prime}}}Y_{\mathbf{1^\prime}}^{(4)} H_{5}
 + \alpha_{u_3}  (T_3T_3)_{\mathbf{1^{\prime\prime}}}Y_{\mathbf{1^\prime}}^{(8)} H_{5}
\,.
\end{eqnarray}
The matrices $M_N$ and ${\cal Y}_5^{\nu}$ are identical with those of the model ${\cal V}_{25}$ as well.
The matrices ${\cal Y}_{\overline{5}}^d$ and ${\cal Y}_5^u$ are the same as those of the model ${\cal I}_{2}$ as well.
The phases of the parameters $\alpha_{N_1}$, $\alpha_{\nu_1}$, $\alpha_{\nu_2}$, $\alpha_{\nu_3}$, $\alpha_{d_1}$, $\alpha_{d_2}$, $\alpha_{d_3}$, $\alpha_{d_1}^\prime$ and $\alpha_{u_3}$ are unphysical, and the remaining parameters $\alpha_{N_2}$, $\alpha_{d_4}$, $\alpha_{d_2}^\prime$, $\alpha_{d_3}^\prime$, $\alpha_{d_4}^\prime$, $\alpha_{u_1}$ and $\alpha_{u_2}$ are complex.
\item{~Model ${\cal V}_{28}$~:~$\begin{aligned}
&(\rho_{N_1},\rho_{N_2},\rho_{N_3},\rho_F,\rho_{T_1},\rho_{T_2},\rho_{T_3})=(\mathbf{1^{\prime}},\mathbf{1^{\prime}},\mathbf{1^{\prime\prime}},\mathbf{3},\mathbf{1^{\prime\prime}},\mathbf{1^{\prime\prime}},\mathbf{1^{\prime}})\,,\\
& (k_{N_1},k_{N_2},k_{N_3},k_{F},k_{T_1},k_{T_2},k_{T_3})=(0,2,0,2,0,2,4)
\end{aligned}$}

The quark and lepton masses are described by the following superpotential:
\begin{eqnarray}
\nonumber {\cal W}_\nu &=&
   \alpha_{N_1} \Lambda (N_1N_3)_{\mathbf{1}}
 + \alpha_{N_2} \Lambda (N_2N_2)_{\mathbf{1^{\prime\prime}}}Y_{\mathbf{1^\prime}}^{(4)}
 + \alpha_{\nu_1} (N_1F)_{\mathbf{3}}Y_{\mathbf{3}}^{(2)} H_{5}
 + \alpha_{\nu_2} (N_2F)_{\mathbf{3}}Y_{\mathbf{3}}^{(4)} H_{5}
\\
\nonumber&&
 + \alpha_{\nu_3} (N_3F)_{\mathbf{3}}Y_{\mathbf{3}}^{(2)} H_{5}
\,,\\
\nonumber {\cal W}_d &=&
   \alpha_{d_1} (FT_1)_{\mathbf{3}}Y_{\mathbf{3}}^{(2)} H_{\overline{5}}
 + \alpha_{d_2} (FT_2)_{\mathbf{3}}Y_{\mathbf{3}}^{(4)} H_{\overline{5}}
 + \alpha_{d_3} (FT_3)_{\mathbf{3}}Y_{\mathbf{3},I}^{(6)} H_{\overline{5}}
 + \alpha_{d_4} (FT_3)_{\mathbf{3}}Y_{\mathbf{3},II}^{(6)} H_{\overline{5}}
\\\nonumber &&
 + \alpha^\prime_{d_1} (FT_1)_{\mathbf{3}}Y_{\mathbf{3}}^{(2)} H_{\overline{45}}
 + \alpha^\prime_{d_2} (FT_2)_{\mathbf{3}}Y_{\mathbf{3}}^{(4)} H_{\overline{45}}
 + \alpha^\prime_{d_3} (FT_3)_{\mathbf{3}}Y_{\mathbf{3},I}^{(6)} H_{\overline{45}}
 + \alpha^\prime_{d_4} (FT_3)_{\mathbf{3}}Y_{\mathbf{3},II}^{(6)} H_{\overline{45}}
\,,\\
{\cal W}_u &=&
   \alpha_{u_1}  (T_1T_3)_{\mathbf{1}}Y_{\mathbf{1}}^{(4)} H_{5}
 + \alpha_{u_2}  (T_2T_3)_{\mathbf{1}}Y_{\mathbf{1}}^{(6)} H_{5}
 + \alpha_{u_3}  (T_3T_3)_{\mathbf{1^{\prime\prime}}}Y_{\mathbf{1^\prime}}^{(8)} H_{5}
\,.
\end{eqnarray}
The matrices $M_N$ and ${\cal Y}_5^{\nu}$ are identical with those of the model ${\cal V}_{25}$ as well.
The matrices ${\cal Y}_{\overline{5}}^d$ and ${\cal Y}_5^u$ are identical with those of the model ${\cal I}_{3}$ as well.
The phases of the parameters $\alpha_{N_1}$, $\alpha_{\nu_1}$, $\alpha_{\nu_2}$, $\alpha_{\nu_3}$, $\alpha_{d_1}$, $\alpha_{d_2}$, $\alpha_{d_3}$, $\alpha_{d_1}^\prime$ and $\alpha_{u_3}$ can be removed by field redefinition, and $\alpha_{N_2}$, $\alpha_{d_4}$, $\alpha_{d_2}^\prime$, $\alpha_{d_3}^\prime$, $\alpha_{d_4}^\prime$, $\alpha_{u_1}$ and $\alpha_{u_2}$ are complex.
\end{itemize}
\subsection{Type-VII}
\begin{itemize}[leftmargin=1.0em]
\item{~Model ${\cal VII}_{1}$~:~$\begin{aligned}
&(\rho_{N_1},\rho_{N_2},\rho_F,\rho_{T_1},\rho_{T_2},\rho_{T_3})=(\mathbf{1},\mathbf{1^{\prime\prime}},\mathbf{3},\mathbf{1^{\prime\prime}},\mathbf{1},\mathbf{1})\,,\\
& (k_{N_1},k_{N_2},k_{F},k_{T_1},k_{T_2},k_{T_3})=(1,3,3,1,1,3)
\end{aligned}$}

The modular invariant superpotentials for quark and lepton masses are of the following form,
\begin{eqnarray}
\nonumber {\cal W}_\nu &=&
   \alpha_{N_1} \Lambda (N_1N_2)_{\mathbf{1^{\prime\prime}}}Y_{\mathbf{1^\prime}}^{(4)}
 + \alpha_{\nu_1} (N_1F)_{\mathbf{3}}Y_{\mathbf{3}}^{(4)} H_{5}
 + \alpha_{\nu_2} (N_2F)_{\mathbf{3}}Y_{\mathbf{3},I}^{(6)} H_{5}
 + \alpha_{\nu_3} (N_2F)_{\mathbf{3}}Y_{\mathbf{3},II}^{(6)} H_{5}
\,,\\
\nonumber {\cal W}_d &=&
   \alpha_{d_1} (FT_1)_{\mathbf{3}}Y_{\mathbf{3}}^{(4)} H_{\overline{5}}
 + \alpha_{d_2} (FT_2)_{\mathbf{3}}Y_{\mathbf{3}}^{(4)} H_{\overline{5}}
 + \alpha_{d_3} (FT_3)_{\mathbf{3}}Y_{\mathbf{3},I}^{(6)} H_{\overline{5}}
 + \alpha_{d_4} (FT_3)_{\mathbf{3}}Y_{\mathbf{3},II}^{(6)} H_{\overline{5}}
\\\nonumber &&
 + \alpha^\prime_{d_1} (FT_1)_{\mathbf{3}}Y_{\mathbf{3}}^{(4)} H_{\overline{45}}
 + \alpha^\prime_{d_2} (FT_2)_{\mathbf{3}}Y_{\mathbf{3}}^{(4)} H_{\overline{45}}
 + \alpha^\prime_{d_3} (FT_3)_{\mathbf{3}}Y_{\mathbf{3},I}^{(6)} H_{\overline{45}}
 + \alpha^\prime_{d_4} (FT_3)_{\mathbf{3}}Y_{\mathbf{3},II}^{(6)} H_{\overline{45}}
\,,\\
{\cal W}_u &=&
   \alpha_{u_1}  (T_1T_3)_{\mathbf{1^{\prime\prime}}}Y_{\mathbf{1^\prime}}^{(4)} H_{5}
 + \alpha_{u_2}  (T_2T_3)_{\mathbf{1}}Y_{\mathbf{1}}^{(4)} H_{5}
 + \alpha_{u_3}  (T_3T_3)_{\mathbf{1}}Y_{\mathbf{1}}^{(6)} H_{5}
\,.
\end{eqnarray}
which gives rise to
\begin{eqnarray}
m_M^{\nu} &=& \Lambda  \left(
\begin{array}{cc}
 0 ~&~ \alpha_{N_1}Y_{\mathbf{1}^\prime}^{(4)} \\
 \alpha_{N_1}Y_{\mathbf{1}^\prime}^{(4)} ~&~ 0 \\
\end{array}
\right)\,,\nonumber\\
{\cal Y}_5^{\nu} &=& \left(
\begin{array}{ccc}
 \alpha_{\nu_1}Y_{\mathbf{3},1}^{(4)} ~&~ \alpha_{\nu_1}Y_{\mathbf{3},3}^{(4)} ~&~ \alpha_{\nu_1}Y_{\mathbf{3},2}^{(4)} \\
 \alpha_{\nu_2}Y_{\mathbf{3}I,2}^{(6)}+\alpha_{\nu_3}Y_{\mathbf{3}II,2}^{(6)} ~&~ \alpha_{\nu_2}Y_{\mathbf{3}I,1}^{(6)}+\alpha_{\nu_3}Y_{\mathbf{3}II,1}^{(6)} ~&~ \alpha_{\nu_2}Y_{\mathbf{3}I,3}^{(6)}+\alpha_{\nu_3}Y_{\mathbf{3}II,3}^{(6)} \\
\end{array}
\right)\,.
\end{eqnarray}
The matrices ${\cal Y}_{\overline{5}}^d$ and ${\cal Y}_5^u$ are in common with the corresponding ones of the models ${\cal I}_{6}$ and ${\cal I}_{1}$ respectively.
The parameters $\alpha_{N_1}$, $\alpha_{\nu_1}$, $\alpha_{\nu_2}$, $\alpha_{d_1}$, $\alpha_{d_2}$, $\alpha_{d_3}$, $\alpha_{d_1}^\prime$ and $\alpha_{u_3}$ can taken to be real without loss of generality, while $\alpha_{\nu_3}$, $\alpha_{d_4}$, $\alpha_{d_2}^\prime$, $\alpha_{d_3}^\prime$, $\alpha_{d_4}^\prime$, $\alpha_{u_1}$ and $\alpha_{u_2}$ are complex.
\item{~Model ${\cal VII}_{2}$~:~$\begin{aligned}
&(\rho_{N_1},\rho_{N_2},\rho_F,\rho_{T_1},\rho_{T_2},\rho_{T_3})=(\mathbf{1^{\prime}},\mathbf{1^{\prime}},\mathbf{3},\mathbf{1^{\prime\prime}},\mathbf{1},\mathbf{1})\,,\\
& (k_{N_1},k_{N_2},k_{F},k_{T_1},k_{T_2},k_{T_3})=(1,3,3,1,1,3)
\end{aligned}$}

We can read out the superpotential for quark and leptons as follows,
\begin{eqnarray}
\nonumber {\cal W}_\nu &=&
   \alpha_{N_1} \Lambda (N_1N_2)_{\mathbf{1^{\prime\prime}}}Y_{\mathbf{1^\prime}}^{(4)}
 + \alpha_{\nu_1} (N_1F)_{\mathbf{3}}Y_{\mathbf{3}}^{(4)} H_{5}
 + \alpha_{\nu_2} (N_2F)_{\mathbf{3}}Y_{\mathbf{3},I}^{(6)} H_{5}
 + \alpha_{\nu_3} (N_2F)_{\mathbf{3}}Y_{\mathbf{3},II}^{(6)} H_{5}
\,,\\
\nonumber {\cal W}_d &=&
   \alpha_{d_1} (FT_1)_{\mathbf{3}}Y_{\mathbf{3}}^{(4)} H_{\overline{5}}
 + \alpha_{d_2} (FT_2)_{\mathbf{3}}Y_{\mathbf{3}}^{(4)} H_{\overline{5}}
 + \alpha_{d_3} (FT_3)_{\mathbf{3}}Y_{\mathbf{3},I}^{(6)} H_{\overline{5}}
 + \alpha_{d_4} (FT_3)_{\mathbf{3}}Y_{\mathbf{3},II}^{(6)} H_{\overline{5}}
\\\nonumber &&
 + \alpha^\prime_{d_1} (FT_1)_{\mathbf{3}}Y_{\mathbf{3}}^{(4)} H_{\overline{45}}
 + \alpha^\prime_{d_2} (FT_2)_{\mathbf{3}}Y_{\mathbf{3}}^{(4)} H_{\overline{45}}
 + \alpha^\prime_{d_3} (FT_3)_{\mathbf{3}}Y_{\mathbf{3},I}^{(6)} H_{\overline{45}}
 + \alpha^\prime_{d_4} (FT_3)_{\mathbf{3}}Y_{\mathbf{3},II}^{(6)} H_{\overline{45}}
\,,\\
{\cal W}_u &=&
   \alpha_{u_1}  (T_1T_3)_{\mathbf{1^{\prime\prime}}}Y_{\mathbf{1^\prime}}^{(4)} H_{5}
 + \alpha_{u_2}  (T_2T_3)_{\mathbf{1}}Y_{\mathbf{1}}^{(4)} H_{5}
 + \alpha_{u_3}  (T_3T_3)_{\mathbf{1}}Y_{\mathbf{1}}^{(6)} H_{5}
\,.
\end{eqnarray}
We can straightforwardly read out the right-handed neutrino mass matrix and the Yukawa matrices as follows,
\begin{eqnarray}
{\cal Y}_5^{\nu} &=& \left(
\begin{array}{ccc}
 \alpha_{\nu_1}Y_{\mathbf{3},3}^{(4)} ~&~ \alpha_{\nu_1}Y_{\mathbf{3},2}^{(4)} ~&~ \alpha_{\nu_1}Y_{\mathbf{3},1}^{(4)} \\
 \alpha_{\nu_2}Y_{\mathbf{3}I,3}^{(6)}+\alpha_{\nu_3}Y_{\mathbf{3}II,3}^{(6)} ~&~ \alpha_{\nu_2}Y_{\mathbf{3}I,2}^{(6)}+\alpha_{\nu_3}Y_{\mathbf{3}II,2}^{(6)} ~&~ \alpha_{\nu_2}Y_{\mathbf{3}I,1}^{(6)}+\alpha_{\nu_3}Y_{\mathbf{3}II,1}^{(6)} \\
\end{array}
\right)\,.
\end{eqnarray}
The $M_N$ matrix is in common with that of the model ${\cal VII}_{1}$.
The ${\cal Y}_{\overline{5}}^d$ matrix is the same as that of the model ${\cal I}_{6}$.
The ${\cal Y}_5^u$ matrix is in common with that of the model ${\cal I}_{1}$.
The phases of the parameters $\alpha_{N_1}$, $\alpha_{\nu_1}$, $\alpha_{\nu_2}$, $\alpha_{d_1}$, $\alpha_{d_2}$, $\alpha_{d_3}$, $\alpha_{d_1}^\prime$ and $\alpha_{u_3}$ can be removed by field redefinition, and $\alpha_{\nu_3}$, $\alpha_{d_4}$, $\alpha_{d_2}^\prime$, $\alpha_{d_3}^\prime$, $\alpha_{d_4}^\prime$, $\alpha_{u_1}$ and $\alpha_{u_2}$ are generically complex.
\item{~Model ${\cal VII}_{3}$~:~$\begin{aligned}
&(\rho_{N_1},\rho_{N_2},\rho_F,\rho_{T_1},\rho_{T_2},\rho_{T_3})=(\mathbf{1^{\prime}},\mathbf{1^{\prime\prime}},\mathbf{3},\mathbf{1^{\prime\prime}},\mathbf{1},\mathbf{1})\,,\\
& (k_{N_1},k_{N_2},k_{F},k_{T_1},k_{T_2},k_{T_3})=(1,3,3,1,1,3)
\end{aligned}$}

We can read out the superpotential for quark and leptons as follows,
\begin{eqnarray}
\nonumber {\cal W}_\nu &=&
   \alpha_{N_1} \Lambda (N_1N_2)_{\mathbf{1}}Y_{\mathbf{1}}^{(4)}
 + \alpha_{\nu_1} (N_1F)_{\mathbf{3}}Y_{\mathbf{3}}^{(4)} H_{5}
 + \alpha_{\nu_2} (N_2F)_{\mathbf{3}}Y_{\mathbf{3},I}^{(6)} H_{5}
 + \alpha_{\nu_3} (N_2F)_{\mathbf{3}}Y_{\mathbf{3},II}^{(6)} H_{5}
\,,\\
\nonumber {\cal W}_d &=&
   \alpha_{d_1} (FT_1)_{\mathbf{3}}Y_{\mathbf{3}}^{(4)} H_{\overline{5}}
 + \alpha_{d_2} (FT_2)_{\mathbf{3}}Y_{\mathbf{3}}^{(4)} H_{\overline{5}}
 + \alpha_{d_3} (FT_3)_{\mathbf{3}}Y_{\mathbf{3},I}^{(6)} H_{\overline{5}}
 + \alpha_{d_4} (FT_3)_{\mathbf{3}}Y_{\mathbf{3},II}^{(6)} H_{\overline{5}}
\\\nonumber &&
 + \alpha^\prime_{d_1} (FT_1)_{\mathbf{3}}Y_{\mathbf{3}}^{(4)} H_{\overline{45}}
 + \alpha^\prime_{d_2} (FT_2)_{\mathbf{3}}Y_{\mathbf{3}}^{(4)} H_{\overline{45}}
 + \alpha^\prime_{d_3} (FT_3)_{\mathbf{3}}Y_{\mathbf{3},I}^{(6)} H_{\overline{45}}
 + \alpha^\prime_{d_4} (FT_3)_{\mathbf{3}}Y_{\mathbf{3},II}^{(6)} H_{\overline{45}}
\,,\\
{\cal W}_u &=&
   \alpha_{u_1}  (T_1T_3)_{\mathbf{1^{\prime\prime}}}Y_{\mathbf{1^\prime}}^{(4)} H_{5}
 + \alpha_{u_2}  (T_2T_3)_{\mathbf{1}}Y_{\mathbf{1}}^{(4)} H_{5}
 + \alpha_{u_3}  (T_3T_3)_{\mathbf{1}}Y_{\mathbf{1}}^{(6)} H_{5}
\,,
\end{eqnarray}
which lead to
\begin{eqnarray}
m_M^{\nu} &=& \Lambda  \left(
\begin{array}{cc}
 0 ~&~ \alpha_{N_1}Y_\mathbf{1}^{(4)} \\
 \alpha_{N_1}Y_\mathbf{1}^{(4)} ~&~ 0 \\
\end{array}
\right)\,,\nonumber\\
{\cal Y}_5^{\nu} &=& \left(
\begin{array}{ccc}
 \alpha_{\nu_1}Y_{\mathbf{3},3}^{(4)} ~&~ \alpha_{\nu_1}Y_{\mathbf{3},2}^{(4)} ~&~ \alpha_{\nu_1}Y_{\mathbf{3},1}^{(4)} \\
 \alpha_{\nu_2}Y_{\mathbf{3}I,2}^{(6)}+\alpha_{\nu_3}Y_{\mathbf{3}II,2}^{(6)} ~&~ \alpha_{\nu_2}Y_{\mathbf{3}I,1}^{(6)}+\alpha_{\nu_3}Y_{\mathbf{3}II,1}^{(6)} ~&~ \alpha_{\nu_2}Y_{\mathbf{3}I,3}^{(6)}+\alpha_{\nu_3}Y_{\mathbf{3}II,3}^{(6)} \\
\end{array}
\right)\,.
\end{eqnarray}
The matrices ${\cal Y}_{\overline{5}}^d$ and ${\cal Y}_5^u$ coincide with those of the model ${\cal VII}_{1}$ as well.
The phases of the parameters $\alpha_{N_1}$, $\alpha_{\nu_1}$, $\alpha_{\nu_2}$, $\alpha_{d_1}$, $\alpha_{d_2}$, $\alpha_{d_3}$, $\alpha_{d_1}^\prime$ and $\alpha_{u_3}$ can be removed by field redefinition, and $\alpha_{\nu_3}$, $\alpha_{d_4}$, $\alpha_{d_2}^\prime$, $\alpha_{d_3}^\prime$, $\alpha_{d_4}^\prime$, $\alpha_{u_1}$ and $\alpha_{u_2}$ are generically complex.
\item{~Model ${\cal VII}_{4}$~:~$\begin{aligned}
&(\rho_{N_1},\rho_{N_2},\rho_F,\rho_{T_1},\rho_{T_2},\rho_{T_3})=(\mathbf{1},\mathbf{1},\mathbf{3},\mathbf{1^{\prime\prime}},\mathbf{1^{\prime\prime}},\mathbf{1^{\prime}})\,,\\
& (k_{N_1},k_{N_2},k_{F},k_{T_1},k_{T_2},k_{T_3})=(0,2,2,0,2,4)
\end{aligned}$}

We can read out the superpotential for quark and leptons as follows,
\begin{eqnarray}
\nonumber {\cal W}_\nu &=&
   \alpha_{N_1} \Lambda (N_1N_1)_{\mathbf{1}}
 + \alpha_{N_2} \Lambda (N_2N_2)_{\mathbf{1}}Y_{\mathbf{1}}^{(4)}
 + \alpha_{\nu_1} (N_1F)_{\mathbf{3}}Y_{\mathbf{3}}^{(2)} H_{5}
 + \alpha_{\nu_2} (N_2F)_{\mathbf{3}}Y_{\mathbf{3}}^{(4)} H_{5}
\,,\\
\nonumber {\cal W}_d &=&
   \alpha_{d_1} (FT_1)_{\mathbf{3}}Y_{\mathbf{3}}^{(2)} H_{\overline{5}}
 + \alpha_{d_2} (FT_2)_{\mathbf{3}}Y_{\mathbf{3}}^{(4)} H_{\overline{5}}
 + \alpha_{d_3} (FT_3)_{\mathbf{3}}Y_{\mathbf{3},I}^{(6)} H_{\overline{5}}
 + \alpha_{d_4} (FT_3)_{\mathbf{3}}Y_{\mathbf{3},II}^{(6)} H_{\overline{5}}
\\\nonumber &&
 + \alpha^\prime_{d_1} (FT_1)_{\mathbf{3}}Y_{\mathbf{3}}^{(2)} H_{\overline{45}}
 + \alpha^\prime_{d_2} (FT_2)_{\mathbf{3}}Y_{\mathbf{3}}^{(4)} H_{\overline{45}}
 + \alpha^\prime_{d_3} (FT_3)_{\mathbf{3}}Y_{\mathbf{3},I}^{(6)} H_{\overline{45}}
 + \alpha^\prime_{d_4} (FT_3)_{\mathbf{3}}Y_{\mathbf{3},II}^{(6)} H_{\overline{45}}
\,,\\
{\cal W}_u &=&
   \alpha_{u_1}  (T_1T_3)_{\mathbf{1}}Y_{\mathbf{1}}^{(4)} H_{5}
 + \alpha_{u_2}  (T_2T_3)_{\mathbf{1}}Y_{\mathbf{1}}^{(6)} H_{5}
 + \alpha_{u_3}  (T_3T_3)_{\mathbf{1^{\prime\prime}}}Y_{\mathbf{1^\prime}}^{(8)} H_{5}
\,.
\end{eqnarray}
The right-handed neutrino mass matrix and the Yukawa coupling matrices are predicted to be
\begin{eqnarray}
m_M^{\nu} &=& \Lambda  \left(
\begin{array}{cc}
 \alpha_{N_1} ~&~ 0 \\
 0 ~&~ \alpha_{N_2}Y_\mathbf{1}^{(4)} \\
\end{array}
\right)\,,\nonumber\\
{\cal Y}_5^{\nu} &=& \left(
\begin{array}{ccc}
 \alpha_{\nu_1}Y_{\mathbf{3},1}^{(2)} ~&~ \alpha_{\nu_1}Y_{\mathbf{3},3}^{(2)} ~&~ \alpha_{\nu_1}Y_{\mathbf{3},2}^{(2)} \\
 \alpha_{\nu_2}Y_{\mathbf{3},1}^{(4)} ~&~ \alpha_{\nu_2}Y_{\mathbf{3},3}^{(4)}  ~&~ \alpha_{\nu_2}Y_{\mathbf{3},2}^{(4)} \\
\end{array}
\right)\,.
\end{eqnarray}
The matrices ${\cal Y}_{\overline{5}}^d$ and ${\cal Y}_5^u$ are the same as those of the model ${\cal I}_{3}$ as well.
The phases of the parameters $\alpha_{N_1}$, $\alpha_{\nu_1}$, $\alpha_{\nu_2}$, $\alpha_{d_1}$, $\alpha_{d_2}$, $\alpha_{d_3}$, $\alpha_{d_1}^\prime$ and $\alpha_{u_3}$ are unphysical, and the other parameters are complex.
\item{~Model ${\cal VII}_{5}$~:~$\begin{aligned}
&(\rho_{N_1},\rho_{N_2},\rho_F,\rho_{T_1},\rho_{T_2},\rho_{T_3})=(\mathbf{1},\mathbf{1^{\prime}},\mathbf{3},\mathbf{1^{\prime\prime}},\mathbf{1^{\prime\prime}},\mathbf{1^{\prime}})\,,\\
& (k_{N_1},k_{N_2},k_{F},k_{T_1},k_{T_2},k_{T_3})=(0,2,2,0,2,4)
\end{aligned}$}

We can read out the superpotentials relevant to quark and lepton masses
\begin{eqnarray}
\nonumber {\cal W}_\nu &=&
   \alpha_{N_1} \Lambda (N_1N_1)_{\mathbf{1}}
 + \alpha_{N_2} \Lambda (N_2N_2)_{\mathbf{1^{\prime\prime}}}Y_{\mathbf{1^\prime}}^{(4)}
 + \alpha_{\nu_1} (N_1F)_{\mathbf{3}}Y_{\mathbf{3}}^{(2)} H_{5}
 + \alpha_{\nu_2} (N_2F)_{\mathbf{3}}Y_{\mathbf{3}}^{(4)} H_{5}
\,,\\
\nonumber {\cal W}_d &=&
   \alpha_{d_1} (FT_1)_{\mathbf{3}}Y_{\mathbf{3}}^{(2)} H_{\overline{5}}
 + \alpha_{d_2} (FT_2)_{\mathbf{3}}Y_{\mathbf{3}}^{(4)} H_{\overline{5}}
 + \alpha_{d_3} (FT_3)_{\mathbf{3}}Y_{\mathbf{3},I}^{(6)} H_{\overline{5}}
 + \alpha_{d_4} (FT_3)_{\mathbf{3}}Y_{\mathbf{3},II}^{(6)} H_{\overline{5}}
\\\nonumber &&
 + \alpha^\prime_{d_1} (FT_1)_{\mathbf{3}}Y_{\mathbf{3}}^{(2)} H_{\overline{45}}
 + \alpha^\prime_{d_2} (FT_2)_{\mathbf{3}}Y_{\mathbf{3}}^{(4)} H_{\overline{45}}
 + \alpha^\prime_{d_3} (FT_3)_{\mathbf{3}}Y_{\mathbf{3},I}^{(6)} H_{\overline{45}}
 + \alpha^\prime_{d_4} (FT_3)_{\mathbf{3}}Y_{\mathbf{3},II}^{(6)} H_{\overline{45}}
\,,\\
{\cal W}_u &=&
   \alpha_{u_1}  (T_1T_3)_{\mathbf{1}}Y_{\mathbf{1}}^{(4)} H_{5}
 + \alpha_{u_2}  (T_2T_3)_{\mathbf{1}}Y_{\mathbf{1}}^{(6)} H_{5}
 + \alpha_{u_3}  (T_3T_3)_{\mathbf{1^{\prime\prime}}}Y_{\mathbf{1^\prime}}^{(8)} H_{5}
\,.
\end{eqnarray}
The right-handed neutrino mass matrix and the Yukawa matrix are given by
\begin{eqnarray}
m_M^{\nu} &=& \Lambda  \left(
\begin{array}{cc}
 \alpha_{N_1} ~&~ 0 \\
 0 ~&~ \alpha_{N_2}Y_{\mathbf{1}^\prime}^{(4)} \\
\end{array}
\right)\,,\nonumber\\
{\cal Y}_5^{\nu} &=& \left(
\begin{array}{ccc}
 \alpha_{\nu_1}Y_{\mathbf{3},1}^{(2)} ~&~ \alpha_{\nu_1}Y_{\mathbf{3},3}^{(2)} ~&~ \alpha_{\nu_1}Y_{\mathbf{3},2}^{(2)} \\
 \alpha_{\nu_2}Y_{\mathbf{3},3}^{(4)}  ~&~ \alpha_{\nu_2}Y_{\mathbf{3},2}^{(4)} ~&~ \alpha_{\nu_2}Y_{\mathbf{3},1}^{(4)} \\
\end{array}
\right)\,.
\end{eqnarray}
The matrices ${\cal Y}_{\overline{5}}^d$ and ${\cal Y}_5^u$ coincide with those of the model ${\cal VII}_{4}$ as well.
The parameters $\alpha_{N_1}$, $\alpha_{\nu_1}$, $\alpha_{\nu_2}$, $\alpha_{d_1}$, $\alpha_{d_2}$, $\alpha_{d_3}$, $\alpha_{d_1}^\prime$ and $\alpha_{u_3}$ are real, and $\alpha_{N_2}$, $\alpha_{d_4}$, $\alpha_{d_2}^\prime$, $\alpha_{d_3}^\prime$, $\alpha_{d_4}^\prime$, $\alpha_{u_1}$ and $\alpha_{u_2}$ are generically complex.
\item{~Model ${\cal VII}_{6}$~:~$\begin{aligned}
&(\rho_{N_1},\rho_{N_2},\rho_F,\rho_{T_1},\rho_{T_2},\rho_{T_3})=(\mathbf{1},\mathbf{1^{\prime}},\mathbf{3},\mathbf{1^{\prime\prime}},\mathbf{1^{\prime\prime}},\mathbf{1^{\prime}})\,,\\
& (k_{N_1},k_{N_2},k_{F},k_{T_1},k_{T_2},k_{T_3})=(2,2,2,0,2,4)
\end{aligned}$}

We can read out the superpotentials relevant to quark and lepton masses
\begin{eqnarray}
\nonumber {\cal W}_\nu &=&
   \alpha_{N_1} \Lambda (N_1N_1)_{\mathbf{1}}Y_{\mathbf{1}}^{(4)}
 + \alpha_{N_2} \Lambda (N_2N_2)_{\mathbf{1^{\prime\prime}}}Y_{\mathbf{1^\prime}}^{(4)}
 + \alpha_{\nu_1} (N_1F)_{\mathbf{3}}Y_{\mathbf{3}}^{(4)} H_{5}
 + \alpha_{\nu_2} (N_2F)_{\mathbf{3}}Y_{\mathbf{3}}^{(4)} H_{5}
\,,\\
\nonumber {\cal W}_d &=&
   \alpha_{d_1} (FT_1)_{\mathbf{3}}Y_{\mathbf{3}}^{(2)} H_{\overline{5}}
 + \alpha_{d_2} (FT_2)_{\mathbf{3}}Y_{\mathbf{3}}^{(4)} H_{\overline{5}}
 + \alpha_{d_3} (FT_3)_{\mathbf{3}}Y_{\mathbf{3},I}^{(6)} H_{\overline{5}}
 + \alpha_{d_4} (FT_3)_{\mathbf{3}}Y_{\mathbf{3},II}^{(6)} H_{\overline{5}}
\\\nonumber &&
 + \alpha^\prime_{d_1} (FT_1)_{\mathbf{3}}Y_{\mathbf{3}}^{(2)} H_{\overline{45}}
 + \alpha^\prime_{d_2} (FT_2)_{\mathbf{3}}Y_{\mathbf{3}}^{(4)} H_{\overline{45}}
 + \alpha^\prime_{d_3} (FT_3)_{\mathbf{3}}Y_{\mathbf{3},I}^{(6)} H_{\overline{45}}
 + \alpha^\prime_{d_4} (FT_3)_{\mathbf{3}}Y_{\mathbf{3},II}^{(6)} H_{\overline{45}}
\,,\\
{\cal W}_u &=&
   \alpha_{u_1}  (T_1T_3)_{\mathbf{1}}Y_{\mathbf{1}}^{(4)} H_{5}
 + \alpha_{u_2}  (T_2T_3)_{\mathbf{1}}Y_{\mathbf{1}}^{(6)} H_{5}
 + \alpha_{u_3}  (T_3T_3)_{\mathbf{1^{\prime\prime}}}Y_{\mathbf{1^\prime}}^{(8)} H_{5}
\,,
\end{eqnarray}
which gives rise to
\begin{eqnarray}
m_M^{\nu} &=& \Lambda  \left(
\begin{array}{cc}
 \alpha_{N_1}Y_\mathbf{1}^{(4)} ~&~ 0 \\
 0 ~&~ \alpha_{N_2}Y_{\mathbf{1}^\prime}^{(4)} \\
\end{array}
\right)\,,\nonumber\\
{\cal Y}_5^{\nu} &=& \left(
\begin{array}{ccc}
 \alpha_{\nu_1}Y_{\mathbf{3},1}^{(4)} ~&~ \alpha_{\nu_1}Y_{\mathbf{3},3}^{(4)} ~&~ \alpha_{\nu_1}Y_{\mathbf{3},2}^{(4)} \\
 \alpha_{\nu_2}Y_{\mathbf{3},3}^{(4)}  ~&~ \alpha_{\nu_2}Y_{\mathbf{3},2}^{(4)} ~&~ \alpha_{\nu_2}Y_{\mathbf{3},1}^{(4)} \\
\end{array}
\right)\,.
\end{eqnarray}
The matrices ${\cal Y}_{\overline{5}}^d$ and ${\cal Y}_5^u$ are in common with the corresponding ones of the model ${\cal VII}_{4}$ as well.
The phases of the parameters $\alpha_{N_1}$, $\alpha_{\nu_1}$, $\alpha_{\nu_2}$, $\alpha_{d_1}$, $\alpha_{d_2}$, $\alpha_{d_3}$, $\alpha_{d_1}^\prime$ and $\alpha_{u_3}$ can be removed by field redefinition, and the other parameters are complex.
\end{itemize}

\section{\label{sec:numerical-results}Numerical results}

We have numerically scanned the parameter space of each model to optimize the agreement between predictions and experimental data. The dynamics determining the VEV of the complex modulus $\tau$ is an open question now, consequently we treat the VEV of $\tau$ as a free parameter to match the experimental data. As explicitly shown in previous section, the phases of some coupling constants can be removed by a field redefinition while others are generally complex. Each model depends on a set of dimensionless input parameters and three overall mass scales for the up type quark, charged lepton and neutrino mass matrices. The dimensionless input parameters include the modulus $\tau$ and the ratios of the coupling constants and they determine the mixing angles, CP violation phases and the fermion mass ratios. As can be seen from Eq.~\eqref{eq:Me-Md}, the down-type quark and charged lepton mass matrices are closely related in
$SU(5)$ and they share a same overall factor. We can use the measured values of $\Delta m_{21}^2$, $m_\tau$ and $m_t$ to determine the overall mass parameters of the neutrino, the charged lepton, the up quark mass matrices.
As a measure of goodness of fit, we use a $\chi^2$ function defined in the usual way,
\begin{eqnarray}
\chi^2_{total}=\sum_i\left(\frac{\xi_i-\overline{\xi_i}}{\sigma_i}\right)^2\,,
\end{eqnarray}
where $\overline{\xi_i}$ and $\sigma_i$ refer to the global best fit values and the $1\sigma$ deviations of the corresponding observables and their values are listed in table~\ref{tab:quark-data}.

The total $\chi^2_{total}$ can be split into the lepton and quark contributions $\chi^2_l$ and $\chi^2_q$ respectively with $\chi^2_{total}=\chi^2_l+\chi^2_q$. We construct the lepton sector $\chi^2_l$ function with the mass ratios $m_e/m_\mu$, $m_\mu/m_\tau$, $\Delta m_{21}^2/\Delta m_{31}^2$ and the lepton mixing parameters $\sin^2\theta_{12}^l$, $\sin^2\theta_{13}^l$, $\sin^2\theta_{23}^l$, $\delta_{CP}^l$. The quark sector $\chi^2_q$ function is constructed from the quark mass ratios $m_u/m_c$, $m_c/m_t$, $m_d/m_s$, $m_s/m_b$ and the quark mixing parameters $\theta_{12}^q$, $\theta_{13}^q$, $\theta_{23}^q$, $\delta_{CP}^q$. The tau mass $m_\tau$ is fixed to its experimental best fit  value to determine the common overall scale of the charged lepton and down quark mass matrices, the contribution from the mass ratio $m_b/m_\tau$ is included into the $\chi^2_q$. Notice that the overall scale factors of the mass matrices don't affect the value of $\chi^2_{total}$. The values of the quark and charged lepton masses and the CKM mixing parameters at the GUT scale are taken from Ref.~\cite{Antusch:2013jca} with $\tan\beta = 10$ and the SUSY breaking scale $M_{\text{SUSY}}=10$ TeV. The experimental values of the neutrino mixing parameters are taken from NuFIT v5.0 with Super-Kamiokanda atmospheric data~\cite{Esteban:2020cvm}. Since the normal ordering (NO) neutrino mass spectrum is slightly preferred over the inverted ordering (IO) masses by the present data~\cite{Esteban:2020cvm}, we focus on the NO in the following.
\begin{table}[t!]
\centering
\begin{tabular}{|c|c|} \hline  \hline
Parameters & Best fit values and $1\sigma$ ranges \\ \hline
$m_u/m_c$ & $(1.9286\pm 0.6017)\times 10^{-3}$ \\
$m_c/m_t$ & $(2.8213 \pm 0.1195)\times 10^{-3}$ \\
$m_d/m_s$ & $(5.0523 \pm 0.6191)\times 10^{-2}$ \\
$m_s/m_b$ & $(1.8241 \pm 0.1005)\times 10^{-2}$ \\
$m_b/m_\tau$ & $(0.7434\pm 0.0090)\times 10^{-2}$ \\ \hline
$\delta^q_{CP}$ & $69.213^{\circ}\pm 3.115^{\circ}$ \\
$\theta^q_{12}$ & $0.22736\pm 0.00073$ \\
$\theta^q_{13}$ & $0.00349 \pm 0.00013$ \\
$\theta^q_{23}$ & $0.04015 \pm 0.00064$ \\ \hline \hline
$m_e/m_\mu $ & $ (4.73689 \pm 0.04019)\times 10^{-3}$ \\
$m_\mu/m_\tau$ & $(5.85684 \pm 0.04654)\times 10^{-2}$ \\\hline
$\sin^2\theta_{12}^l$ & $0.304^{+0.012}_{-0.012}$ \\
$\sin^2\theta_{23}^l$ & $0.573^{+0.016}_{-0.020}$ \\
$\sin^2\theta_{13}^l$ & $0.02238^{+0.00062}_{-0.00063}$ \\
$\delta^l_{CP}/^\circ$ & $197_{-24}^{+27}$ \\
$\frac{\Delta m_{21}^2}{10^{-5}{\rm eV}^2}$ & $7.42^{+0.21}_{-0.20}$ \\
$\frac{\Delta m_{31}^2}{10^{-3}{\rm eV}^2}$ & $2.517^{+0.026}_{-0.028}$ \\
\hline \hline
$m_t$ & $87.4555\pm2.0893 ~\text{GeV}$ \\
$m_\tau$ & $1.30234\pm0.0068~\text{GeV}$ \\ \hline\hline
\end{tabular}
\caption{\label{tab:quark-data}The best fit values and $1\sigma$ errors of the mass ratios and mixing parameters of quarks and leptons. The values of the quark and charged lepton masses and the CKM parameters at the GUT scale are taken from Ref.~~\cite{Antusch:2013jca} with SUSY breaking scale $M_{\text{SUSY}}=10$ TeV and $\tan\beta=10$, $\bar{\eta}_b=0.09375$~\cite{Antusch:2013jca}. The lepton mixing parameters are taken from~\cite{Esteban:2020cvm} for normal ordering neutrino masses. }
\end{table}

For any given values of the input parameters of a model, we diagonalize the mass matrices to extract the lepton and quark masses and mixing matrices, and then the value of $\chi^2_{total}$ is calculated. The absolute value of all dimensionless parameters are treated as random numbers varying freely between $0$ and $10^8$ and their phases are distributed uniformly in the region $[0,2\pi]$, while the VEV of the modulus $\tau$ is limited in the fundamental region $\mathcal{D}=\left\{\tau| \texttt{Im}(\tau)>0, |\texttt{Re}(\tau)|\leq\frac{1}{2}, |\tau|\geq1\right\}$. We numerically search for the global minimum of $\chi^2_{total}$ by using the minimization
algorithms incorporated in the CERN developed package \texttt{TMinuit} to optimize the values of the input parameters~\cite{minuit}. We have scanned all models with number of parameters (including the real and imaginary part of $\tau$) less than $25$. Requiring $\chi^2_{total}<100$, we find $17$ type-I, $2$ type-II, $28$ type-V and $6$ type-VII models are viable.
The representations and weight assignments of the matter fields for these models are summarized in table~\ref{tab:models-summary}. Moreover, we display the best fit values of the free parameters and the predictions for fermion masses and flavor mixing parameters as well as the minimal values of $\chi^2_l$, $\chi^2_q$ and $\chi^2_{total}$ in tables~\ref{tab:fitI1}-\ref{tab:fitVII2}. The results of leptonic PMNS parameters and the quark CKM parameters are extracted in the standard way as defined in the Particle Data Group (PDG). For instance, the PDG parametrization of the lepton PMNS mixing matrix is as follow,
\begin{equation}
\label{eq:UPMNS-param}U_{PMNS}=\left(\begin{array}{ccc}
c^{l}_{12} c^{l}_{13} ~& s^{l}_{12} c^{l}_{13} ~& s^{l}_{13} e^{-i \delta^{l}_{C P}} \\
-s^{l}_{12} c^{l}_{23}-c^{l}_{12} s^{l}_{13} s^{l}_{23} e^{i \delta^{l}_{C P}} ~& c^{l}_{12} c^{l}_{23}-s^{l}_{12} s^{l}_{13} s^{l}_{23} e^{i \delta^{l}_{C P}} ~& c^{l}_{13} s^{l}_{23} \\
s^{l}_{12} s^{l}_{23}-c^{l}_{12} s^{l}_{13} c^{l}_{23} e^{i \delta^{l}_{C P}} ~& -c_{12} s^{l}_{23}-s^{l}_{12} s^{l}_{13} c^{l}_{23} e^{i \delta^{l}_{C P}} ~& c^{l}_{13} c^{l}_{23}
\end{array}\right) Q\,,
\end{equation}
with $c^{l}_{ij}=\cos \theta^{l}_{ij}, s^{l}_{ij}=\sin^{l}\theta_{ij}, \delta^{l}_{CP}$ is the Dirac CP violating phase, and $Q=\operatorname{diag}(1, e^{i \frac{\alpha_{21}}{2}}, e^{i \frac{\alpha_{31}}{2}})$ is the Majorana phase matrix factor where $\alpha_{21}$ and $\alpha_{31}$ are the so-called Majorana CP phases. If the lightest neutrino is massless $m_1=0$ or $m_3=0$, $\alpha_{31}$ is unphysical and the only physical Majorana phase is $\phi=\alpha_{21}$. The quark CKM mixing matrix is parameterized similar to Eq.~\eqref{eq:UPMNS-param} without the Majorana phase matrix $Q$.

As can be seen from tables~\ref{tab:fitI1} to \ref{tab:fitVII2}, there are 13 models ${\cal I}_{4}$, ${\cal I}_{5}$, ${\cal I}_{7}$, ${\cal I}_{9}$, ${\cal I}_{10}$, ${\cal I}_{11}$, ${\cal I}_{12}$, ${\cal I}_{15}$, ${\cal V}_{5}$, ${\cal V}_{6}$, ${\cal V}_{7}$, ${\cal V}_{18}$ and ${\cal V}_{25}$ in good agreement with the experiment data at $1\sigma$ level, and 13 models
${\cal I}_{6}$, ${\cal I}_{13}$, ${\cal I}_{16}$, ${\cal I}_{17}$, ${\cal V}_{9}$, ${\cal V}_{11}$, ${\cal V}_{12}$, ${\cal V}_{15}$, ${\cal V}_{16}$, ${\cal V}_{17}$, ${\cal V}_{19}$, ${\cal V}_{20}$ and ${\cal V}_{26}$
are compatible with data at $3\sigma$ level. All these models depend on 24 real parameters including the real and imaginary part of $\tau$. In order accommodate the quark and lepton mass hierarchies, we see that hierarchical values of the coupling constants are necessary, and this is expected to be naturally explained by the weighton mechanism~\cite{King:2020qaj}. It is remarkable that the model $\mathcal{I}_1$ has only 18 real parameters including $\texttt{Re}(\tau)$ and $\texttt{Im}(\tau)$ and the predictions are in qualitative agreement with the observations, although $\sin^2\theta^{l}_{23}$ and $\delta^q_{CP}$ are slightly above the $3\sigma$ allowed regions and the charm quark mass $m_c$ is a bit smaller and the up quark mass is vanishing with $m_u=0$. In all the 22 parameters models $\mathcal{I}_2$, $\mathcal{I}_3$, $\mathcal{V}_1$, $\mathcal{V}_2$, $\mathcal{VII}_1$, $\mathcal{VII}_2$, $\mathcal{VII}_3$ together with the models ${\cal V}_{21}$, ${\cal V}_{22}$, ${\cal V}_{23}$, ${\cal V}_{24}$, ${\cal V}_{27}$, ${\cal V}_{28}$, $\mathcal{VII}_4$, $\mathcal{VII}_5$ and $\mathcal{VII}_6$, the top left $2\times2$ block of the Yukawa coupling matrix ${\cal Y}_5^u$ is zero such that we have $m_u=0$ which is compatible with the tiny despite non-vanishing up quark mass. Moreover, there are only two right-handed neutrinos in the type-VII models, and the lightest neutrino is massless.

\begin{figure}[hptb!]
\centering
\includegraphics[width=0.8\textwidth]{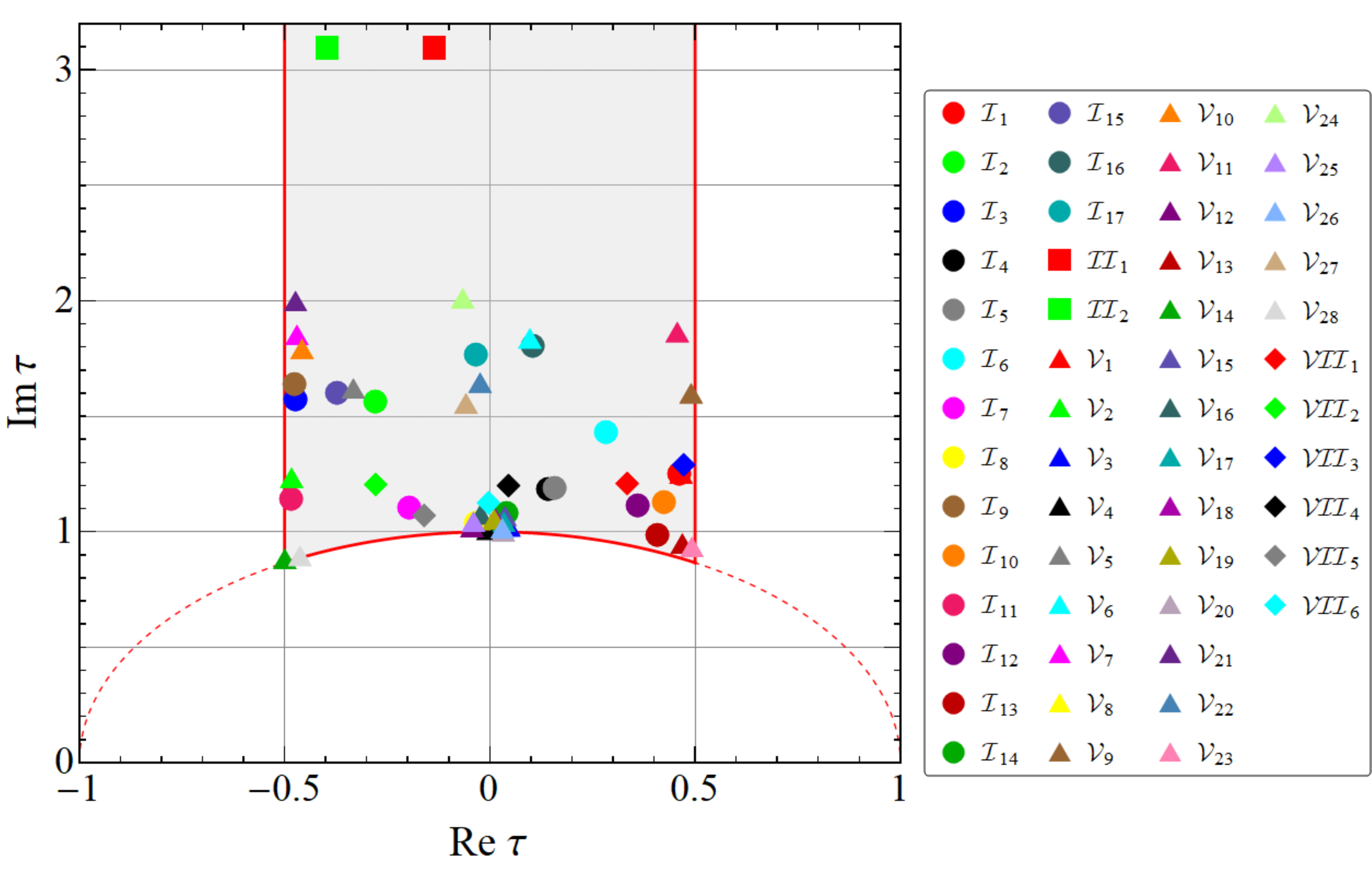}
\caption{\label{fig:tauBF}The best fit values of the complex modulus $\tau$ for the viable models summarized in table~\ref{tab:models-summary}. }
\end{figure}

\begin{figure}[hptb!]
\centering
\includegraphics[width=0.999\textwidth]{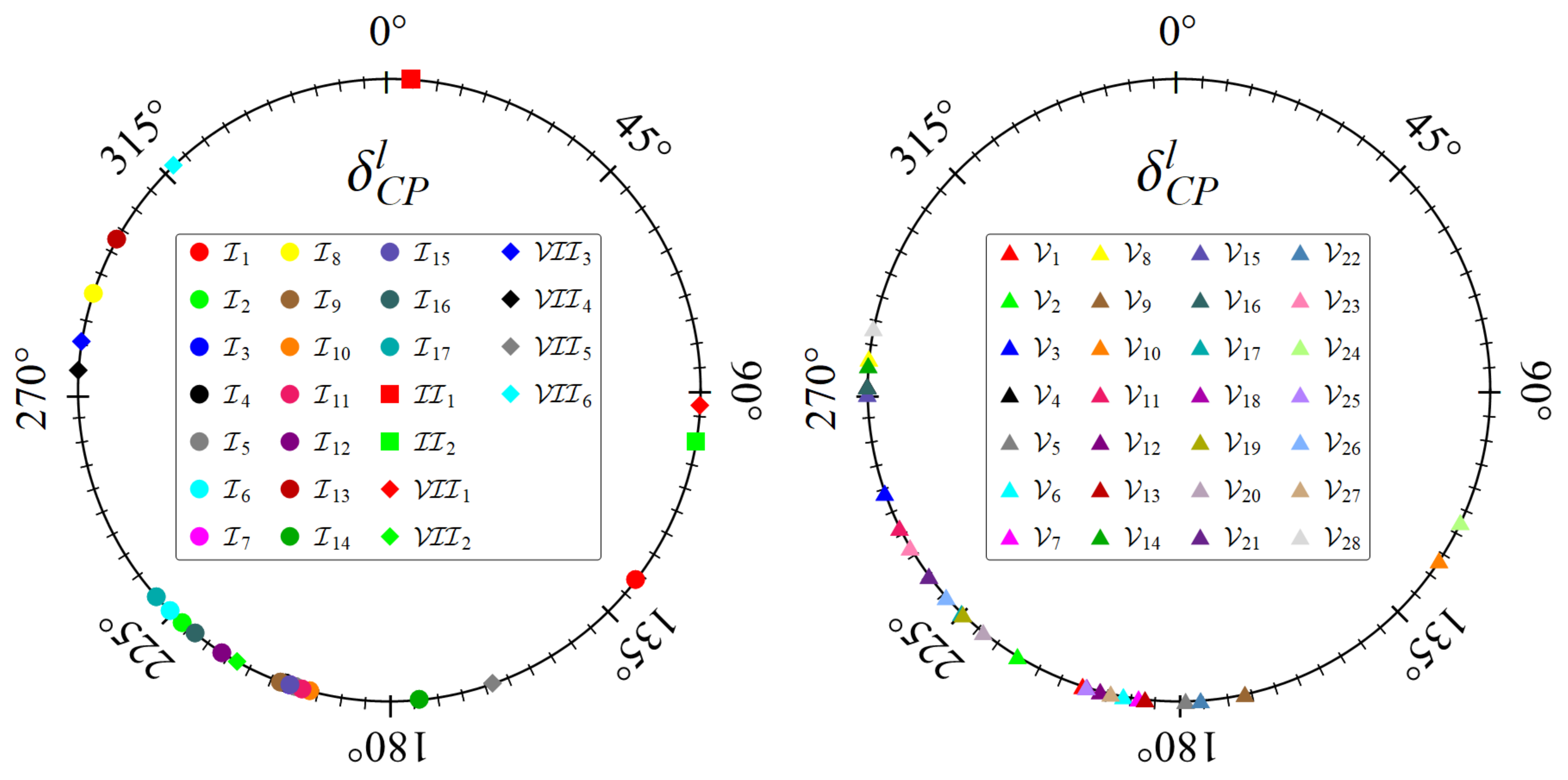}
\caption{\label{fig:deltaCP-127-5}The predictions for the Dirac CP phase $\delta^l_{CP}$. The left panel is for the models of type-I, type-II and type-VII, and the right is for the type-V models.}
\end{figure}

\begin{figure}[t!]
\centering
\includegraphics[width=0.8\textwidth]{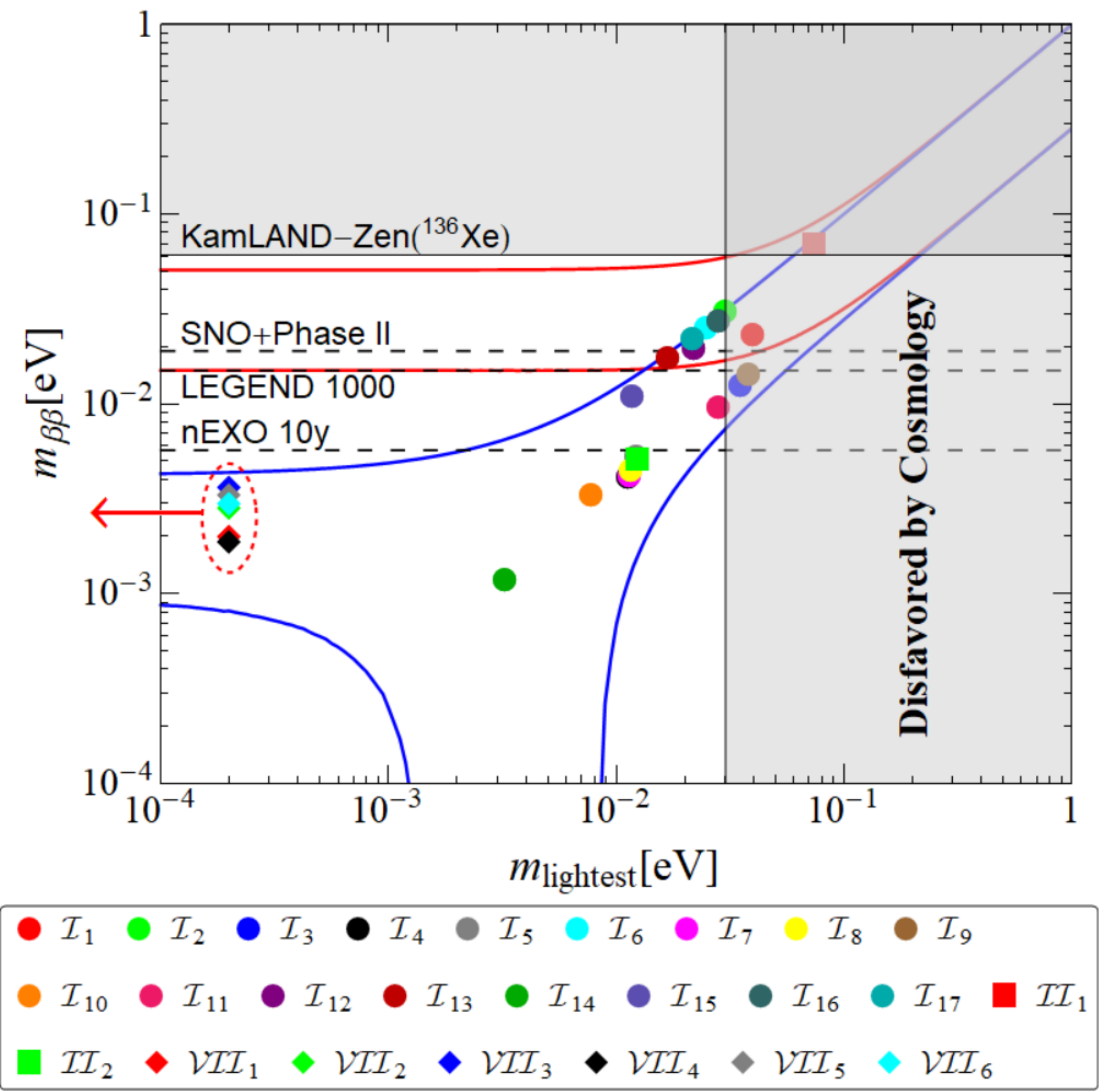}
\caption{\label{fig:mee127}The predictions for the effective mass of the neutrinoless double beta decay for the models of type I, type II and type VII. Notice that the lightest neutrino is massless for type VII models since only two right-handed neutrinos are introduced. The blue (red) lines denote the most general allowed regions for NO (IO) where the neutrino oscillation parameters are freely varied in their $3\sigma$ regions~\cite{Esteban:2020cvm}. The vertical grey exclusion band denotes the bound on the lightest neutrino mass coming from the cosmological data $\Sigma_{i}m_{i}<0.120\,\text{eV}$ at $95\%$ confidence level obtained by the Planck collaboration~\cite{Aghanim:2018eyx}. The current experimental bound from KamLAND-Zen~\cite{KamLAND-Zen:2016pfg} and the estimated experimental sensitivities of future 0$\nu\beta\beta$ experiments are indicated by the horizontal lines. }
\end{figure}

\begin{figure}[t!]
\centering
\includegraphics[width=0.8\textwidth]{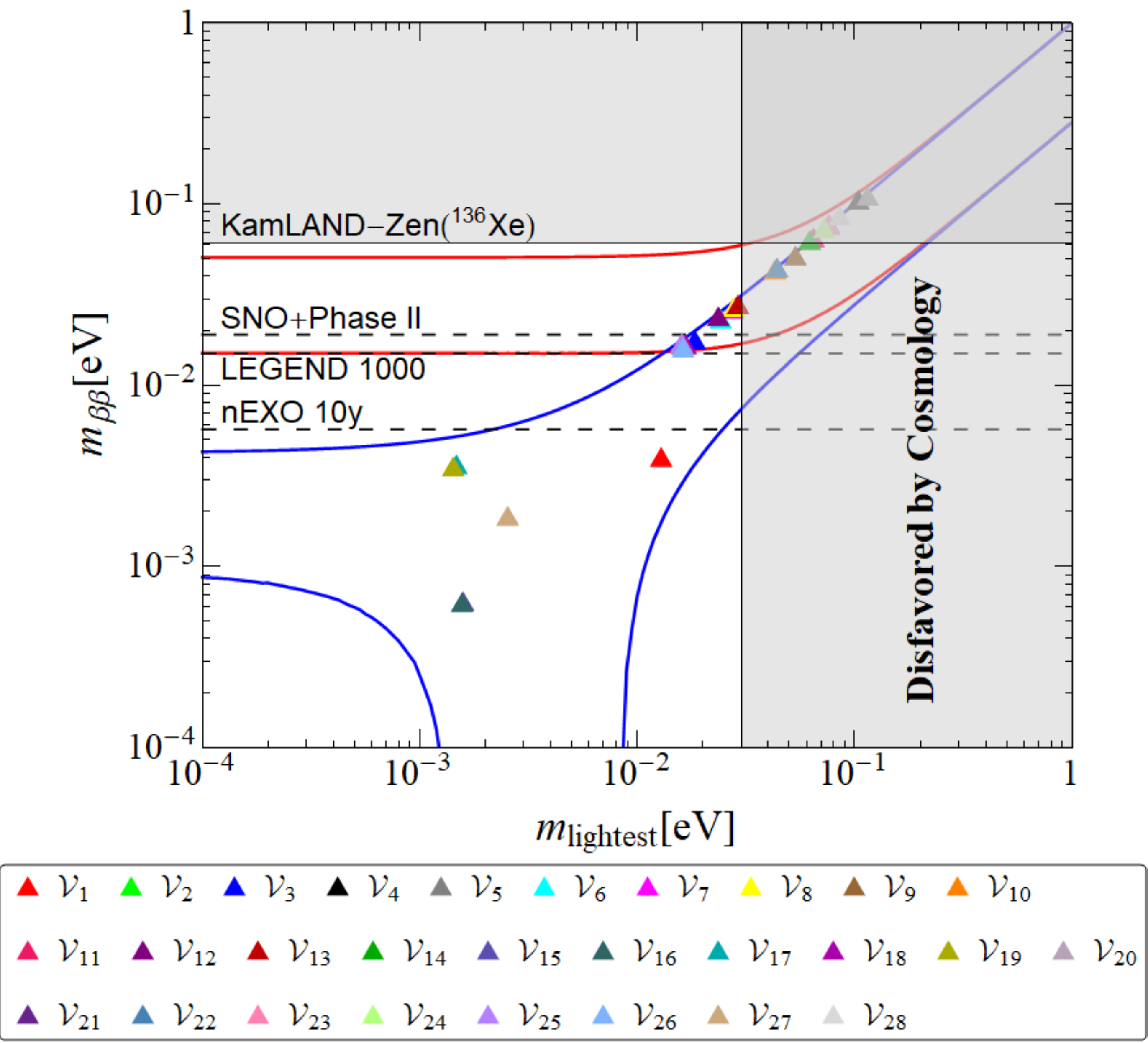}
\caption{\label{fig:mee5}The predictions for the effective mass of the neutrinoless double beta decay for the type V models. We adopt the same convention as figure~\ref{fig:mee127}. }
\end{figure}

We plot the best fit value of the modulus $\tau$ in  figure~\ref{fig:tauBF}, it can be seen that $\tau$ tends to distribute around the boundary of the fundamental domain $\mathcal{D}$, in particular some are clustered close to the self-dual point $\tau=i$ where $S$ is unbroken. The value of the Dirac CP phase $\delta^{l}_{CP}$ is still unknown, the current and upcoming long-baseline experiments will be able to place important constraints. As can be seen from figure~\ref{fig:deltaCP-127-5}, the predictions for $\delta^{l}_{CP}$ mostly lie in the region of $[\pi,3\pi/2]$ and they could be tested at future experiments. The most sensitive probe to whether neutrinos are Dirac or Majorana states is the neutrinoless
double beta decay ($0\nu\beta\beta$): $(A, Z)\rightarrow(A, Z+2)+e^{-}+e^{-}$. The decay amplitude is proportional to the effective Majorana mass
\begin{eqnarray}
m_{\beta\beta}=\left|m_1\cos^2\theta_{12}\cos^2\theta_{13}+m_2\sin^2\theta_{12}\cos^2\theta_{13}e^{i\alpha_{21}}+m_3\sin^2\theta_{13}e^{i(\alpha_{31}-2\delta_{CP})}\right|\,,
\end{eqnarray}
which reduces to
\begin{eqnarray}
m_{\beta\beta}=\left|m_2\sin^2\theta_{12}\cos^2\theta_{13}e^{i\phi}+m_3\sin^2\theta_{13}e^{-2i\delta_{CP}}\right|\,,
\end{eqnarray}
for $m_1=0$. The current most stringent limit is $m_{\beta\beta}<61$ meV from KamLAND-Zen ~\cite{KamLAND-Zen:2016pfg}. From the predicted values of lepton mixing angles and neutrino masses, one can easily determine the effective mass $m_{\beta\beta}$. The predictions for $m_{\beta\beta}$ of the type-I, II and VII models are shown in figure~\ref{fig:mee127} and those of type-V models are displayed in figure~\ref{fig:mee5}. There are many experiments which are in various stages of planning and construction. the new generation of $0\nu\beta\beta$ decay experiments will significantly increase the sensitivity to this rare process such that a considerable amount of our models are within the reach in foreseeable future.

\clearpage

\section{\label{sec:conclusion}Conclusion}

GUTs are an important framework to understand the origin of the observed patterns of fermion masses and flavor mixing. The quark and leptons fields in each generation are embedded into a GUT multiplet so that they are closely related. It is usually assumed that the three generations of fermions are related by certain family symmetry to address the flavor mixing structure of quarks and leptons. The role of modular invariance as flavor symmetry can overcome the drawback of the conventional discrete flavor symmetry models. The modular symmetry is broken by the VEV of the complex modulus $\tau$, and the Yukawa couplings are modular forms. Flavor models based on modular symmetry can be highly predictive, and the neutrino masses and lepton mixing mixing can be predicted in terms of few input parameters. Inspired by the success in the lepton sector, the modular symmetry has been extended to the quark sector~\cite{Okada:2018yrn,Okada:2019uoy,King:2020qaj,Lu:2019vgm,Okada:2020rjb,Liu:2020akv,Yao:2020zml}. In the present work, we imposed $\Gamma_3\cong A_4$ modular symmetry on the $SU(5)$ GUT to give a unified description of both quarks and leptons.

In order to account for the mass difference of charged leptons and down-type quarks, we introduced the Higgs multiplet $H_{\overline{45}}$ besides $H_{5}$ and $H_{\overline{5}}$. All the three Higgs multiplets $H_{5}$, $H_{\overline{5}}$ and $H_{\overline{45}}$ were assumed to be invariant under $A_4$ with zero modular weight. The neutrino masses are generated by the type-I seesaw mechanism, and both scenarios with three and two right-handed neutrinos were considered. The matter fields $N$, $\overline{F}$ and $T$ transform as either triplets $\mathbf{3}$ or singlets $\mathbf{1}$, $\mathbf{1}'$, $\mathbf{1}''$ under $A_4$. We have given the most general form of the quark and lepton mass matrices for different possible representation assignments, as shown in section~\ref{sec:lassification}. All possible models were classified according to the transformation properties of the matter fields, and we found there are five types of models for three right-handed neutrinos and two types of models for two right-handed neutrinos. If $N$, $\overline{F}$ and $T$ are all singlets of $A_4$, the Yukawa superpotential would be less restricted by modular symmetry and generally more free parameters would be involved, hence we did not consider these cases in the present work since we were interested in the simplest possibilities.

We have performed a numerical scan over the parameter space of each model, and searched for the minimum of the $\chi^2_{total}$ function to optimize the agreement between predictions and data.
For normal ordering neutrino masses, intensive numerical calculations revealed that there are 17 type-I models, 2 type-II models, 28 type-V and 6 type-VII models with $\chi^2_{total}<100$ with the number of real free parameters not larger than 24 including the real and imaginary part of $\tau$. Out of the 54 considered models, we found that 12 type-I and 14 type-V models are in agreement with the experimental data at $3\sigma$ level, as can be seen from tables~\ref{tab:fitI1} to \ref{tab:fitVII2}. All those phenomenological viable examples have 24 parameters.

The predictions for the leptonic CP violating Dirac phase, the lightest neutrino mass and the neutrinoless double beta decay parameter as shown in the figures are observed to cover a wide range of possible values, but appear to be clustered around particular regions.
Consequently the forthcoming generation of neutrino oscillation experiments sensitive to the leptonic CP phase, in conjunction with the upcoming neutrinoless double beta decay experiments, will together be able to discriminate between the various types of models based on
$\Gamma_3\cong A_4$ modular symmetry with $SU(5)$ GUTs considered here.

\section*{Acknowledgements}

PC and GJD are grateful to Dr. Chang-Yuan Yao for his kind help on numerical analysis. PC is supported by the National Natural Science Foundation of China under Grant Nos 11847240. GJD is supported by the National Natural Science Foundation of China under Grant Nos 11975224, 11835013, 11947301, 12047502.
SFK acknowledges the STFC Consolidated Grant ST/L000296/1 and the European Union's Horizon 2020 Research and Innovation programme under Marie Sk\l{}odowska-Curie grant agreement
HIDDeN European ITN project (H2020-MSCA-ITN-2019//860881-HIDDeN).

\begin{appendix}

\setcounter{equation}{0}
\renewcommand{\theequation}{\thesection.\arabic{equation}}

\section{\label{sec:appendix-best-fit}Results of the best fit for the benchmark models  }

In this Appendix, we collect the best fit values of the input parameters and the corresponding predictions for the quark and lepton masses and mixing parameters for the phenomenologically viable models listed in table~\ref{tab:models-summary}. Here we have considered normal ordering neutrino masses which is slightly preferred by the experimental data~\cite{Esteban:2020cvm}.

\begin{table}[h]
 \centering
\begin{tabular}{|c|c|c|c|}
 \hline
 Model & $\mathcal{I}_{1}$ & $\mathcal{I}_{2}$ & $\mathcal{I}_{3}$
 \\\hline\hline$\texttt{Re}(\tau)$ & $0.4623$ & $-0.2778$ & $-0.4732$\\
$\texttt{Im}(\tau)$ & $1.249$ & $1.560$ & $1.574$\\
$|\alpha_{\nu_2}/\alpha_{\nu_1}|$ & $1.216$ & $2.497$ & $1.961$\\
$\arg(\alpha_{\nu_2}/\alpha_{\nu_1})/\pi$ & $1.513$ & $1.913$ & $1.634$\\
$\alpha_{d_1}/\alpha_{d_3}$ & $0.0275$ & $0.00253$ & $0.0148$\\
$\alpha_{d_2}/\alpha_{d_3}$ & $0.465$ & $2.089\times 10^{-6}$ & $0.0844$\\
$|\alpha_{d_4}/\alpha_{d_3}|$ & ---  & $2.731$ & $7.592$\\
\hline\hline
\end{tabular}
\caption{\label{tab:fitI1}The best fit values of the free parameters and the corresponding predictions for lepton and quark mixing
parameters and fermion masses for the phenomenologically viable models of Type-I.}
\end{table}
\begin{table}[h]
 \centering
 \begin{tabular}{|c|c|c|c|c|}
 \hline
 Model & $\mathcal{I}_{4}$ & $\mathcal{I}_{5}$ & $\mathcal{I}_{6}$ & $\mathcal{I}_{7}$
 \\\hline\hline$\texttt{Re}(\tau)$ & $0.1427$ & $0.1601$ & $0.2830$ & $-0.1956$\\
$\texttt{Im}(\tau)$ & $1.184$ & $1.188$ & $1.429$ & $1.103$\\
$|\alpha_{\nu_2}/\alpha_{\nu_1}|$ & $2.949$ & $2.937$ & $2.609$ & $3.051$\\
$\arg(\alpha_{\nu_2}/\alpha_{\nu_1})/\pi$ & $4.567\times 10^{-8}$ & $0.00486$ & $0.0791$ & $1.999$\\
$\alpha_{d_1}/\alpha_{d_3}$ & $0.0523$ & $0.0839$ & $0.00416$ & $0.220$\\
$\alpha_{d_2}/\alpha_{d_3}$ & $0.00504$ & $0.00463$ & $0.0794$ & $0.0213$\\
$|\alpha_{d_4}/\alpha_{d_3}|$ & $0.620$ & $2.136$ & $1.068$ & $4.410$\\
\hline\hline
\end{tabular}
\caption{\label{tab:fitI2}The best fit values of the free parameters and the corresponding predictions for lepton and quark mixing
parameters and fermion masses for the phenomenologically viable models of Type-I (continuation of table~\ref{tab:fitI1}).}
\end{table}
\begin{table}[h]
 \centering
 \begin{tabular}{|c|c|c|c|c|c|}
 \hline
 Model & $\mathcal{I}_{8}$ & $\mathcal{I}_{9}$ & $\mathcal{I}_{10}$ & $\mathcal{I}_{11}$ & $\mathcal{I}_{12}$
 \\\hline\hline$\texttt{Re}(\tau)$ & $-0.03236$ & $-0.4738$ & $0.4264$ & $-0.4828$ & $0.3621$\\
$\texttt{Im}(\tau)$ & $1.036$ & $1.637$ & $1.125$ & $1.143$ & $1.113$\\
$|\alpha_{N_2}/\alpha_{N_1}|$ & $1.740$ & ---  & ---  & ---  & --- \\
$\arg(\alpha_{N_2}/\alpha_{N_1})/\pi$ & $0.662$ & ---  & ---  & ---  & --- \\
$|\alpha_{N_3}/\alpha_{N_1}|$ & $0.821$ & ---  & ---  & ---  & --- \\
$\arg(\alpha_{N_3}/\alpha_{N_1})/\pi$ & $0.621$ & ---  & ---  & ---  & --- \\
$|\alpha_{\nu_2}/\alpha_{\nu_1}|$ & $1.717$ & $1.930$ & $0.683$ & $2.273$ & $3.224$\\
\hline\hline
\end{tabular}
\caption{\label{tab:fitI3}The best fit values of the free parameters and the corresponding predictions for lepton and quark mixing
parameters and fermion masses for the phenomenologically viable models of Type-I (continuation of table~\ref{tab:fitI2}).}
\end{table}
\begin{table}[h]
 \centering
 \begin{tabular}{|c|c|c|c|c|c|}
 \hline
 Model & $\mathcal{I}_{13}$ & $\mathcal{I}_{14}$ & $\mathcal{I}_{15}$ & $\mathcal{I}_{16}$ & $\mathcal{I}_{17}$
 \\\hline\hline$\texttt{Re}(\tau)$ & $0.4092$ & $0.04232$ & $-0.3716$ & $0.1069$ & $-0.03262$\\
$\texttt{Im}(\tau)$ & $0.984$ & $1.081$ & $1.602$ & $1.804$ & $1.763$\\
$|\alpha_{N_1}/\alpha_{N_3}|$ & $6.71\times 10^{-4}$ & $7.575\times 10^{-5}$ & $0.800$ & $3.402\times 10^{-4}$ & $3.246\times 10^{-4}$\\
$\arg(\alpha_{N_1}/\alpha_{N_3})/\pi$ & $1.241$ & $1.569$ & $0.677$ & $0.883$ & $1.204$\\
$|\alpha_{N_2}/\alpha_{N_3}|$ & $1.432$ & $1.765$ & $1.346$ & $0.852$ & $1.570$\\
$\arg(\alpha_{N_2}/\alpha_{N_3})/\pi$ & $0.124$ & $1.997$ & $0.680$ & $1.694$ & $0.868$\\
$|\alpha_{\nu_2}/\alpha_{\nu_1}|$ & $2.149$ & $2.181$ & $2.782$ & $1.542$ & $1.534$\\
\hline\hline
\end{tabular}
\caption{\label{tab:fitI4}The best fit values of the free parameters and the corresponding predictions for lepton and quark mixing
parameters and fermion masses for the phenomenologically viable models of Type-I (continuation of table~\ref{tab:fitI3}).}
\end{table}
\begin{table}[h]
 \centering
 \begin{tabular}{|c|c|c|}
 \hline
 Model & $\mathcal{II}_{1}$ & $\mathcal{II}_{2}$
 \\\hline\hline$\texttt{Re}(\tau)$ & $-0.1342$ & $-0.3947$\\
$\texttt{Im}(\tau)$ & $3.095$ & $3.092$\\
$|\alpha_{\nu_2}/\alpha_{\nu_1}|$ & $0.326$ & $0.378$\\
$\arg(\alpha_{\nu_2}/\alpha_{\nu_1})/\pi$ & $0.368$ & $4.329\times 10^{-4}$\\
$|\alpha_{\nu_3}/\alpha_{\nu_1}|$ & $1.685$ & $0.0148$\\
$\arg(\alpha_{\nu_3}/\alpha_{\nu_1})/\pi$ & $1.992$ & $1.721$\\
$\alpha_{d_1}/\alpha_{d_3}$ & $17.909$ & $0.0538$\\
\hline\hline
\end{tabular}
\caption{\label{tab:fitII1}The best fit values of the free parameters and the corresponding predictions for lepton and quark mixing
parameters and fermion masses for the phenomenologically viable models of Type-II.}
\end{table}
\begin{table}[h]
 \centering
 \begin{tabular}{|c|c|c|c|c|c|}
 \hline
 Model & $\mathcal{V}_{1}$ & $\mathcal{V}_{2}$ & $\mathcal{V}_{3}$ & $\mathcal{V}_{4}$ & $\mathcal{V}_{5}$
 \\\hline\hline$\texttt{Re}(\tau)$ & $0.4670$ & $-0.4818$ & $0.04772$ & $-0.003098$ & $-0.3313$\\
$\texttt{Im}(\tau)$ & $1.250$ & $1.232$ & $1.021$ & $1.007$ & $1.618$\\
$|\alpha_{N_2}/\alpha_{N_1}|$ & $171.272$ & $5.059$ & $1.917\times 10^7$ & $7.571\times 10^5$ & $1.181\times 10^7$\\
$\arg(\alpha_{N_2}/\alpha_{N_1})/\pi$ & $0.203$ & $0.723$ & $0.157$ & $1.146$ & $1.987$\\
$|\alpha_{N_3}/\alpha_{N_1}|$ & $8.270$ & $94.999$ & $9.\times 10^5$ & $5.692\times 10^5$ & $4.117\times 10^6$\\
$\arg(\alpha_{N_3}/\alpha_{N_1})/\pi$ & $1.857$ & $1.081$ & $1.156$ & $1.120$ & $1.924$\\
$\alpha_{\nu_2}/\alpha_{\nu_1}$ & $7.255$ & $2.597$ & $8.209\times 10^4$ & $1042.230$ & $5765.608$\\
\hline\hline
\end{tabular}
\caption{\label{tab:fitV1}The best fit values of the free parameters and the corresponding predictions for lepton and quark mixing
parameters and fermion masses for the phenomenologically viable models of Type-V.}
\end{table}
\begin{table}[h]
 \centering
 \begin{tabular}{|c|c|c|c|c|c|}
 \hline
 Model & $\mathcal{V}_{6}$ & $\mathcal{V}_{7}$ & $\mathcal{V}_{8}$ & $\mathcal{V}_{9}$ & $\mathcal{V}_{10}$
 \\\hline\hline$\texttt{Re}(\tau)$ & $0.09874$ & $-0.4687$ & $-0.02455$ & $0.4915$ & $-0.4562$\\
$\texttt{Im}(\tau)$ & $1.835$ & $1.850$ & $1.049$ & $1.598$ & $1.790$\\
$|\alpha_{N_2}/\alpha_{N_1}|$ & $1619.521$ & $2.779\times 10^4$ & $7.979\times 10^4$ & $4.37\times 10^7$ & $2.863\times 10^4$\\
$\arg(\alpha_{N_2}/\alpha_{N_1})/\pi$ & $1.994$ & $1.908$ & $0.998$ & $1.031$ & $1.002$\\
$|\alpha_{N_3}/\alpha_{N_1}|$ & $8701.305$ & $2.797\times 10^4$ & $1.405\times 10^4$ & $9.499\times 10^6$ & $1562.126$\\
$\arg(\alpha_{N_3}/\alpha_{N_1})/\pi$ & $0.755$ & $1.711$ & $1.645$ & $0.0975$ & $0.211$\\
$\alpha_{\nu_2}/\alpha_{\nu_1}$ & $32.356$ & $374.097$ & $255.583$ & $6894.423$ & $178.981$\\
\hline\hline
\end{tabular}
\caption{\label{tab:fitV2}The best fit values of the free parameters and the corresponding predictions for lepton and quark mixing
parameters and fermion masses for the phenomenologically viable models of Type-V (continuation of table~\ref{tab:fitV1}).}
\end{table}
\begin{table}[h]
 \centering
 \begin{tabular}{|c|c|c|c|c|}
 \hline
 Model & $\mathcal{V}_{11}$ & $\mathcal{V}_{12}$ & $\mathcal{V}_{13}$ & $\mathcal{V}_{14}$
 \\\hline\hline$\texttt{Re}(\tau)$ & $0.4578$ & $-0.04235$ & $0.4701$ & $-0.4983$\\
$\texttt{Im}(\tau)$ & $1.863$ & $1.018$ & $0.947$ & $0.881$\\
$|\alpha_{N_2}/\alpha_{N_1}|$ & $4.766\times 10^5$ & $67.594$ & $3040.620$ & $855.157$\\
$\arg(\alpha_{N_2}/\alpha_{N_1})/\pi$ & $0.992$ & $0.497$ & $1.868$ & $1.025$\\
$|\alpha_{N_3}/\alpha_{N_1}|$ & $2.698\times 10^7$ & $1.746\times 10^4$ & $5427.280$ & $2436.021$\\
$\arg(\alpha_{N_3}/\alpha_{N_1})/\pi$ & $1.859$ & $0.187$ & $1.937$ & $0.164$\\
$\alpha_{\nu_2}/\alpha_{\nu_1}$ & $707.032$ & $777.226$ & $1801.840$ & $544.650$\\
\hline\hline
\end{tabular}
\caption{\label{tab:fitV3}The best fit values of the free parameters and the corresponding predictions for lepton and quark mixing
parameters and fermion masses for the phenomenologically viable models of Type-V (continuation of table~\ref{tab:fitV2}).}
\end{table}
\begin{table}[h]
 \centering
 \begin{tabular}{|c|c|c|c|c|}
 \hline
 Model & $\mathcal{V}_{15}$ & $\mathcal{V}_{16}$ & $\mathcal{V}_{17}$ & $\mathcal{V}_{18}$
 \\\hline\hline$\texttt{Re}(\tau)$ & $0.03809$ & $-0.02129$ & $0.03559$ & $0.01418$\\
$\texttt{Im}(\tau)$ & $1.067$ & $1.074$ & $1.041$ & $1.056$\\
$|\alpha_{N_2}/\alpha_{N_1}|$ & $2.837\times 10^5$ & $1.745\times 10^7$ & $3.18\times 10^6$ & $3.461\times 10^7$\\
$\arg(\alpha_{N_2}/\alpha_{N_1})/\pi$ & $0.886$ & $0.884$ & $1.090$ & $1.010$\\
$\alpha_{\nu_2}/\alpha_{\nu_1}$ & $346.937$ & $1.324\times 10^4$ & $1.129\times 10^4$ & $5535.626$\\
$\alpha_{\nu_3}/\alpha_{\nu_1}$ & $2274.630$ & $3644.926$ & $180.796$ & $1.08\times 10^4$\\
$\alpha_{d_1}/\alpha_{d_3}$ & $0.0718$ & $0.0107$ & $0.00398$ & $0.00370$\\
\hline\hline
\end{tabular}
\caption{\label{tab:fitV4}The best fit values of the free parameters and the corresponding predictions for lepton and quark mixing
parameters and fermion masses for the phenomenologically viable models of Type-V (continuation of table~\ref{tab:fitV3}).}
\end{table}
\begin{table}[h]
 \centering
 \begin{tabular}{|c|c|c|c|c|c|}
 \hline
 Model & $\mathcal{V}_{19}$ & $\mathcal{V}_{20}$ & $\mathcal{V}_{21}$ & $\mathcal{V}_{22}$ & $\mathcal{V}_{23}$
 \\\hline\hline$\texttt{Re}(\tau)$ & $0.01168$ & $0.03374$ & $-0.4722$ & $-0.02250$ & $0.4935$\\
$\texttt{Im}(\tau)$ & $1.052$ & $1.000$ & $1.997$ & $1.644$ & $0.932$\\
$|\alpha_{N_2}/\alpha_{N_1}|$ & $3.413\times 10^7$ & $1.429\times 10^6$ & $1.445\times 10^7$ & $2.521\times 10^5$ & $1.38\times 10^7$\\
$\arg(\alpha_{N_2}/\alpha_{N_1})/\pi$ & $1.087$ & $1.734$ & $1.142$ & $1.773$ & $0.912$\\
$\alpha_{\nu_2}/\alpha_{\nu_1}$ & $7261.654$ & $2467.673$ & $3033.538$ & $422.269$ & $3529.148$\\
$\alpha_{\nu_3}/\alpha_{\nu_1}$ & $3004.292$ & $1032.908$ & $5167.630$ & $719.326$ & $4260.711$\\
$\alpha_{d_1}/\alpha_{d_3}$ & $5.973\times 10^{-6}$ & $1.462\times 10^{-5}$ & $0.00428$ & $0.00272$ & $4.274\times 10^{-6}$\\
\hline\hline
\end{tabular}
\caption{\label{tab:fitV5}The best fit values of the free parameters and the corresponding predictions for lepton and quark mixing
parameters and fermion masses for the phenomenologically viable models of Type-V (continuation of table~\ref{tab:fitV4}).}
\end{table}
\begin{table}[h]
 \centering
 \begin{tabular}{|c|c|c|c|c|c|}
 \hline
 Model & $\mathcal{V}_{24}$ & $\mathcal{V}_{25}$ & $\mathcal{V}_{26}$ & $\mathcal{V}_{27}$ & $\mathcal{V}_{28}$
 \\\hline\hline$\texttt{Re}(\tau)$ & $-0.06476$ & $-0.03872$ & $0.03142$ & $-0.05690$ & $-0.4614$\\
$\texttt{Im}(\tau)$ & $2.009$ & $1.042$ & $1.013$ & $1.554$ & $0.894$\\
$|\alpha_{N_2}/\alpha_{N_1}|$ & $2.321\times 10^6$ & $6260.908$ & $423.935$ & $2.772\times 10^4$ & $1.419\times 10^4$\\
$\arg(\alpha_{N_2}/\alpha_{N_1})/\pi$ & $1.825$ & $0.166$ & $1.054$ & $1.999$ & $1.834$\\
$\alpha_{\nu_2}/\alpha_{\nu_1}$ & $1345.488$ & $5562.698$ & $585.205$ & $9996.838$ & $8367.220$\\
$\alpha_{\nu_3}/\alpha_{\nu_1}$ & $1882.619$ & $8929.744$ & $1441.682$ & $4377.767$ & $5285.115$\\
$\alpha_{d_1}/\alpha_{d_3}$ & $1.065\times 10^{-4}$ & $0.00622$ & $6.\times 10^{-5}$ & $6.512\times 10^{-4}$ & $1.667\times 10^{-6}$\\
\hline\hline
\end{tabular}
\caption{\label{tab:fitV6}The best fit values of the free parameters and the corresponding predictions for lepton and quark mixing
parameters and fermion masses for the phenomenologically viable models of Type-V (continuation of table~\ref{tab:fitV5}).}
\end{table}
\begin{table}[h]
 \centering
 \begin{tabular}{|c|c|c|c|}
 \hline
 Model & $\mathcal{VII}_{1}$ & $\mathcal{VII}_{2}$ & $\mathcal{VII}_{3}$
 \\\hline\hline$\texttt{Re}(\tau)$ & $0.3358$ & $-0.2769$ & $0.4734$\\
$\texttt{Im}(\tau)$ & $1.207$ & $1.203$ & $1.288$\\
$|\alpha_{\nu_3}/\alpha_{\nu_2}|$ & $0.739$ & $0.580$ & $2.329$\\
$\arg(\alpha_{\nu_3}/\alpha_{\nu_2})/\pi$ & $1.992$ & $1.062$ & $0.889$\\
$\alpha_{d_1}/\alpha_{d_3}$ & $0.0544$ & $0.00557$ & $0.00769$\\
$\alpha_{d_2}/\alpha_{d_3}$ & $0.00406$ & $0.0476$ & $0.101$\\
$|\alpha_{d_4}/\alpha_{d_3}|$ & $0.861$ & $1.328$ & $1.200$\\
\hline\hline
\end{tabular}
\caption{\label{tab:fitVII1}The best fit values of the free parameters and the corresponding predictions for lepton and quark mixing
parameters and fermion masses for the phenomenologically viable models of Type-VII.}
\end{table}
\begin{table}[h]
 \centering
 \begin{tabular}{|c|c|c|c|}
 \hline
 Model & $\mathcal{VII}_{4}$ & $\mathcal{VII}_{5}$ & $\mathcal{VII}_{6}$
 \\\hline\hline$\texttt{Re}(\tau)$ & $0.04682$ & $-0.1587$ & $-0.001939$\\
$\texttt{Im}(\tau)$ & $1.198$ & $1.068$ & $1.123$\\
$|\alpha_{N_2}/\alpha_{N_1}|$ & $3.397\times 10^4$ & $1441.443$ & $4902.592$\\
$\arg(\alpha_{N_2}/\alpha_{N_1})/\pi$ & $0.143$ & $0.164$ & $0.0237$\\
$\alpha_{\nu_2}/\alpha_{\nu_1}$ & $101.846$ & $19.052$ & $103.807$\\
$\alpha_{d_1}/\alpha_{d_3}$ & $0.00964$ & $0.00815$ & $0.00759$\\
$\alpha_{d_2}/\alpha_{d_3}$ & $0.0187$ & $0.0538$ & $0.0444$\\
\hline\hline
\end{tabular}
\caption{\label{tab:fitVII2}The best fit values of the free parameters and the corresponding predictions for lepton and quark mixing
parameters and fermion masses for the phenomenologically viable models of Type-VII (continuation of table~\ref{tab:fitVII1}).}
\end{table}

\end{appendix}

\providecommand{\href}[2]{#2}\begingroup\raggedright\endgroup


\begin{thebibliography}{10}

\bibitem{Glashow:1961tr}
S.~L. Glashow, ``{Partial Symmetries of Weak Interactions},''
  \href{http://dx.doi.org/10.1016/0029-5582(61)90469-2}{{\em Nucl. Phys.}
  {\bfseries 22} (1961) 579--588}.

\bibitem{Zyla:2020zbs}
{\bfseries Particle Data Group} Collaboration, P.~Zyla {\em et~al.}, ``{Review
  of Particle Physics},'' \href{http://dx.doi.org/10.1093/ptep/ptaa104}{{\em
  PTEP} {\bfseries 2020} (2020) 083C01}.

\bibitem{Esteban:2020cvm}
I.~Esteban, M.~C. Gonzalez-Garcia, M.~Maltoni, T.~Schwetz, and A.~Zhou, ``{The
  fate of hints: updated global analysis of three-flavor neutrino
  oscillations},'' \href{http://dx.doi.org/10.1007/JHEP09(2020)178}{{\em JHEP}
  {\bfseries 09} (2020) 178}, \href{http://arxiv.org/abs/2007.14792}{{\ttfamily
  arXiv:2007.14792 [hep-ph]}}.

\bibitem{King:2013eh}
S.~F. King and C.~Luhn, ``{Neutrino Mass and Mixing with Discrete Symmetry},''
  \href{http://dx.doi.org/10.1088/0034-4885/76/5/056201}{{\em Rept. Prog.
  Phys.} {\bfseries 76} (2013) 056201},
  \href{http://arxiv.org/abs/1301.1340}{{\ttfamily arXiv:1301.1340 [hep-ph]}}.

\bibitem{Feruglio:2017spp}
F.~Feruglio, {\em {Are neutrino masses modular forms?}},
  \href{http://dx.doi.org/10.1142/9789813238053_0012}{pp.~227--266}.
\newblock 2019.
\newblock \href{http://arxiv.org/abs/1706.08749}{{\ttfamily arXiv:1706.08749
  [hep-ph]}}.

\bibitem{Liu:2019khw}
X.-G. Liu and G.-J. Ding, ``{Neutrino Masses and Mixing from Double Covering of
  Finite Modular Groups},''
  \href{http://dx.doi.org/10.1007/JHEP08(2019)134}{{\em JHEP} {\bfseries 1908}
  (2019) 134}, \href{http://arxiv.org/abs/1907.01488}{{\ttfamily
  arXiv:1907.01488 [hep-ph]}}.

\bibitem{Kobayashi:2018vbk}
T.~Kobayashi, K.~Tanaka, and T.~H. Tatsuishi, ``{Neutrino mixing from finite
  modular groups},'' \href{http://dx.doi.org/10.1103/PhysRevD.98.016004}{{\em
  Phys.Rev.} {\bfseries D98} (2018) 016004},
  \href{http://arxiv.org/abs/1803.10391}{{\ttfamily arXiv:1803.10391
  [hep-ph]}}.

\bibitem{Kobayashi:2018wkl}
T.~Kobayashi, Y.~Shimizu, K.~Takagi, M.~Tanimoto, T.~H. Tatsuishi, and
  H.~Uchida, ``{Finite modular subgroups for fermion mass matrices and
  baryon/lepton number violation},''
  \href{http://dx.doi.org/10.1016/j.physletb.2019.05.034}{{\em Phys.Lett.}
  {\bfseries B794} (2019) 114--121},
  \href{http://arxiv.org/abs/1812.11072}{{\ttfamily arXiv:1812.11072
  [hep-ph]}}.

\bibitem{Kobayashi:2019rzp}
T.~Kobayashi, Y.~Shimizu, K.~Takagi, M.~Tanimoto, and T.~H. Tatsuishi,
  ``{Modular $S_3$-invariant flavor model in SU(5) grand unified theory},''
  \href{http://dx.doi.org/10.1093/ptep/ptaa055}{{\em PTEP} {\bfseries 2020}
  (2020) 053B05}, \href{http://arxiv.org/abs/1906.10341}{{\ttfamily
  arXiv:1906.10341 [hep-ph]}}.

\bibitem{Okada:2019xqk}
H.~Okada and Y.~Orikasa, ``{Modular $S_3$ symmetric radiative seesaw model},''
  \href{http://dx.doi.org/10.1103/PhysRevD.100.115037}{{\em Phys.Rev.}
  {\bfseries D100} (2019) 115037},
  \href{http://arxiv.org/abs/1907.04716}{{\ttfamily arXiv:1907.04716
  [hep-ph]}}.

\bibitem{Criado:2018thu}
J.~C. Criado and F.~Feruglio, ``{Modular Invariance Faces Precision Neutrino
  Data},'' \href{http://dx.doi.org/10.21468/SciPostPhys.5.5.042}{{\em SciPost
  Phys.} {\bfseries 5} (2018) 042},
  \href{http://arxiv.org/abs/1807.01125}{{\ttfamily arXiv:1807.01125
  [hep-ph]}}.

\bibitem{Kobayashi:2018scp}
T.~Kobayashi, N.~Omoto, Y.~Shimizu, K.~Takagi, M.~Tanimoto, and T.~H.
  Tatsuishi, ``{Modular A$_{4}$ invariance and neutrino mixing},''
  \href{http://dx.doi.org/10.1007/JHEP11(2018)196}{{\em JHEP} {\bfseries 1811}
  (2018) 196}, \href{http://arxiv.org/abs/1808.03012}{{\ttfamily
  arXiv:1808.03012 [hep-ph]}}.

\bibitem{deAnda:2018ecu}
F.~J. de~Anda, S.~F. King, and E.~Perdomo, ``{$SU(5)$ grand unified theory with
  $A_4$ modular symmetry},''
  \href{http://dx.doi.org/10.1103/PhysRevD.101.015028}{{\em Phys.Rev.}
  {\bfseries D101} (2020) 015028},
  \href{http://arxiv.org/abs/1812.05620}{{\ttfamily arXiv:1812.05620
  [hep-ph]}}.

\bibitem{Okada:2018yrn}
H.~Okada and M.~Tanimoto, ``{CP violation of quarks in $A_4$ modular
  invariance},'' \href{http://dx.doi.org/10.1016/j.physletb.2019.02.028}{{\em
  Phys.Lett.} {\bfseries B791} (2019) 54--61},
  \href{http://arxiv.org/abs/1812.09677}{{\ttfamily arXiv:1812.09677
  [hep-ph]}}.

\bibitem{Novichkov:2018yse}
P.~Novichkov, S.~Petcov, and M.~Tanimoto, ``{Trimaximal Neutrino Mixing from
  Modular A4 Invariance with Residual Symmetries},''
  \href{http://dx.doi.org/10.1016/j.physletb.2019.04.043}{{\em Phys.Lett.}
  {\bfseries B793} (2019) 247--258},
  \href{http://arxiv.org/abs/1812.11289}{{\ttfamily arXiv:1812.11289
  [hep-ph]}}.

\bibitem{Nomura:2019jxj}
T.~Nomura and H.~Okada, ``{A modular $A_4$ symmetric model of dark matter and
  neutrino},'' \href{http://dx.doi.org/10.1016/j.physletb.2019.134799}{{\em
  Phys.Lett.} {\bfseries B797} (2019) 134799},
  \href{http://arxiv.org/abs/1904.03937}{{\ttfamily arXiv:1904.03937
  [hep-ph]}}.

\bibitem{Okada:2019uoy}
H.~Okada and M.~Tanimoto, ``{Towards unification of quark and lepton flavors in
  $A_4$ modular invariance},''
  \href{http://arxiv.org/abs/1905.13421}{{\ttfamily arXiv:1905.13421
  [hep-ph]}}.

\bibitem{Nomura:2019yft}
T.~Nomura and H.~Okada, ``{A two loop induced neutrino mass model with modular
  $A_4$ symmetry},'' \href{http://arxiv.org/abs/1906.03927}{{\ttfamily
  arXiv:1906.03927 [hep-ph]}}.

\bibitem{Ding:2019zxk}
G.-J. Ding, S.~F. King, and X.-G. Liu, ``{Modular A$_{4}$ symmetry models of
  neutrinos and charged leptons},''
  \href{http://dx.doi.org/10.1007/JHEP09(2019)074}{{\em JHEP} {\bfseries 1909}
  (2019) 074}, \href{http://arxiv.org/abs/1907.11714}{{\ttfamily
  arXiv:1907.11714 [hep-ph]}}.

\bibitem{Okada:2019mjf}
H.~Okada and Y.~Orikasa, ``{A radiative seesaw model in modular $A_4$
  symmetry},'' \href{http://arxiv.org/abs/1907.13520}{{\ttfamily
  arXiv:1907.13520 [hep-ph]}}.

\bibitem{Nomura:2019lnr}
T.~Nomura, H.~Okada, and O.~Popov, ``{A modular $A_4$ symmetric scotogenic
  model},'' \href{http://dx.doi.org/10.1016/j.physletb.2020.135294}{{\em
  Phys.Lett.} {\bfseries B803} (2020) 135294},
  \href{http://arxiv.org/abs/1908.07457}{{\ttfamily arXiv:1908.07457
  [hep-ph]}}.

\bibitem{Kobayashi:2019xvz}
T.~Kobayashi, Y.~Shimizu, K.~Takagi, M.~Tanimoto, and T.~H. Tatsuishi, ``{$A_4$
  lepton flavor model and modulus stabilization from $S_4$ modular symmetry},''
  \href{http://dx.doi.org/10.1103/PhysRevD.101.039904}{{\em Phys.Rev.}
  {\bfseries D100} (2019) 115045},
  \href{http://arxiv.org/abs/1909.05139}{{\ttfamily arXiv:1909.05139
  [hep-ph]}}.

\bibitem{Asaka:2019vev}
T.~Asaka, Y.~Heo, T.~H. Tatsuishi, and T.~Yoshida, ``{Modular $A_4$ invariance
  and leptogenesis},'' \href{http://dx.doi.org/10.1007/JHEP01(2020)144}{{\em
  JHEP} {\bfseries 2001} (2020) 144},
  \href{http://arxiv.org/abs/1909.06520}{{\ttfamily arXiv:1909.06520
  [hep-ph]}}.

\bibitem{Gui-JunDing:2019wap}
G.-J. Ding, S.~F. King, X.-G. Liu, and J.-N. Lu, ``{Modular S$_{4}$ and A$_{4}$
  symmetries and their fixed points: new predictive examples of lepton
  mixing},'' \href{http://dx.doi.org/10.1007/JHEP12(2019)030}{{\em JHEP}
  {\bfseries 1912} (2019) 030},
  \href{http://arxiv.org/abs/1910.03460}{{\ttfamily arXiv:1910.03460
  [hep-ph]}}.

\bibitem{Zhang:2019ngf}
D.~Zhang, ``{A modular $A_4$ symmetry realization of two-zero textures of the
  Majorana neutrino mass matrix},''
  \href{http://dx.doi.org/10.1016/j.nuclphysb.2020.114935}{{\em Nucl.Phys.}
  {\bfseries B952} (2020) 114935},
  \href{http://arxiv.org/abs/1910.07869}{{\ttfamily arXiv:1910.07869
  [hep-ph]}}.

\bibitem{Nomura:2019xsb}
T.~Nomura, H.~Okada, and S.~Patra, ``{An Inverse Seesaw model with
  $A_4$-modular symmetry},'' \href{http://arxiv.org/abs/1912.00379}{{\ttfamily
  arXiv:1912.00379 [hep-ph]}}.

\bibitem{Wang:2019xbo}
X.~Wang, ``{Lepton flavor mixing and CP violation in the minimal type-(I+II)
  seesaw model with a modular $A_4$ symmetry},''
  \href{http://dx.doi.org/10.1016/j.nuclphysb.2020.115105}{{\em Nucl.Phys.}
  {\bfseries B957} (2020) 115105},
  \href{http://arxiv.org/abs/1912.13284}{{\ttfamily arXiv:1912.13284
  [hep-ph]}}.

\bibitem{Kobayashi:2019gtp}
T.~Kobayashi, T.~Nomura, and T.~Shimomura, ``{Type II seesaw models with
  modular $A_4$ symmetry},''
  \href{http://dx.doi.org/10.1103/PhysRevD.102.035019}{{\em Phys.Rev.}
  {\bfseries D102} (2020) 035019},
  \href{http://arxiv.org/abs/1912.00637}{{\ttfamily arXiv:1912.00637
  [hep-ph]}}.

\bibitem{King:2020qaj}
S.~J. King and S.~F. King, ``{Fermion mass hierarchies from modular
  symmetry},'' \href{http://dx.doi.org/10.1007/JHEP09(2020)043}{{\em JHEP}
  {\bfseries 2009} (2020) 043},
  \href{http://arxiv.org/abs/2002.00969}{{\ttfamily arXiv:2002.00969
  [hep-ph]}}.

\bibitem{Ding:2020yen}
G.-J. Ding and F.~Feruglio, ``{Testing Moduli and Flavon Dynamics with Neutrino
  Oscillations},'' \href{http://dx.doi.org/10.1007/JHEP06(2020)134}{{\em JHEP}
  {\bfseries 2006} (2020) 134},
  \href{http://arxiv.org/abs/2003.13448}{{\ttfamily arXiv:2003.13448
  [hep-ph]}}.

\bibitem{Okada:2020rjb}
H.~Okada and M.~Tanimoto, ``{Quark and lepton flavors with common modulus
  $\tau$ in $A_4$ modular symmetry},''
  \href{http://arxiv.org/abs/2005.00775}{{\ttfamily arXiv:2005.00775
  [hep-ph]}}.

\bibitem{Nomura:2020opk}
T.~Nomura and H.~Okada, ``{A linear seesaw model with $A_4$-modular flavor and
  local $U(1)_{B-L}$ symmetries},''
  \href{http://arxiv.org/abs/2007.04801}{{\ttfamily arXiv:2007.04801
  [hep-ph]}}.

\bibitem{Okada:2020brs}
H.~Okada and M.~Tanimoto, ``{Spontaneous CP violation by modulus $\tau$ in
  $A_4$ model of lepton flavors},''
  \href{http://arxiv.org/abs/2012.01688}{{\ttfamily arXiv:2012.01688
  [hep-ph]}}.

\bibitem{Yao:2020qyy}
C.-Y. Yao, J.-N. Lu, and G.-J. Ding, ``{Modular Invariant $A_{4}$ Models for
  Quarks and Leptons with Generalized CP Symmetry},''
  \href{http://arxiv.org/abs/2012.13390}{{\ttfamily arXiv:2012.13390
  [hep-ph]}}.

\bibitem{Feruglio:2021dte}
F.~Feruglio, V.~Gherardi, A.~Romanino, and A.~Titov, ``{Modular Invariant
  Dynamics and Fermion Mass Hierarchies around $\tau = i$},''
  \href{http://arxiv.org/abs/2101.08718}{{\ttfamily arXiv:2101.08718
  [hep-ph]}}.

\bibitem{Penedo:2018nmg}
J.~Penedo and S.~Petcov, ``{Lepton Masses and Mixing from Modular $S_4$
  Symmetry},'' \href{http://dx.doi.org/10.1016/j.nuclphysb.2018.12.016}{{\em
  Nucl.Phys.} {\bfseries B939} (2019) 292--307},
  \href{http://arxiv.org/abs/1806.11040}{{\ttfamily arXiv:1806.11040
  [hep-ph]}}.

\bibitem{Novichkov:2018ovf}
P.~Novichkov, J.~Penedo, S.~Petcov, and A.~Titov, ``{Modular S$_{4}$ models of
  lepton masses and mixing},''
  \href{http://dx.doi.org/10.1007/JHEP04(2019)005}{{\em JHEP} {\bfseries 1904}
  (2019) 005}, \href{http://arxiv.org/abs/1811.04933}{{\ttfamily
  arXiv:1811.04933 [hep-ph]}}.

\bibitem{deMedeirosVarzielas:2019cyj}
I.~de~Medeiros~Varzielas, S.~F. King, and Y.-L. Zhou, ``{Multiple modular
  symmetries as the origin of flavor},''
  \href{http://dx.doi.org/10.1103/PhysRevD.101.055033}{{\em Phys.Rev.}
  {\bfseries D101} (2020) 055033},
  \href{http://arxiv.org/abs/1906.02208}{{\ttfamily arXiv:1906.02208
  [hep-ph]}}.

\bibitem{Kobayashi:2019mna}
T.~Kobayashi, Y.~Shimizu, K.~Takagi, M.~Tanimoto, and T.~H. Tatsuishi, ``{New
  $A_4$ lepton flavor model from $S_4$ modular symmetry},''
  \href{http://dx.doi.org/10.1007/JHEP02(2020)097}{{\em JHEP} {\bfseries 2002}
  (2020) 097}, \href{http://arxiv.org/abs/1907.09141}{{\ttfamily
  arXiv:1907.09141 [hep-ph]}}.

\bibitem{King:2019vhv}
S.~F. King and Y.-L. Zhou, ``{Trimaximal TM$_1$ mixing with two modular $S_4$
  groups},'' \href{http://dx.doi.org/10.1103/PhysRevD.101.015001}{{\em
  Phys.Rev.} {\bfseries D101} (2020) 015001},
  \href{http://arxiv.org/abs/1908.02770}{{\ttfamily arXiv:1908.02770
  [hep-ph]}}.

\bibitem{Criado:2019tzk}
J.~C. Criado, F.~Feruglio, and S.~J. King, ``{Modular Invariant Models of
  Lepton Masses at Levels 4 and 5},''
  \href{http://dx.doi.org/10.1007/JHEP02(2020)001}{{\em JHEP} {\bfseries 2002}
  (2020) 001}, \href{http://arxiv.org/abs/1908.11867}{{\ttfamily
  arXiv:1908.11867 [hep-ph]}}.

\bibitem{Wang:2019ovr}
X.~Wang and S.~Zhou, ``{The minimal seesaw model with a modular S$_{4}$
  symmetry},'' \href{http://dx.doi.org/10.1007/JHEP05(2020)017}{{\em JHEP}
  {\bfseries 2005} (2020) 017},
  \href{http://arxiv.org/abs/1910.09473}{{\ttfamily arXiv:1910.09473
  [hep-ph]}}.

\bibitem{Wang:2020dbp}
X.~Wang, ``{Dirac neutrino mass models with a modular $S_4$ symmetry},''
  \href{http://dx.doi.org/10.1016/j.nuclphysb.2020.115247}{{\em Nucl.Phys.}
  {\bfseries B962} (2021) 115247},
  \href{http://arxiv.org/abs/2007.05913}{{\ttfamily arXiv:2007.05913
  [hep-ph]}}.

\bibitem{Novichkov:2018nkm}
P.~Novichkov, J.~Penedo, S.~Petcov, and A.~Titov, ``{Modular A$_{5}$ symmetry
  for flavour model building},''
  \href{http://dx.doi.org/10.1007/JHEP04(2019)174}{{\em JHEP} {\bfseries 1904}
  (2019) 174}, \href{http://arxiv.org/abs/1812.02158}{{\ttfamily
  arXiv:1812.02158 [hep-ph]}}.

\bibitem{Ding:2019xna}
G.-J. Ding, S.~F. King, and X.-G. Liu, ``{Neutrino mass and mixing with $A_5$
  modular symmetry},''
  \href{http://dx.doi.org/10.1103/PhysRevD.100.115005}{{\em Phys.Rev.}
  {\bfseries D100} (2019) 115005},
  \href{http://arxiv.org/abs/1903.12588}{{\ttfamily arXiv:1903.12588
  [hep-ph]}}.

\bibitem{Ding:2020msi}
G.-J. Ding, S.~F. King, C.-C. Li, and Y.-L. Zhou, ``{Modular Invariant Models
  of Leptons at Level 7},''
  \href{http://dx.doi.org/10.1007/JHEP08(2020)164}{{\em JHEP} {\bfseries 2008}
  (2020) 164}, \href{http://arxiv.org/abs/2004.12662}{{\ttfamily
  arXiv:2004.12662 [hep-ph]}}.

\bibitem{Lu:2019vgm}
J.-N. Lu, X.-G. Liu, and G.-J. Ding, ``{Modular symmetry origin of texture
  zeros and quark lepton unification},''
  \href{http://dx.doi.org/10.1103/PhysRevD.101.115020}{{\em Phys.Rev.}
  {\bfseries D101} (2020) 115020},
  \href{http://arxiv.org/abs/1912.07573}{{\ttfamily arXiv:1912.07573
  [hep-ph]}}.

\bibitem{Liu:2020akv}
X.-G. Liu, C.-Y. Yao, and G.-J. Ding, ``{Modular Invariant Quark and Lepton
  Models in Double Covering of $S_4$ Modular Group},''
  \href{http://arxiv.org/abs/2006.10722}{{\ttfamily arXiv:2006.10722
  [hep-ph]}}.

\bibitem{Novichkov:2020eep}
P.~Novichkov, J.~Penedo, and S.~Petcov, ``{Double Cover of Modular $S_4$ for
  Flavour Model Building},'' \href{http://arxiv.org/abs/2006.03058}{{\ttfamily
  arXiv:2006.03058 [hep-ph]}}.

\bibitem{Wang:2020lxk}
X.~Wang, B.~Yu, and S.~Zhou, ``{Double Covering of the Modular $A^{}_5$ Group
  and Lepton Flavor Mixing in the Minimal Seesaw Model},''
  \href{http://arxiv.org/abs/2010.10159}{{\ttfamily arXiv:2010.10159
  [hep-ph]}}.

\bibitem{Liu:2020msy}
X.-G. Liu, C.-Y. Yao, B.-Y. Qu, and G.-J. Ding, ``{Half-integral weight modular
  forms and application to neutrino mass models},''
  \href{http://dx.doi.org/10.1103/PhysRevD.102.115035}{{\em Phys. Rev. D}
  {\bfseries 102} no.~11, (2020) 115035},
  \href{http://arxiv.org/abs/2007.13706}{{\ttfamily arXiv:2007.13706
  [hep-ph]}}.

\bibitem{Yao:2020zml}
C.-Y. Yao, X.-G. Liu, and G.-J. Ding, ``{Fermion Masses and Mixing from Double
  Cover and Metaplectic Cover of $A_5$ Modular Group},''
  \href{http://arxiv.org/abs/2011.03501}{{\ttfamily arXiv:2011.03501
  [hep-ph]}}.

\bibitem{Novichkov:2019sqv}
P.~P. Novichkov, J.~T. Penedo, S.~T. Petcov, and A.~V. Titov, ``{Generalised CP
  Symmetry in Modular-Invariant Models of Flavour},''
  \href{http://dx.doi.org/10.1007/JHEP07(2019)165}{{\em JHEP} {\bfseries 07}
  (2019) 165},
\href{http://arxiv.org/abs/1905.11970}{{\ttfamily arXiv:1905.11970 [hep-ph]}}.

\bibitem{Baur:2019kwi}
A.~Baur, H.~P. Nilles, A.~Trautner, and P.~K.~S. Vaudrevange, ``{Unification of
  Flavor, CP, and Modular Symmetries},''
\href{http://arxiv.org/abs/1901.03251}{{\ttfamily arXiv:1901.03251 [hep-th]}}.

\bibitem{Baur:2019iai}
A.~Baur, H.~P. Nilles, A.~Trautner, and P.~K. Vaudrevange, ``{A String Theory
  of Flavor and $\mathcal{CP}$},''
  \href{http://dx.doi.org/10.1016/j.nuclphysb.2019.114737}{{\em Nucl. Phys. B}
  {\bfseries 947} (2019) 114737},
  \href{http://arxiv.org/abs/1908.00805}{{\ttfamily arXiv:1908.00805
  [hep-th]}}.

\bibitem{Ding:2020zxw}
G.-J. Ding, F.~Feruglio, and X.-G. Liu, ``{Automorphic Forms and Fermion
  Masses},'' \href{http://dx.doi.org/10.1007/JHEP01(2021)037}{{\em JHEP}
  {\bfseries 01} (2021) 037}, \href{http://arxiv.org/abs/2010.07952}{{\ttfamily
  arXiv:2010.07952 [hep-th]}}.

\bibitem{Georgi:1974sy}
H.~Georgi and S.~L. Glashow, ``{Unity of All Elementary Particle Forces},''
  \href{http://dx.doi.org/10.1103/PhysRevLett.32.438}{{\em Phys. Rev. Lett.}
  {\bfseries 32} (1974) 438--441}.

\bibitem{King:2017guk}
S.~F. King, ``{Unified Models of Neutrinos, Flavour and CP Violation},''
  \href{http://dx.doi.org/10.1016/j.ppnp.2017.01.003}{{\em Prog. Part. Nucl.
  Phys.} {\bfseries 94} (2017) 217--256},
  \href{http://arxiv.org/abs/1701.04413}{{\ttfamily arXiv:1701.04413
  [hep-ph]}}.

\bibitem{Ma:2001dn}
E.~Ma and G.~Rajasekaran, ``{Softly broken A(4) symmetry for nearly degenerate
  neutrino masses},'' \href{http://dx.doi.org/10.1103/PhysRevD.64.113012}{{\em
  Phys. Rev. D} {\bfseries 64} (2001) 113012},
  \href{http://arxiv.org/abs/hep-ph/0106291}{{\ttfamily arXiv:hep-ph/0106291}}.

\bibitem{Bjorkeroth:2015ora}
F.~Bj\"orkeroth, F.~J. de~Anda, I.~de~Medeiros~Varzielas, and S.~F. King,
  ``{Towards a complete A$_{4} \times$ SU(5) SUSY GUT},''
  \href{http://dx.doi.org/10.1007/JHEP06(2015)141}{{\em JHEP} {\bfseries 06}
  (2015) 141}, \href{http://arxiv.org/abs/1503.03306}{{\ttfamily
  arXiv:1503.03306 [hep-ph]}}.

\bibitem{Du:2020ylx}
X.~Du and F.~Wang, ``{SUSY Breaking Constraints on Modular flavor $S_3$
  Invariant $SU(5)$ GUT Model},''
  \href{http://arxiv.org/abs/2012.01397}{{\ttfamily arXiv:2012.01397
  [hep-ph]}}.

\bibitem{Zhao:2021jxg}
Y.~Zhao and H.-H. Zhang, ``{Adjoint SU(5) GUT model with Modular $S_4$
  Symmetry},'' \href{http://arxiv.org/abs/2101.02266}{{\ttfamily
  arXiv:2101.02266 [hep-ph]}}.

\bibitem{Georgi:1979df}
H.~Georgi and C.~Jarlskog, ``{A New Lepton - Quark Mass Relation in a Unified
  Theory},'' \href{http://dx.doi.org/10.1016/0370-2693(79)90842-6}{{\em
  Phys.Lett.} {\bfseries B86} (1979) 297--300}.

\bibitem{Minkowski:1977sc}
P.~Minkowski, ``{$\mu \to e\gamma$ at a Rate of One Out of $10^{9}$ Muon
  Decays?},'' \href{http://dx.doi.org/10.1016/0370-2693(77)90435-X}{{\em Phys.
  Lett. B} {\bfseries 67} (1977) 421--428}.

\bibitem{Yanagida:1979as}
T.~Yanagida, ``{Horizontal gauge symmetry and masses of neutrinos},'' {\em
  Conf. Proc. C} {\bfseries 7902131} (1979) 95--99.

\bibitem{GellMann:1980vs}
M.~Gell-Mann, P.~Ramond, and R.~Slansky, ``{Complex Spinors and Unified
  Theories},'' {\em Conf. Proc. C} {\bfseries 790927} (1979) 315--321,
  \href{http://arxiv.org/abs/1306.4669}{{\ttfamily arXiv:1306.4669 [hep-th]}}.

\bibitem{Mohapatra:1979ia}
R.~N. Mohapatra and G.~Senjanovic, ``{Neutrino Mass and Spontaneous Parity
  Nonconservation},'' \href{http://dx.doi.org/10.1103/PhysRevLett.44.912}{{\em
  Phys. Rev. Lett.} {\bfseries 44} (1980) 912}.

\bibitem{Schechter:1980gr}
J.~Schechter and J.~W.~F. Valle, ``{Neutrino Masses in SU(2) x U(1)
  Theories},'' \href{http://dx.doi.org/10.1103/PhysRevD.22.2227}{{\em Phys.
  Rev. D} {\bfseries 22} (1980) 2227}.

\bibitem{King:1999mb}
S.~F. King, ``{Large mixing angle MSW and atmospheric neutrinos from single
  right-handed neutrino dominance and U(1) family symmetry},''
  \href{http://dx.doi.org/10.1016/S0550-3213(00)00109-7}{{\em Nucl. Phys. B}
  {\bfseries 576} (2000) 85--105},
  \href{http://arxiv.org/abs/hep-ph/9912492}{{\ttfamily arXiv:hep-ph/9912492}}.

\bibitem{Frampton:2002qc}
P.~H. Frampton, S.~L. Glashow, and T.~Yanagida, ``{Cosmological sign of
  neutrino CP violation},''
  \href{http://dx.doi.org/10.1016/S0370-2693(02)02853-8}{{\em Phys. Lett. B}
  {\bfseries 548} (2002) 119--121},
  \href{http://arxiv.org/abs/hep-ph/0208157}{{\ttfamily arXiv:hep-ph/0208157}}.

\bibitem{Antusch:2013jca}
S.~Antusch and V.~Maurer, ``{Running quark and lepton parameters at various
  scales},'' \href{http://dx.doi.org/10.1007/JHEP11(2013)115}{{\em JHEP}
  {\bfseries 1311} (2013) 115},
  \href{http://arxiv.org/abs/1306.6879}{{\ttfamily arXiv:1306.6879 [hep-ph]}}.

\bibitem{minuit}
\url{https://seal.web.cern.ch/seal/snapshot/work-packages/mathlibs/minuit/}.

\bibitem{KamLAND-Zen:2016pfg}
{\bfseries KamLAND-Zen} Collaboration, A.~Gando {\em et~al.}, ``{Search for
  Majorana Neutrinos near the Inverted Mass Hierarchy Region with
  KamLAND-Zen},'' \href{http://dx.doi.org/10.1103/PhysRevLett.117.082503}{{\em
  Phys. Rev. Lett.} {\bfseries 117} no.~8, (2016) 082503},
  \href{http://arxiv.org/abs/1605.02889}{{\ttfamily arXiv:1605.02889
  [hep-ex]}}. [Addendum: Phys.Rev.Lett. 117, 109903 (2016)].

\bibitem{Aghanim:2018eyx}
{\bfseries Planck} Collaboration, N.~Aghanim {\em et~al.}, ``{Planck 2018
  results. VI. Cosmological parameters},''
  \href{http://dx.doi.org/10.1051/0004-6361/201833910}{{\em Astron. Astrophys.}
  {\bfseries 641} (2020) A6}, \href{http://arxiv.org/abs/1807.06209}{{\ttfamily
  arXiv:1807.06209 [astro-ph.CO]}}.

\end{thebibliography}
\end{document}